\documentclass[9pt,a4paper]{article}

\usepackage{amsmath}
\usepackage{amssymb}
\usepackage{mathrsfs}
\usepackage{amsthm}
\usepackage{graphicx}
\usepackage{color}
\usepackage{float}
\usepackage{subfigure}
\usepackage{framed}
\usepackage{ulem}
\usepackage{array}
\usepackage{makecell}
\restylefloat{table}
\usepackage{xcolor}
\usepackage[export]{adjustbox}
\usepackage[utf8]{inputenc}

\usepackage{setspace}

\def\N{\mathbb N}

\theoremstyle{definition}
	\newtheorem{definition}{Definition}[section]
	\newtheorem{remark}{Remark}

    \newtheorem*{conjecture}{Conjecture}
	\newtheorem{lemma}[definition]{Lemma}
	
	\newtheorem{theorem}[definition]{Theorem}
\theoremstyle{remark}

	\title{Rogue waves and their patterns in the vector nonlinear Schr\"{o}dinger equation}
	
	\author{Guangxiong Zhang$^1$,   Peng Huang$^1$, Bao-Feng Feng$^{*2}$, Chengfa Wu$^{*1,3}$
\\
\\
 $^1$Institute for Advanced Study, Shenzhen University, \\ Shenzhen 518060, People's Republic of China
\\
 $^2$School of Mathematical and Statistical Sciences, \\ The University of
Texas Rio Grande Valley Edinburg,
\\ Edinburg, TX 78541-2999, USA
\\
$^3$College of Mathematics and Statistics, Shenzhen University, \\ Shenzhen, 518060, People's Republic of China
\\
}

\date{\today}

\begin{document}

\maketitle

%\graphicspath{ {./RootStructurePicture/}}

\graphicspath{ {./ComparisonPicture3com/},{./ComparisonPicture4com/},{./RootStructurePicture/} }

%\graphicspath{ {./Figure4/} }
%\graphicspath{ {./Comparison picture 3com/}, {./Comparison picture 4com/}  }

\maketitle

\begin{figure}[b]
\rule[-2.5truemm]{5cm}{0.1truemm}\\[2mm]
{\footnotesize  *Corresponding authors. Email address: baofeng.feng@utrgv.edu (B. F. Feng), cfwu@szu.edu.cn (C. F. Wu).}

\end{figure}

\begin{abstract}
In this paper, we study the general rogue wave solutions and their patterns in the vector (or $M$-component) nonlinear Schr\"{o}dinger (NLS) equation. By applying the Kadomtsev-Petviashvili hierarchy reduction method, we derived an explicit solution for the rogue wave expressed by $\tau$ functions that are determinants of $K\times K$ block matrices ($K=1,2,\cdots, M$)  with an index jump of $M+1$. Patterns of the rogue waves for $M=3,4$ and $K=1$ are thoroughly investigated. We find that when a specific internal parameter is large enough, the wave patterns are linked to the root structures of generalized Wronskian-Hermite polynomial hierarchy in contrast with rogue wave patterns of the scalar NLS equation, the Manakov system and many others. Moreover, the generalized Wronskian-Hermite polynomial hierarchy includes the Yablonskii-Vorob'ev polynomial hierarchy and Okamoto polynomial hierarchies as special cases, which have been used to describe the rogue wave patterns of the scalar NLS equation and the Manakov system, respectively. As a result,  we extend the most recent results by Yang {\it et al.} for the scalar NLS equation and the Manakov system. It is noted that the case $M=3$ displays a new feature different from the previous results. The predicted rogue wave patterns are compared with the ones of the true solutions for both cases of $M=3,4$. An excellent agreement is achieved.
%and they are in accordance with each other.

\vspace{1em}

{\bf Keywords}: Kadomtsev-Petviashvili hierarchy reduction method, vector nonlinear Schr\"{o}dinger equation, rogue waves, pattern formation, Wronskian-Hermite polynomials

\end{abstract}

%\newpage

\section{Introduction}

Rogue waves have been known in the maritime community as part of folklore for centuries. Notable features of such waves include sudden emergence, abnormally large amplitude, and disappearance without any trace. These characteristics indicate that rogue waves may result in tremendous impacts on their surrounding environment and have been associated with many maritime disasters \cite{dysthe2008oceanic}. Systematic studies on rouge waves started only after the first verified measurement of an extreme water wave on 1995 \cite{haver2004possible}. Remarkably,
research on rogue waves has developed considerably since 2007, following the discovery of rouge waves in optical fibres \cite{solli2007optical}, which has attracted much interest in both optics and hydrodynamics. Since then, there has been an explosion of studies to explore rogue waves extended to other physical systems, such as superfluid helium \cite{ganshin2008observation}, Bose-Einstein condensates \cite{bludov2009matter}, capillary waves \cite{shats2010capillary}, and plasmas \cite{bailung2011observation}.

In optics and hydrodynamics, the mathematical models governing wave propagation can be derived from Maxwell's equations and the Euler equations, respectively  \cite{dudley2019rogue}. Under further assumptions, the nonlinear Schr\"odinger (NLS) equation, which describes the evolution of slowly varying wave packets in nonlinear wave systems, can be reduced from both of these two models \cite{ablowitz1991solitons}. Owing to its integrability, the NLS equation has been widely studied \cite{besse2002order,eliasson2010kam,kato1987nonlinear,pelinovsky2004parametric,weinstein1982nonlinear,yang2016normal} and shown to admit a number of analytic solutions.  In particular, one of its rational solutions, namely the Peregrine soliton \cite{peregrine1983water}, is widely regarded as the prototype of rogue waves. In the past two decades, mathematical study on rogue waves has attracted much attention, and various higher-order rogue wave solutions of the NLS equation have been constructed \cite{akhmediev2009rogue,dubard2010multi,kedziora2011circular,guo2012nonlinear,ohta2012general}. It is worth noting that these solutions in turn have facilitated the experimental studies of rogue waves. On the other hand, explicit rogue wave solutions have been derived in various integrable equations, such as the derivative NLS equation \cite{yang2020rogue}, the Yajima-Oikawa equation \cite{chen2015rational,chen2018general}, the three-wave equation \cite{yang2021general},  the Manakov system \cite{baronio2014vector,chen2015vector},
the Sasa-Satsuma equation \cite{feng2022higher,wu2022general}, and many others. Besides, rogue waves of infinite order have been uncovered \cite{bilman2020extreme,bilman2022broader} by making use of the Riemann-Hilbert approach \cite{yang2010nonlinear}, while rogue waves on the periodic background \cite{chen2018rogue,chen2019rogue,feng2020multi} have also been explored. Large-order asymptotics for solitons \cite{bilman2019large} and \cite{bilman2019robust} rogue waves of the NLS equation were analyzed by using the inverse-scattering transform method.

In addition to their physical significance, rouge waves may exhibit extremely regular and symmetric patterns, which are intriguing and can provide critical information for predicting subsequent rogue waves from previous ones. For instance, circular rogue wave clusters of the NLS equation were reported in \cite{kedziora2011circular} by using Darboux transformation and numerical simulations. Soon after this phenomenon was confirmed analytically in \cite{he2013generating}, a systematic classification of the NLS rogue wave patterns was obtained in \cite{kedziora2013classifying} according to the order of rogue waves and the parameter shifts involved in the Akhmediev breathers in the rogue-wave limit. Moreover, this study reveals various highly symmetric geometric structures of rogue waves under certain choices of parameters, including triangles, pentagons, heptagons, and nonagons. A even more remarkable observation is that, which was first shown in \cite{yang2021rogue}, the distribution of rogue waves for specific choices of parameters looks very similar to another independent object, that is, the root structure of the Yablonskii-Vorob'ev polynomial hierarchy, which is closely related to rational solutions of the Painlev\'{e} II hierarchy \cite{clarkson2003second}. When specific internal parameter is large enough, the deep connection between these two objects has been established analytically in \cite{yang2021rogue}, which is a remarkable progress in the study of rogue waves. Following this work, it is found that \cite{yang2021universal} such patterns are universal, as rogue waves of many other integrable equations demonstrate similar patterns, such as the Boussinesq equation and the Manakov system, as long as the Schur polynomials involved in the $\tau$ functions have index jumps of two. Beyond that, Yang and Yang \cite{yang2022rogue} very recently discovered that other rogue wave patterns exist when the index jumps are three, and these patterns are characterized by root structures of Okamoto polynomial hierarchies.

Inspired by the works in \cite{yang2021rogue,yang2021universal,yang2022rogue}, some natural problems arise.
\begin{itemize}
 \item Can we construct the rogue wave solution in $M$-component NLS equation?

  \item What about the patterns of rogue waves for $M=3,4$ or even the general case?  Are these patterns related to some orthogonal polynomial hierarchy?

\end{itemize}

The main objective of this paper is to solve the problems listed above by considering the vector NLS equation
\begin{equation}\label{vector NLS}
  \mathrm{i}  u_{j,t} + u_{j,xx} +\left(\sum_{k=1}^M \sigma_k\left|u_k\right|^2\right) u_j=0, \quad j=1,2, \cdots, M,
\end{equation}
where $M$ is a positive integer and $\sigma_k = \pm 1$. For $M=2$, it is known as the Manakov system \cite{manakov1974theory}, which is a model that governs soliton propagation through optical fiber arrays  \cite{akhmediev1997nonlinear,kang1996observation,kanna2001exact}. Our results consist of two ingredients. First, we will apply the Kadomtsev-Petviashvili hierarchy reduction technique to derive rogue wave solutions of the vector NLS equation whose $\tau$ functions are represented by determinants of $K\times K$ block matrices ($K=1,2,\cdots, M$)  with index jumps of $M+1$.
The crucial point of this part is to solve a system of algebraic equations (see Lemma \ref{multiple roots} and its proof). Then we will study the rogue wave patterns for $M=3,4$ and $K=1$. We find that when a specific internal parameter is large enough, these patterns are connected to the root structure of a new orthogonal polynomial hierarchy, which is called {\it generalized Wronskian-Hermite polynomials} (see Section \ref{section-Preliminaries}.2). Moreover, we notice that the Yablonskii-Vorob'ev polynomial hierarchy and Okamoto polynomial hierarchies are special cases of the generalized Wronskian-Hermite polynomials. Accordingly, our results have unified rogue wave patterns of the scalar NLS equation and the vector NLS equation \eqref{vector NLS} for $M=2,3,4$. In addition, we find that the proof in the inner region for $M=3$ is different from other cases. In the proofs of the inner region of the scalar NLS equation and the Manakov system, one can perform row and column operations to reduce the $\tau$ functions into determinants of block matrices with lower triangular matrices at the (1,1) entry whose elements on the diagonal are all $1$. Then the sizes of the determinants can be decreased; hence these waves can be approximated by possible lower-order rogue waves in the inner region. However, although the waves for $M=3$ can also be approximated by possible lower-order rogue waves in the inner region, it turns out that in certain cases the sizes of the determinants remain unchanged after row and column operations, as the (1,1) entries of the block matrices are no longer triangular. The predicted rogue wave patterns are compared with actual ones, and excellent agreement is achieved.

%This can be explained as follows.

The structure of this paper can now be explained. Section \ref{section-Preliminaries} presents some preliminary results that will be used in the subsequent discussions. We first provide explicit rogue wave solutions with index jumps of $M+1$ of the vector NLS equation, and it is shown that these solutions are expressed by $K\times K$ block matrices ($K=1,2,\cdots, M$). This is followed by an introduction to the generalized Wronskian-Hermite polynomials and the study of their root structures. Then rogue wave patterns for the three- and four-component NLS equations under the condition that specific parameters are large enough are stated in Section \ref{section-Rogue wave patterns}, which form the main results of this paper. Section \ref{section-comparison} is devoted to comparing predicted and actual rogue wave patterns, while the proofs of the main results are provided in Section \ref{section-Proof of the main results}. We summarize the main results of this paper in Section \ref{section-Conclusion}. Finally, the proof of Lemma \ref{multiple roots}, which involves the study of multiple roots of some rational function and plays a pivotal role in this paper, and derivations of rogue wave solutions in the vector NLS equation are given in Appendices A and B respectively, while the results on root structures of the generalized Wronskian-Hermite polynomials of jump $k=4,5$ are proved in Appendix C.

\section{Preliminaries} \label{section-Preliminaries}

\subsection{Rogue wave solutions of the vector nonlinear Schr\"odinger equation}

This section presents rogue wave solutions of the vector NLS equation \eqref{vector NLS}, which possesses an infinite dimensional algebra of non-commutative symmetries \cite{kodama2001symmetry}. We note that these solutions have been studied before \cite{baronio2012solutions,chen2015vector,ling2014high,mu2015dynamics,rao2019vector,zhang2021multi}. In particular, vector Peregrine solitons were found by applying the loop group theory in \cite{zhang2021multi}, in which the authors proposed the problem of whether patterns of these rogue waves are related to Yablonskii-Vorob'ev polynomial hierarchy. This problem was later confirmed for the Manakov system  \cite{yablonskii1959rational}, which has been taken as an example to show that universal rogue wave patterns associated with the Yablonskii-Vorob'ev polynomial hierarchy exist in integrable systems. Very recently, new patterns of another class of (degenerate) rogue waves of the Manakov system have been obtained by Yang and Yang \cite{yang2022rogue} through establishing the connection between these waves and the Okamoto polynomial hierarchies. A remarkable feature of these new patterns is that, unlike previous patterns, the transformations between the locations of fundamental rogue waves and zeros of the Okamoto polynomial hierarchies are nonlinear, thereby leading to deformations of rogue patterns. Inspired by these studies, we will extend the results in  \cite{yang2022rogue} and solve the problem mentioned above in \cite{zhang2021multi} for $M=3,4$ by studying patterns of degenerate rogue waves of the vector NLS equation \eqref{vector NLS}.
To this end, we introduce some notations and lemma that will be needed.

 The Schur polynomials $S_n(\boldsymbol{x})$ are defined by
	$$
	\sum_{n=0}^{\infty}S_n(\boldsymbol{x})\lambda^n=\exp\left(\sum_{k=1}^{\infty}x_k\lambda^k\right),
	$$
	where $\boldsymbol{x}=(x_1,x_2,\cdots)$. To be more specific, we have
\begin{equation}\label{Schur polynomials}
  S_{0}(\boldsymbol{x})=1, \quad S_{1}(\boldsymbol{x})=x_{1}, \quad S_{2}(\boldsymbol{x})=\frac{1}{2} x_{1}^{2}+x_{2},  \ldots, \quad S_{j}(\boldsymbol{x})=\sum_{l_{1}+2 l_{2}+\cdots+m l_{m}=j}\left(\prod_{i=1}^{m} \frac{x_{i}^{l_{i}}}{l_{i} !}\right).
\end{equation}
Further, we define $S_j(\boldsymbol{x})\equiv0$ for $ j<0 $.

%We also need the following lemma.

\begin{lemma} \label{multiple roots}
Let $M$ be a positive integer and $\lambda_1>0, r_j \not = 0,k_j$ be real constants, $j=1,2,\dots, M$, where the $k_j$'s are distinct. %, and $k_i \not = k_j$ for
Let $\mathcal{R}_M(z)$ be a rational function defined by
\begin{equation}\label{dimension reduction VS rat}
\mathcal{R}_M(z)= \sum_{j=1}^M\frac{r_j }{(z+  k_j)^2} + 2.
\end{equation}
%where $r_j \not = 0,  j=1,2,\dots, M$.
Then $\mathcal{R}_M(z)=0$ has a pair of complex conjugate roots with nonzero imaginary parts of multiplicity $M$
\begin{equation}%\label{}
     \lambda_1  \cos[ \pi /(M+1)] -  k_1  \pm  \mathrm{i}  \lambda_1 \sin[ \pi /(M+1)],
\end{equation}
if the parameters  $r_j,k_j$, $j=2,\dots, M$, satisfy the conditions
\begin{eqnarray*}
  k_j  &=& k_1 +  \lambda_1 \left( \sin[  \pi /(M+1)]  \cot[j \pi /(M+1)] -  \cos[ \pi /(M+1)]\right),
\end{eqnarray*}
and
\begin{eqnarray}%\label{}
  r_j
  = 2 (-1)^{j+1}  \prod_{\substack{i=1 \\ i \not=j}}^M (k_j-k_i)^{-1}  \left(\lambda_1 \frac{\sin[  \pi /(M+1)] }{\sin[j \pi /(M+1)]} \right)^{M+1}.
\end{eqnarray}

\end{lemma}

\begin{remark}
The equation $\mathcal{R}_M(z)=0$ may have real roots of multiplicity $M$ as well. For instance,  the equation
\begin{equation*}
    \frac{162}{(z+2)^2}-\frac{256}{(z+3)^2}-\frac{16}{(z+1)^2}+2=0
\end{equation*}
has a real root $1$ of multiplicity $3$, while the equation
\begin{equation*}
    -\frac{1024}{(z+2)^2}+\frac{3125}{(z+3)^2}-\frac{2592}{(z+4)^2}+\frac{81}{(z+1)^2}+2=0
\end{equation*}
has a real root $2$ of multiplicity $4$.
Nevertheless, this case will not occur in our subsequent discussions.

\end{remark}

We will provide the proof of Lemma \ref{multiple roots} in Appendix A. Next, we define the functions $\mathcal{G}_M(p)$ and $p(\kappa)$ respectively by
%\begin{equation}
%\end{equation}
%be the function defined by
\begin{eqnarray}
\mathcal{G}_M(p) &=& \sum_{j=1}^M\frac{\sigma_j \rho_j^2}{p-\mathrm{i} k_j} +2 p, \label{definition of G(p))} %\quad  M=3,4,
\\
   \mathcal{G}_M(p(\kappa)) &=& \dfrac{\mathcal{G}_M(p(0))}{M+1} \sum_{n=1}^{M+1} \exp\left(\exp\left(\dfrac{2 n \pi \mathrm{i}}{M+1}\right) \kappa\right), \nonumber
   \\
   &=& \dfrac{\mathcal{G}_M(p(0))}{M+1} \sum_{n=1}^{M+1} \exp\left({\cos{\left(\dfrac{2 n \pi }{M+1}\right)} \kappa}\right) \cos\left(\sin \left(\dfrac{2 n \pi }{M+1}\right)\kappa\right), \label{definition of p(kappa))}
\end{eqnarray}
where $\rho_j >0, k_j$ are real constants,  $j=1,2,\cdots,M$. We may deduce from Lemma \ref{multiple roots} that if   $\sigma_j, \rho_j $ and $ k_j, j=1,2,\cdots,M$, satisfy the constraints
\begin{equation}\label{contraints of multiple zeros}
\begin{aligned}
 k_j  &= k_1 +  \lambda_1 \left( \sin[  \pi /(M+1)]  \cot[j \pi /(M+1)] -  \cos[ \pi /(M+1)]\right),\\
 \sigma_j\rho_j^2
  &= 2 (-1)^{j+1}  \prod_{\substack{i=1 \\ i \not=j}}^M (k_j-k_i)^{-1}  \left(\lambda_1 \frac{\sin[  \pi /(M+1)] }{\sin[j \pi /(M+1)]} \right)^{M+1},
  \end{aligned}
\end{equation}
%\begin{eqnarray*}
%  k_j  &=& k_1 +  \lambda_1 \left( \sin[  \pi /(M+1)]  \cot[j \pi /(M+1)] -  \cos[ \pi /(M+1)]\right),
%\end{eqnarray*}
%and
%\begin{eqnarray}%\label{}
%  \sigma_j\rho_j^2
%  = 2 (-1)^{j+1}  \prod_{\substack{i=1 \\ i \not=j}}^M (k_j-k_i)^{-1}  \left(\lambda_1 \frac{\sin[  \pi /(M+1)] }{\sin[j \pi /(M+1)]} \right)^{M+1} .
%\end{eqnarray}
then the algebraic equation
\begin{equation}
\mathcal{G}_M^{\prime}(p)=0
\end{equation}
has a pair of non-imaginary roots of multiplicity $M$ given by
\begin{equation}%\label{}
      \pm  \lambda_1 \sin[ \pi /(M+1)]-\mathrm{i} \lambda_1  \cos[ \pi /(M+1)] + \mathrm{i} k_1 .
\end{equation}

\begin{theorem} \label{RW solutions of vector NLS}
% Let  $\sigma_j=1 $, $j=1,2,\cdots,M$, and $\mathcal{G}_M(p), p(\kappa)$ be functions defined by \eqref{definition of G(p))} and \eqref{definition of p(kappa))} respectively, where $M$ is a positive integer.
 Let $M$ be a positive integer,  $\rho_j >0, k_j$ be real constants, and $\sigma_j=1 $, where $j=1,2,\cdots,M$. Assume  $ \rho_j $ and $ k_j$ are given by \eqref{contraints of multiple zeros}. Let $\mathcal{G}_M(p), p(\kappa)$ be functions defined by \eqref{definition of G(p))} and \eqref{definition of p(kappa))} respectively.
 % and $\rho_j >0, k_j$ be real constants, where $j=1,2,\cdots,M$. Let
%$\mathcal{G}_M(p)$ be the function defined by
%\begin{equation}
%\mathcal{G}_M(p)= \sum_{j=1}^M\frac{\sigma_j \rho_j^2}{p-\mathrm{i} k_j} +2 p, \quad  M \in \mathbb{Z^+}
%\end{equation}
%and $p(\kappa)$ be the function defined by
%\begin{equation}
%   \mathcal{G}_M(p(\kappa)) = \dfrac{\mathcal{G}_M(p(0))}{M+1} \sum_{n=1}^{M+1} \exp\left(\exp\left(\dfrac{2 n \pi \mathrm{i}}{M+1}\right) \kappa\right),
%\end{equation}
%or
%\begin{equation}
%   \mathcal{G}_M(p(\kappa)) = \dfrac{\mathcal{G}_M(p(0))}{M+1} \sum_{n=1}^{M+1} \exp\left({\cos{\left(\dfrac{2 n \pi }{M+1}\right)} \kappa}\right) \cos\left(\sin \left(\dfrac{2 n \pi }{M+1}\right)\kappa\right).
%\end{equation}
% where
% \begin{equation}
% \lambda_n=e^{\dfrac{2 n \pi \mathrm{i}}{M+1}}, \quad n=1,2,\cdots,M+1.
% \end{equation}
Let $\boldsymbol{x}_{I}^{\pm}=\left(x_{1,I}^{\pm}, x_{2,I}^{\pm}, \cdots\right), I = 1, 2, \dots, M$, and $\boldsymbol{s}=\left(s_{1}, s_{2}, \cdots\right)$ be the vectors defined by
%\begin{equation}
\begin{eqnarray}
&&x_{i,I}^{+}=\alpha_i x+\beta_i \mathrm{i} t+ \sum_{j=1}^{M}n_j \theta_{ij} + a_{i,I}, \label{values of x+} \\
&&x_{i,I}^{-}=\alpha_i^* x-\beta_i^* \mathrm{i} t-\sum_{j=1}^{M}n_j \theta_{ij}^* +a_{i,I}^*,\\
&&\ln \left[\frac{1}{\kappa}\left(\frac{p_{0}+p_{0}^{*}}{p_{1}}\right)\left(\frac{p(\kappa)-p_{0}}{p(\kappa)+p_{0}^{*}}\right)\right]=\sum_{r=1}^{\infty} s_{r} \kappa^{r}, \label{sr}
\end{eqnarray}
%\end{equation}
where the asterisk `$*$' represents complex conjugation, $p_0=p(0),~p_1=p^{\prime}(0)$, the $a_{i,I}$'s are arbitrary constants, and $\alpha_i,~\beta_i,~\theta_{ij},~j = 1, 2, \dots, M$, are defined by the expansions
$$
p(\kappa)-p_{0}=\sum_{r=1}^{\infty} \alpha_{r} \kappa^{r}, \quad p^{2}(\kappa)-p_{0}^{2}=\sum_{r=1}^{\infty} \beta_{r} \kappa^{r},\quad\ln \frac{p(\kappa)-\mathrm{i} k_{j}}{p_{0}-\mathrm{i} k_{j}}=\sum_{r=1}^{\infty} \theta_{rj} \kappa^{r}.
$$
%If   $\rho_j $ and $ k_j, j=1,2,\cdots,M$, satisfy the constraints
%\\
%$\cdots$
%\\
%then the algebraic equation
%\begin{equation}
%\mathcal{G}_M^{\prime}(p)=0
%\end{equation}
%has a pair of non-imaginary roots of multiplicity $M$.
In this case, the $M$-component NLS equation \eqref{vector NLS} admits $\mathcal{N}$-th order rogue wave solutions
\begin{equation} \label{}
 u_{j,\mathcal{N}} = \frac{g_{j,\mathcal{N}}}{f_{\mathcal{N}}}  e^{\mathrm{i}\left(k_j x+w_j t\right)} , \quad j=1,2,\cdots,M,
\end{equation}
where %all of $\rho_{i}$, $k_{i}$ and $w_{i}$ are arbitrary real constants, but under the following constraints
%{\color{red}
\begin{equation} \label{definition of N-theorem}
w_j=\sum_{i=1}^{M}\sigma_i \rho_i^2 - k_j^2, \quad \mathcal{N} = \left(N_1, N_2, \ldots, N_M\right),
\end{equation}%}
with $N_j~(j=1,2,\cdots,M)$ being nonnegative integers,
and $f$ and $g_{j}$ are given by
\begin{equation}
f_{\mathcal{N}}=\tau_{\mathbf{n}_0}, \quad g_{j,\mathcal{N}}=\tau_{\mathbf{n}_j}
\end{equation}
with
$$
\mathbf{n}_0=(0,0, \ldots, 0) \in \mathbb{R}^M, \quad \mathbf{n}_j=\sum_{l=1}^M \delta_{j l} \boldsymbol{e}_l,
$$
$\boldsymbol{e}_l$ being the standard unit vector in $\mathbb{R}^M$ and $\delta_{j l}$ being the Kronecker delta. Here, $\tau_{\mathbf{n}}$ is given by the following $K \times K \, (K=1,2,\cdots,M)$ block determinant%{\color{red}
\begin{equation} \label{tau-block matrix-theorem}
\tau_{\mathbf{n}}=\det\left(\begin{array}{llll}
\tau_{\mathbf{n}}^{[I_1,I_1]} & \tau_{\mathbf{n}}^{[I_1,I_2]}&\cdots&\tau_{\mathbf{n}}^{[I_1,I_K]} \\
\tau_{\mathbf{n}}^{[I_2,I_1]} & \tau_{\mathbf{n}}^{[I_2,I_2]}&\cdots&\tau_{\mathbf{n}}^{[I_2,I_K]} \\ \vdots & \vdots& \ddots &\vdots \\\tau_{\mathbf{n}}^{[I_K,I_1]} & \tau_{\mathbf{n}}^{[I_K,I_2]}&\cdots&\tau_{\mathbf{n}}^{[I_K,I_K]}
\end{array}\right)_{N \times N},
\end{equation}
where
\begin{eqnarray}
&&\mathbf{n}=\left(n_1, n_2, \ldots, n_M\right), \quad 1\leq I_1<I_2<\cdots<I_K\leq M,
\\
&& \tau_{\mathbf{n}}^{[I, J]}=\left(m_{(M+1) i-I,(M+1) j-J}^{(\mathbf{n},I, J)}\right)_{1 \leq i \leq N_I, 1 \leq j \leq N_J}, \quad 1 \leq I, J \leq M,  \label{tau-entry of block matrix-theorem}
\end{eqnarray}
$n_1, n_2, \ldots, n_M $ are integers,  $ I_j, N_{I_j} \, (j=1,2,\cdots,K)$  are positive integers with $N_{I_1}+N_{I_2}+\cdots+N_{I_K}=N$ and $N_l = 0$ for $l \in \{1,2,\cdots,M\}  \backslash \{I_1,I_2,\cdots,I_K\}$, and the corresponding matrix elements of \eqref{tau-entry of block matrix-theorem} are defined by
%{\color{red}
\begin{eqnarray} \label{matrix elments-Mcom}
   % m_{i, j}^{(\mathbf{n},I,J)}&=&\sum_{v=0}^{\min (i,
    m_{i, j}^{(\mathbf{n},I, J)}&=&\sum_{v=0}^{\min (i, j)}\left[\frac{\left|p_{1}\right|^{2}}{\left(p_{0}+p_{0}^{*}\right)^{2}}\right]^{v} S_{i-v}\left(\boldsymbol{x}_I^{+}(\mathbf{n})+v \boldsymbol{s}\right) S_{j-v}\left(\boldsymbol{x}_J^{-}(\mathbf{n})+v \boldsymbol{s}^{*}\right).  \label{matrix elments-Mcom}
\end{eqnarray}
%}
\end{theorem}

We provide the proof of Theorem \ref{RW solutions of vector NLS} in Appendix B.

\begin{remark}
The rogue wave solutions to the vector NLS equation \eqref{vector NLS} in Theorem \ref{RW solutions of vector NLS} are represented by $\tau$ functions which have matrix elements expressed by Schur polynomials with index jumps of $M+1$. %, while the index jumps of Schur polynomials corresponding to the four-component NLS equation are four.
These rogue waves exist only when $\mathcal{G}_M^{\prime}(p)=0$ has non-imaginary roots of order $M$. This indicates that the scalar NLS equation and the Manakov system have no such kind of rogue waves with index jumps of $\mathcal{J} \geq 4$. In addition,  when $\mathcal{G}_M^{\prime}(p)=0$ has simple or double roots, the vector NLS equations would possess similar types of rogue waves as the scalar NLS equation or the Manakov system, where the index jumps of the corresponding Schur polynomials are 2 or 3. In such cases, the rogue wave patterns are similar to those of the scalar NLS equation  \cite{yang2021rogue} or the Manakov system \cite{yang2022rogue} and hence we will not present these rogue waves.
\end{remark}

\begin{remark}
We restrict the study of rogue wave patterns to the cases of $M=3,4$. Then the $p(\kappa)$ introduced in \eqref{definition of p(kappa))} can be expressed by
% Note that $\mathcal{G}_M(p)$ are the functions defined by
\begin{equation}
\mathcal{G}_M(p(\kappa))= \begin{cases}
\dfrac{\mathcal{G}_3(p(0))}{4}\left( e^\kappa+ e^{-\kappa}+2\cos \kappa\right),  & M=3, \\ \dfrac{\mathcal{G}_4(p(0))}{5}\left(e^\kappa+2 e^{\left(-\frac{\sqrt{5}}{4}-\frac{1}{4}\right) \kappa}\left(e^{\frac{\sqrt{5} \kappa}{2}} \cos \left(\sqrt{\frac{\sqrt{5}}{8}+\frac{5}{8} }\kappa\right)+\cos \left(\sqrt{\frac{5}{8}-\frac{\sqrt{5}}{8} }\kappa\right)\right)\right), \quad & M=4.\end{cases}
\end{equation}
%{\color{red} We note that there are two irreducible free parameters in the expressions of the roots in \eqref{three-order roots} and \eqref{four-order roots}.}
It is also clear that there are other parameter choices for $\mathcal{G}_M^{\prime}(p)=0 \, (M=3,4)$ to have a pair of non-imaginary roots of order $M$, on account of the symmetry of the vector NLS equation \eqref{vector NLS}. Let $(i,j,l)$ be any permutation of the set $\{1, 2, 3\}$, then the condition  \eqref{contraints of multiple zeros} can be replaced by
\begin{equation} \label{triple root condition}
  \rho_i=\rho_j=\sqrt{2}\rho_l=2|k_l-k_i| =2|k_l-k_j|\neq0, \quad  k_i \not = k_j. %\quad \text{ for } M=3,
\end{equation}
Similarly, assume $(i,j,l,m)$ is any permutation of the set $\{1, 2, 3, 4\}$, then the condition  \eqref{contraints of multiple zeros} can be replaced by
\begin{eqnarray} \label{quadruple root condition}
 2\rho_i^2=(3-\sqrt{5})\rho_j^2 = (3-\sqrt{5})\rho_l^2=2\rho_m^2
=(6-2\sqrt{5})(k_j-k_i)^2 =4(k_l-k_i)^2 = (6+2\sqrt{5})(k_m-k_i)^2\neq0.  %\quad   \text{ for }  M=4,
\end{eqnarray}
Under these conditions, the roots are
\begin{equation}
 p_0=
 \begin{cases}
\pm \dfrac{\rho_i}{2}+ \mathrm{i} k_l, \quad \text{ for } M=3,
 \\
 \pm \dfrac{1}{4} \sqrt{5+\sqrt{5}} \rho_i +\mathrm{i} \left( k_i -\dfrac{1}{4} \sqrt{3-\sqrt{5}} \rho_i \right), \quad \text{ for } M=4.
 \end{cases}
\end{equation}

%\begin{eqnarray} \label{quadruple root condition}
%&& 2\rho_1^2=(3-\sqrt{5})\rho_2^2 = (3-\sqrt{5})\rho_3^2=2\rho_4^2 \nonumber
%\\
%&=&(6-2\sqrt{5})(k_2-k_1)^2 =4(k_3-k_1)^2 = (6+2\sqrt{5})(k_4-k_1)^2\neq0,  \quad   \text{ for }  M=4,
%\end{eqnarray}

%{\color{blue} Other cases such that $\mathcal{G}_3^{\prime}(p)=0$ and $\mathcal{G}_4^{\prime}(p)=0$ have non-imaginary roots of order 3 and 4.}
\end{remark}

\begin{remark}
%$K \in \{1,2,\dots, M\}$
When $K=1$, %there exists $I_1$ such that $N_K = N$,
the $\tau$ functions are comprised of determinants of single block matrices, i.e.,
%{\color{red}
\begin{eqnarray*}
\tau_{\mathbf{n}} &=& \det_{1 \leq i, j \leq N} \left(m_{(M+1) i-I_1, (M+1) j-I_1}^{(\mathbf{n},I_1, I_1)}\right), \quad 1 \leq I_1 \leq M,
%\\
%\tau_{n, k,l}&=&\left(m_{4 i-I, 4 j-I}^{(n, k, l, I, I)}\right)_{1 \leq i, j \leq N}, \quad 1 \leq I \leq 3,
%\\
%\tau_{n, k,l, r}&=&\left(m_{5 i-J, 5 j-J}^{(n, k, l, r, J, J)}\right)_{1 \leq i,   \leq j \leq N}, \quad 1 \leq  J \leq 4,
\end{eqnarray*}
%}
where $m_{i, j}^{(\mathbf{n},I, J)}$ is given by \eqref{matrix elments-Mcom}. %and \eqref{matrix elments-4com}.
In this case, we define the rogue wave solutions in Theorem \ref{RW solutions of vector NLS} to be the $I_1$-th type, $1 \leq I_1 \leq M$, and simply denote $\boldsymbol{x}_{I}^{\pm}$ by $\boldsymbol{x}^{\pm}$ by ignoring the dependence on $I.$
\end{remark}

%{\color{red}
\begin{remark}
By rewriting $\tau_{\mathbf{n}}$ into a
larger determinant similar to \cite{ohta2012general}, we can show that the degrees of the polynomials $\tau_{\mathbf{n}} $ for $ M=3,4$  with respect to $x$ and $t$ in Theorem \ref{RW solutions of vector NLS} are
\begin{equation}
\deg(\tau_{\mathbf{n}})=%\left\{
\begin{cases}
3\left(N_1^2+N_2^2+N_3^2\right)-2(N_1N_2+N_1N_3+N_2N_3)
+\left(3N_1+N_2-N_3\right), \quad M=3, \\
5\left(N_1^2+N_2^2+N_3^2+N_4^2\right)- (N_1+N_2+N_3+N_4)^2+4 N_1+ 2 N_2- 2 N_4, \quad M=4,
\end{cases}
%\right.
\end{equation}
where $N_j \, (j=1,2,\dots, M)$ are non-negative integers such that $N_1+N_2+\cdots+N_M=N$. We note that, when $N_I = 0,$ it means that the block matrices $\tau_{\mathbf{n}}^{[I,I_l]}$ and $\tau_{\mathbf{n}}^{[I_l,I]} \, (l=1,2,\dots, M)$ do not appear in \eqref{tau-block matrix-theorem}.

\end{remark}

\begin{remark} \label{Values of s_r}
It can be calculated that
\begin{equation}
%\left\{
\begin{aligned}
s_1=s_2=s_3=s_5=s_6=s_7=s_9=s_{10}=s_{11}=0, \quad \text{ when } M=3, %\\
%s_1=s_2=s_3=s_4=s_6=s_7=s_8=s_9=s_{11}=0, \quad M=4
\end{aligned}
%\right.
\end{equation}
and
\begin{equation}
%\left\{
\begin{aligned}
%s_1=s_2=s_3=s_5=s_6=s_7=s_9=s_{10}=s_{11}=0, \quad M=3, \\
s_1=s_2=s_3=s_4=s_6=s_7=s_8=s_9=s_{11}=0, \quad \text{ when } M=4,
\end{aligned}
%\right.
\end{equation}
in \eqref{sr}, but we do not know whether or not $s_i=0$ holds for all $i \in \N$ such that $i \not \equiv 0 \mod{(M+1)}$ when $ M=3,4$. We also note that $x_{r,I}^{\pm}$ can be removed from the solution when $r \equiv 0 \mod (M+1)$, by using the technique developed in \cite{yang2021rogue}.
\end{remark}

%}

\subsection{Generalized Wronskian-Hermite polynomials}
Hermite polynomials are a sequence of classical orthogonal polynomials, and they arise in many areas of mathematics, such as probability, combinatorics, random matrix theory, etc. Like other orthogonal polynomials, Hermite polynomials can be defined from various viewpoints. It is also worth noting that there are two different standardizations in common use. However, it turns out neither of them is convenient for the analysis of wave patterns. Instead, we will introduce a slightly different definition \cite{yang2022pattern}. Let $p_j(z)$ be Schur polynomials defined by
\begin{equation}\label{polynomial p}
  \sum_{j=0}^{\infty} p_j(z) \epsilon^j=\exp \left(z \epsilon+\epsilon^2\right),
\end{equation}
with $p_j(z)\equiv0$ for $j<0$. Then it can be shown that the polynomials $p_j(z)$ are related to Hermite polynomials via certain rescaling. % that depends on the choice of Hermite polynomials.

Next, we introduce Wronskian-Hermite polynomials, which have appeared in the study of certain monodromy-free Schr\"{o}dinger operators \cite{oblomkov1999monodromy}. Let $N$ be a positive integer and  $\Lambda=\left(n_1, n_2, \ldots, n_N\right)$, where $\left\{n_i\right\}$ are distinct positive integers such that $n_1<n_2<\cdots<$ $n_N$, then the Wronskian-Hermite polynomial $W_{\Lambda}(z)$ is defined as
\begin{equation} \label{Wronskian-Hermite-det form}
  W_{\Lambda}(z)=\left|\begin{array}{cccc}
p_{n_1}(z) & p_{n_1-1}(z) & \cdots & p_{n_1-N+1}(z) \\
p_{n_2}(z) & p_{n_2-1}(z) & \cdots & p_{n_2-N+1}(z) \\
\vdots & \vdots & \vdots & \vdots \\
p_{n_N}(z) & p_{n_N-1}(z) & \cdots & p_{n_N-N+1}(z)
\end{array}\right|.
\end{equation}
Note from \eqref{polynomial p} that  $p_{k+1}^{\prime}(z)=p_k(z)$. This implies that the Wronskian-Hermite polynomial $W_{\Lambda}(z)$ can be rewritten as
\begin{equation}\label{Wronskian-Hermite-Wronskian form}
 W_{\Lambda}(z)=\operatorname{Wronskian}\left[p_{n_1}(z), p_{n_2}(z), \ldots, p_{n_N}(z)\right].
\end{equation}
 In particular, when the indices $\left(n_1, n_2, \ldots, n_N\right)$ are consecutive, these polynomials are called generalized Hermite polynomials, which are closely related to rational solutions of the fourth Painlev\'{e} equation \cite{clarkson2003second}. % $\left(\mathrm{P}_{\mathrm{IV}}\right)$ (Clarkson 2003).

The Yablonskii-Vorob'ev polynomials \cite{yablonskii1959rational,vorob1965rational} and Okamoto polynomials \cite{okamoto1986studies} are another two important classes of special polynomials, and as shown in \cite{clarkson2003second,yang2022rogue}, they can be generalized to hierarchies that have close connections with rogue wave patterns of certain integrable systems \cite{yang2021universal,yang2022rogue}. It turns out that the Wronskian-Hermite polynomials can be generalized in a similar way.
Let $p_j^{[m]}(z)$, where $m>1$ is a positive integer, be Schur polynomials defined by
\begin{equation}\label{polynomial p}
  \sum_{j=0}^{\infty} p_j^{[m]}(z) \epsilon^j=\exp \left(z \epsilon+\epsilon^m\right),
\end{equation}
with $p_j^{[m]}(z)\equiv0$ for $j<0$. Then the generalized Wronskian-Hermite polynomials are defined by %the Wronskians
\begin{equation}   \label{Wronskian-Hermite hierarchy}
  W_{\Lambda}^{[m]}(z)
 = \left|\begin{array}{cccc}
p_{n_1}^{[m]}(z) & p_{n_1-1}^{[m]}(z) & \cdots & p_{n_1-N+1}^{[m]}(z) \\
p_{n_2}^{[m]}(z) & p_{n_2-1}^{[m]}(z) & \cdots & p_{n_2-N+1}^{[m]}(z) \\
\vdots & \vdots & \vdots & \vdots \\
p_{n_N}^{[m]}(z) & p_{n_N-1}^{[m]}(z) & \cdots & p_{n_N-N+1}^{[m]}(z)
\end{array}\right|. %                      \nonumber
%\\
%&=&\operatorname{Wronskian}\left[p_{n_1}(z), p_{n_2}(z), \ldots, p_{n_N}(z)\right].
\end{equation}
In particular, these polynomials are called generalized Wronskian-Hermite polynomials of jump $k>0$ if %the differences of indices between two consecutive rows are $k$, i.e.,
$n_{j+1}-n_j = k, j =1,2,\dots, N-1$. Further, when $n_1=l,$ where $1\leq l < k$, %(what happens when $l=k$?)}
 we denote the generalized Wronskian-Hermite polynomial of jump $k>0$ by $W_{N}^{[m,k,l]}(z)$, i.e.,
 \begin{equation} \label{Wronskian-Hermite with jumps}
  W_{N}^{[m,k,l]}(z)= c_N^{[m,k,l]}\left|\begin{array}{cccc}
p_{l}^{[m]}(z) & p_{l-1}^{[m]}(z) & \cdots & p_{l-N+1}^{[m]}(z) \\
p_{l+k}^{[m]}(z) & p_{l+k-1}^{[m]}(z) & \cdots & p_{l+k-N+1}^{[m]}(z) \\
\vdots & \vdots & \vdots & \vdots \\
p_{l+k(N-1)}^{[m]}(z) & p_{l+k(N-1)-1}^{[m]}(z) & \cdots & p_{l+k(N-1)-N+1}^{[m]}(z)
\end{array}\right|,
\end{equation}
where %{\color{red}
%$$c_N^{[m,k,l]}  =k^{-N(N-1)}\dfrac{l!(k+l)!(2k+l)!\cdots[(N-1)k+l]!}{0!1!2!3!\cdots(N-1)!}.$$}
%{\color{red}
\begin{eqnarray}
  c_N^{[m,k,l]}  &=& \left(\prod_{n=1}^N\gamma_n !\right)  \bigg/ \left[\prod_{1 \leq i<j \leq N}\left(\gamma_j-\gamma_i\right) \right], \label{def of cN} \\
   \gamma_n &=& l+(n-1)k, \quad 1 \leq n \leq N.
\end{eqnarray}
%}
For the convenience of later use, we have multiplied a constant $c_N^{[m,k,l]}$ in \eqref{Wronskian-Hermite with jumps}, which makes  $W_{N}^{[m,k,l]}(z)$ a monic polynomial.

Since we can deduce from \eqref{polynomial p} that $(p_{j+1}^{[m]})'(z) = p_j^{[m]}(z)$, $W_{N}^{[m,k,l]}(z)$ can be rewritten as
\begin{equation}\label{Wronskian-Hermite-hierarchy Wronskian form}
 W_{N}^{[m,k,l]}(z)=\operatorname{Wronskian}\left[p_{l}(z), p_{l+k}(z), \ldots, p_{l+k(N-1)}(z)\right].
\end{equation}
If we take $m=2,k=4$, then the first few $W_{N}^{[2,4,l]}(z) \, (N,l=1,2,3)$ are
%\begin{eqnarray*}
%% \nonumber to remove numbering (before each equation)
%  W_{1}^{[2,4,1]}(z) &=& \\
%  W_{2}^{[2,4,1]}(z) &=&   \\
%  W_{3}^{[2,4,1]}(z) &=&   \\
%  W_{1}^{[2,4,2]}(z) &=&   \\
%  W_{2}^{[2,4,2]}(z) &=&   \\
%  W_{3}^{[2,4,2]}(z) &=&   \\
%  W_{1}^{[2,4,3]}(z) &=&   \\
%  W_{2}^{[2,4,3]}(z) &=&   \\
%  W_{3}^{[2,4,3]}(z) &=&
%\end{eqnarray*}
\begin{flalign*}
    && W_1^{[2,4,1]}(z) & = z, & \\
    && W_2^{[2,4,1]}(z) & = z^3 \left(z^2+10\right), & \\
    && W_3^{[2,4,1]}(z) & = z^6 \left(z^6+42 z^4+540 z^2+2520\right), & \\
    && W_1^{[2,4,2]}(z) & = z^2+2, & \\
    && W_2^{[2,4,2]}(z) & = z \left(z^6+18 z^4+60 z^2+120\right), & \\
    && W_3^{[2,4,2]}(z) & = z^3 \left(z^{12}+60 z^{10}+1260 z^8+12000 z^6+54000 z^4+181440 z^2+302400\right), & \\
    && W_1^{[2,4,3]}(z) & = z(z^2+6), & \\
    && W_2^{[2,4,3]}(z) & = z^3 \left(z^6+30 z^4+252 z^2+840\right), & \\
    && W_3^{[2,4,3]}(z) & = z^6 \left(z^{12}+84 z^{10}+2700 z^8+43680 z^6+388080 z^4+1995840 z^2+4656960\right). &
\end{flalign*}
We remark that, when $k=2$, the generalized Wronskian-Hermite polynomials $W_{N}^{[2m+1,2,1]}(z)$ are related to the
Yablonskii-Vorob'ev polynomials through some rescaling. In addition, $W_{N}^{[m,3,1]}(z)$ and $W_{N}^{[m,3,2]}(z)$ are multiples of the Okamoto polynomial hierarchies of $Q_N^{[m]}(z)$ and $R_N^{[m]}(z)$ respectively \cite{yang2022rogue}. In other words, the Yablonskii-Vorob'ev polynomial hierarchy and the Okamoto polynomial hierarchies are special cases of the generalized Wronskian-Hermite polynomials.
%{\color{red} It is noted that when $m =0 \bmod k $, then $W_N^{[m,k,l]}(z)=???, l =1,2,\dots, k$. }
%{\color{red} It is noted that when $m \equiv 0 \bmod k $, then $W_N^{[m,k,l]}(z)=z^{{\theta}_k}, l =1,2,\dots, k-1$, }
%where
%$$  {\theta}_4=N (3N-3+2l)/2,
%\quad
%  {\theta}_5=N (2N-2+l).
%$$

As we will see in the subsequent sections, rogue wave patterns of the vector NLS equation \eqref{vector NLS} are asymptotically determined by the distribution of zeros of the generalized Wronskian-Hermite polynomials. Root structures of certain special cases of the generalized Wronskian-Hermite polynomials have been obtained in previous studies, such as the  Yablonskii-Vorob'ev polynomial hierarchy \cite{clarkson2003second,balogh2016hankel,fukutani2000special,taneda2000remarks,buckingham2014large} and the Okamoto polynomials hierarchies \cite{clarkson2003fourth,kametaka1983poles,fukutani2000special}. For instance, it has been shown that all nonzero roots of the Yablonskii-Vorob'ev polynomials and the Okamoto polynomials $Q_N^{[1]}(z)$ and $R_N^{[1]}(z)$ are simple \cite{fukutani2000special,kametaka1983poles}. Despite that, as far as we know, root structures for higher members of generalized Wronskian-Hermite polynomials have not been studied yet.

Now we discuss root structures of the generalized Wronskian-Hermite polynomials of jump $4$ and $5$, which will be used in later studies on rogue wave patterns. Let $N_0$ be the remainder of $N$ divided by $m$, i.e.,
$$
N_0 \equiv N \bmod m \quad {\text or } \quad N = k m + N_0,
$$
where $k$ is a nonnegative integer, and we denote $[a]$ by the largest integer less than or equal to a real number $a$. Then our results can be summarized as follows.

%{\color{red} When $\Gamma = 0$?}

\begin{theorem} \label{root sturcture of jump 4}
The generalized Wronskian-Hermite polynomials $W_N^{[m,4,l]}$ of jump $4$ are monic with degree $   N (3N-3+2l)/2,$ and has the form
\begin{equation} \label{root sturcture of jump 4-simple form}
W_N^{[m,4,l]}=z^{\Gamma} w_N^{[m,4,l]}(\zeta), \quad
\zeta = z^m,
\end{equation}
where $w_N^{[m,4,l]}{(\zeta)}$ is a monic polynomial with real coefficients, $w_N^{[m,4,l]}{(0)} \not = 0$, and $\Gamma$ is the multiplicity of the zero root given by % and we will give the its fomulas when $k=4,5$ and $1\leq l\leq k-1$.
%For the case of $m= 1 $ or $3 \mod 4$, %($m\mod k= 2$ is a singular case that our metheod does't work)
%Moreover, we have
\begin{equation}
\Gamma=\dfrac{3}{2}\left(N_1^2+N_2^2+N_3^2\right)-(N_1N_2+N_1N_3+N_2N_3)
+\dfrac{1}{2}\left(3N_1+N_2-N_3\right)
\end{equation}
with the values of $N_1, N_2 $ and $N_3$ characterized as follows.
\begin{itemize}
  \item When $m \equiv 1 \mod 4$, we have

  \begin{equation*}
l=3: \left(N_1, N_2,N_3\right)= \begin{cases}\left(N_0, 0,0\right), & 0 \leq N_0 \leq\left[\frac{m}{4}\right]
\\ \left(\left[\frac{m}{4}\right], N_0-\left[\frac{m}{4}\right],0\right), & \left[\frac{m}{4}\right]+1 \leq N_0 \leq 2\left[\frac{m}{4}\right]
\\ \left(\left[\frac{m}{4}\right],\left[\frac{m}{4}\right], N_0-2\left[\frac{m}{4}\right]\right), &  2\left[\frac{m}{4}\right]+1 \leq N_0 \leq 3\left[\frac{m}{4}\right]
\\
\left(m-1-N_0, m-1-N_0,m-1-N_0\right), & \text 3\left[\frac{m}{4}\right]+1 \leq N_0 \leq m-1\end{cases}
\end{equation*}
\begin{equation*}
l=2: \left(N_1, N_2,N_3\right)=
\begin{cases}
\left(0, N_0,0\right), &   0 \leq N_0 \leq\left[\frac{m}{4}\right]
\\
\left(0,\left[\frac{m}{4}\right], N_0-\left[\frac{m}{4}\right]\right), &  \left[\frac{m}{4}\right]+1 \leq N_0 \leq 2\left[\frac{m}{4}\right]
\\
\left(\left[\frac{m}{4}\right]-1,\left[\frac{m}{4}\right]-1, N_0-2\left[\frac{m}{4}\right]-1\right), & 2\left[\frac{m}{4}\right]+1 \leq N_0 \leq 3\left[\frac{m}{4}\right]
\\
\left(m-1-N_0, m-1-N_0,m-1-N_0\right), &   3\left[\frac{m}{4}\right]+1 \leq N_0 \leq m-1\end{cases}
\end{equation*}
\begin{equation*}
l=1: \left(N_1, N_2,N_3\right)=
\begin{cases}
\left(0,0,N_0\right), &   0 \leq N_0 \leq\left[\frac{m}{4}\right]
\\
\left(\left[\frac{m}{4}\right]-1, N_0-\left[\frac{m}{4}\right]-1,0\right), &  \left[\frac{m}{4}\right]+1 \leq N_0 \leq 2\left[\frac{m}{4}\right]+1
\\
\left(\left[\frac{m}{4}\right]-1,\left[\frac{m}{4}\right], N_0-2\left[\frac{m}{4}\right]-1\right), &  2\left[\frac{m}{4}\right]+2 \leq N_0 \leq 3\left[\frac{m}{4}\right]+1
\\
\left(m-1-N_0, m-N_0,m-N_0\right), &  3\left[\frac{m}{4}\right]+2 \leq N_0 \leq m-1.
\end{cases}
\end{equation*}

 \item  When $m \equiv 2 \mod 4$, we have
 \begin{equation*}
l=3: \left(N_1, N_2,N_3\right)= \begin{cases}\left(\frac{k m}{2}+N_0, 0,\frac{k m}{2}\right), &   0 \leq N_0 \leq\left[\frac{m}{4}\right]
\\ \left(\frac{k m}{2}+\left[\frac{m}{4}\right], 0,\frac{k m}{2} +N_0 - \left[\frac{m}{4}\right]\right), & \left[\frac{m}{4}\right]+1 \leq N_0 \leq 3\left[\frac{m}{4}\right]+1
\\ \left(\frac{k m}{2}+N_0-2\left[\frac{m}{4}\right]-1,0,\frac{k m}{2}+2\left[\frac{m}{4}\right]+1\right), &   3\left[\frac{m}{4}\right]+2 \leq N_0 \leq m-1
\end{cases}
\end{equation*}
\begin{equation*}
l=2: \left(N_1, N_2,N_3\right)= \begin{cases}\left(\frac{k m}{2}-1, 0,\frac{k m}{2}+N_0\right), &   0 \leq N_0 \leq\left[\frac{m}{4}\right]
\\ \left(\frac{k m}{2}+N_0-1-\left[\frac{m}{4}\right], 0,\frac{k m}{2} + \left[\frac{m}{4}\right]\right), & \left[\frac{m}{4}\right]+1 \leq N_0 \leq 3\left[\frac{m}{4}\right]+1
\\ \left(\frac{k m}{2}+2\left[\frac{m}{4}\right],0, \frac{k m}{2}+N_0-2\left[\frac{m}{4}\right]-1\right), &   3\left[\frac{m}{4}\right]+2 \leq N_0 \leq m-1
\end{cases}
\end{equation*}
\begin{equation*}
l=1: \left(N_1, N_2,N_3\right)= \begin{cases}\left(\frac{k m}{2}, 0,\frac{k m}{2}+N_0\right), &   0 \leq N_0 \leq\left[\frac{m}{4}\right]+1
\\ \left(\frac{k m}{2} +N_0 - \left[\frac{m}{4}\right]-1,0,\frac{k m}{2}+\left[\frac{m}{4}\right]+1\right), & \left[\frac{m}{4}\right]+2 \leq N_0 \leq 3\left[\frac{m}{4}\right]+2
\\ \left(\frac{k m}{2}+2\left[\frac{m}{4}\right]+1,0,\frac{k m}{2}+N_0-2\left[\frac{m}{4}\right]-1\right), &   3\left[\frac{m}{4}\right]+3 \leq N_0 \leq m-1.
\end{cases}
\end{equation*}

  \item  When $m \equiv 3 \mod 4$, we have
\begin{equation*}
l=3: \left(N_1, N_2,N_3\right)= \begin{cases}\left(N_0, 0,0\right), &   0 \leq N_0 \leq\left[\frac{m}{4}\right]
\\ \left(N_0-1-\left[\frac{m}{4}\right], \left[\frac{m}{4}\right],0\right), & \left[\frac{m}{4}\right]+1 \leq N_0 \leq 2\left[\frac{m}{4}\right]+1
\\ \left(N_0-2\left[\frac{m}{4}\right]-2,\left[\frac{m}{4}\right], \left[\frac{m}{4}\right]\right), &   2\left[\frac{m}{4}\right]+2 \leq N_0 \leq 3\left[\frac{m}{4}\right]+2
\\ \left(m-1-N_0, m-1-N_0,m-1-N_0\right), &  3\left[\frac{m}{4}\right]+3 \leq N_0 \leq m-1\end{cases}
\end{equation*}
\begin{equation*}
l=2: \left(N_1, N_2,N_3\right)= \begin{cases}\left(0,N_0,0\right), &  0 \leq N_0 \leq\left[\frac{m}{4}\right]+1
\\ \left(N_0-\left[\frac{m}{4}\right]-1, \left[\frac{m}{4}\right]+1,0\right), &  \left[\frac{m}{4}\right]+2 \leq N_0 \leq 2\left[\frac{m}{4}\right]+1
\\ \left(N_0-2\left[\frac{m}{4}\right]-2,\left[\frac{m}{4}\right], \left[\frac{m}{4}\right]+1\right), &  2\left[\frac{m}{4}\right]+2 \leq N_0 \leq 3\left[\frac{m}{4}\right]+2
\\ \left(m-1-N_0, m-1-N_0,m-N_0\right), &   3\left[\frac{m}{4}\right]+3 \leq N_0 \leq m-1\end{cases}
\end{equation*}
\begin{equation*}
l=1: \left(N_1, N_2,N_3\right)=
\begin{cases}\left(0,0,N_0\right), &   0 \leq N_0 \leq\left[\frac{m}{4}\right]+1
\\ \left(0,N_0-\left[\frac{m}{4}\right]-1, \left[\frac{m}{4}\right]+1\right), &  \left[\frac{m}{4}\right]+2 \leq N_0 \leq 2\left[\frac{m}{4}\right]+2
\\ \left(N_0-2\left[\frac{m}{4}\right]-2,\left[\frac{m}{4}\right]+1, \left[\frac{m}{4}\right]+1\right), &   2\left[\frac{m}{4}\right]+3 \leq N_0 \leq 3\left[\frac{m}{4}\right]+2
\\ \left(m-1-N_0, m-1-N_0,m-N_0\right), &  3\left[\frac{m}{4}\right]+3 \leq N_0 \leq m-1.
\end{cases}
\end{equation*}

  %\item
\end{itemize}

\end{theorem}

\begin{theorem} \label{root sturcture of jump 5}

The generalized Wronskian-Hermite polynomials $W_N^{[m,5,l]}$ of jump $5$ are monic with degree $   N (2N-2+l),$ and has the form
\begin{equation} \label{root sturcture of jump 5-simple form}
W_N^{[m,5,l]}=z^{\Gamma} w_N^{[m,5,l]}(\zeta), \quad
\zeta = z^m,
\end{equation}
where $w_N^{[m,5,l]}{(\zeta)}$ is a monic polynomial with real coefficients, $w_N^{[m,5,l]}{(0)} \not = 0$, and $\Gamma$ is the multiplicity of the zero root given by % and we will give the its fomulas when $k=4,5$ and $1\leq l\leq k-1$.
%For the case of $m= 1, 2, 3, 4 \mod 5$, %($m\mod k= 2$ is a singular case that our metheod does't work)
%we have
\begin{equation} \label{gamma-4com}
\Gamma=\frac{5}{2}\left(N_1^2+N_2^2+N_3^2+N_4^2\right)-\frac{1}{2}(N_1+N_2+N_3+N_4)^2
+2N_1+N_2-N_4.
\end{equation}
The values of $N_1, N_2,N_3,N_4$ can be characterized in a similar way as Theorem \ref{root sturcture of jump 4} (the details are provided in Lemma \ref{values of N_i-root structure} of Appendix C).

\end{theorem}

We provide the proofs of Theorems \ref{root sturcture of jump 4} and \ref{root sturcture of jump 5} in Appendix C.

\begin{remark} %\label{root sturcture of jump 5}
We note that Theorems \ref{root sturcture of jump 4} and \ref{root sturcture of jump 5} provide the multiplicities of the zero root of the generalized Wronskian-Hermite polynomials $W_N^{[m,4,l]}$ and $W_N^{[m,5,l]}$ respectively, and as we will see subsequently, these multiplicities are essential in the analysis of rogue wave patterns in the inner region when specific parameters are very large (see Theorems \ref{Rogue wave patterns-3com} and \ref{Rogue wave patterns-4com}). It is also clear that the roots of $W_N^{[m,4,l]}$ are distributed symmetrically on some circles in the sense that if $z_0$ is a root of $W_N^{[m,4,l]}$, then so is $z_0 \exp(2 k \pi \mathrm{i}/m)$, where $k=0,1,\dots, m-1$.

\begin{figure}[htb]
	\centering
	\renewcommand\arraystretch{0.5}
	\setlength\tabcolsep{0pt}
	
	\begin{tabular}{m{0.6cm}<{\centering}m{2.9cm}<{\centering}m{2.9cm}<{\centering}m{2.9cm}<{\centering}m{2.9cm}<{\centering}m{2.9cm}<{\centering}m{0.6cm}<{\centering}r}
	&\textbf{$m=2$}  & \textbf{$m=3$} & \textbf{$m=5$}& \textbf{$m=6$}&\textbf{$m=7$}&\\
	\ \rotatebox{90}{\textbf{$N=2$}}& \includegraphics[height=27mm,width=27mm]{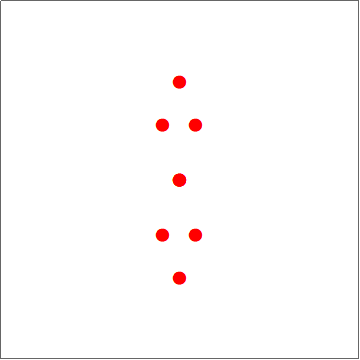}&\includegraphics[height=27mm,width=27mm]{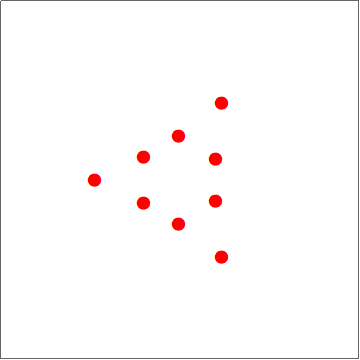}&\includegraphics[height=27mm,width=27mm]{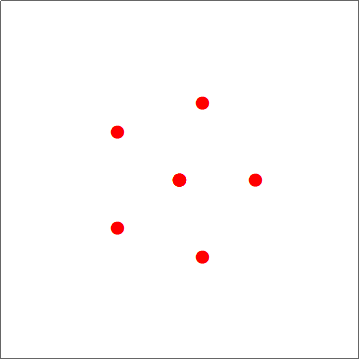}&\includegraphics[height=27mm,width=27mm]{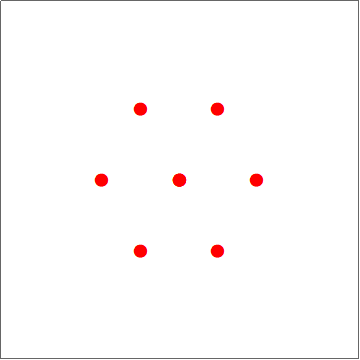}&\includegraphics[height=27mm,width=27mm]{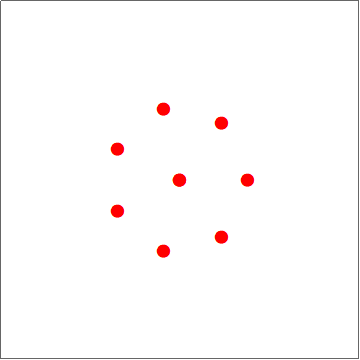}&\rotatebox{270}{\textbf{Im(z)}}\\
	\ \rotatebox{90}{\textbf{$N=3$}} &\includegraphics[height=27mm,width=27mm]{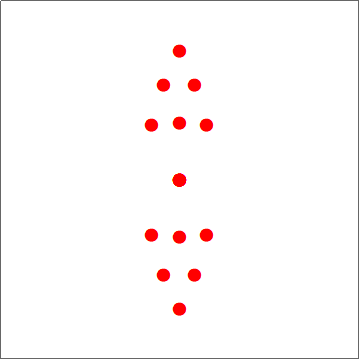}&\includegraphics[height=27mm,width=27mm]{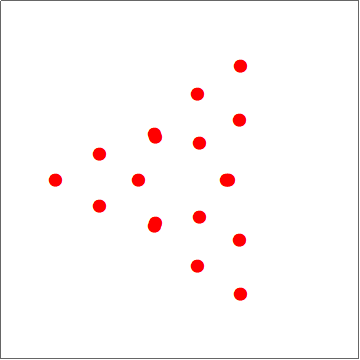}&\includegraphics[height=27mm,width=27mm]{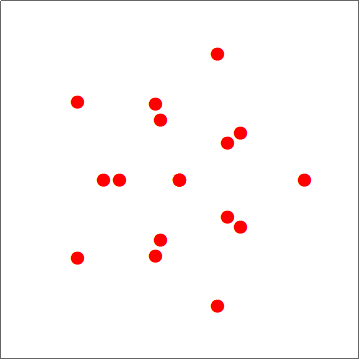}&\includegraphics[height=27mm,width=27mm]{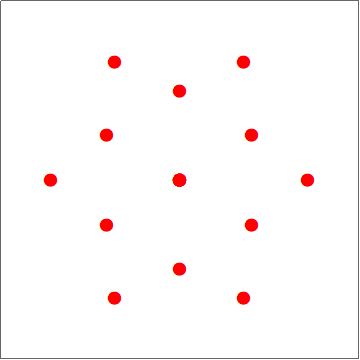}&\includegraphics[height=27mm,width=27mm]{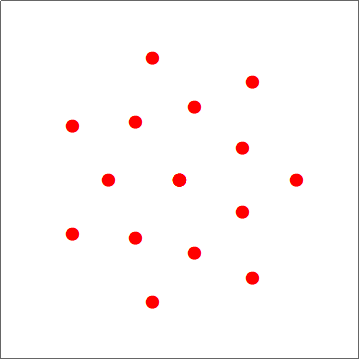}&\rotatebox{270}{\textbf{Im(z)}}\\
	\ \rotatebox{90}{\textbf{$N=4$}} &\includegraphics[height=27mm,width=27mm]{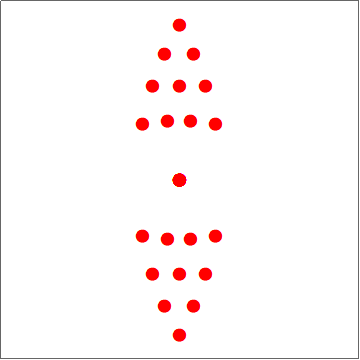}&\includegraphics[height=27mm,width=27mm]{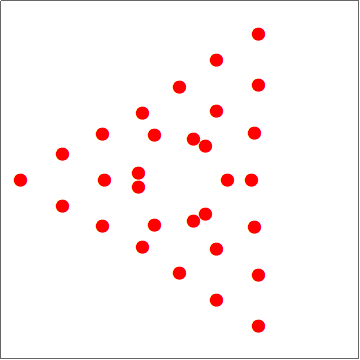}&\includegraphics[height=27mm,width=27mm]{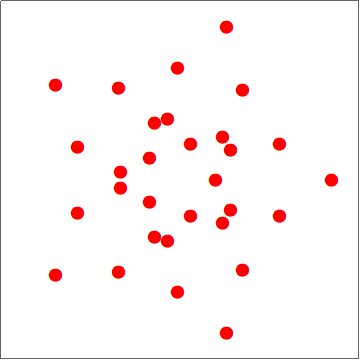}&\includegraphics[height=27mm,width=27mm]{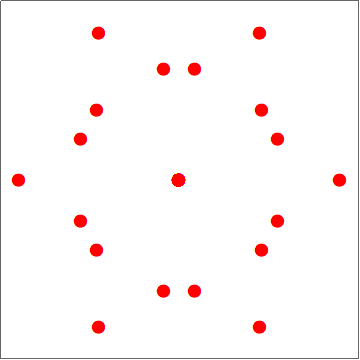}&\includegraphics[height=27mm,width=27mm]{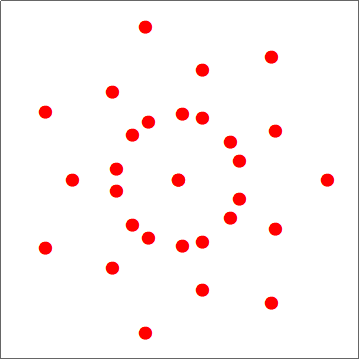}&\rotatebox{270}{\textbf{Im(z)}} \\
	\ \rotatebox{90}{\textbf{$N=5$}} &\includegraphics[height=27mm,width=27mm]{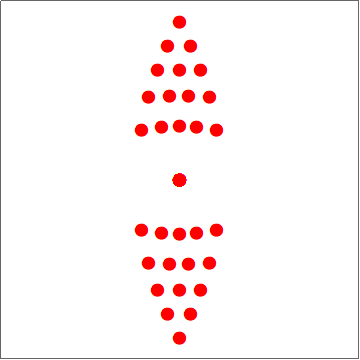}&\includegraphics[height=27mm,width=27mm]{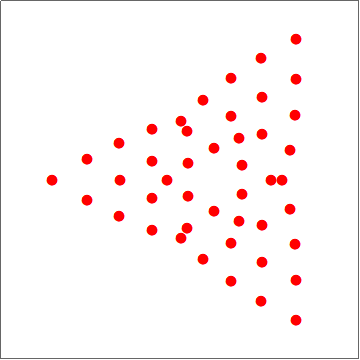}&\includegraphics[height=27mm,width=27mm]{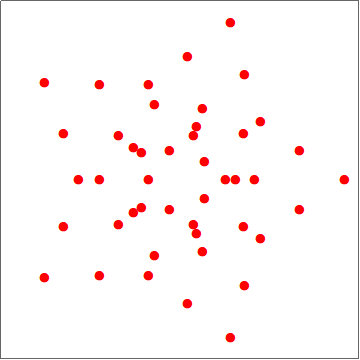}&\includegraphics[height=27mm,width=27mm]{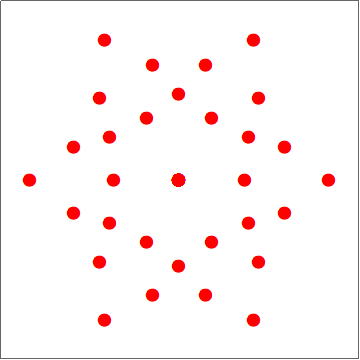}&\includegraphics[height=27mm,width=27mm]{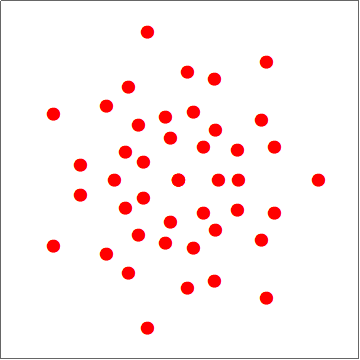}& \rotatebox{270}{\textbf{Im(z)}}\\
	&\textbf{Re(z)} & \textbf{Re(z)} & \textbf{Re(z)} & \textbf{Re(z)}& \textbf{Re(z)}
	\end{tabular}
	
	\caption{Plots of the roots of the polynomials $W_N^{[m, 4, 3]}(z)$ for $2 \leq N \leq 5$ and $m=2,3,5,6,7$.}
	
	\label{roots-43}
\end{figure}

\begin{figure}[htb]
	\centering
	\renewcommand\arraystretch{0.5}
	\setlength\tabcolsep{0pt}
	
	\begin{tabular}{m{0.6cm}<{\centering}m{2.9cm}<{\centering}m{2.9cm}<{\centering}m{2.9cm}<{\centering}m{2.9cm}<{\centering}m{2.9cm}<{\centering}m{0.6cm}<{\centering}r}
	&\textbf{$m=2$}  & \textbf{$m=3$} & \textbf{$m=5$}& \textbf{$m=6$}&\textbf{$m=7$} &\\
	\ \rotatebox{90}{\textbf{$N=2$}} & \includegraphics[height=27mm,width=27mm]{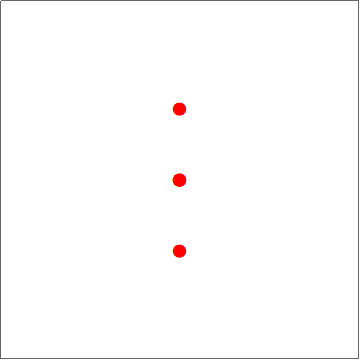}&\includegraphics[height=27mm,width=27mm]{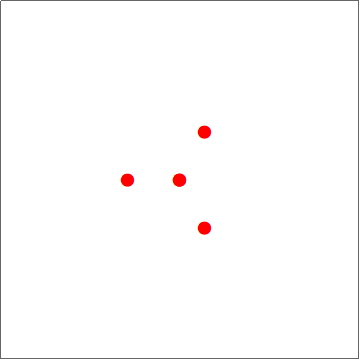}&\includegraphics[height=27mm,width=27mm]{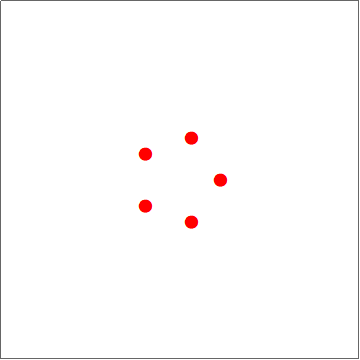}&  &  &\rotatebox{270}{\textbf{Im(z)}}\\
	\ \rotatebox{90}{\textbf{$N=3$}} &\includegraphics[height=27mm,width=27mm]{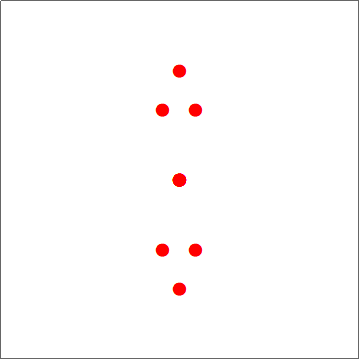}&\includegraphics[height=27mm,width=27mm]{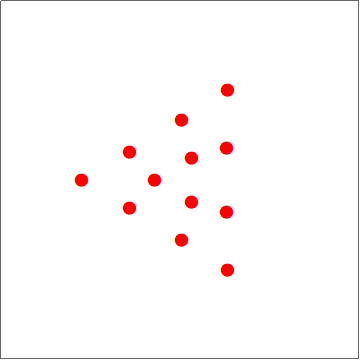}&\includegraphics[height=27mm,width=27mm]{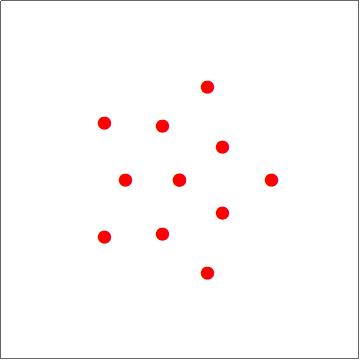}&\includegraphics[height=27mm,width=27mm]{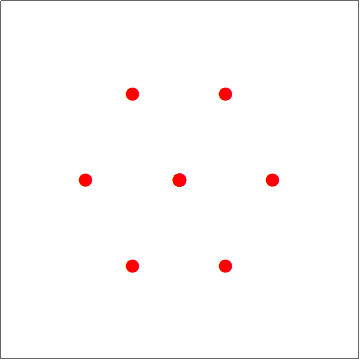}&\includegraphics[height=27mm,width=27mm]{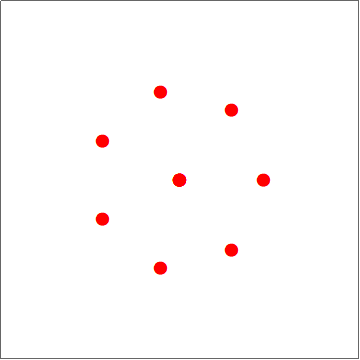}&\rotatebox{270}{\textbf{Im(z)}} \\
	\ \rotatebox{90}{\textbf{$N=4$}} &\includegraphics[height=27mm,width=27mm]{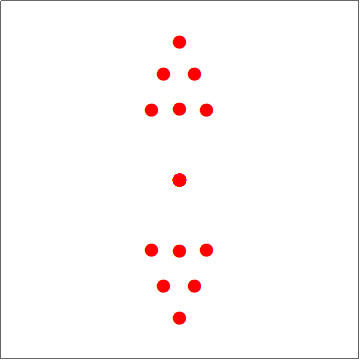}&\includegraphics[height=27mm,width=27mm]{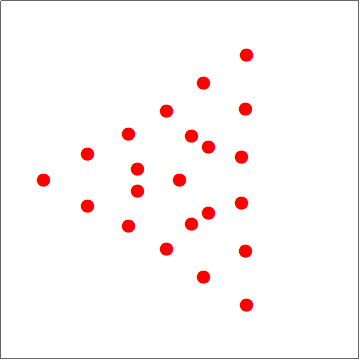}&\includegraphics[height=27mm,width=27mm]{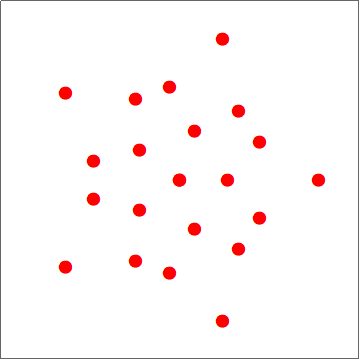}&\includegraphics[height=27mm,width=27mm]{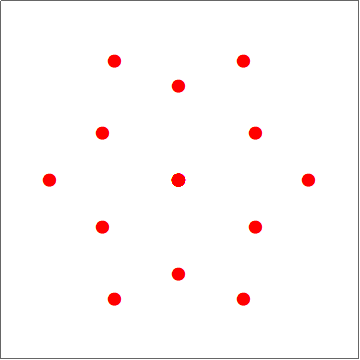}&\includegraphics[height=27mm,width=27mm]{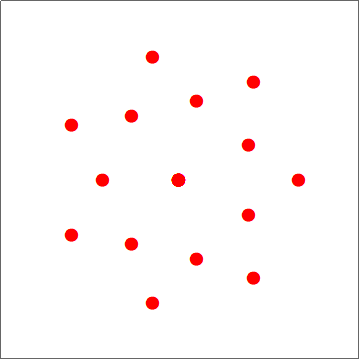}&\rotatebox{270}{\textbf{Im(z)}} \\
	\ \rotatebox{90}{\textbf{$N=5$}}&\includegraphics[height=27mm,width=27mm]{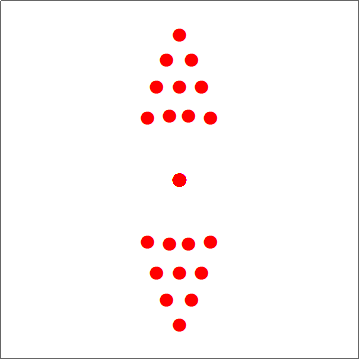}&\includegraphics[height=27mm,width=27mm]{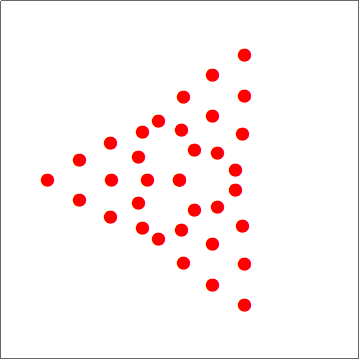}&\includegraphics[height=27mm,width=27mm]{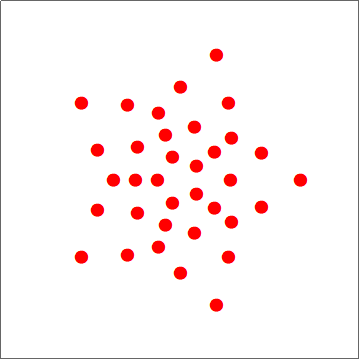}&\includegraphics[height=27mm,width=27mm]{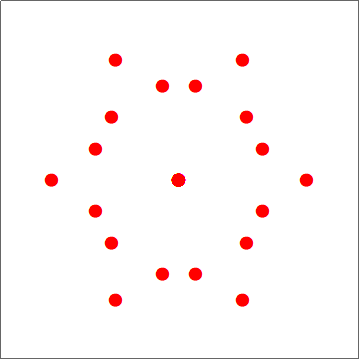}&\includegraphics[height=27mm,width=27mm]{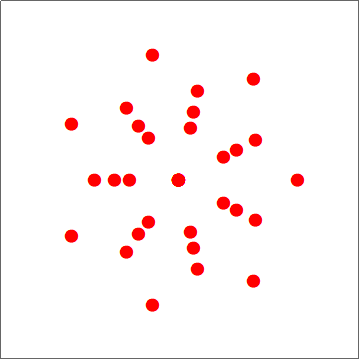}&\rotatebox{270}{\textbf{Im(z)}} \\
	&\textbf{Re(z)} & \textbf{Re(z)} & \textbf{Re(z)} & \textbf{Re(z)}& \textbf{Re(z)}
	\end{tabular}
	
	\caption{Plots of the roots of the polynomials $W_N^{[m, 4, 1]}(z)$ for $2 \leq N \leq 5$ and $m=2,3,5,6,7$.}
	
	\label{roots-41}
\end{figure}

\begin{figure}[htb]
	\centering
	\renewcommand\arraystretch{0.5}
	\setlength\tabcolsep{0pt}
	
	\begin{tabular}{m{0.6cm}<{\centering}m{2.8cm}<{\centering}m{2.8cm}<{\centering}m{2.8cm}<{\centering}m{2.8cm}<{\centering}m{2.8cm}<{\centering}m{0.6cm}<{\centering}r}
	&\textbf{$m=2$}  & \textbf{$m=3$} & \textbf{$m=4$}& \textbf{$m=6$}&\textbf{$m=7$} &\\
	\ \rotatebox{90}{\textbf{$N=2$}} & \includegraphics[height=27mm,width=27mm]{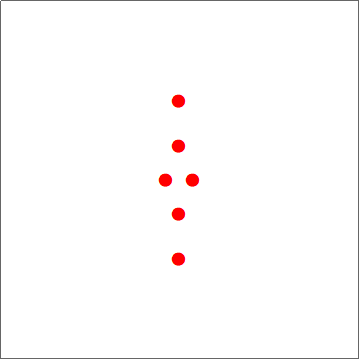}&\includegraphics[height=27mm,width=27mm]{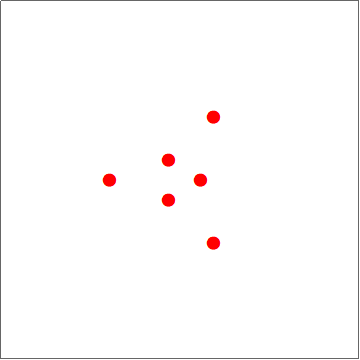}&\includegraphics[height=27mm,width=27mm]{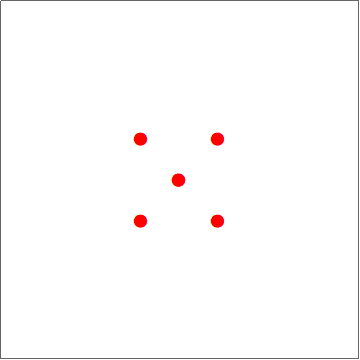}& \includegraphics[height=27mm,width=27mm]{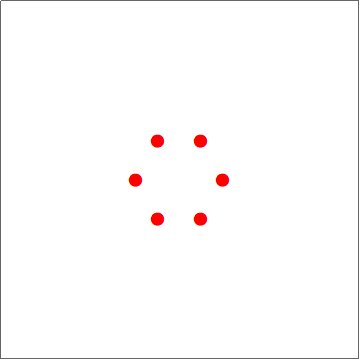} &  &\rotatebox{270}{\textbf{Im(z)}}\\
	\ \rotatebox{90}{\textbf{$N=3$}} &\includegraphics[height=27mm,width=27mm]{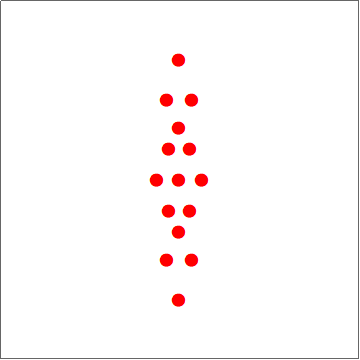}&\includegraphics[height=27mm,width=27mm]{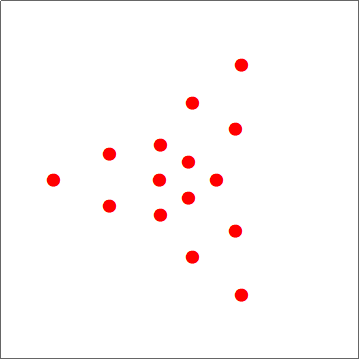}&\includegraphics[height=27mm,width=27mm]{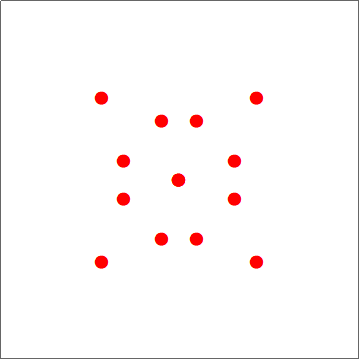}&\includegraphics[height=27mm,width=27mm]{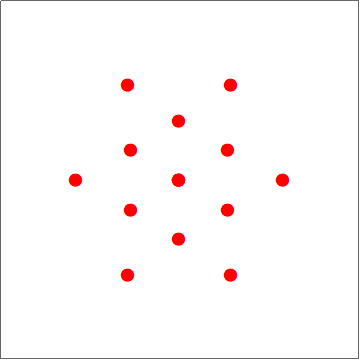}&\includegraphics[height=27mm,width=27mm]{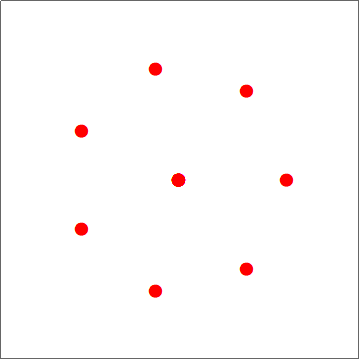}&\rotatebox{270}{\textbf{Im(z)}} \\
	\ \rotatebox{90}{\textbf{$N=4$}} &\includegraphics[height=27mm,width=27mm]{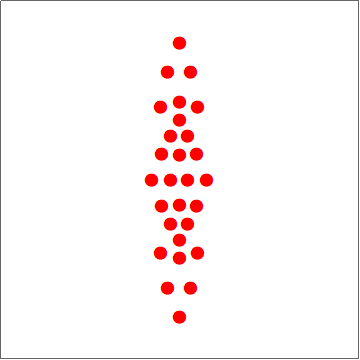}&\includegraphics[height=27mm,width=27mm]{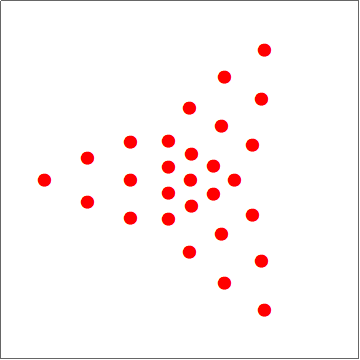}&\includegraphics[height=27mm,width=27mm]{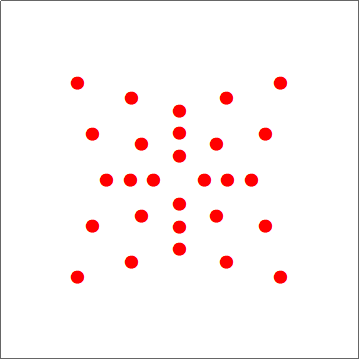}&\includegraphics[height=27mm,width=27mm]{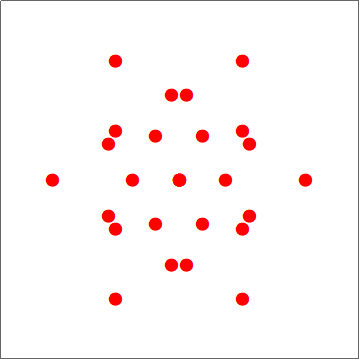}&\includegraphics[height=27mm,width=27mm]{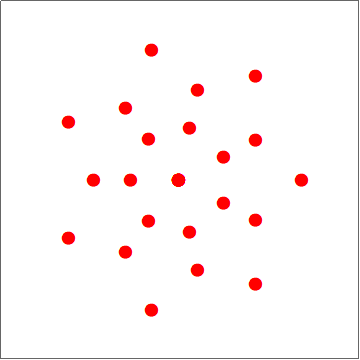}&\rotatebox{270}{\textbf{Im(z)}} \\
	\ \rotatebox{90}{\textbf{$N=5$}}&\includegraphics[height=27mm,width=27mm]{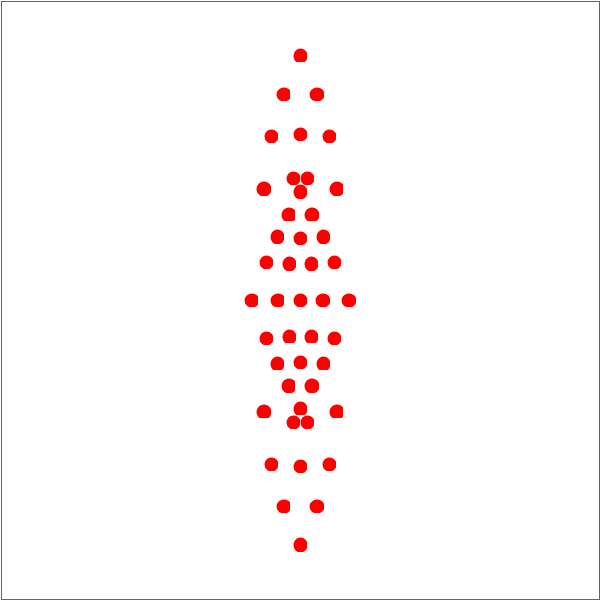}&\includegraphics[height=27mm,width=27mm]{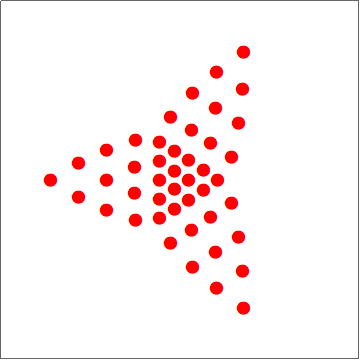}&\includegraphics[height=27mm,width=27mm]{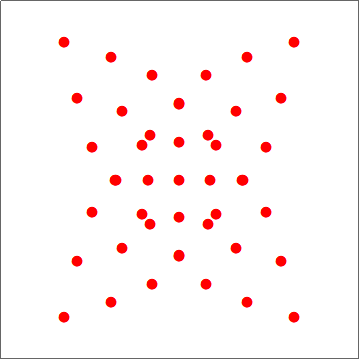}&\includegraphics[height=27mm,width=27mm]{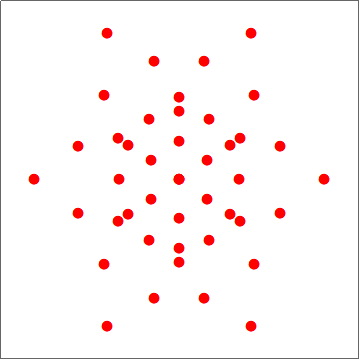}&\includegraphics[height=27mm,width=27mm]{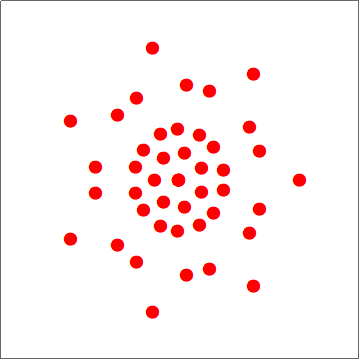}&\rotatebox{270}{\textbf{Im(z)}} \\
	&\textbf{Re(z)} & \textbf{Re(z)} & \textbf{Re(z)} & \textbf{Re(z)}& \textbf{Re(z)}
	\end{tabular}
	
	\caption{Plots of the roots of the polynomials $W_N^{[m, 5, 1]}(z)$ for $2 \leq N \leq 5$ and $m=2,3,5,6,7$.}
	
	\label{roots-51}
\end{figure}

% {\color{red} When $\lambda = 0$?}

In the analytical study of rogue wave patterns, a crucial assumption is that all the nonzero roots of the corresponding generalized Wronskian-Hermite polynomials $W_N^{[m,k,l]}$ are simple \cite{yang2021rogue,yang2021universal,yang2022rogue}. This assumption has been proved for Yablonskii-Vorob'ev polynomials $W_{N}^{[3,2,1]}(z)$ and Okamoto polynomials $W_{N}^{[2,3,1]}(z)$ \cite{fukutani2000special} and $W_{N}^{[2,3,2]}(z)$ \cite{fukutani2000special,kametaka1983poles}. Nevertheless, this assumption has not been verified for the general case. Since our results will also rely on this assumption, we propose a conjecture similar to those in \cite{yang2021rogue,yang2021universal,yang2022rogue}.

\begin{conjecture}
All nonzero roots of the generalized Wronskian-Hermite polynomials $W_N^{[m,k,l]}$ are simple for any integers $N\geq 1, m\geq 1, k \geq 2, 1 \leq l \leq k-1$.

\end{conjecture}

Although we are not able to prove this conjecture, we have verified it numerically for a variety of special cases which include all the particular generalized Wronskian-Hermite polynomials $W_N^{[m,k,l]}$ that will be involved in this paper.

The distribution of roots of the Yablonskii-Vorob'ev polynomial hierarchies and Okamoto polynomial hierarchies demonstrates highly regular and symmetric structures \cite{clarkson2003second,clarkson2003fourth}. The original Yablonskii-Vorob'ev polynomials form approximately equilateral triangles, while the higher members of the Yablonskii-Vorob'ev polynomial hierarchy form various shapes, such as pentagons, septagons, nonagons and
undecagons, etc., depending on the values of $m$ \cite{yang2022rogue}. The roots of Okamoto polynomials exhibit completely different structures compared with Yablonskii-Vorob'ev polynomials. Both of the $W_{N}^{[2,3,1]}(z)$ and $W_{N}^{[2,3,2]}(z)$ have similar root structures as theses roots are located on two ``triangles" except that $W_{N}^{[2,3,1]}(z)$ has an extra row of roots on a straight line between these two triangles. Here, we use ``triangles" because the edges of these triangles are curved rather than straight lines. A natural question is what characteristics the root structures of the generalized Wronskian-Hermite polynomials would exhibit. To this end, we plot the roots of $W_{N}^{[m,4,l]}(z)$ and $W_{N}^{[m,5,l]}(z)$ in Figs. \ref{roots-43}-\ref{roots-41} and \ref{roots-51}, respectively.
% Add a few sentences to describe the features of figures.}

\end{remark}

%\newpage

\section{Rogue wave patterns of the three- and four-component nonlinear Schr\"odinger equation} \label{section-Rogue wave patterns}

\begin{theorem} \label{Rogue wave patterns-3com}
Let $p_0, p_1,  \theta_{1n}, \rho_n, k_n, w_n \, (n=1,2,3)$ be the same as in Theorem \ref{RW solutions of vector NLS}.
Assume that $|a_m|\gg1$ and all other parameters are $O(1)$ in the $i$-th type $\mathcal{N}_i$-th order rogue waves ($  i =1,2,3$)
\begin{equation}\label{RW solution-3com-(1,1) type}
  u_{1, \mathcal{N}_i}(x, t), \quad u_{2, \mathcal{N}_i}(x, t), \quad u_{3, \mathcal{N}_i }(x, t),
\end{equation}
of the three-component nonlinear Schr\"odinger equation, where
$$\displaystyle{\mathcal{N}_i=N \sum_{j=1}^3 \delta_{ij} \boldsymbol{e}_j},$$
 $N$ is a positive integer, $\boldsymbol{e}_j$ is the standard unit vector in $\mathbb{R}^3$ %, i.e.,
% \begin{equation}%\label{}
%   \boldsymbol{e}_j = ???
% \end{equation}
  and  $\delta_{ij}$ is the Kronecker delta. We also assume that all non-zero roots of the generalized Wronskian-Hermite polynomials $W_N^{[m,4,4-i]}$ of jump $4$ are simple.
  Then, we have the following results concerning the asymptotics of the rogue waves \eqref{RW solution-3com-(1,1) type}.

 \begin{itemize}
   \item [(1)]  In the outer region on the $(x,t)$ plane, when $\sqrt{x^{2}+t^{2}}=O\left(\left|a_m\right|^{1 /m}\right)$, the $\mathcal{N}_i$-th order rouge waves separate into $N(3N+5-2i)/2-\Gamma$ fundamental rouge waves, where $\Gamma$ is given in \eqref{root sturcture of jump 4-simple form}. These fundamental rouge waves are
\begin{eqnarray}
&\hat{u}_1(x, t)=\rho_1 e^{\mathrm{i}\left(k_1 x+\omega_1 t\right)}  \dfrac{\left[p_1 x+2 p_0 p_1(\mathrm{i} t)+\theta_{11}\right]\left[p_1^* x-2 p_0^* p_1^*(\mathrm{i} t)-\theta_{11}^*\right]+\left|h_0\right|^2}{\left|p_1 x+2 p_0 p_1(\mathrm{i} t)\right|^2+\left|h_0\right|^2}, \label{fundmental_u1_3com} \\
&\hat{u}_2(x, t)=\rho_2  e^{\mathrm{i}\left(k_2 x+\omega_2 t\right)} \dfrac{\left[p_1 x+2 p_0 p_1(\mathrm{i} t)+\theta_{12}\right]\left[p_1^* x-2 p_0^* p_1^*(\mathrm{i} t)-\theta_{12}^*\right]+\left|h_0\right|^2}{\left|p_1 x+2 p_0 p_1(\mathrm{i} t)\right|^2+\left|h_0\right|^2}, \label{fundmental_u2_3com} \\
&\hat{u}_3(x, t)=\rho_3 e^{\mathrm{i}\left(k_3 x+\omega_3 t\right)} \dfrac{\left[p_1 x+2 p_0 p_1(\mathrm{i} t)+\theta_{13}\right]\left[p_1^* x-2 p_0^* p_1^*(\mathrm{i} t)-\theta_{13}^*\right]+\left|h_0\right|^2}{\left|p_1 x+2 p_0 p_1(\mathrm{i} t)\right|^2+\left|h_0\right|^2}, \label{fundmental_u3_3com}
\end{eqnarray}
where $|h_0|^2=\left|p_1\right|^2/\left(p_0+p_0^*\right)^2$, and their positions $(\hat{x}_0, \hat{t}_0)$ are given by
\begin{eqnarray} \label{prediction of locations of RW-3com}
\hat{x}_0 =\dfrac{1}{\Re\left(p_0\right)} \Re\left[\dfrac{p_0^*}{p_1}\left(z_0 a_m^{1 / m}- \Delta_i\right)\right],
\quad
\hat{t}_0 =\dfrac{1}{2 \Re\left(p_0\right)} \Im\left[\dfrac{1}{p_1}\left(z_0 a_m^{1 / m}-\Delta_i\right)\right],
\end{eqnarray}
where $z_0$ is any one of the non-zero simple roots of $W_{N}^{[m,4,4-i]}(z)$, $\Delta_i$ is a $z_0$-dependent $O(1)$ quantity, and $\left(\Re,\Im\right)$ refer to the real and imaginary parts of a complex number, respectively. The approximation error here is $ O\left(|a_{m}|^{-1 /m}\right)$.
In other words, when $|a_m|\gg1 $ and $\left(x-\hat{x}_0\right)^2+\left(t-\hat{t}_0\right)^2=O(1)$, we have the following asymptotics
\begin{eqnarray}
&u_{n, \mathcal{N}_i}(x, t)=\hat{u}_n\left(x-\hat{x}_0, t-\hat{t}_0\right) +O\left(\left|a_m\right|^{-1 / m}\right), \quad n =1,2,3.
%\\
%&u_{1, \mathcal{N}_i}(x, t)=\hat{u}_1\left(x-\hat{x}_0, t-\hat{t}_0\right) +O\left(\left|a_m\right|^{-1 / m}\right), \\
%&u_{2, \mathcal{N}_i}(x, t)=\hat{u}_2\left(x-\hat{x}_0, t-\hat{t}_0\right) +O\left(\left|a_m\right|^{-1 / m}\right) ,\\
%&u_{3, \mathcal{N}_i}(x, t)=\hat{u}_3 \left(x-\hat{x}_0, t-\hat{t}_0\right) +O\left(\left|a_m\right|^{-1 / m}\right).
\end{eqnarray}

   \item [(2)]  In the inner
	region, where $x^2+t^2=O(1)$,	 if zero is a root of the generalized Wronskian-Hermite polynomials $W_N^{[m,4,4-i]}(z)$, then $\left[u_{1,\mathcal{N}_i}(x, t), u_{2,\mathcal{N}_i}(x,
	t),u_{3,\mathcal{N}_i}(x, t)\right]$ is approximately a lower
	$\widehat{\mathcal{N}}_i$-th order rogue wave $$u_{1, \widehat{\mathcal{N}}_i}(x,
	t), \quad u_{2, \widehat{\mathcal{N}}_i}(x, t), \quad
	u_{3, \widehat{\mathcal{N}}_i}(x, t)$$
  where  $\widehat{\mathcal{N}}_i=\displaystyle{\sum_{j=1}^{3}N_{j,4-i}\boldsymbol{e}_j}$ and $N_{j,l}$ refers to the value of $N_j$ against $l \in \{1,2,3\}$ given in Theorem \ref{root sturcture of jump 4}.  Moreover, the internal parameters
  \begin{eqnarray*}
  \left(\hat{a}_{1,n}, \hat{a}_{2,n}, \hat{a}_{3,n}, \hat{a}_{5,n}, \hat{a}_{6,n}\ldots, \hat{a}_{4 N_{n,4-i}-n,n}\right), \quad n=1,2,3,
  \end{eqnarray*}
  in this lower-order rogue waves are related to those in the original rogue wave as follows.
  \begin{itemize}
    \item For $ m \equiv 1 $  or $  3 \mod 4,$ we have
    	$$
	\hat{a}_{j, 1}=\hat{a}_{j, 2}=\hat{a}_{j, 3}=a_j+\displaystyle{\left(N-\displaystyle{\sum_{n=1}^3N_{n,4-i}}\right)} s_j, \quad j=1,2,3,5,6,7 \cdots.
	$$

    \item  For $m \equiv 2 \mod 4$, we have
    \begin{equation*}
\hat{a}_{j, 1}=\hat{a}_{j, 3}=
  \begin{cases}
  a_j+\displaystyle{\left(N-\displaystyle{\sum_{n=1}^3N_{n,4-i}}\right)} s_j, \quad \text{ if } \quad j=1,2,3,5, \cdots,m-1,m+1,\cdots,
  \\
  \displaystyle{\left(N-\displaystyle{\sum_{n=1}^3N_{n,4-i}}\right)} s_j, \quad \text{ if } \quad j=m.
  \end{cases}
\end{equation*}
  \end{itemize}
Here, $s_j$ is defined in Theorem \ref{RW solutions of vector NLS}. % and $\bar{N}_i=\displaystyle{\sum_{n=1}^3N_{n,4-i}}$.
The approximation error of this lower-order rogue wave  is $O\left(\left|a_m\right|^{-1}\right)$. In other words, when $\left|a_m\right| \gg 1$ and $x^2+t^2=$ $O(1)$, we have
\begin{eqnarray*}
    && u_{n,\mathcal{N}_i}\left(x, t ; a_2, a_3, a_5, a_6, \cdots\right) \\
   &=& u_{n,\widehat{\mathcal{N}}_i}\left(x, t ; \hat{a}_{j, 1}, \hat{a}_{j, 2},\hat{a}_{j, 3}, j=1,2,3,5,6 \ldots\right)+O\left(\left|a_m\right|^{-1}\right), \quad n=1,2,3.
\end{eqnarray*}
	If zero is not a root of $W_N^{[m,4,4-i]}(z)$, the solution $$\left[u_{1,\mathcal{N}_i}(x, t), \quad u_{2,\mathcal{N}_i}(x,
	t),  \quad u_{3,\mathcal{N}_i}(x, t)\right]$$ is approximately the constant background $$\left[\rho_1 e^{\mathrm{i}\left(k_1 x+\omega_1 t\right)}, \quad \rho_2 e^{\mathrm{i}\left(k_3 x+\omega_2 t\right)}, \quad  \rho_3 e^{\mathrm{i}\left(k_3 x+\omega_3 t\right)}\right].$$ %when $|a_m| \gg 1$.
 \end{itemize}

\end{theorem}

\begin{theorem} \label{Rogue wave patterns-4com}
Let $p_0, p_1,  \theta_{1n}, \rho_n, k_n, w_n \, (n=1,2,3,4)$ be the same as in Theorem \ref{RW solutions of vector NLS}.
Assume that $|a_m|\gg1$ and all other parameters are $O(1)$ in the $i$-th type $\mathcal{N}_i$-th order rogue waves ($  i =1,2,3,4$)
\begin{equation}\label{RW solution-4com-(1,1) type}
  u_{1, \mathcal{N}_i}(x, t), \quad u_{2, \mathcal{N}_i}(x, t), \quad u_{3, \mathcal{N}_i }(x, t), \quad u_{4, \mathcal{N}_i }(x, t)
\end{equation}
of the four-component nonlinear Schr\"odinger equation, where
$$\displaystyle{\mathcal{N}_i=N \sum_{j=1}^4 \delta_{ij} \boldsymbol{e_j}},$$
 $N$ is a positive integer, $\boldsymbol{e_j}$ is the standard unit vector in $\mathbb{R}^4$ and $\delta_{ij}$ is the Kronecker delta. We also assume that all non-zero roots of the generalized Wronskian-Hermite polynomials $W_N^{[m,5,5-i]}$ of jump $5$ are simple.
  Then, we have the following results concerning the asymptotics of the rogue waves \eqref{RW solution-4com-(1,1) type}.

 \begin{itemize}
   \item [(1)]  In the outer region on the $(x,t)$ plane, when $\sqrt{x^{2}+t^{2}}=O\left(\left|a_m\right|^{1 /m}\right)$, the $N$-th order rouge waves separate into $N (2N+3-i)-\Gamma$ fundamental rouge waves, where $\Gamma$ is given in \eqref{root sturcture of jump 5-simple form}. These fundamental rouge waves are
\begin{eqnarray}
&\bar{u}_1(x, t)=\rho_1 e^{\mathrm{i}\left(k_1 x+\omega_1 t\right)}  \dfrac{\left[p_1 x+2 p_0 p_1(\mathrm{i} t)+\theta_{11}\right]\left[p_1^* x-2 p_0^* p_1^*(\mathrm{i} t)-\theta_{11}^*\right]+\left|h_0\right|^2}{\left|p_1 x+2 p_0 p_1(\mathrm{i} t)\right|^2+\left|h_0\right|^2}, \\
&\bar{u}_2(x, t)=\rho_2  e^{\mathrm{i}\left(k_2 x+\omega_2 t\right)} \dfrac{\left[p_1 x+2 p_0 p_1(\mathrm{i} t)+\theta_{12}\right]\left[p_1^* x-2 p_0^* p_1^*(\mathrm{i} t)-\theta_{12}^*\right]+\left|h_0\right|^2}{\left|p_1 x+2 p_0 p_1(\mathrm{i} t)\right|^2+\left|h_0\right|^2},\\
&\bar{u}_3(x, t)=\rho_3 e^{\mathrm{i}\left(k_3 x+\omega_3 t\right)} \dfrac{\left[p_1 x+2 p_0 p_1(\mathrm{i} t)+\theta_{13}\right]\left[p_1^* x-2 p_0^* p_1^*(\mathrm{i} t)-\theta_{13}^*\right]+\left|h_0\right|^2}{\left|p_1 x+2 p_0 p_1(\mathrm{i} t)\right|^2+\left|h_0\right|^2},\\
&\bar{u}_4(x, t)=\rho_4 e^{\mathrm{i}\left(k_4 x+\omega_4 t\right)} \dfrac{\left[p_1 x+2 p_0 p_1(\mathrm{i} t)+\theta_{14}\right]\left[p_1^* x-2 p_0^* p_1^*(\mathrm{i} t)-\theta_{14}^*\right]+\left|h_0\right|^2}{\left|p_1 x+2 p_0 p_1(\mathrm{i} t)\right|^2+\left|h_0\right|^2},
\end{eqnarray}
where $|h_0|^2=\left|p_1\right|^2/\left(p_0+p_0^*\right)^2$, and their positions $(\bar{x}_0, \bar{t}_0)$ are given by
\begin{eqnarray} \label{prediction of locations of RW-4com}
\bar{x}_0 &=\dfrac{1}{\Re\left(p_0\right)} \Re\left[\dfrac{p_0^*}{p_1}\left(z_0 a_m^{1 / m}-\bar{\Delta}_i\right)\right],
\quad
\bar{t}_0 &=\dfrac{1}{2 \Re\left(p_0\right)} \Im\left[\dfrac{1}{p_1}\left(z_0 a_m^{1 / m}-\bar{\Delta}_i\right)\right],
\end{eqnarray}
where $z_0$ is any one of the non-zero simple roots of  $W_{N}^{[m,5,5-i]}(z)$, $\bar{\Delta}_i$ is a $z_0$-dependent $O(1)$ quantity. %, and %$\left(\Re,\Im\right)$ refer to the real and imaginary parts of the complex numbers, respectively.
The approximation error here is $ O\left(|a_{m}|^{-1 /m}\right)$.
In other words, when $|a_m|\gg1 $ and $\left(x-\bar{x}_0\right)^2+\left(t-\bar{t}_0\right)^2=O(1)$, we have the following asymptotics
\begin{eqnarray}
&u_{n, \mathcal{N}_i}(x, t)=\bar{u}_n\left(x-\bar{x}_0, t-\bar{t}_0\right) +O\left(\left|a_m\right|^{-1 / m}\right), \quad n=1,2,3,4. %\\
%&u_{1, \mathcal{N}_i}(x, t)=\bar{u}_1\left(x-\bar{x}_0, t-\bar{t}_0\right) +O\left(\left|a_m\right|^{-1 / m}\right), \\
%&u_{2, \mathcal{N}_i}(x, t)=\bar{u}_2\left(x-\bar{x}_0, t-\bar{t}_0\right) +O\left(\left|a_m\right|^{-1 / m}\right) ,\\
%&u_{3, \mathcal{N}_i}(x, t)=\bar{u}_3 \left(x-\bar{x}_0, t-\bar{t}_0\right) +O\left(\left|a_m\right|^{-1 / m}\right), \\
%&u_{4, \mathcal{N}_i}(x, t)=\bar{u}_4 \left(x-\bar{x}_0, t-\bar{t}_0\right) +O\left(\left|a_m\right|^{-1 / m}\right).
\end{eqnarray}

   \item [(2)]  In the inner
	region, where $x^2+t^2=O(1)$,	 if zero is a root of the generalized Wronskian-Hermite polynomials $W_N^{[m,5,5-i]}(z)$, then $\left[u_{1,\mathcal{N}_i}(x, t), u_{2,\mathcal{N}_i}(x,
	t),u_{3,\mathcal{N}_i}(x, t),u_{4,\mathcal{N}_i}(x, t)\right]$ is approximately a lower
	$\overline{\mathcal{N}}_i$-th order rogue wave $$u_{1, \overline{\mathcal{N}}_i}(x,
	t), u_{2, \overline{\mathcal{N}}_i}(x, t),
	u_{3, \overline{\mathcal{N}}_i}(x, t),
	u_{4, \overline{\mathcal{N}}_i}(x, t)$$
  where  $\overline{\mathcal{N}}_i=\displaystyle{\sum_{j=1}^{4}N_{j,5-i}\boldsymbol{e}_j}$ and $N_{j,l}$ refers to the value of $N_j$ against $l \in \{1,2,3,4\}$  are given in Theorem \ref{root sturcture of jump 5}.  Moreover, the internal parameters
  \begin{eqnarray*}
  \left(\bar{a}_{1,n}, \bar{a}_{2,n}, \bar{a}_{3,n}, \bar{a}_{4,n}, \bar{a}_{6,n}\ldots, \bar{a}_{5 N_{n,5-i}-n,n}\right), \quad n=1,2,3,4, %\\
  %  \left(\hat{a}_{1,1}, \hat{a}_{2,1}, \hat{a}_{3,1}, \hat{a}_{4,1}, \hat{a}_{6,1}\ldots, \hat{a}_{5 N_{1,5-i}-1,1}\right) \\
%    \left(\hat{a}_{1,2}, \hat{a}_{2,2}, \hat{a}_{3,2}, \hat{a}_{4,2}, \hat{a}_{6,2}\ldots, \hat{a}_{5 N_{2,5-i}-2,2}\right) \\
%    \left(\hat{a}_{1,3}, \hat{a}_{2,3}, \hat{a}_{3,3}, \hat{a}_{4,3}, \hat{a}_{6,3}\ldots, \hat{a}_{5 N_{3,5-i}-3,3}\right) \\
%    \left(\hat{a}_{1,4}, \hat{a}_{2,4}, \hat{a}_{3,4}, \hat{a}_{4,4}, \hat{a}_{6,4}\ldots, \hat{a}_{5 N_{4,5-i}-4,4}\right)
  \end{eqnarray*}
  in this lower-order rogue waves are related to those in the original rogue wave by
	$$
	\bar{a}_{j, 1}=\bar{a}_{j, 2}=\bar{a}_{j, 3}=\bar{a}_{j, 4}=a_j+\displaystyle{\left(N-\displaystyle{\sum_{n=1}^4N_{n,5-i}}\right)} s_j, \quad j=1,2,3,4,6,7 \cdots,
	$$
	where $s_j$ is defined in Theorem \ref{RW solutions of vector NLS}. % and $\overline{N}_i=\displaystyle{\sum_{n=1}^4N_{n,5-i}}$. %and numerically given in {\color{red}$\cdots$}, $\Sigma_i^{4}=N_{1,5-i}+N_{2,5-i}+N_{3,5-i}+N_{4,5-i}$.
The approximation error of this lower-order rogue wave  is $O\left(\left|a_m\right|^{-1}\right)$. In other words, when $\left|a_m\right| \gg 1$ and $x^2+t^2=$ $O(1)$, we have
\begin{eqnarray*}
    && u_{n,\mathcal{N}_i}\left(x, t ; a_2, a_3, a_4, a_6, \cdots\right) \\
   &=& u_{n,\overline{\mathcal{N}}_i}\left(x, t ; \bar{a}_{j, 1}, \bar{a}_{j, 2},\bar{a}_{j, 3}, \bar{a}_{j, 4}, j=1,2,3,4,6,7 \ldots\right)+O\left(\left|a_m\right|^{-1}\right), \quad n=1,2,3,4.
\end{eqnarray*}
	If zero is not a root of $W_N^{[m,5,5-i]}(z)$, the solution
$$
\left[u_{1,\mathcal{N}_i}(x, t), \quad u_{2,\mathcal{N}_i}(x,
	t), \quad u_{3,\mathcal{N}_i}(x, t), \quad u_{4,\mathcal{N}_i}(x, t)\right]
$$ is approximately the constant background
$$
\left[\rho_1 e^{\mathrm{i}\left(k_1 x+\omega_1 t\right)}, \quad \rho_2 e^{\mathrm{i}\left(k_3 x+\omega_2 t\right)}, \quad \rho_3 e^{\mathrm{i}\left(k_3 x+\omega_3 t\right)}, \quad \rho_4 e^{\mathrm{i}\left(k_4 x+\omega_4 t\right)} \right].
$$ %when $|a_m| \gg 1$.
 \end{itemize}

\end{theorem}

%\newpage

\section{Comparison between predicted and true rogue wave patterns} \label{section-comparison}

\subsection{Comparison in the three-component NLS equation}
In this subsection, we compare our predictions of rouge wave patterns in Theorem \ref{Rogue wave patterns-3com} with true rouge waves of the three-component NLS equation. It is noted that the predicted $i$-th type $|u_{n, \mathcal{N}_i}(x, t)|, n=1,2,3,$ from Theorem \ref{Rogue wave patterns-3com} can be divided into a simple form
\begin{equation}
|u_{n, \mathcal{N}_i}(x, t)|=|u_{n, \widehat{\mathcal{N}}_i}(x, t)|+\sum_{j=1}^{N_p}\left(|\hat{u}_n\left(x-\hat{x}_0^{(j)}, t-\hat{t}_0^{(j)}\right)|-\rho_n\right), \quad n =1,2,3,
\end{equation}
where $u_{n, \widehat{\mathcal{N}}_i}(x, t)$ is a lower-order rogue wave of the three-component NLS equation with all its internal parameters set to $0$, $\widehat{\mathcal{N}}_i=(N_1, N_2, N_3)$ is given by Theorems \ref{Rogue wave patterns-3com} and \ref{root sturcture of jump 4}, $\hat{u}_n\left(x, t\right), n=1,2,3,$ is the fundamental rogue wave of the three-component NLS equation, whose predicted location $\left(\hat{x}_0^{(j)}, \hat{t}_0^{(j)}\right)$ can be obtained from \eqref{prediction of locations of RW-3com}, and $N_p$ is the number of fundamental rogue waves given in Theorem \ref{Rogue wave patterns-3com}. To analyze the triple root case, we choose the background wavenumbers $k_2 =-k_1 = 1$ and $ k_3 = 0$, which gives $\rho_1= \rho_2=\rho_3^2=2$ by \eqref{triple root condition}. Further, we select $p_0=1$, $p_1=-1/\sqrt[4]{3}$ and $p_2=1/\sqrt{3}$ for the subsequent analysis.
\subsubsection{First-type rogue waves of the three-component NLS equation }
We start with $(2,0,0)$-th order rogue wave solutions. Moreover, we let one of the internal parameters $(a_2, a_3, a_5, a_6, a_7)$ be large and set others to $0$. We note that $a_1$ can be set to 0 by normalization, and $a_4$ is a parameter that can be removed. Then, the very large parameter is one of
\begin{equation}
    a_2=30,\quad a_3=100,\quad a_5=1200,\quad a_6=3000,\quad a_7=7000.
\end{equation}
According to Theorem \ref{Rogue wave patterns-3com}, the position $(\hat{x}_0, \hat{t}_0)$ of each fundamental rogue wave corresponding to
$$
u_{1, \mathcal{N}_1}(x, t), \quad u_{2, \mathcal{N}_1}(x, t), \quad u_{3, \mathcal{N}_1 }(x, t)
$$
can be predicted by equation \eqref{prediction of locations of RW-3com}. The lower $(N_1, N_2, N_3)$-th order rogue wave would appear in the inner region, and the value of $(N_1, N_2, N_3)$ can be obtained from Theorems \ref{Rogue wave patterns-3com} and \ref{root sturcture of jump 4}. In our prediction, the $(N_1, N_2, N_3)$ values for these five rogue solutions are
$$
(N_1,N_2,N_3) = (1,0,1), \quad  (0,0,0), \quad (1,1,0), \quad (1,0,1), \quad (0,1,0),
$$
respectively. Note that $(0,0,0)$ means no lower-order rogue wave exists in the inner region. Because of our choice of parameters $a_m$ and the value of $s_j$ shown in Remark \ref{Values of s_r}, the internal parameters in these predicted lower $(N_1, N_2, N_3)$-th order rogue waves of the inner region are all zero.
%all selected to be zero.

\begin{figure}[h]
\centering
\renewcommand\arraystretch{0.5}
\setlength\tabcolsep{0pt}
\resizebox{\linewidth}{!}{
\begin{tabular}{m{0.6cm}<{\centering}m{4.1cm}<{\centering}m{4.1cm}<{\centering}m{4.05cm}<{\centering}m{4.1cm}<{\centering}m{4.05cm}<{\centering}m{0.6cm}<{\centering}r}
&\textbf{$a_2=30$}  & \textbf{$a_3= 100$} & \textbf{$a_5=1200 $}& \textbf{$a_6=3000 $}& \textbf{$a_7=7000 $}&\\
\rotatebox{90}{\text{ predicted locations }} & \includegraphics[height=40mm,width=40mm]{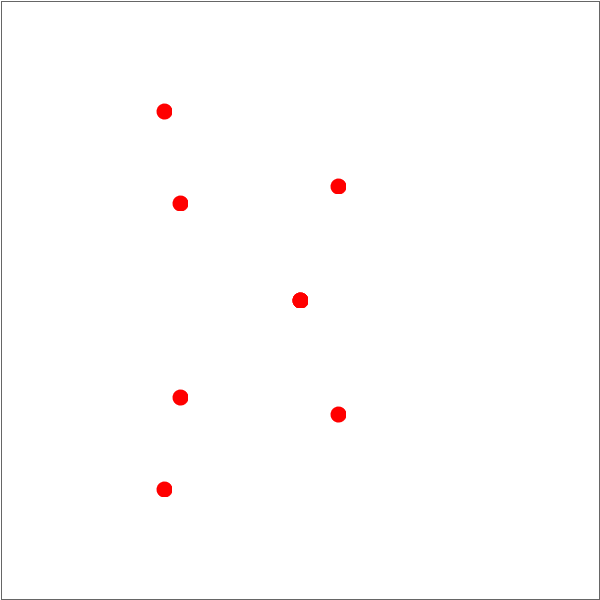}&\includegraphics[height=40mm,width=40mm]{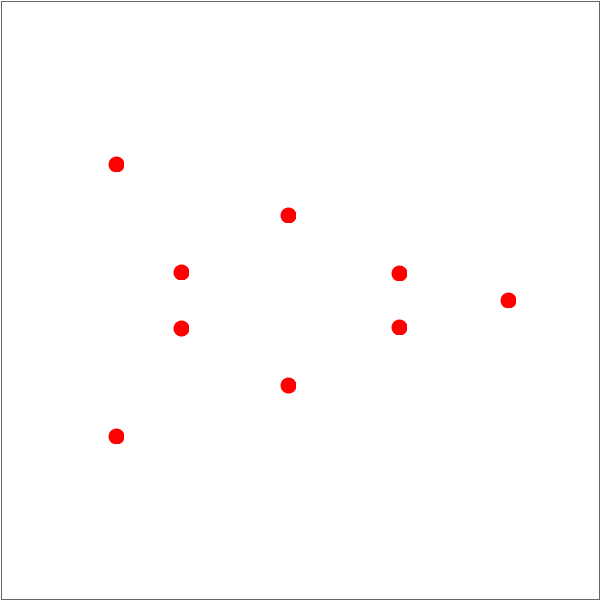}&\includegraphics[height=40mm,width=40mm]{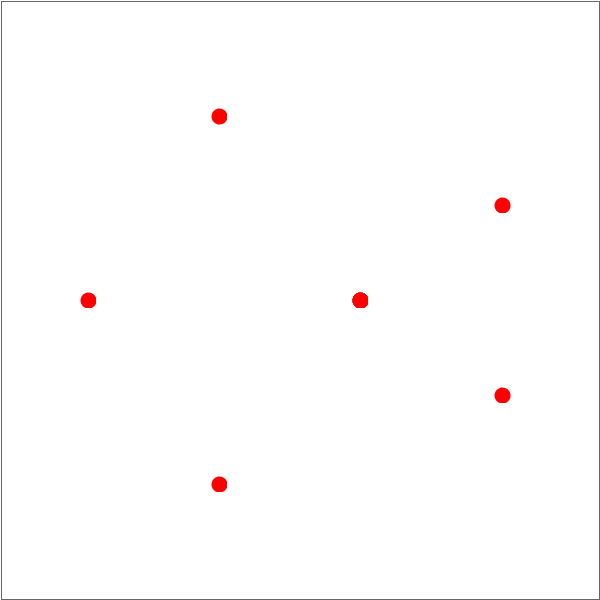}&\includegraphics[height=40mm,width=40mm]{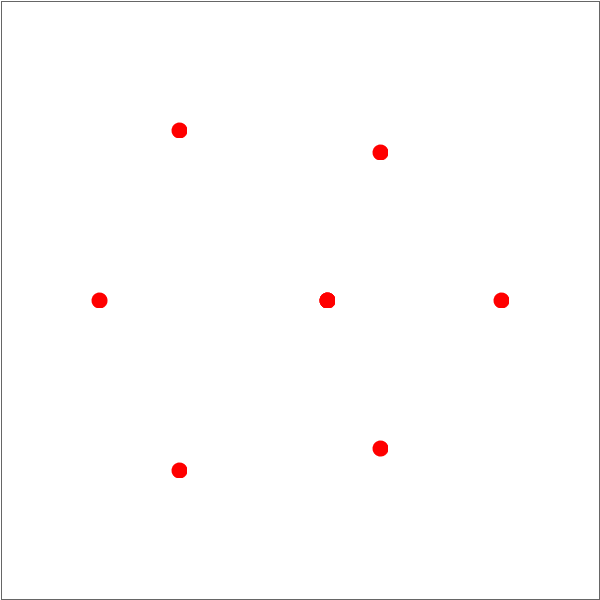}&\includegraphics[height=40mm,width=40mm]{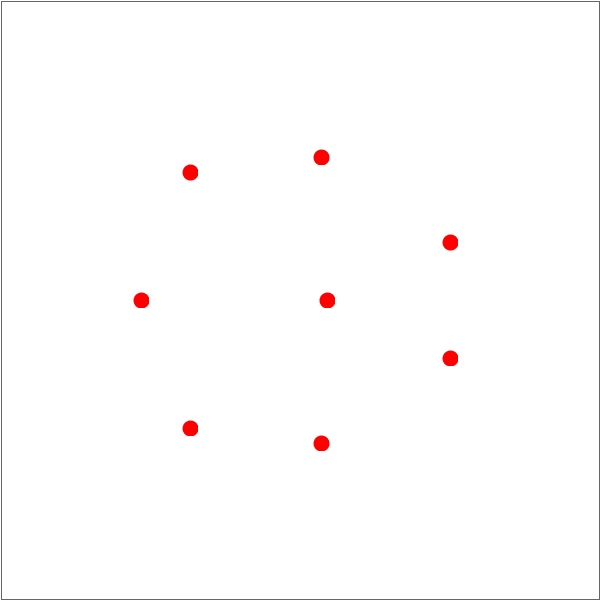}&\rotatebox{0}{\textbf{$t$}}\\
\rotatebox{0}{\textbf{$\left|u_1\right|$}} & \includegraphics[height=40mm,width=40mm]{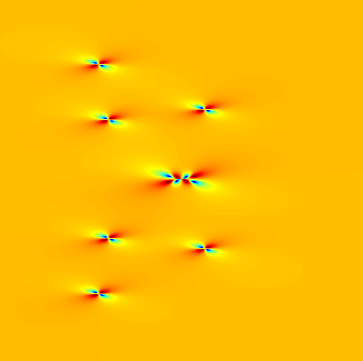}&\includegraphics[height=40mm,width=40mm]{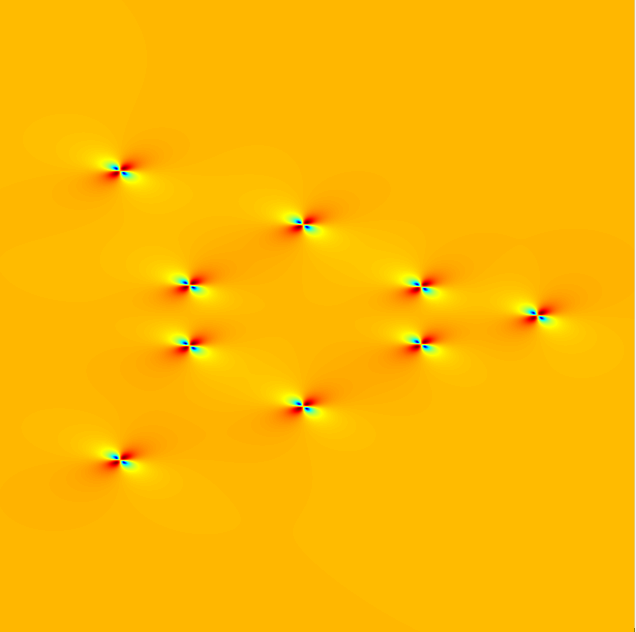}&\includegraphics[height=40mm,width=40mm]{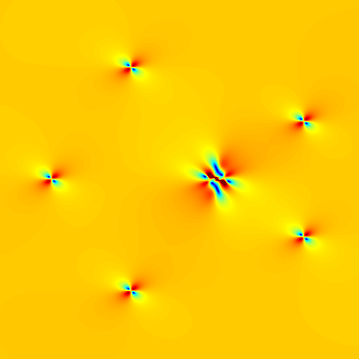}&\includegraphics[height=40mm,width=40mm]{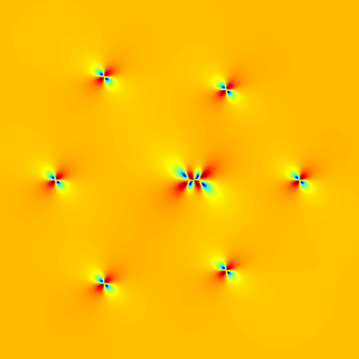}&\includegraphics[height=40mm,width=40mm]{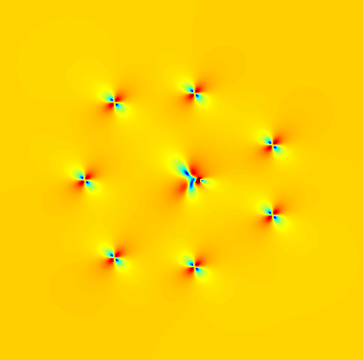}&\rotatebox{0}{\textbf{$t$}}\\
\rotatebox{0}{\textbf{$\left|u_2\right|$}} & \includegraphics[height=40mm,width=40mm]{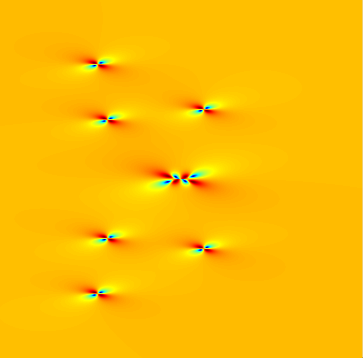}&\includegraphics[height=40mm,width=40mm]{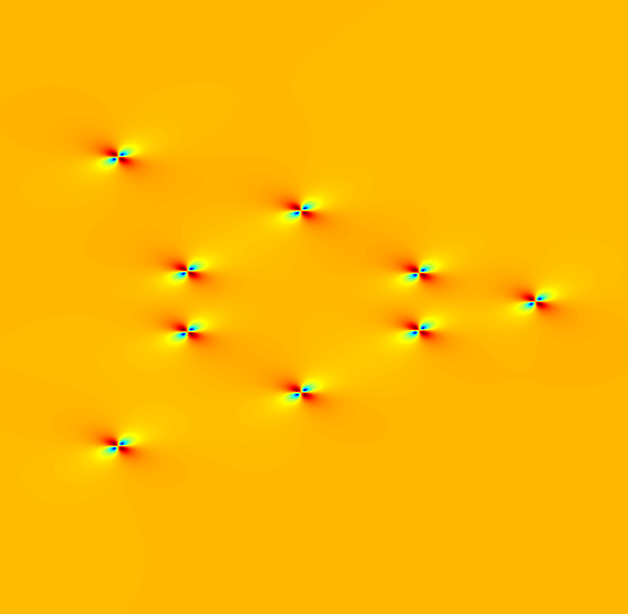}&\includegraphics[height=40mm,width=40mm]{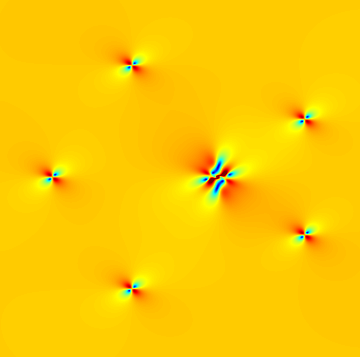}&\includegraphics[height=40mm,width=40mm]{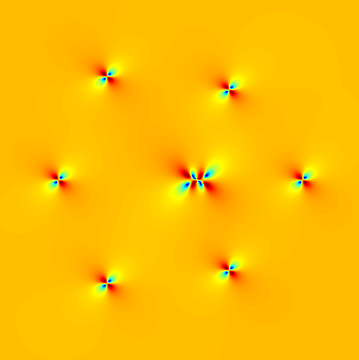}&\includegraphics[height=40mm,width=40mm]{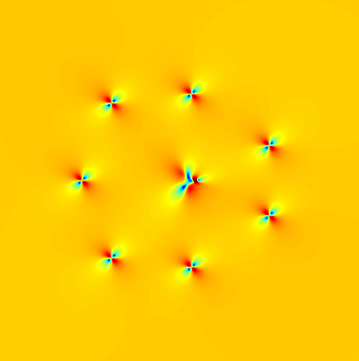}&\rotatebox{0}{\textbf{$t$}}\\
\rotatebox{0}{\textbf{$\left|u_3\right|$}} & \includegraphics[height=40mm,width=40mm]{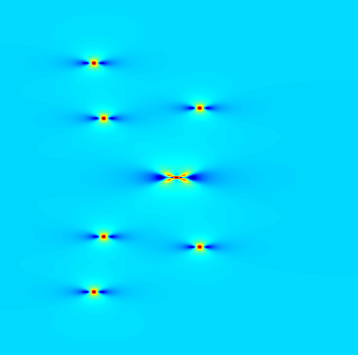}&\includegraphics[height=40mm,width=40mm]{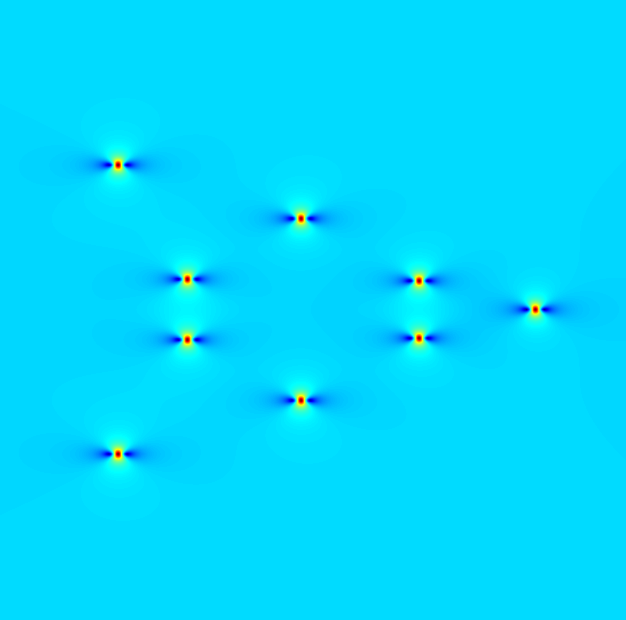}&\includegraphics[height=40mm,width=40mm]{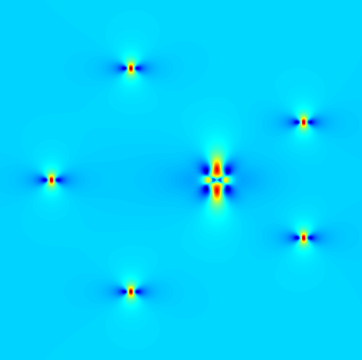}&\includegraphics[height=40mm,width=40mm]{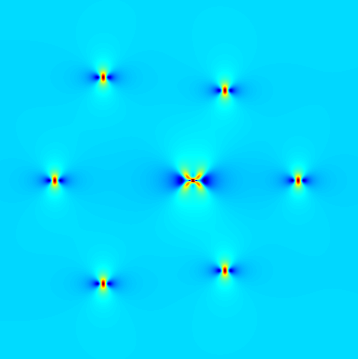}&\includegraphics[height=40mm,width=40mm]{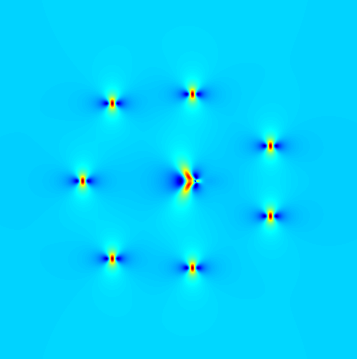}&\rotatebox{0}{\textbf{$t$}}\\
& $x$ & $x$ & $x$ & $x$ & $x$
\end{tabular}
}
\caption{Predicted $1$st type $(2,0,0)$-th order rogue waves of the three-component NLS from Theorem \ref{Rogue wave patterns-3com}. Each column depicts rogue waves with a single large parameter $a_m$, whose value is indicated on top, and all other internal parameters are set to zero. Top row: predicted $\left(\hat{x}_0, \hat{t}_0\right)$ locations by formulae \eqref{prediction of locations of RW-3com}. Second row: predicted $\left|u_1(x, t)\right|$.  Third row: predicted $\left|u_2(x, t)\right|$.  Bottom row: predicted $\left|u_3(x, t)\right|$. First column: the $(x, t)$ intervals are $-21 \leq x \leq 21$, $-25 \leq t \leq 25$. Second column: the $(x, t)$ intervals are $-30 \leq x \leq 30$, $-25 \leq t \leq 25$. Third column: the $(x, t)$ intervals are $-30 \leq x \leq 20$, $-16 \leq t \leq 16$. Fourth column: the $(x, t)$ intervals are $-30 \leq x \leq 25$, $-15 \leq t \leq 15$. Fifth column: the $(x, t)$ intervals are $-30 \leq x \leq 25$, $-15 \leq t \leq 15$.}
\label{1st-type_RW_N2_P}
\end{figure}

For $[u_{1, \mathcal{N}_1}(x, t), u_{2, \mathcal{N}_1}(x, t), u_{3, \mathcal{N}_1 }(x, t)]$, their corresponding predicted rogue wave patterns are illustrated in the last three rows of Fig. \ref{1st-type_RW_N2_P}, with the first row being the locations of predicted rogue waves.
%Each column is separately when the parameter $(a_2, a_3,a_5, a_6, a_7)$ is large.
These predicted rogue waves are generated in the following way. We first replace each non-center dot, which is the non-zero root, in the first row of Fig. \ref{1st-type_RW_N2_P} by a fundamental rogue wave according to \eqref{fundmental_u1_3com}-\eqref{fundmental_u3_3com}. Then the center dot is replaced by a lower $(N_1, N_2, N_3)$-th order rogue wave with all internal parameters set to zero.
%Then, the predicted rogue patterns are shown in the last three rows of Fig.\ref{1st-type_RW_N2_P}.

It can be seen from Fig. \ref{1st-type_RW_N2_P} that the large-$a_2$ solution displays a skewed double-triangle, corresponding to the double-triangle root structure of $W_2^{[2,4,3]}(z)$. The large-$a_3$ solution exhibits a skewed triple-triangle, corresponding to the triple-triangle root structure of $W_2^{[3,4,3]}(z)$. The large-$a_5$ solution displays a deformed pentagon, corresponding to the pentagon-shaped root structure of $W_2^{[5,4,3]}(z)$. The large-$a_6$ solution exhibits a deformed hexagon, corresponding to the hexagon-shaped root structure of $W_2^{[6,4,3]}(z)$. The large-$a_7$ solution displays a deformed heptagon, corresponding to the heptagon-shaped root structure of $W_2^{[7,4,3]}(z)$. It seems that triple-triangle is a new type of pattern compared with those of the NLS equation \cite{yang2021general} and Manakov system \cite{yang2022pattern}.

\begin{figure}[h]
\centering
\renewcommand\arraystretch{0.5}
\setlength\tabcolsep{0pt}
\resizebox{\linewidth}{!}{
\begin{tabular}{m{0.6cm}<{\centering}m{4.1cm}<{\centering}m{4.1cm}<{\centering}m{4.05cm}<{\centering}m{4.1cm}<{\centering}m{4.05cm}<{\centering}m{0.6cm}<{\centering}r}
&\textbf{$a_2=30$}  & \textbf{$a_3= 100$} & \textbf{$a_5=1200 $}& \textbf{$a_6=3000 $}& \textbf{$a_7=7000 $}&\\
%\rotatebox{90}{\text{ predicted locations }} & \includegraphics[height=40mm,width=40mm]{RW32211.png}&\includegraphics[height=40mm,width=40mm]{RW32311.png}&\includegraphics[height=40mm,width=40mm]{RW32511.png}&\includegraphics[height=40mm,width=40mm]{RW32611.png}&\includegraphics[height=40mm,width=40mm]{RW32711.png}&\rotatebox{0}{\textbf{$t$}}\\
\rotatebox{0}{\textbf{$\left|u_1\right|$}} & \includegraphics[height=40mm,width=40mm]{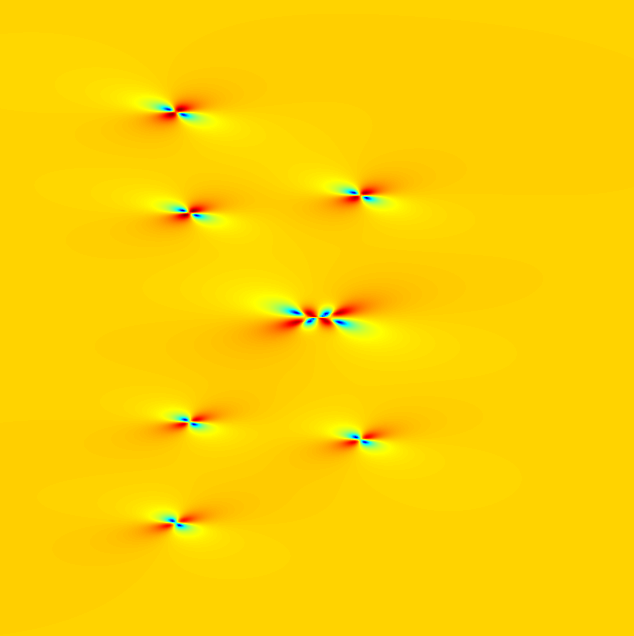}&\includegraphics[height=40mm,width=40mm]{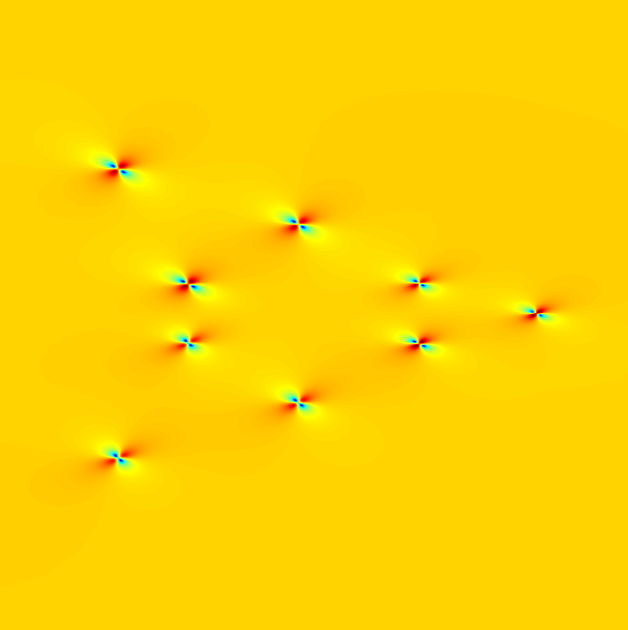}&\includegraphics[height=40mm,width=40mm]{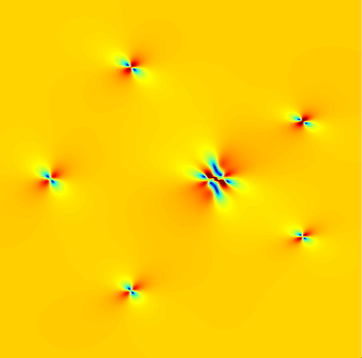}&\includegraphics[height=40mm,width=40mm]{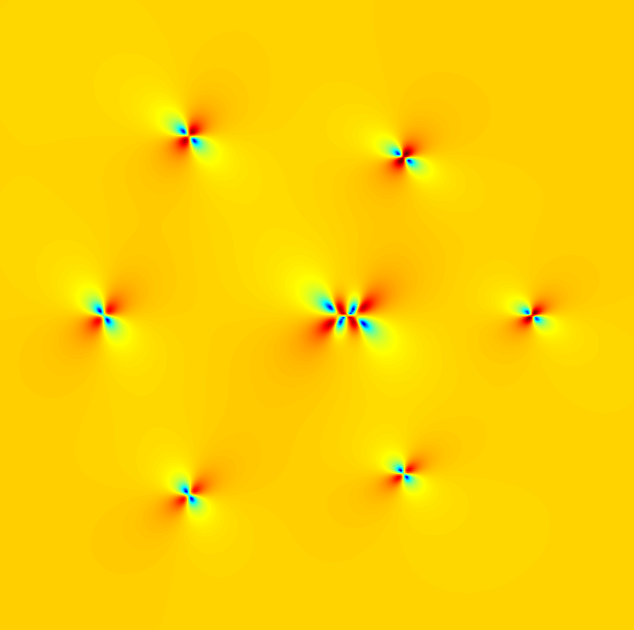}&\includegraphics[height=40mm,width=40mm]{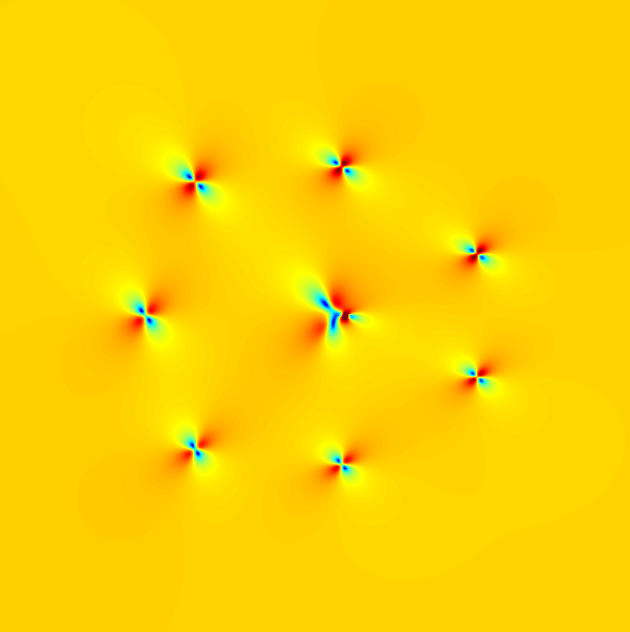}&\rotatebox{0}{\textbf{$t$}}\\
\rotatebox{0}{\textbf{$\left|u_2\right|$}} & \includegraphics[height=40mm,width=40mm]{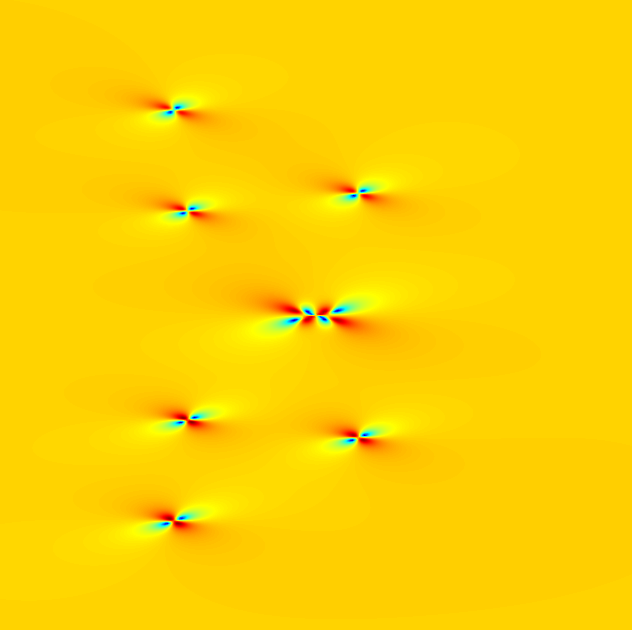}&\includegraphics[height=40mm,width=40mm]{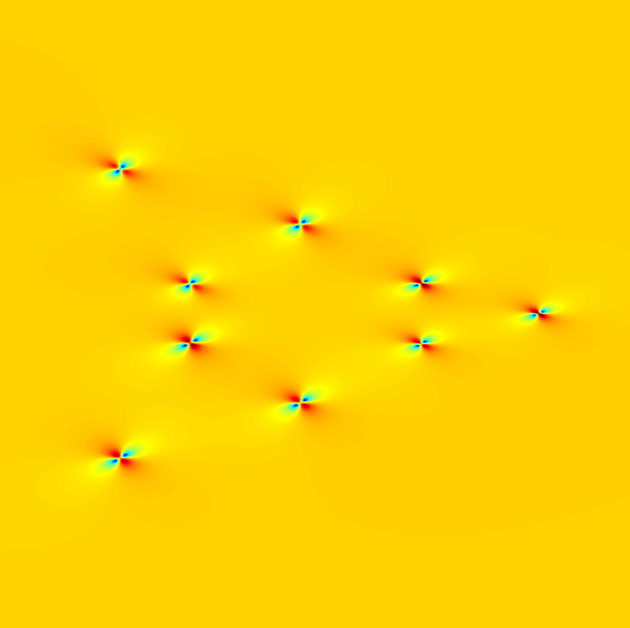}&\includegraphics[height=40mm,width=40mm]{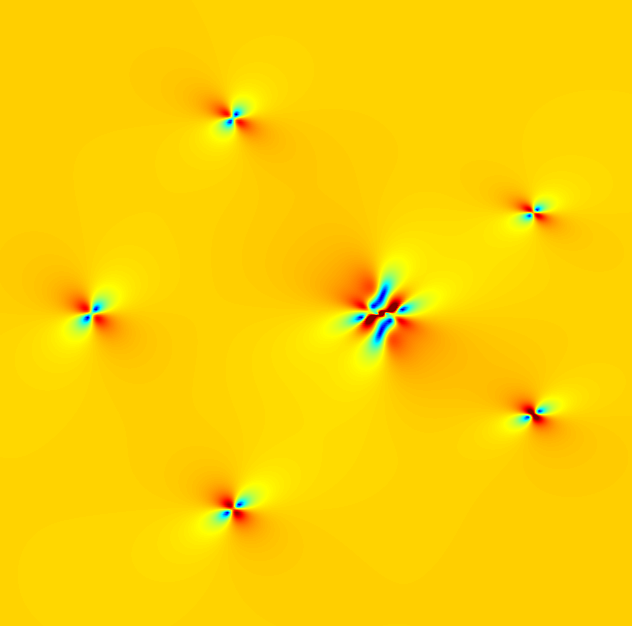}&\includegraphics[height=40mm,width=40mm]{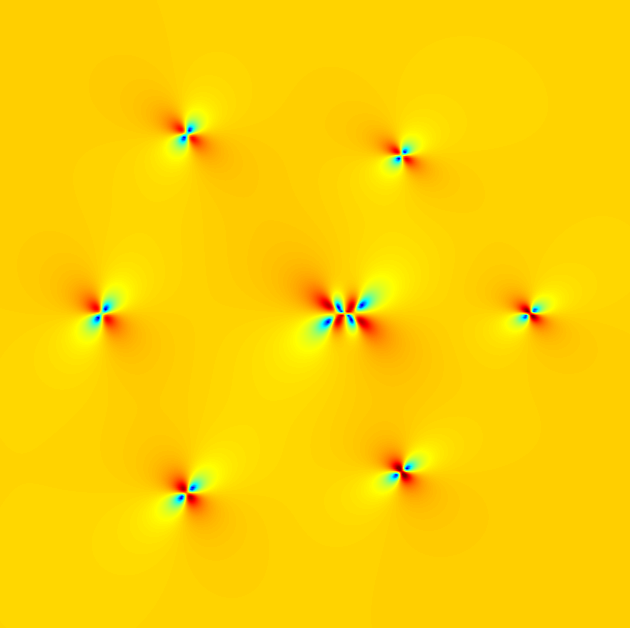}&\includegraphics[height=40mm,width=40mm]{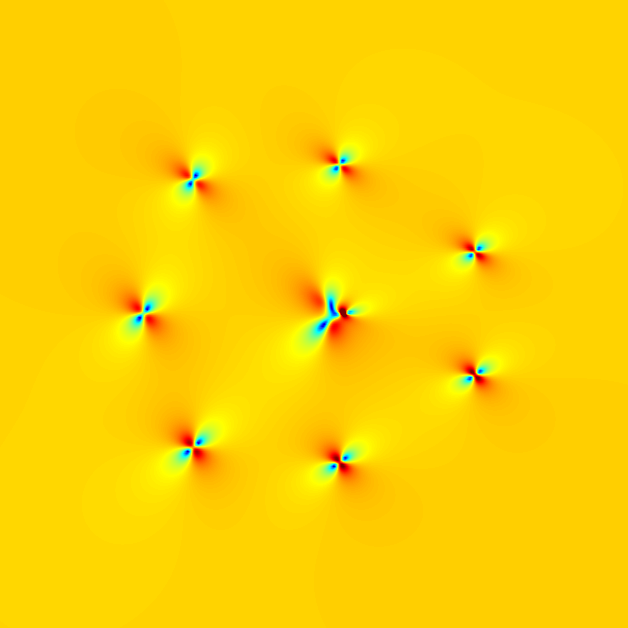}&\rotatebox{0}{\textbf{$t$}}\\
\rotatebox{0}{\textbf{$\left|u_3\right|$}} & \includegraphics[height=40mm,width=40mm]{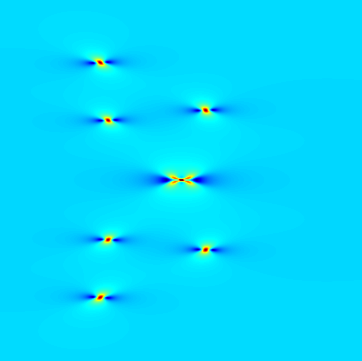}&\includegraphics[height=40mm,width=40mm]{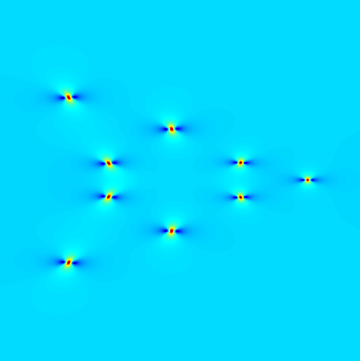}&\includegraphics[height=40mm,width=40mm]{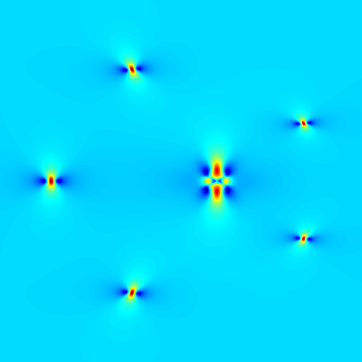}&\includegraphics[height=40mm,width=40mm]{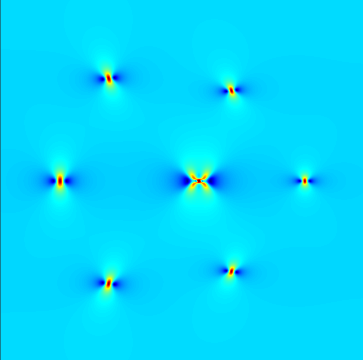}&\includegraphics[height=40mm,width=40mm]{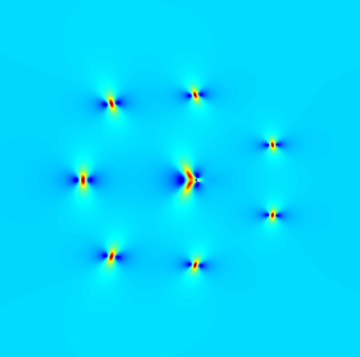}&\rotatebox{0}{\textbf{$t$}}\\
& $x$ & $x$ & $x$ & $x$ & $x$
\end{tabular}
}
\caption{True $1$st type $(2,0,0)$-th order rogue waves of the three-component NLS with the same parameters as Fig. \ref{1st-type_RW_N2_P}.
% Each column corresponds to rogue waves with a single large parameter $a_m$, whose value is indicated on top, and all other internal parameters are set to be zero. Top row: true $\left|u_1(x, t)\right|$. Middle row:true $\left|u_2(x, t)\right|$. Bottom row: true $\left|u_3(x, t)\right|$.
The $(x, t)$ interval for each column is the same as the corresponding column in Fig. \ref{1st-type_RW_N2_P}.
% First column: the $(x, t)$ intervals are $-21 \leq x \leq 21$, $-25 \leq t \leq 25$. Second column: the $(x, t)$ intervals are $-30 \leq x \leq 30$, $-25 \leq t \leq 25$. Third column: the $(x, t)$ intervals are $-30 \leq x \leq 20$, $-16 \leq t \leq 16$. Fourth column: the $(x, t)$ intervals are $-30 \leq x \leq 25$, $-15 \leq t \leq 15$. Fifth column: the $(x, t)$ intervals are $-30 \leq x \leq 25$, $-15 \leq t \leq 15$.
}
\label{1st-type_RW_N2_T}
\end{figure}

By comparison of the true rogue waves to the predicted ones (see Figs. \ref{1st-type_RW_N2_P} and \ref{1st-type_RW_N2_T}), we can observe that each of the rogue waves matches perfectly in terms of position and rogue wave shape. %Although there are a few rogue waves that are slightly misplaced, this is due to our approximation error.
Notice that the predicted pattern looks very different from the root structure of $W_2^{[m,4,3]}(z)$. This is due to the term $\Delta_1$ leading to a nonlinear transformation from the root structure. When $|a_m|$ is set to be very large, the term $\Delta_1$ can be neglected, and the patterns become much closer to certain linear transformations of the root structure of $W_2^{[m,4,3]}(z)$.

\begin{figure}[h]
\centering
\renewcommand\arraystretch{0.5}
\setlength\tabcolsep{0pt}
\resizebox{\linewidth}{!}{
\begin{tabular}{m{0.6cm}<{\centering}m{4.1cm}<{\centering}m{0.6cm}<{\centering}m{4.1cm}<{\centering}m{0.6cm}<{\centering}m{4.1cm}<{\centering}m{0.6cm}<{\centering}m{4.1cm}<{\centering}m{0.6cm}<{\centering}r}
\rotatebox{0}{\textbf{$\left|u_3\right|$}}&\includegraphics[height=40mm,width=40mm]{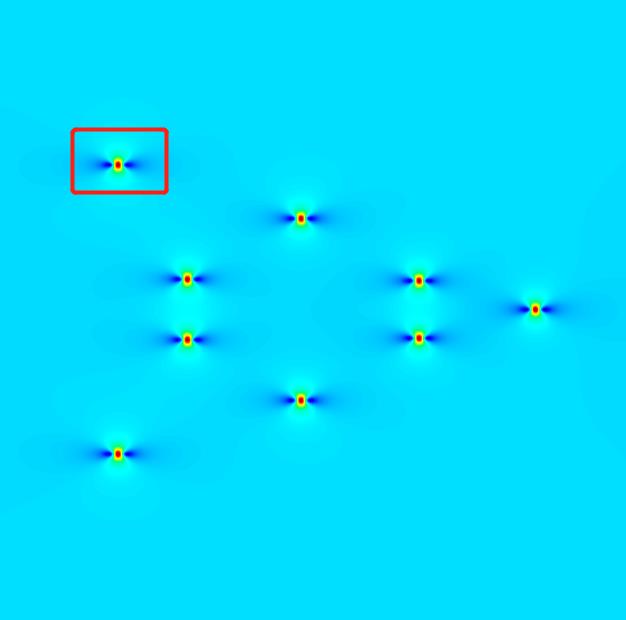}&\rotatebox{0}{\textbf{$t$}} &\includegraphics[height=40mm,width=40mm]{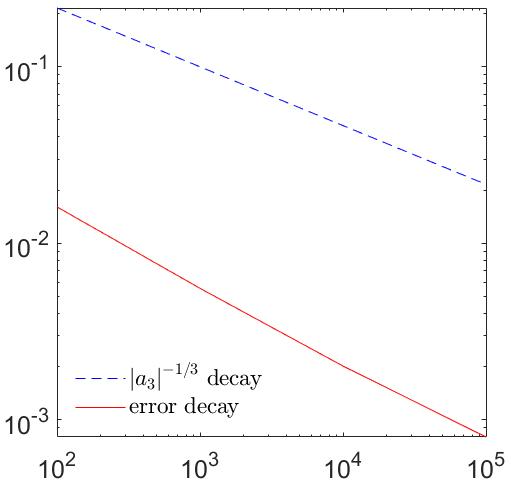}&\rotatebox{0}{\textbf{$\left|u_3\right|$}}&\includegraphics[height=40mm,width=40mm]{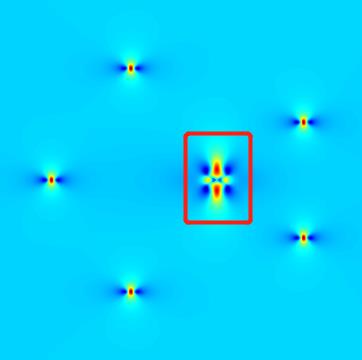}&\rotatebox{0}{\textbf{$t$}}&\includegraphics[height=40mm,width=40mm]{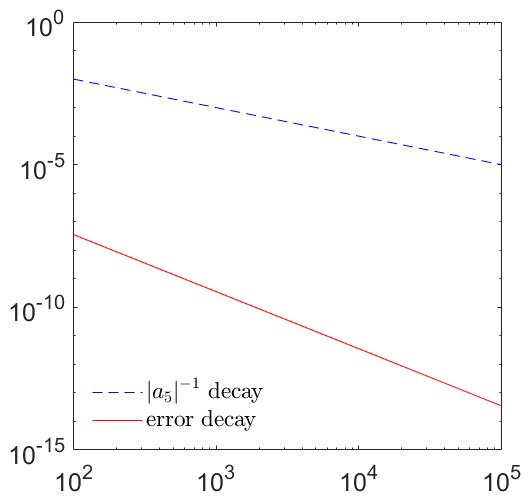}&&\\
& $x$ & & & & $x$ & &\\
& $(a)$ & & $(b)$ & & $(c)$ & & $(d)$
\end{tabular}
}
\caption{Decay of errors in our predictions of Theorem \ref{Rogue wave patterns-3com} for the outer and inner regions of the $1$st type $(2, 0, 0)$-th order rogue waves in the three-component NLS with various large real values of $a_3$ or $a_5$, while other internal parameters are set to zero.  (a) $|u_3(x, t)|$ of the true
rogue wave with $a_3 = 100$. (b) Decay of error versus $a_3$ for the outer fundamental rogue wave marked by the red box, together with the $|a_3|^{-1/3}$ decay for comparison.  (c) $|u_3(x, t)|$ of the true
rogue wave with $a_5 = 1000$. (d) Decay of error versus $a_5$ for the lower order rogue wave marked by the red box, together with the $|a_5|^{-1}$ decay for comparison. }
\label{decay of error 3com_1st}
\end{figure}

Apart from the above observations, we can also qualitatively compare the differences between predicted and true rogue waves. To illustrate this, we choose the $1$st type $(2, 0, 0)$-th order rogue waves, then select various large real values of $a_3$ to analyze errors in the outer region and various large real values of $a_5$ to analyze errors in the inner region.
Referring to the work of Yang and Yang \cite{yang2021rogue}, we define
\begin{equation}
    \text{error of Peregrine location}=\sqrt{\left(\hat{x}_0-x_0\right)^2+\left(\hat{t}_0-t_0\right)^2},
\end{equation}
and
\begin{equation}
    \text{error of inner region}=\left|u_{1, \mathcal{N}_1 }(x, t)-\hat{u}_{1,\widehat{\mathcal{N}}_1}(x, t)\right|_{x=t=0},
\end{equation}
where $(x_0, t_0)$ is the location where each rogue wave reaches maximum modulus value and $(\hat{x}_0, \hat{t}_0)$ is the predicted location of each fundamental rogue wave. These errors and decay rate of $|a_3|^{-1/3}$ and $|a_5|^{-1}$ for large-$a_3$ and large-$a_5$ solutions are plotted in Fig. \ref{decay of error 3com_1st}. As can be seen, the results of the numerical analysis also match very well.

\subsubsection{Second-type rogue waves of the three-component NLS equation}
In this case, we mainly carry out the second-type rogue waves in detail by taking $N_2 = 3$, i.e., $(N_1, N_2, N_3)=(0,3,0)$. %In this case, we choose third-order rogue waves, i.e., $N = 3$.
For brevity, we only let one of the internal parameters $(a_2, a_3, a_5, a_6, a_7)$ be large, and the others are set to $0$. The very large parameter is one of
\begin{equation}
    a_2=30,\quad a_3=400,\quad a_5=4800,\quad a_6=3000,\quad a_7=7000.
\end{equation}
According to Theorem \ref{Rogue wave patterns-3com}, the position $(\hat{x}_0, \hat{t}_0)$ of each fundamental rogue wave
$$
u_{1, \mathcal{N}_2}(x, t), \quad u_{2, \mathcal{N}_2}(x, t), \quad u_{3, \mathcal{N}_2 }(x, t)
$$
can be predicted by equation \eqref{prediction of locations of RW-3com}. The lower $(N_1, N_2, N_3)$-th order rogue wave would appear in the inner region. The values of $(N_1,N_2,N_3)$ can be deduced by Theorems \ref{Rogue wave patterns-3com} and \ref{root sturcture of jump 4}. In our prediction, the $(N_1,N_2,N_3)$ values for these five rogue solutions are
$$
(N_1,N_2,N_3) = (1,0,1), \quad  (0,0,0),  \quad  (0,0,0), \quad  (1,0,1),  \quad  (1,2,0),
$$
respectively.  Note that $(0,0,0)$ means no lower-order rogue wave exists in the inner region. On account of our choice of parameters $a_m$ and the value of $s_j$ shown in Remark \ref{Values of s_r}, the internal parameters in these predicted lower $(N_1, N_2, N_3)$-th order rogue waves of the inner region are all chosen to be zero.

\begin{figure}[h]
\centering
\renewcommand\arraystretch{0.5}
\setlength\tabcolsep{0pt}
\resizebox{\linewidth}{!}{
\begin{tabular}{m{0.6cm}<{\centering}m{4.1cm}<{\centering}m{4.1cm}<{\centering}m{4.05cm}<{\centering}m{4.1cm}<{\centering}m{4.05cm}<{\centering}m{0.6cm}<{\centering}r}
&\textbf{$a_2=30$}  & \textbf{$a_3= 400$} & \textbf{$a_5=4800 $}& \textbf{$a_6=3000 $}& \textbf{$a_7=7000 $}&\\
\rotatebox{90}{\text{ predicted locations }} & \includegraphics[height=40mm,width=40mm]{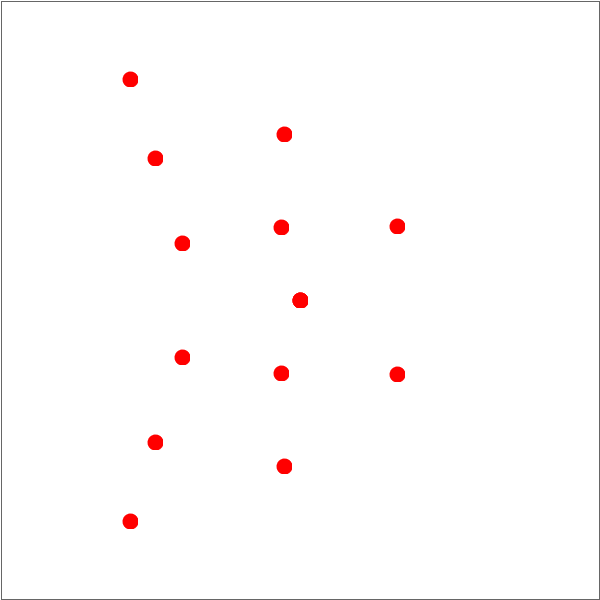}&\includegraphics[height=40mm,width=40mm]{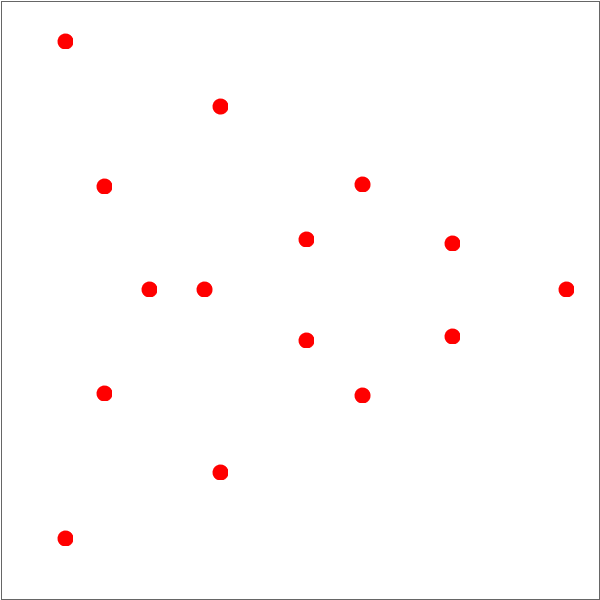}&\includegraphics[height=40mm,width=40mm]{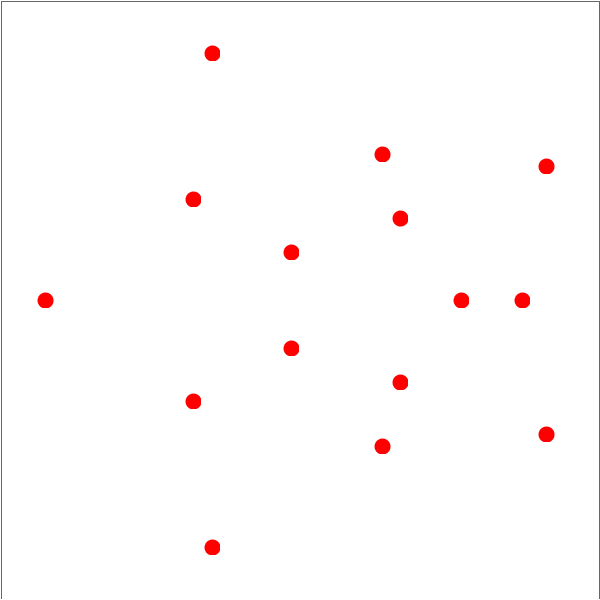}&\includegraphics[height=40mm,width=40mm]{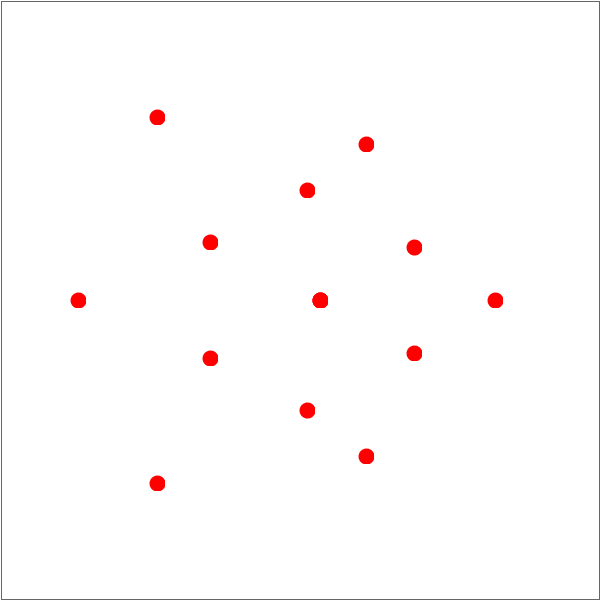}&\includegraphics[height=40mm,width=40mm]{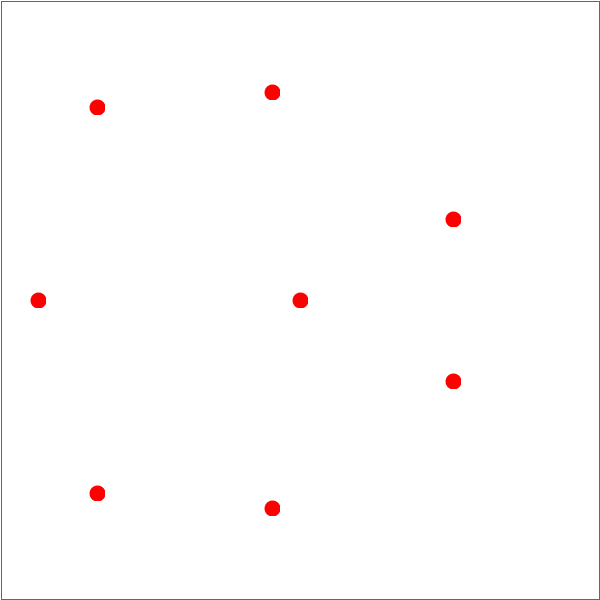}&\rotatebox{0}{\textbf{$t$}}\\
\rotatebox{0}{\textbf{$\left|u_1\right|$}} & \includegraphics[height=40mm,width=40mm]{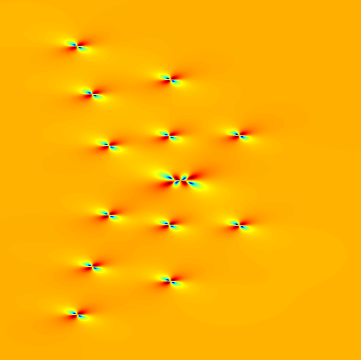}&\includegraphics[height=40mm,width=40mm]{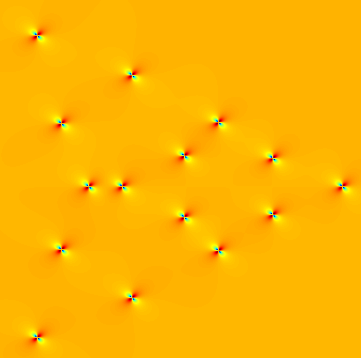}&\includegraphics[height=40mm,width=40mm]{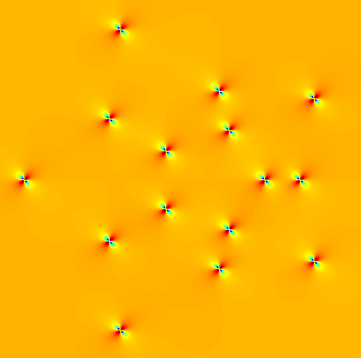}&\includegraphics[height=40mm,width=40mm]{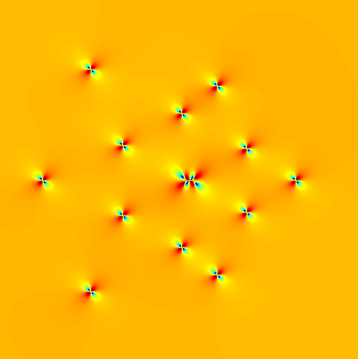}&\includegraphics[height=40mm,width=40mm]{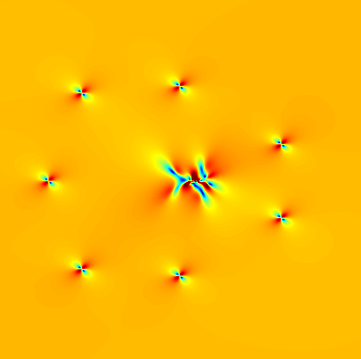}&\rotatebox{0}{\textbf{$t$}}\\
\rotatebox{0}{\textbf{$\left|u_2\right|$}} & \includegraphics[height=40mm,width=40mm]{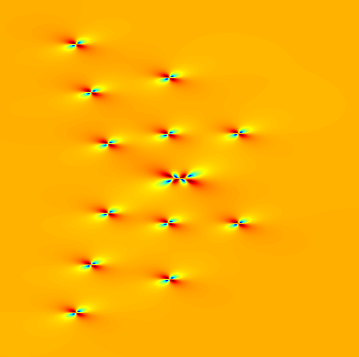}&\includegraphics[height=40mm,width=40mm]{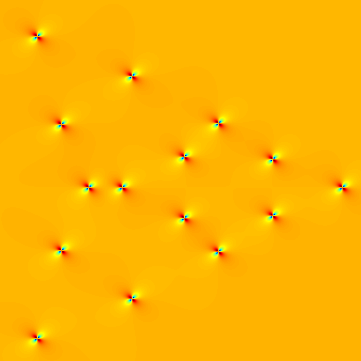}&\includegraphics[height=40mm,width=40mm]{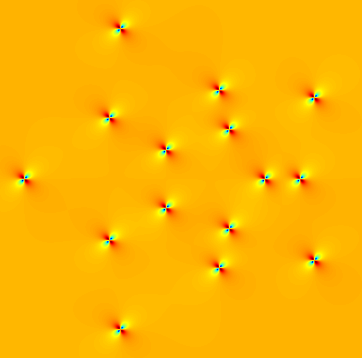}&\includegraphics[height=40mm,width=40mm]{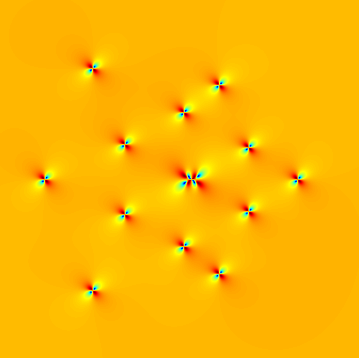}&\includegraphics[height=40mm,width=40mm]{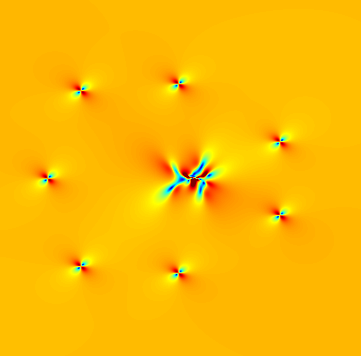}&\rotatebox{0}{\textbf{$t$}}\\
\rotatebox{0}{\textbf{$\left|u_3\right|$}} & \includegraphics[height=40mm,width=40mm]{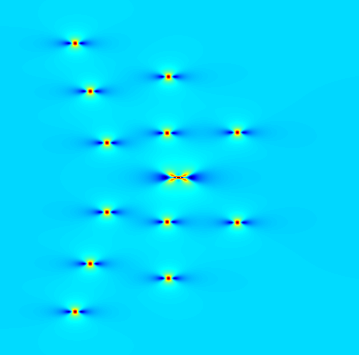}&\includegraphics[height=40mm,width=40mm]{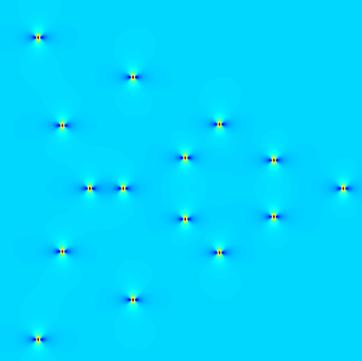}&\includegraphics[height=40mm,width=40mm]{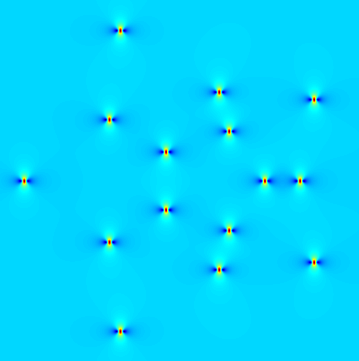}&\includegraphics[height=40mm,width=40mm]{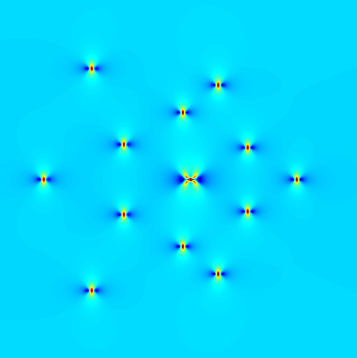}&\includegraphics[height=40mm,width=40mm]{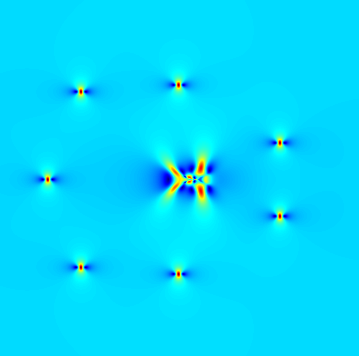}&\rotatebox{0}{\textbf{$t$}}\\
& $x$ & $x$ & $x$ & $x$ & $x$
\end{tabular}
}
\caption{Predicted $2$nd type $(0,3,0)$-th order rogue waves of the three-component NLS from Theorem \ref{Rogue wave patterns-3com}. Each column depicts rogue waves with a single large parameter $a_m$, whose value is indicated on top, and all other internal parameters are set to zero. Top row: predicted $\left(\hat{x}_0, \hat{t}_0\right)$ locations by formulae \eqref{prediction of locations of RW-3com}. Second row: predicted $\left|u_1(x, t)\right|$.  Third row: predicted $\left|u_2(x, t)\right|$.  Bottom row: predicted $\left|u_3(x, t)\right|$. First column: the $(x, t)$ intervals are $-25 \leq x \leq 25$, $-26 \leq t \leq 26$. Second column: the $(x, t)$ intervals are $-45 \leq x \leq 51$, $-29 \leq t \leq 27$. Third column: the $(x, t)$ intervals are $-47 \leq x \leq 31$, $-22 \leq t \leq 22$. Fourth column: the $(x, t)$ intervals are $-40 \leq x \leq 35$, $-20 \leq t \leq 20$. Fifth column: the $(x, t)$ intervals are $-35 \leq x \leq 30$, $-20 \leq t \leq 20$.}
\label{2st-type_RW_N3_P}
\end{figure}

For $[u_{1, \mathcal{N}_2}(x, t), u_{2, \mathcal{N}_2}(x, t), u_{3, \mathcal{N}_2 }(x, t)]$, their corresponding lower-order rogue wave patterns are shown in the last three rows of Fig. \ref{2st-type_RW_N3_P}, with the first row being the predicted locations of the rogue waves. %Each column is when one of $(a_2, a_3,a_5, a_6, a_7)$ is large, respectively.
As seen in Fig. \ref{2st-type_RW_N3_P}, solutions in the first column are skewed double-triangles, while solutions from the second to the fifth columns are skewed triple-triangles, pentagons,  hexagons and heptagons respectively.

\begin{figure}[h]
\centering
\renewcommand\arraystretch{0.5}
\setlength\tabcolsep{0pt}
\resizebox{\linewidth}{!}{
\begin{tabular}{m{0.6cm}<{\centering}m{4.1cm}<{\centering}m{4.1cm}<{\centering}m{4.05cm}<{\centering}m{4.1cm}<{\centering}m{4.05cm}<{\centering}m{0.6cm}<{\centering}r}
&\textbf{$a_2=30$}  & \textbf{$a_3= 400$} & \textbf{$a_5=4800 $}& \textbf{$a_6=3000 $}& \textbf{$a_7=7000 $}&\\
%\rotatebox{90}{\text{ predicted locations }} & \includegraphics[height=40mm,width=40mm]{RW33221.png}&\includegraphics[height=40mm,width=40mm]{RW33321.png}&\includegraphics[height=40mm,width=40mm]{RW33521.png}&\includegraphics[height=40mm,width=40mm]{RW33621.png}&\includegraphics[height=40mm,width=40mm]{RW33721.png}&\rotatebox{0}{\textbf{$t$}}\\
\rotatebox{0}{\textbf{$\left|u_1\right|$}} & \includegraphics[height=40mm,width=40mm]{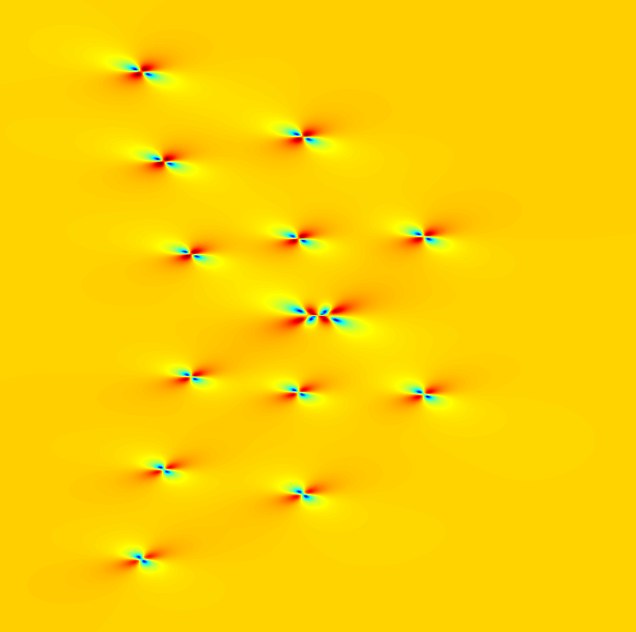}&\includegraphics[height=40mm,width=40mm]{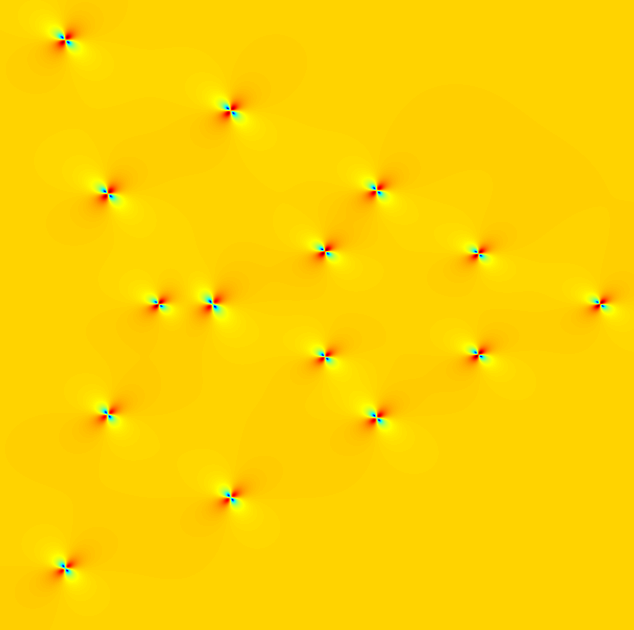}&\includegraphics[height=40mm,width=40mm]{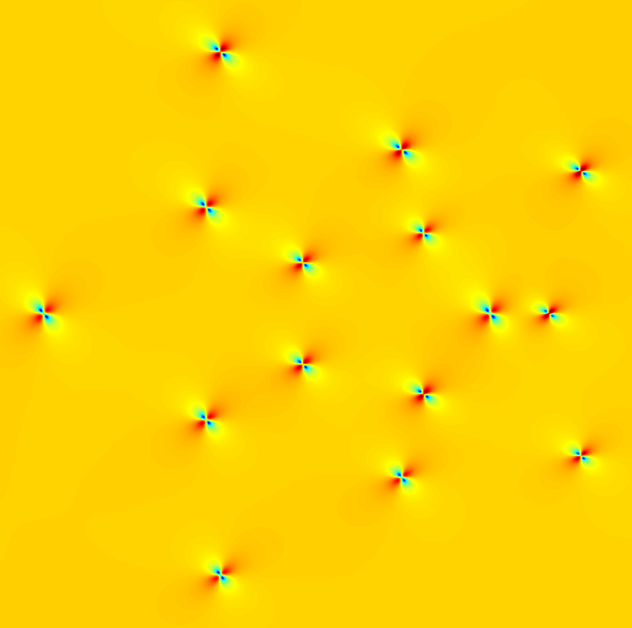}&\includegraphics[height=40mm,width=40mm]{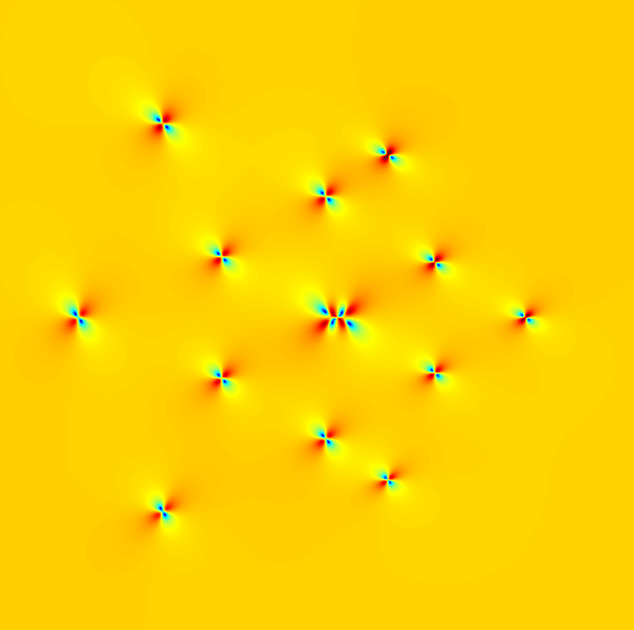}&\includegraphics[height=40mm,width=40mm]{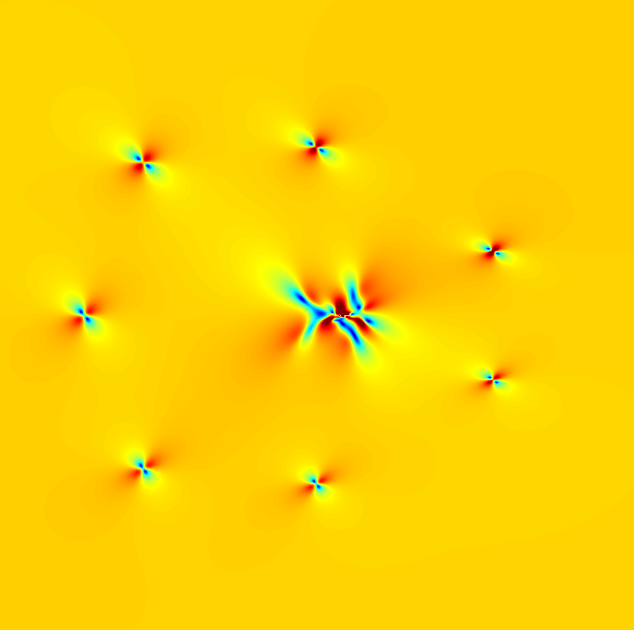}&\rotatebox{0}{\textbf{$t$}}\\
\rotatebox{0}{\textbf{$\left|u_2\right|$}} & \includegraphics[height=40mm,width=40mm]{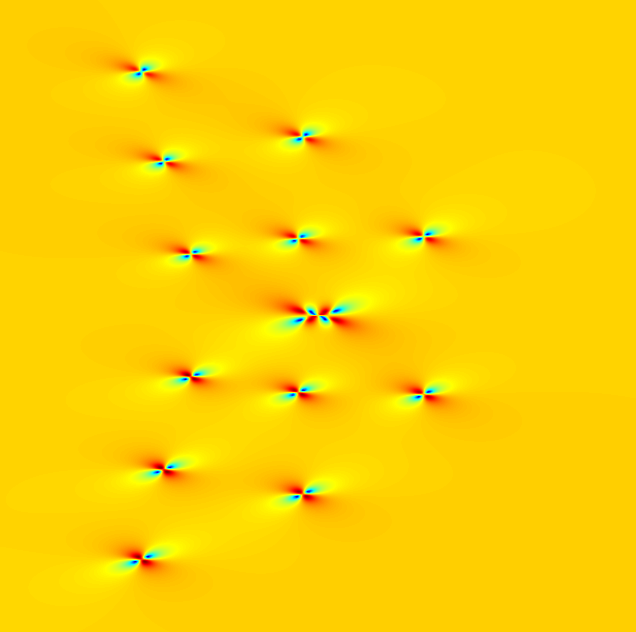}&\includegraphics[height=40mm,width=40mm]{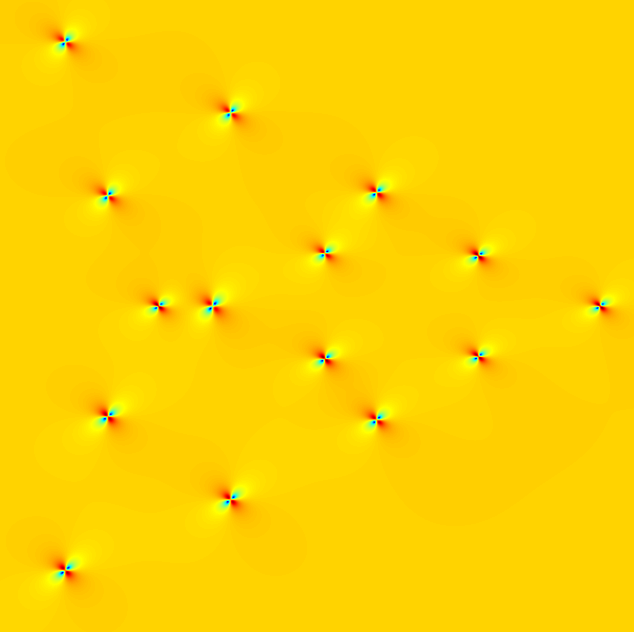}&\includegraphics[height=40mm,width=40mm]{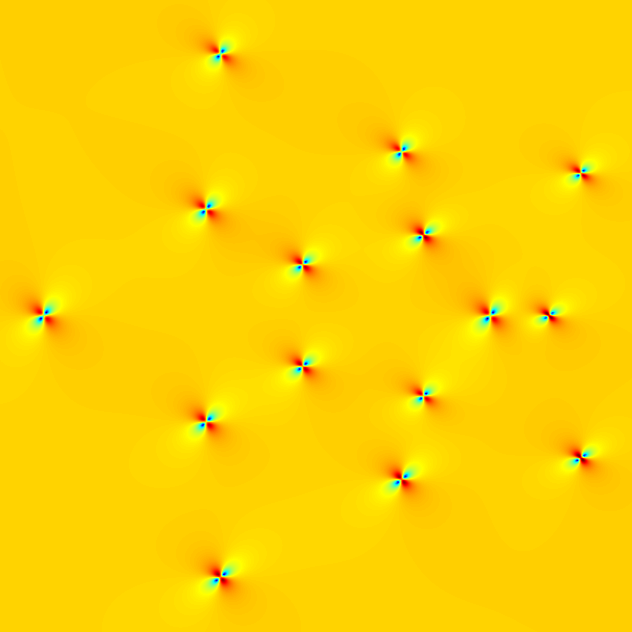}&\includegraphics[height=40mm,width=40mm]{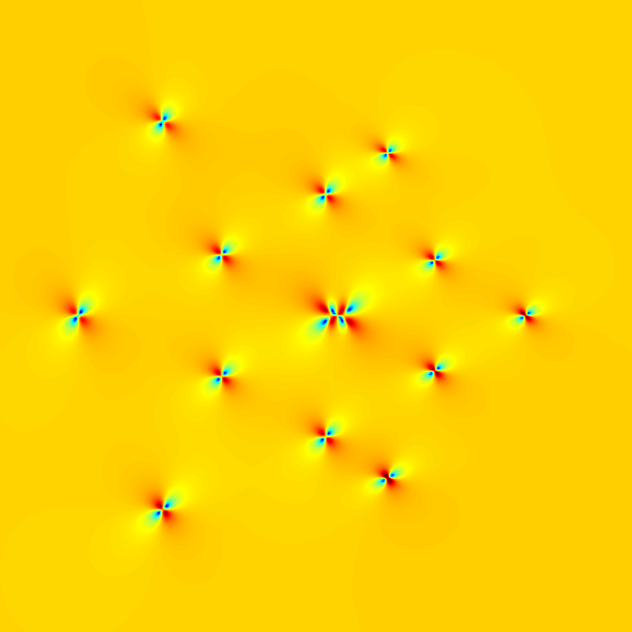}&\includegraphics[height=40mm,width=40mm]{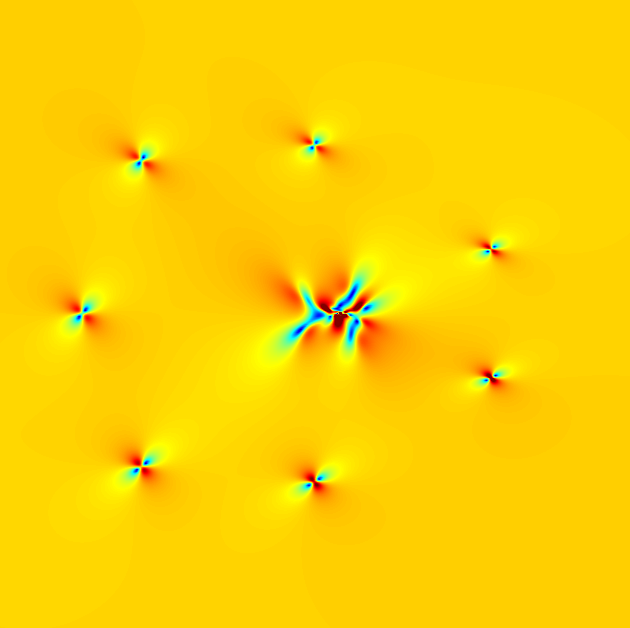}&\rotatebox{0}{\textbf{$t$}}\\
\rotatebox{0}{\textbf{$\left|u_3\right|$}} & \includegraphics[height=40mm,width=40mm]{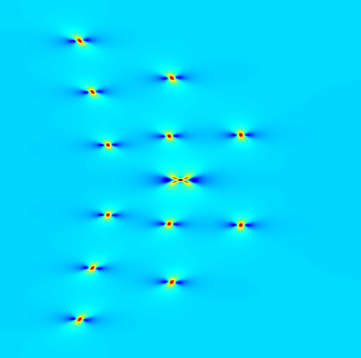}&\includegraphics[height=40mm,width=40mm]{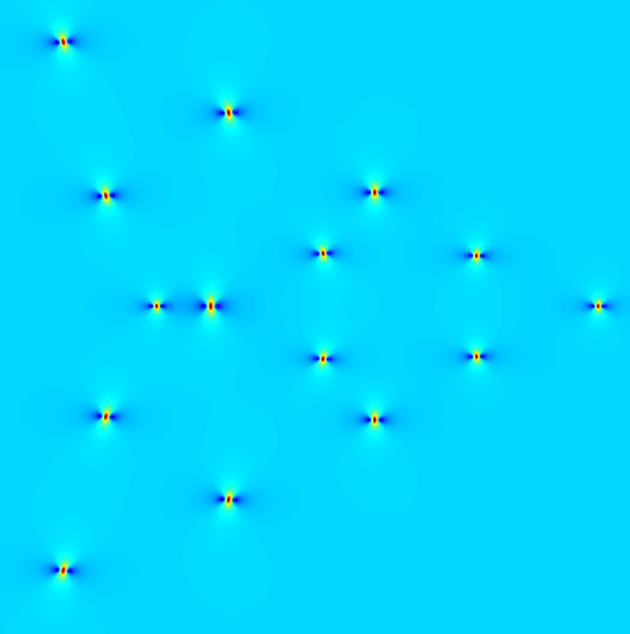}&\includegraphics[height=40mm,width=40mm]{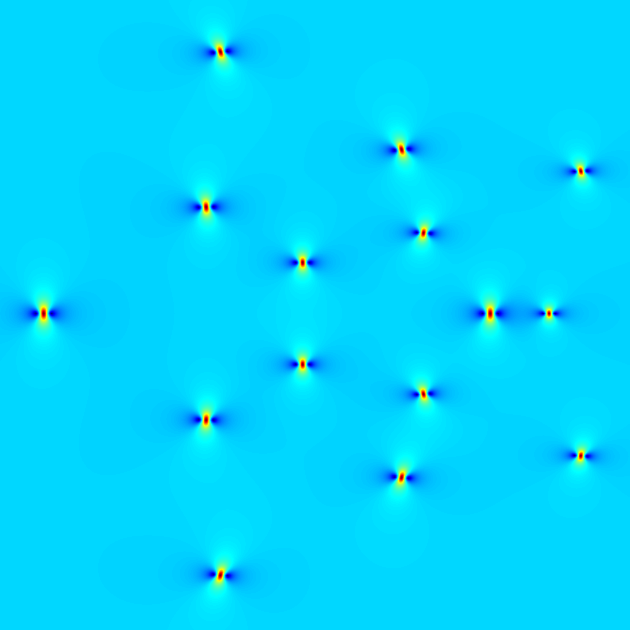}&\includegraphics[height=40mm,width=40mm]{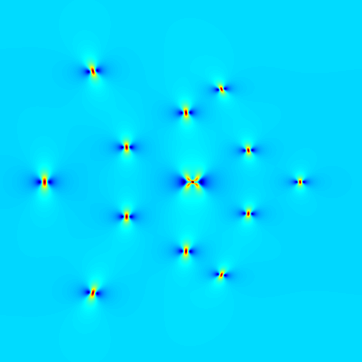}&\includegraphics[height=40mm,width=40mm]{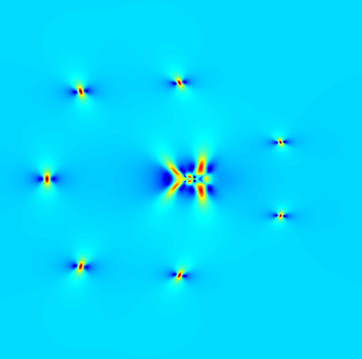}&\rotatebox{0}{\textbf{$t$}}\\
& $x$ & $x$ & $x$ & $x$ & $x$
\end{tabular}
}
\caption{True $2$nd type $(0,3,0)$-th order rogue waves of the three-component NLS with the same parameters as Fig. \ref{2st-type_RW_N3_P}.
% Each column corresponds to rogue waves with a single large parameter $a_m$, whose value is indicated on top, and all other internal parameters are set to zero. Top row: true $\left|u_1(x, t)\right|$. Middle row:true $\left|u_2(x, t)\right|$. Bottom row: true $\left|u_3(x, t)\right|$.
The $(x, t)$ interval for each column is the same as the corresponding column in Fig. \ref{2st-type_RW_N3_P}.
% First column: the $(x, t)$ intervals are $-25 \leq x \leq 25$, $-26 \leq t \leq 26$. Second column: the $(x, t)$ intervals are $-45 \leq x \leq 51$, $-29 \leq t \leq 27$. Third column: the $(x, t)$ intervals are $-47 \leq x \leq 31$, $-22 \leq t \leq 22$. Fourth column: the $(x, t)$ intervals are $-40 \leq x \leq 35$, $-20 \leq t \leq 20$. Fifth column: the $(x, t)$ intervals are $-35 \leq x \leq 30$, $-20 \leq t \leq 20$.
}
\label{2st-type_RW_N3_T}
\end{figure}

Comparing the true rogue waves with predicted ones (see Figs. \ref{2st-type_RW_N3_P} and \ref{2st-type_RW_N3_T}), we can observe that each of the rogue waves strikingly matches in position and rogue wave shape. Not only that, but it is also numerically demonstrated that the actual and predicted results match very well. Since they are very similar to the previous error analysis, we omit the details.
% The results of the numerical analysis also match very well, but we omit the details because they are very similar to the previous error analysis.

\subsubsection{Third-type rogue waves of the three-component NLS equation}
In this case, we choose $(0,0,4)$-th order rogue wave solutions. We only set one of the internal parameters $(a_2, a_3, a_5, a_6, a_7)$ to be large, and the remaining parameters are set to $0$. The very large parameter is one of
\begin{equation}
    a_2=100,\quad a_3=200,\quad a_5=1500,\quad a_6=5000,\quad a_7=10000.
\end{equation}
According to Theorem \ref{Rogue wave patterns-3com}, the position $(\hat{x}_0, \hat{t}_0)$ of each fundamental rogue wave
$$
u_{1, \mathcal{N}_3}(x, t), \quad u_{2, \mathcal{N}_3}(x, t), \quad u_{3, \mathcal{N}_3 }(x, t)
$$
can be predicted by equation \eqref{prediction of locations of RW-3com}. The lower $(N_1, N_2, N_3)$-th order rogue wave would appear in the inner region. The value of $(N_1, N_2, N_3)$ can be obtained from Theorems \ref{Rogue wave patterns-3com} and \ref{root sturcture of jump 4}. In our prediction, the $(N_1,N_2,N_3)$ values for these five rogue wave solutions are
$$
(N_1,N_2,N_3) = (2,0,2), \quad (0,0,1), \quad (0,1,1), \quad (2,0,2),  \quad (0,2,2),
$$
respectively. We remark that $(0,0,1)$ means there is only a fundamental rogue wave in the inner region. The internal parameters in
these predicted lower $(N_1, N_2, N_3)$-th order rogue waves of the inner region are all zero, due to our choice of parameters $a_m$ and the value of $s_j$ shown in Remark \ref{Values of s_r}.

\begin{figure}[h]
\centering
\renewcommand\arraystretch{0.5}
\setlength\tabcolsep{0pt}
\resizebox{\linewidth}{!}{
\begin{tabular}{m{0.6cm}<{\centering}m{4.1cm}<{\centering}m{4.1cm}<{\centering}m{4.05cm}<{\centering}m{4.1cm}<{\centering}m{4.05cm}<{\centering}m{0.6cm}<{\centering}r}
&\textbf{$a_2=100$}  & \textbf{$a_3= 200$} & \textbf{$a_5=1500 $}& \textbf{$a_6=5000 $}& \textbf{$a_7=10000 $}&\\
\rotatebox{90}{\text{ predicted locations }} & \includegraphics[height=40mm,width=40mm]{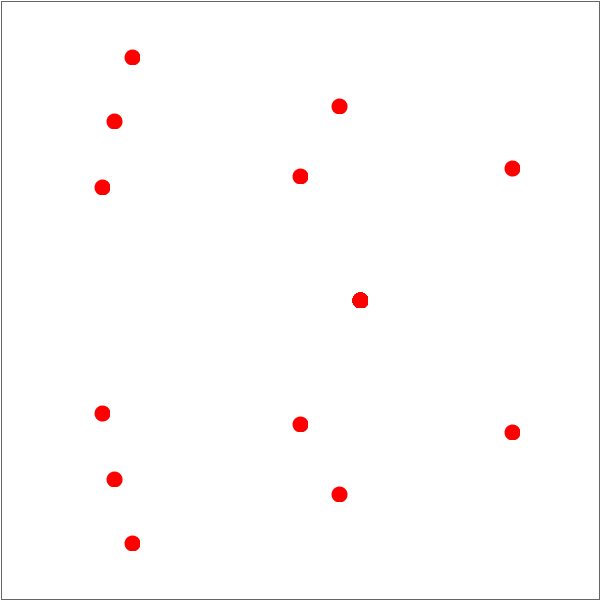}&\includegraphics[height=40mm,width=40mm]{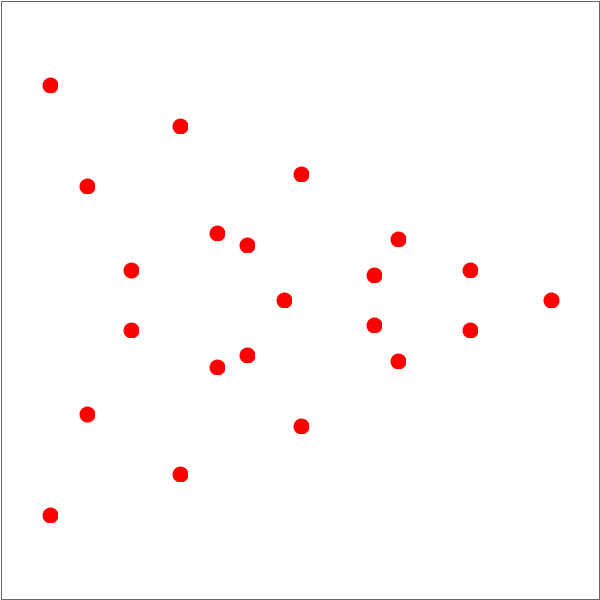}&\includegraphics[height=40mm,width=40mm]{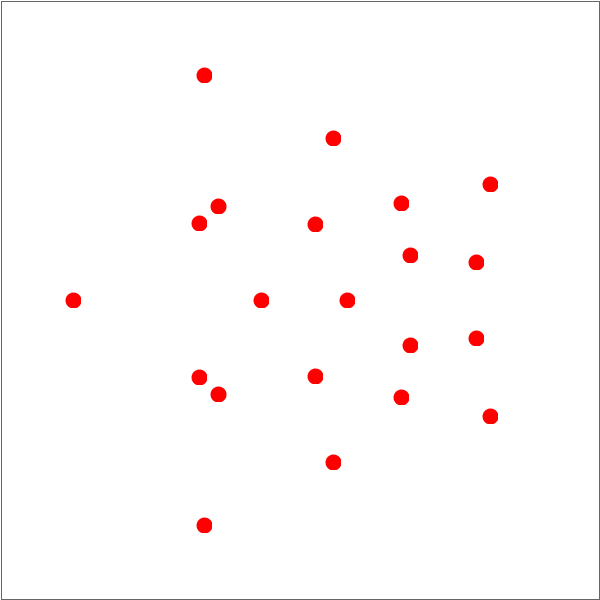}&\includegraphics[height=40mm,width=40mm]{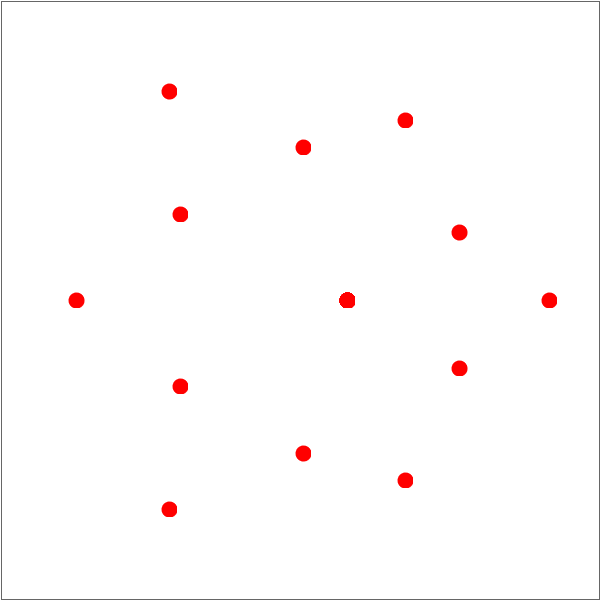}&\includegraphics[height=40mm,width=40mm]{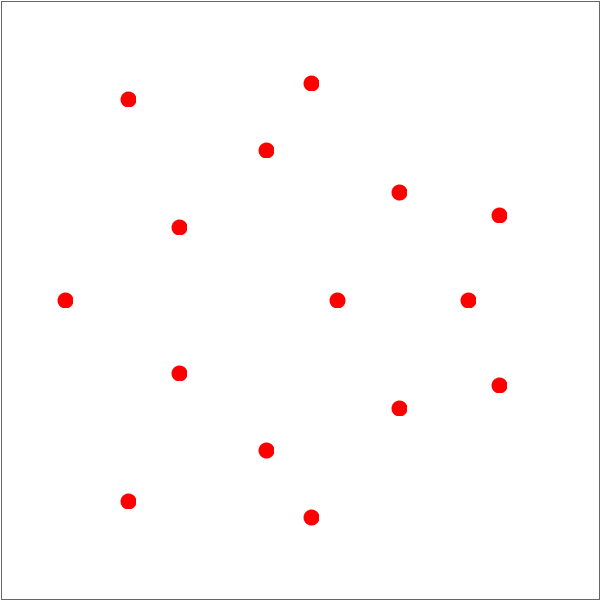}&\rotatebox{0}{\textbf{$t$}}\\
\rotatebox{0}{\textbf{$\left|u_1\right|$}} & \includegraphics[height=40mm,width=40mm]{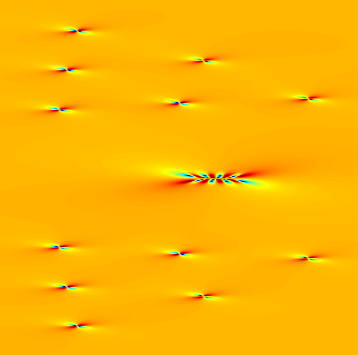}&\includegraphics[height=40mm,width=40mm]{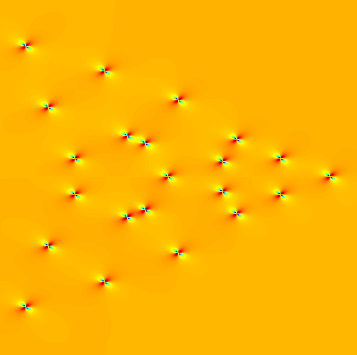}&\includegraphics[height=40mm,width=40mm]{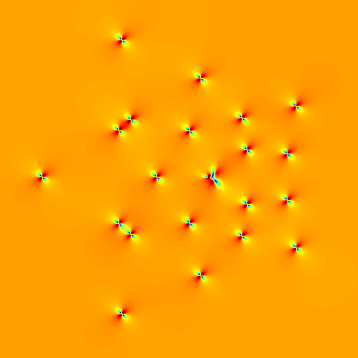}&\includegraphics[height=40mm,width=40mm]{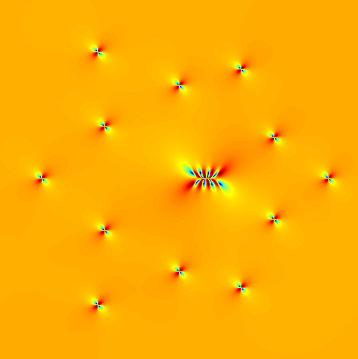}&\includegraphics[height=40mm,width=40mm]{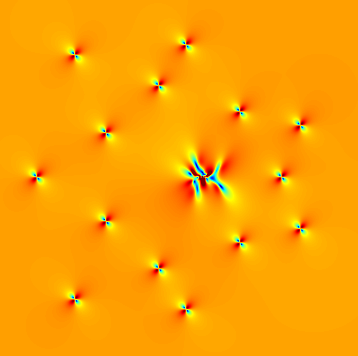}&\rotatebox{0}{\textbf{$t$}}\\
\rotatebox{0}{\textbf{$\left|u_2\right|$}} & \includegraphics[height=40mm,width=40mm]{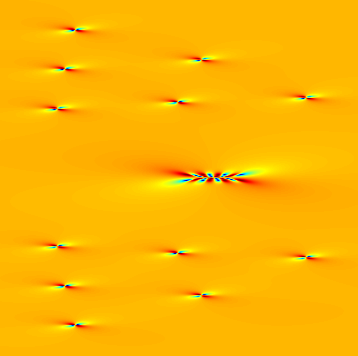}&\includegraphics[height=40mm,width=40mm]{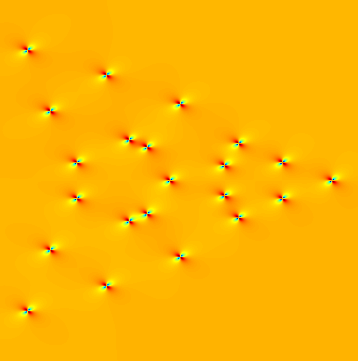}&\includegraphics[height=40mm,width=40mm]{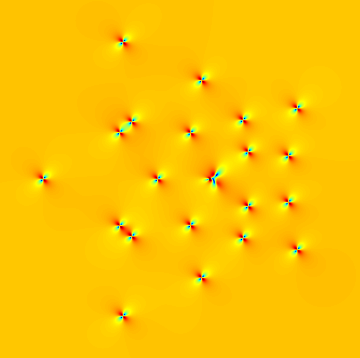}&\includegraphics[height=40mm,width=40mm]{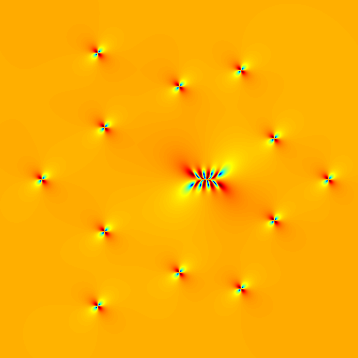}&\includegraphics[height=40mm,width=40mm]{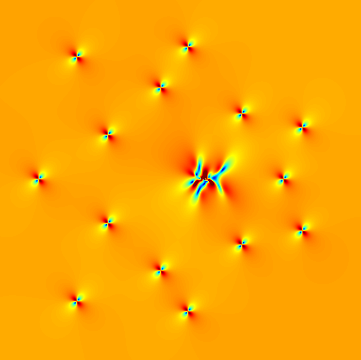}&\rotatebox{0}{\textbf{$t$}}\\
\rotatebox{0}{\textbf{$\left|u_3\right|$}} & \includegraphics[height=40mm,width=40mm]{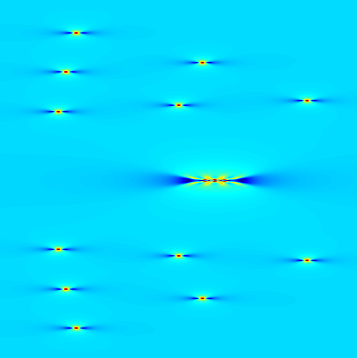}&\includegraphics[height=40mm,width=40mm]{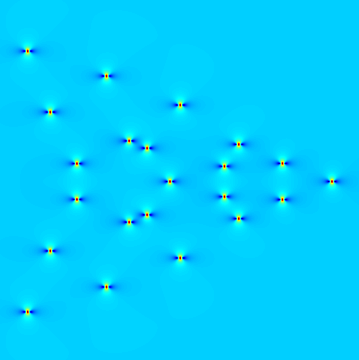}&\includegraphics[height=40mm,width=40mm]{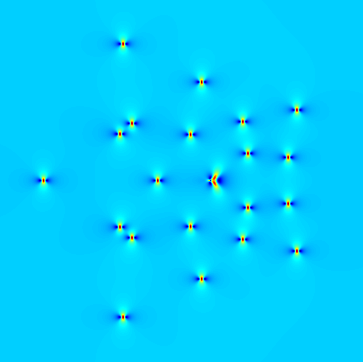}&\includegraphics[height=40mm,width=40mm]{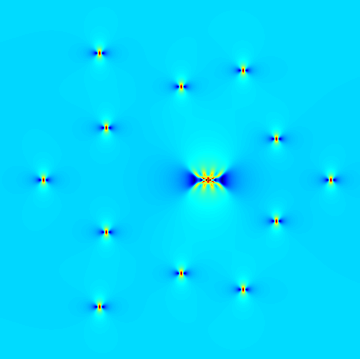}&\includegraphics[height=40mm,width=40mm]{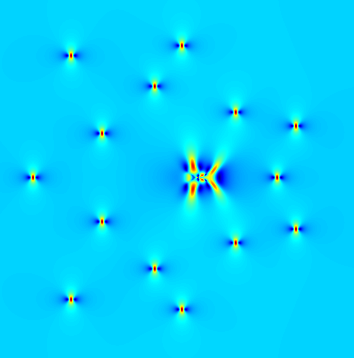}&\rotatebox{0}{\textbf{$t$}}\\
& $x$ & $x$ & $x$ & $x$ & $x$
\end{tabular}
}
\caption{Predicted $3$rd type $(0,0,4)$-th order rogue waves of the three-component NLS from Theorem \ref{Rogue wave patterns-3com}. Each column corresponds to rogue waves with a single large parameter $a_m$, whose value is indicated on top, and all other internal parameters are set to zero. Top row: predicted $\left(\hat{x}_0, \hat{t}_0\right)$ locations by formulae \eqref{prediction of locations of RW-3com}. Second row: predicted $\left|u_1(x, t)\right|$.  Third row: predicted $\left|u_2(x, t)\right|$.  Bottom row: predicted $\left|u_3(x, t)\right|$. First column: the $(x, t)$ intervals are $-30 \leq x \leq 20$, $-50 \leq t \leq 50$. Second column: the $(x, t)$ intervals are $-45 \leq x \leq 50$, $-32 \leq t \leq 32$. Third column: the $(x, t)$ intervals are $-55 \leq x \leq 40$, $-25 \leq t \leq 25$. Fourth column: the $(x, t)$ intervals are $-55 \leq x \leq 40$, $-25 \leq t \leq 25$. Fifth column: the $(x, t)$ intervals are $-45 \leq x \leq 35$, $-40 \leq t \leq 20$.}
\label{3rd-type_RW_N4_P}
\end{figure}

For $[u_{1, \mathcal{N}_3}(x, t), u_{2, \mathcal{N}_3}(x, t), u_{3, \mathcal{N}_3 }(x, t)]$, their corresponding predicted rogue wave patterns are shown in the last three rows of Fig. \ref{3rd-type_RW_N4_P}, with the first row being the locations of the rogue waves.
%Each column is when one of $(a_2, a_3,a_5, a_6, a_7)$ is large, respectively.
It can be seen from Fig. \ref{3rd-type_RW_N4_P} that
the large-$a_3$ solution exhibits a skewed triple-triangle, corresponding to the triple-triangle root structure of $W_4^{[3,4,1]}(z)$. The large-$a_5$ solution displays a deformed pentagon, corresponding to the pentagon-shaped root structure of $W_4^{[5,4,1]}(z)$. The large-$a_6$ solution exhibits a deformed hexagon, corresponding to the hexagon-shaped root structure of $W_4^{[6,4,1]}(z)$.  The large-$a_7$ solution exhibits a deformed heptagon, corresponding to the heptagon-shaped root structure of $W_4^{[7,4,1]}(z)$.

\begin{figure}[h]
\centering
\renewcommand\arraystretch{0.5}
\setlength\tabcolsep{0pt}
\resizebox{\linewidth}{!}{
\begin{tabular}{m{0.6cm}<{\centering}m{4.1cm}<{\centering}m{4.1cm}<{\centering}m{4.05cm}<{\centering}m{4.1cm}<{\centering}m{4.05cm}<{\centering}m{0.6cm}<{\centering}r}
&\textbf{$a_2=100$}  & \textbf{$a_3= 200$} & \textbf{$a_5=1500 $}& \textbf{$a_6=5000 $}& \textbf{$a_7=10000 $}&\\
%\rotatebox{90}{\text{ predicted locations }} & \includegraphics[height=40mm,width=40mm]{RW34231.png}&\includegraphics[height=40mm,width=40mm]{RW34331.png}&\includegraphics[height=40mm,width=40mm]{RW34531.png}&\includegraphics[height=40mm,width=40mm]{RW34631.png}&\includegraphics[height=40mm,width=40mm]{RW34731.png}&\rotatebox{0}{\textbf{$t$}}\\
\rotatebox{0}{\textbf{$\left|u_1\right|$}} & \includegraphics[height=40mm,width=40mm]{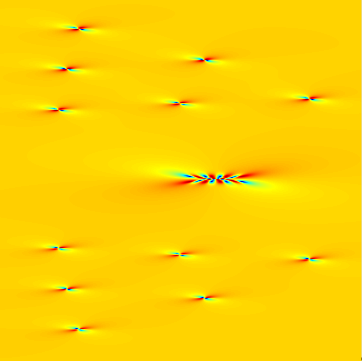}&\includegraphics[height=40mm,width=40mm]{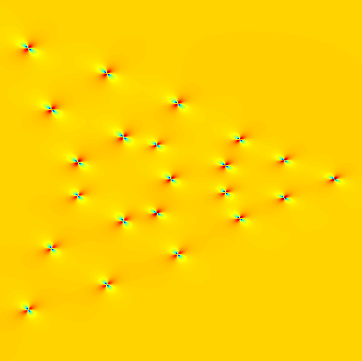}&\includegraphics[height=40mm,width=40mm]{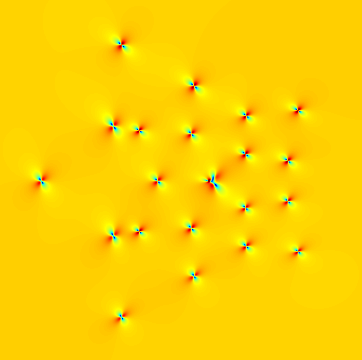}&\includegraphics[height=40mm,width=40mm]{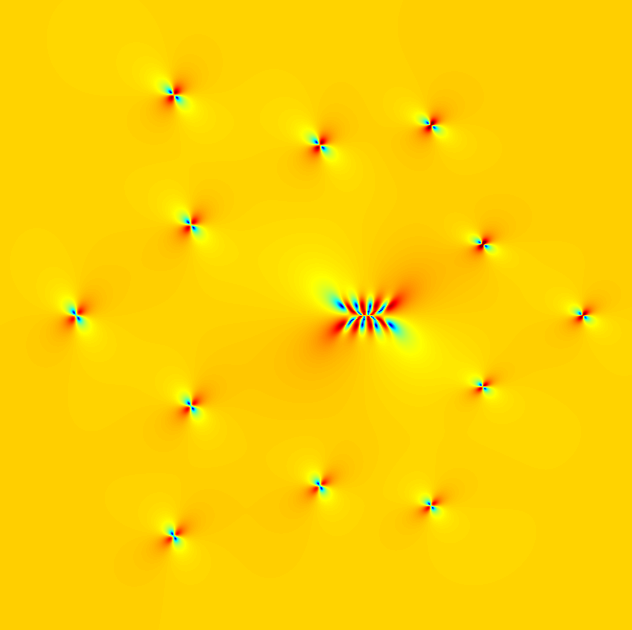}&\includegraphics[height=40mm,width=40mm]{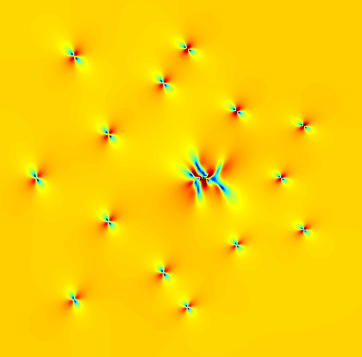}&\rotatebox{0}{\textbf{$t$}}\\
\rotatebox{0}{\textbf{$\left|u_2\right|$}} & \includegraphics[height=40mm,width=40mm]{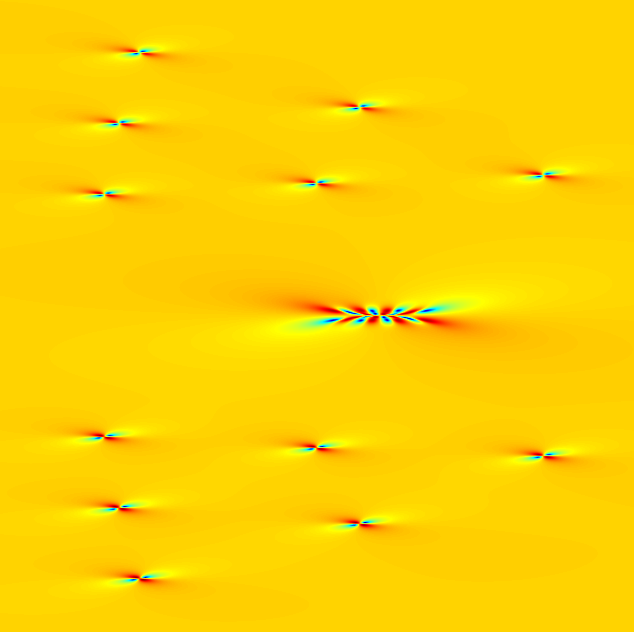}&\includegraphics[height=40mm,width=40mm]{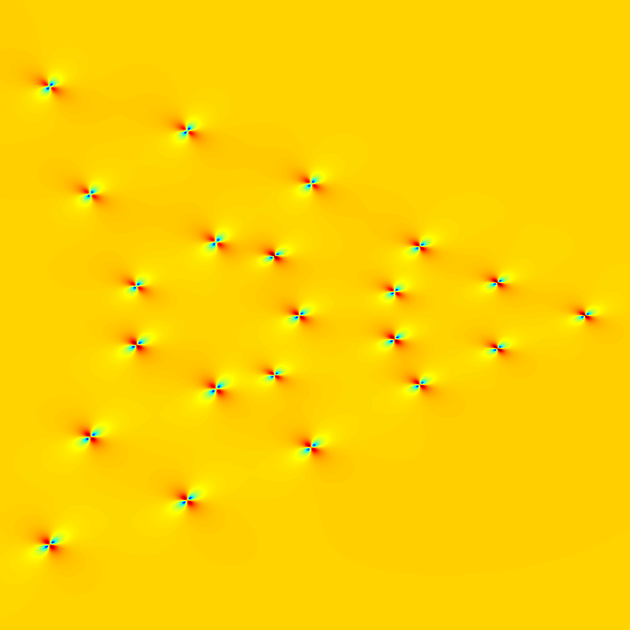}&\includegraphics[height=40mm,width=40mm]{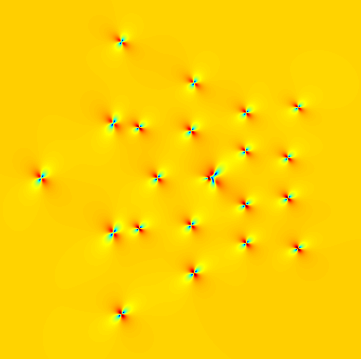}&\includegraphics[height=40mm,width=40mm]{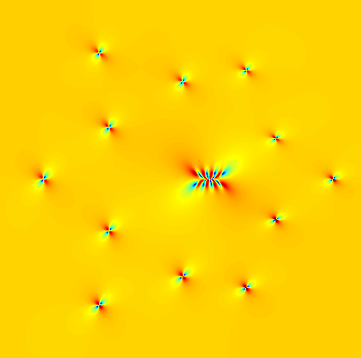}&\includegraphics[height=40mm,width=40mm]{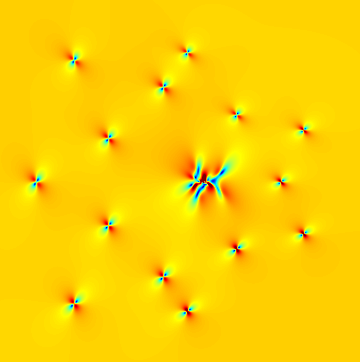}&\rotatebox{0}{\textbf{$t$}}\\
\rotatebox{0}{\textbf{$\left|u_3\right|$}} & \includegraphics[height=40mm,width=40mm]{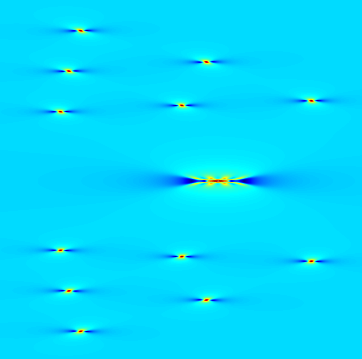}&\includegraphics[height=40mm,width=40mm]{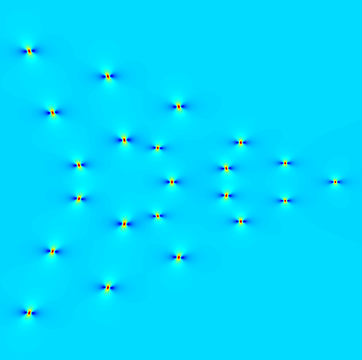}&\includegraphics[height=40mm,width=40mm]{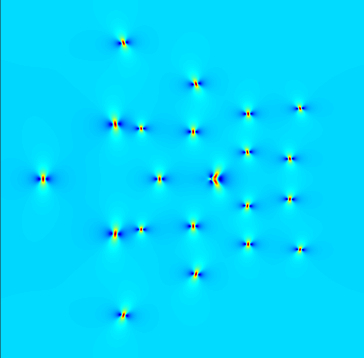}&\includegraphics[height=40mm,width=40mm]{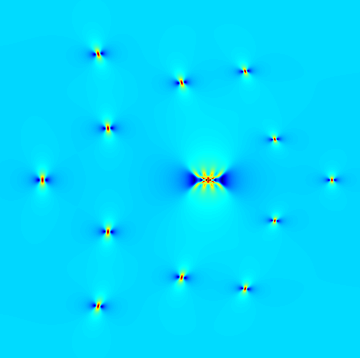}&\includegraphics[height=40mm,width=40mm]{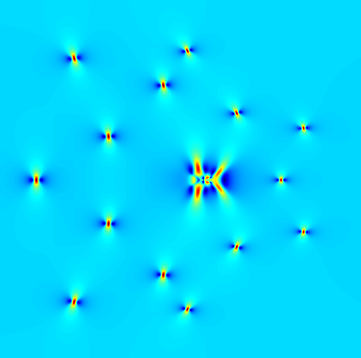}&\rotatebox{0}{\textbf{$t$}}\\
& $x$ & $x$ & $x$ & $x$ & $x$
\end{tabular}
}
\caption{True $3$rd type $(0,0,4)$-th order rogue waves of the three-component NLS with the same parameters as Fig. \ref{3rd-type_RW_N4_P}.
% Each column corresponds to a rogue wave with a single large parameter $a_m$, whose value is indicated on top, and all other internal parameters are set to be zero.  Top row: true $\left|u_1(x, t)\right|$. Middle row:true $\left|u_2(x, t)\right|$. Bottom row: true $\left|u_3(x, t)\right|$.
The $(x, t)$ interval for each column is the same as the corresponding column in Fig. \ref{3rd-type_RW_N4_P}.
% First column: the $(x, t)$ intervals are $-30 \leq x \leq 20$, $-50 \leq t \leq 50$. Second column: the $(x, t)$ intervals are $-45 \leq x \leq 50$, $-32 \leq t \leq 32$. Third column: the $(x, t)$ intervals are $-55 \leq x \leq 40$, $-25 \leq t \leq 25$. Fourth column: the $(x, t)$ intervals are $-55 \leq x \leq 40$, $-25 \leq t \leq 25$. Fifth column: the $(x, t)$ intervals are $-45 \leq x \leq 35$, $-40 \leq t \leq 20$.
}
\label{3rd-type_RW_N4_T}
\end{figure}

Comparing the actual rogue waves with the predicted ones (see Figs. \ref{3rd-type_RW_N4_P} and \ref{3rd-type_RW_N4_T}), we can observe that each of the rogue waves matches perfectly in position and rogue wave shape. Moreover, one can further compare them numerically. The results also support our prediction, and since they are very similar to previous analysis, the details are omitted.

\subsubsection{Effect of parameters on the rogue wave shapes}
We first represent the complex parameter $a_m$ as $a_m=|a_m| \exp{\left(\mathrm{i}\vartheta_m \right)} $. In what follows, we will discuss the effects of the modulus $|a_m|$ and the argument $\vartheta_m$  on the shapes of rogue waves.

\begin{figure}[h]
    \centering
    \renewcommand\arraystretch{0.5}
    \setlength\tabcolsep{0pt}
    %\resizebox{\linewidth}{!}{
    \begin{tabular}{m{0.4cm}<{\centering}m{4.05cm}<{\centering}m{4.05cm}<{\centering}m{4.05cm}<{\centering}m{4.05cm}<{\centering}m{0.6cm}<{\centering}r}
    &\textbf{$\psi_3=-\pi/3$}  & \textbf{$\psi_3=0$} & \textbf{$\psi_3=\pi/3$}&\textbf{$\psi_3=\pi$}&\\
    \rotatebox{90}{\text{predict location}} & \includegraphics[height=40mm,width=40mm]{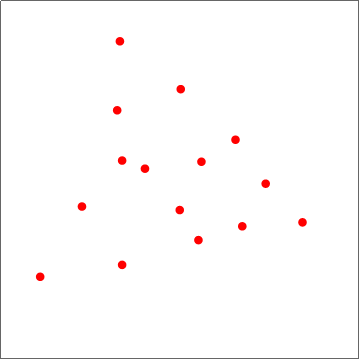}&\includegraphics[height=40mm,width=40mm]{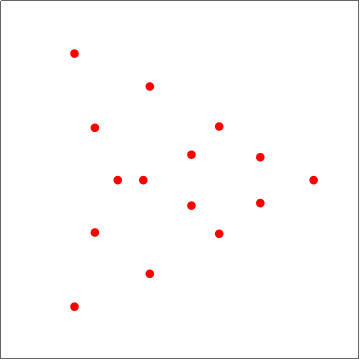}&\includegraphics[height=40mm,width=40mm]{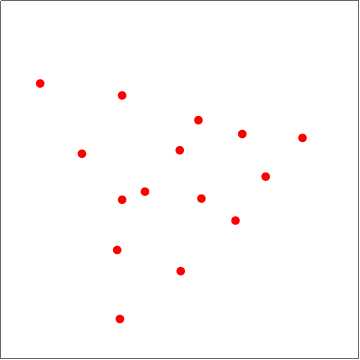}&\includegraphics[height=40mm,width=40mm]{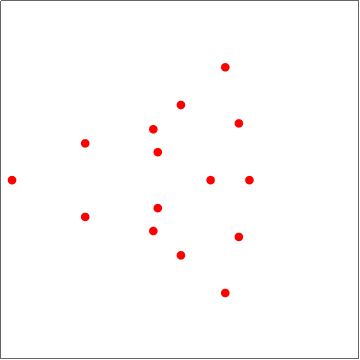}&\rotatebox{0}{\textbf{$t$}}\\
    \rotatebox{90}{\text{true $\left|u_3\right|$}} & \includegraphics[height=40mm,width=40mm]{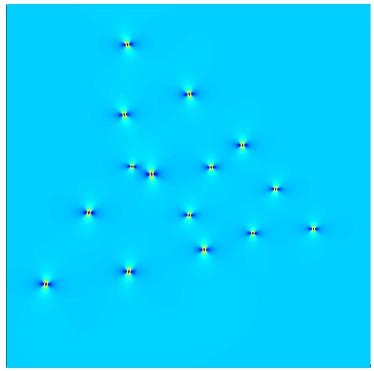}&\includegraphics[height=40mm,width=40mm]{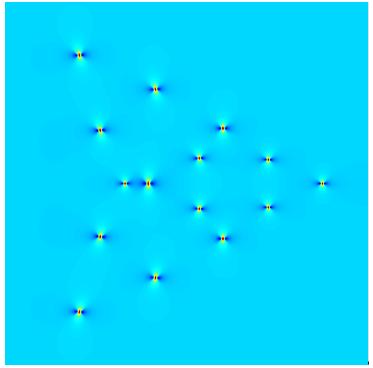}&\includegraphics[height=40mm,width=40mm]{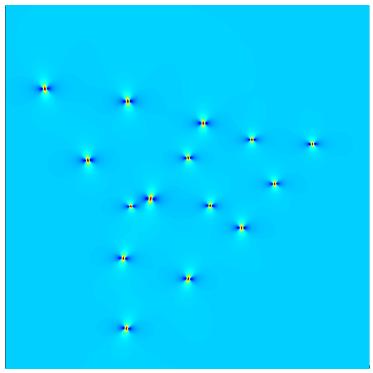}&\includegraphics[height=40mm,width=40mm]{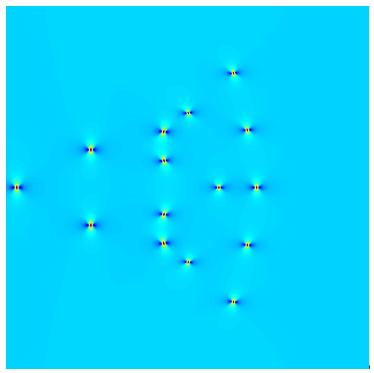}&\rotatebox{0}{\textbf{$t$}}\\
    & $x$ & $x$ & $x$ & $x$
    \end{tabular}
    %}
    \caption{Effect of the argument in $a_3$ on orientations of the $2$nd type $(0,3,0)$-th order rogue waves of the three-component NLS equation. Each column represents a rogue wave with a different value of $a_3$ with the same modulus $300$ but a different argument, and all other internal parameters are set to zero. The $(x, t)$ intervals are $-55 \leq x \leq 55$ and $-30 \leq t \leq 30$.}
    \label{effect of theta_3com_n3}
\end{figure}

To illustrate the effect of the changes of $\vartheta_m$, we consider the $2$nd type $(0,3,0)$-th order rogue waves of the three-component NLS equation and set $|a_3|=300$ while the remaining parameters $a_m$ are set to 0. %Corresponding figures are shown in Fig. 2.
Here we simply choose four values of $a_3$, namely, $$(300\exp{(-\pi\mathrm{i}/3)},\quad 300,\quad 300\exp{(\pi\mathrm{i}/3)},\quad 300\exp{(\pi\mathrm{i}})),$$ then the corresponding predicted and true rogue wave patterns are shown in Fig. \ref{effect of theta_3com_n3}.
It can be seen that the orientation of the rogue wave pattern is changed when we vary the values of $\vartheta_3$. In fact, this can be seen from Theorem \ref{Rogue wave patterns-3com} as well. In addition, we find that the orientation of the rogue wave pattern is obtained by rotating angle arg$(-a_m)/m$ of the root structure of $W_3^{[3,4,2]}(z)$, where ``arg" represents the argument of a complex number.

%As predicted, when the argument of $a_2$ changes, they rotate by the corresponding angles.

%\begin{figure}[htp]
%\centering
%\renewcommand\arraystretch{0.5}
%\setlength\tabcolsep{0pt}
%%\resizebox{\linewidth}{!}{
%\begin{tabular}{m{0.5cm}<{\centering}m{4.05cm}<{\centering}m{4.05cm}<{\centering}m{4.05cm}<{\centering}m{4.05cm}<{\centering}m{0.6cm}<{\centering}r}
%&\textbf{$\psi_3=-\pi/3$}  & \textbf{$\psi_3=0$} & \textbf{$\psi_3=\pi/3$}&\textbf{$\psi_3=\pi$}&\\
%\rotatebox{90}{\text{predict location}} & \includegraphics[height=40mm,width=40mm]{3RW_n3_RotNeg0.33P.png}&\includegraphics[height=40mm,width=40mm]{3RW_n3_a3_300_0_P.png}&\includegraphics[height=40mm,width=40mm]{3RW_n3_a3_300_0.33_P.png}&\includegraphics[height=40mm,width=40mm]{3RW_n3_a3_300_1_P.png}&\rotatebox{0}{\textbf{$t$}}\\
%\rotatebox{90}{\text{true $\left|u_{3}\right|$}} & \includegraphics[height=40mm,width=40mm]{3RW_n3_a3_300_neg0.33_T.png}&\includegraphics[height=40mm,width=40mm]{3RW_n3_a3_300_0_T.png}&\includegraphics[height=40mm,width=40mm]{3RW_n3_a3_300_0.33_T.png}&\includegraphics[height=40mm,width=40mm]{3RW_n3_a3_300_1_T.png}&\rotatebox{0}{\textbf{$t$}}\\
%& $x$ & $x$ & $x$
%\end{tabular}
%%} 3RW_n3_RotNeg0.33P
%\caption{Effect of the argument in $a_3$ on orientations of the $2$nd type $(0,3,0)$-th order rogue waves of the three-component NLS equation. Each column represents a rogue wave with a different value of $a_3$ with the same modulus $300$ but a different argument, and all other internal parameters are set to zero. The $(x, t)$ intervals are $-55 \leq x \leq 55$ and $-30 \leq t \leq 30$.}
%\label{effect of theta_3com_n3}
%\end{figure}

\begin{figure}[h]
\centering
\renewcommand\arraystretch{0.5}
\setlength\tabcolsep{0pt}
%\resizebox{\linewidth}{!}{
\begin{tabular}{m{0.5cm}<{\centering}m{4.1cm}<{\centering}m{4.05cm}<{\centering}m{4.05cm}<{\centering}m{0.6cm}<{\centering}r}
&\textbf{$a_2=20$}  & \textbf{$a_2= 100$} & \textbf{$a_2=500 $}&\\
\rotatebox{90}{\text{predict location}} & \includegraphics[height=40mm,width=40mm]{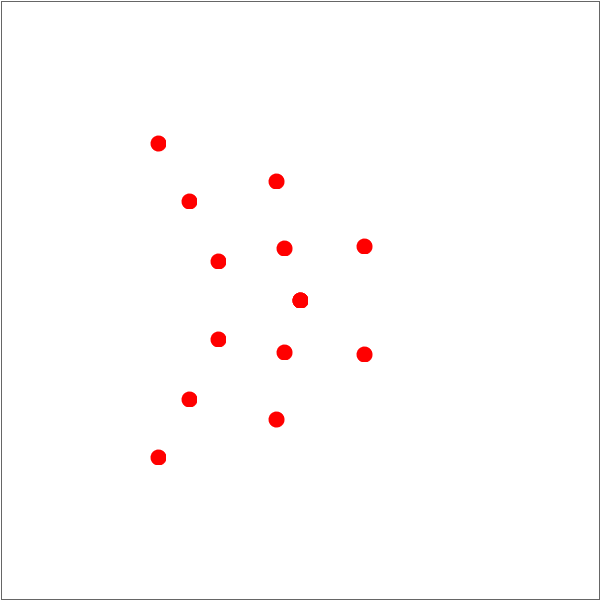}&\includegraphics[height=40mm,width=40mm]{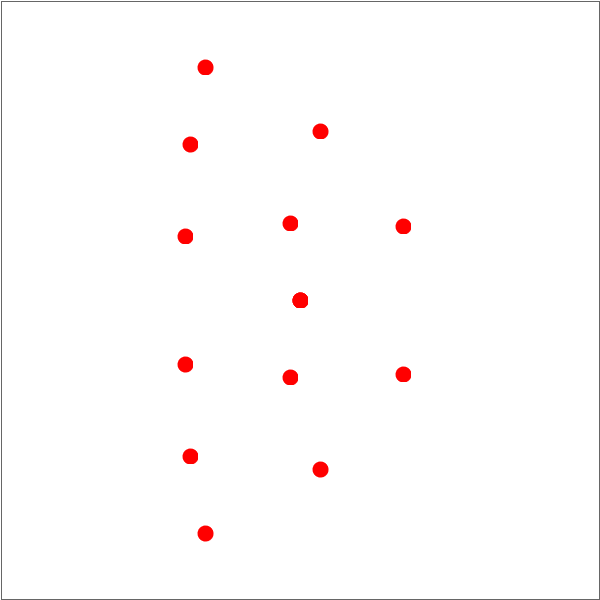}&\includegraphics[height=40mm,width=40mm]{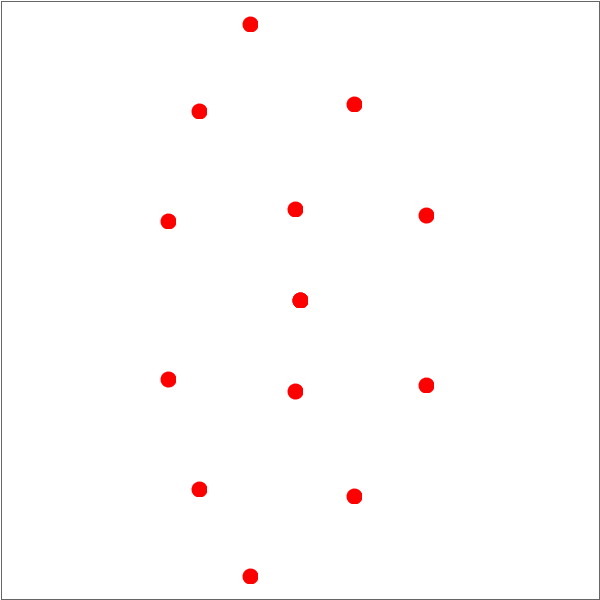}&\rotatebox{0}{\textbf{$t$}}\\
\rotatebox{90}{\text{true $\left|u_3\right|$}} & \includegraphics[height=40mm,width=40mm]{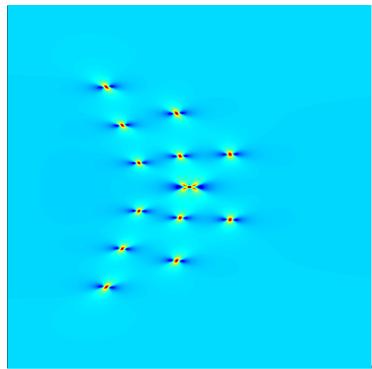}&\includegraphics[height=40mm,width=40mm]{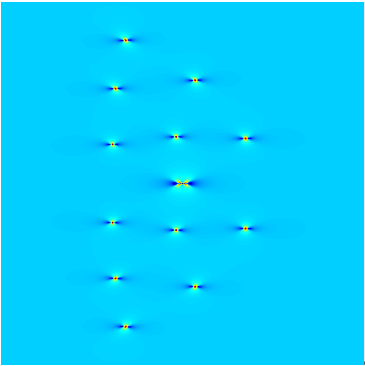}&\includegraphics[height=40mm,width=40mm]{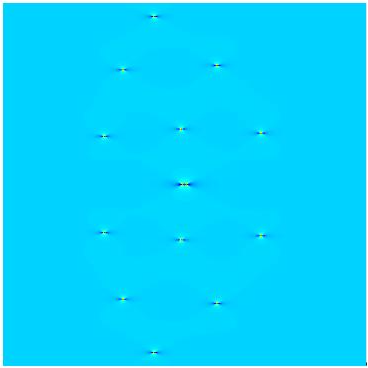}&\rotatebox{0}{\textbf{$t$}}\\
& $x$ & $x$ & $x$
\end{tabular}
%}
\caption{Effect of the size $a_2$ on shapes of the $2$nd type $(0,3,0)$-th order rogue waves of the three-component NLS equation. Each column represents a rogue wave with different values of $a_2$ indicated on top, and all other internal parameters are set to zero. Top row: prediction location of $\left|u_3(x, t)\right|$. Bottom row: true $\left|u_3(x, t)\right|$. First column: the $(x, t)$ intervals are $-30 \leq x,t \leq 30$. Second column: the $(x, t)$ intervals are $-45 \leq x,t \leq 45$. Third column: the $(x, t)$ intervals are $-85 \leq x,t \leq 85$.}
\label{effect of size_3com_n2}
\end{figure}

To study the effect of changes in $|a_m|$ on rogue wave patterns, we choose $2$nd type $(0,3,0)$-th order rogue waves. Set the argument of $a_2$ to $0$ while the remaining parameters $a_m$ are chosen to be 0.
%Corresponding figures are shown in Fig. 3.
The corresponding predicted and true rogue wave patterns are depicted in Fig. \ref{effect of size_3com_n2}.
It can be observed that the shape of the rogue wave pattern becomes closer and closer to the linear transformation of the roots of $W_3^{[2,4,2]}(z)$ when the modulus of $a_2$ gets larger. Specifically, when $a_2=20$, both predicted and true rogue wave patterns look very irregular, especially those located on the negative $x$-axis. However, further increasing the value of $a_2$, the distortion will gradually be weakened. For $a_2 =500$, the shape of rogue waves is very close to a linear transformation of the roots of $W_3^{[2,4,2]}(z)$.
The reason is that $\Delta_2$ is a  $z_0$-dependent $O(1)$ quantity. Specifically, when $a_2$ takes a very large value, the value of $\Delta_2$ in \eqref{prediction of locations of RW-3com} can be ignored and the predicted rogue wave can be obtained approximately by some linear transformation from the root structure of $W_3^{[2,4,2]}(z)$.

%\newpage

\subsection{Comparison in the four-component NLS equation}
In this subsection, we compare our predicted rouge wave patterns in Theorem \ref{Rogue wave patterns-4com} with true rouge waves of the four-component NLS equation. For the quadruple root case, we choose background wavenumbers $k_1=-k_4= \left(\sqrt{5}-1\right)/4$ and $k_2=-k_3=-\left(\sqrt{5}+1\right)^2 /8$, which implies $\rho_1=\rho_4=\sqrt{2}$ and $ \rho_2=\rho_3=\sqrt{\sqrt{5}+3}$
according to \eqref{quadruple root condition}. In this circumstance, we select $p_0=\sqrt{ \left(\sqrt{5}+5\right)/2}/2$ and  $p_1=\sqrt[10]{(11 \sqrt{5}+25)/4608}$. %for the later analysis.

\subsubsection{Third-type rogue waves of the four-component NLS equation}
In this case, we focus on $(0,0,2,0)$-th order rogue wave solution. For brevity, we only consider the first four irreducible parameters $(a_2,a_3,a_4,a_6)$, and set the rest of the parameters to $0$. Then, for the internal parameters $(a_2, a_3, a_4, a_6)$, we let one of them be large and set the others to $0$. The very large parameter is one of
\begin{equation}
    a_2=50,\quad a_3=200,\quad a_4=500,\quad a_6=5000.
\end{equation}
According to Theorem \ref{Rogue wave patterns-4com}, the position $(\bar{x}_0, \bar{t}_0)$ of each fundamental rogue wave
$$
u_{1, \mathcal{N}_1}(x, t), \quad u_{2, \mathcal{N}_1}(x, t), \quad u_{3, \mathcal{N}_1 }(x, t), \quad u_{4, \mathcal{N}_1 }(x, t)
$$
can be predicted by \eqref{prediction of locations of RW-4com}. The $(N_1, N_2, N_3, N_4)$-th order rogue wave would appear in the inner region, and the $(N_1, N_2, N_3, N_4)$ values for these five rogue wave solutions are obtained from Theorems \ref{Rogue wave patterns-4com} and \ref{root sturcture of jump 5} as
$$
(N_1,N_2,N_3,N_4) = (0,0,0,0),\quad (0,0,1,1), \quad(0,1,1,0), \quad(0,0,1,1),
$$
Note that $(0,0,0,0)$ means no lower-order rogue wave exists in the center region. Owing to our choice of parameters $a_m$ and the value of $s_j$ shown in Remark \ref{Values of s_r}, the internal parameters in these predicted lower $(N_1, N_2, N_3, N_4)$-th order rogue waves of the center region are all chosen to be zero.

\begin{figure}[htp]
\centering
\renewcommand\arraystretch{0.5}
\setlength\tabcolsep{0pt}
\resizebox{\linewidth}{!}{
\begin{tabular}{m{0.6cm}<{\centering}m{4.1cm}<{\centering}m{4.1cm}<{\centering}m{4.1cm}<{\centering}m{4.1cm}<{\centering}m{0.6cm}<{\centering}r}
&\textbf{$a_2=50$}  & \textbf{$a_3= 200$} & \textbf{$a_4=500 $}& \textbf{$a_6=5000 $}&\\
\rotatebox{90}{\text{ predicted locations }} & \includegraphics[height=40mm,width=40mm]{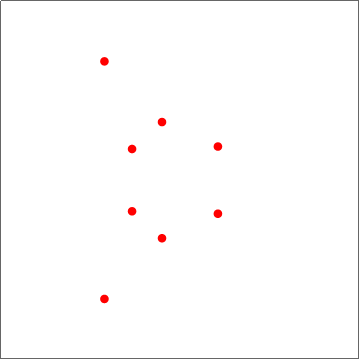}&\includegraphics[height=40mm,width=40mm]{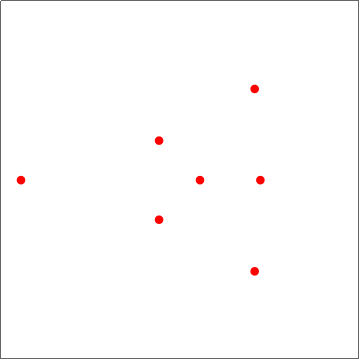}&\includegraphics[height=40mm,width=40mm]{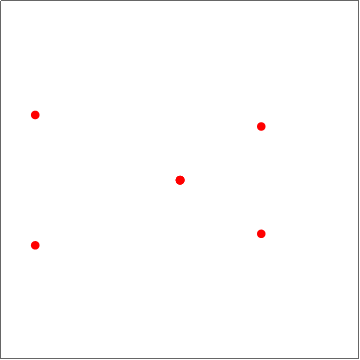}&\includegraphics[height=40mm,width=40mm]{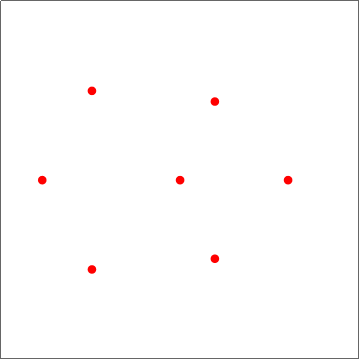}&\rotatebox{0}{\textbf{$t$}}\\
\textbf{$\left|u_1\right|$} & \includegraphics[height=40mm,width=40mm]{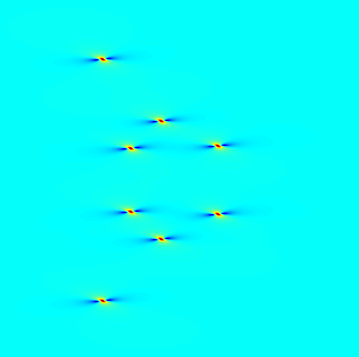}&\includegraphics[height=40mm,width=40mm]{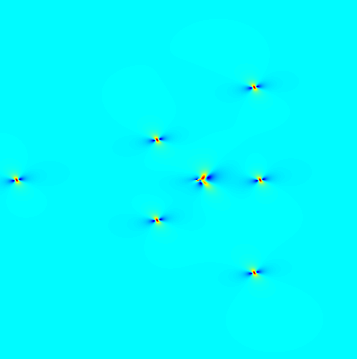}&\includegraphics[height=40mm,width=40mm]{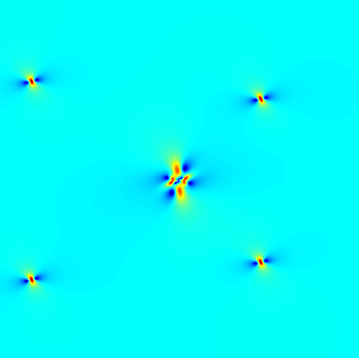}&\includegraphics[height=40mm,width=40mm]{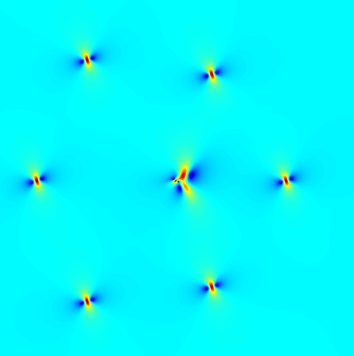}&\rotatebox{0}{\textbf{$t$}}\\
\textbf{$\left|u_2\right|$} & \includegraphics[height=40mm,width=40mm]{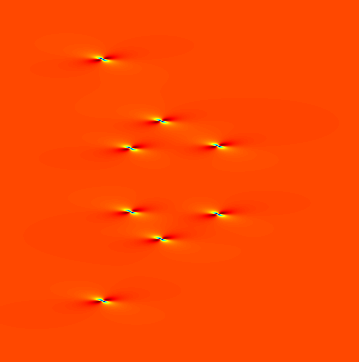}&\includegraphics[height=40mm,width=40mm]{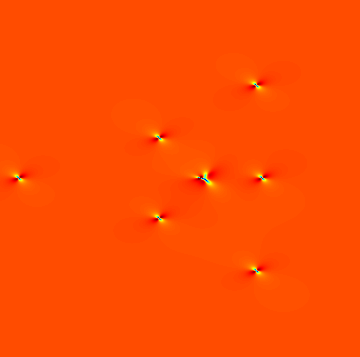}&\includegraphics[height=40mm,width=40mm]{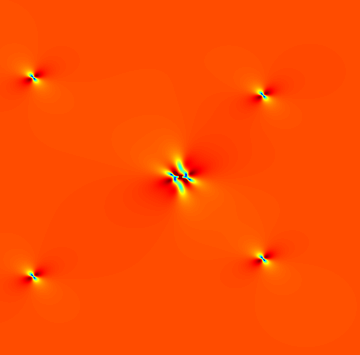}&\includegraphics[height=40mm,width=40mm]{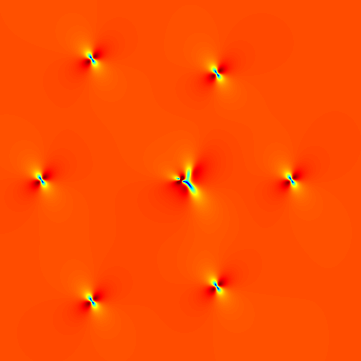}&\rotatebox{0}{\textbf{$t$}}\\
\textbf{$\left|u_3\right|$} & \includegraphics[height=40mm,width=40mm]{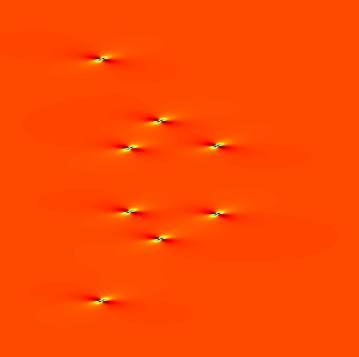}&\includegraphics[height=40mm,width=40mm]{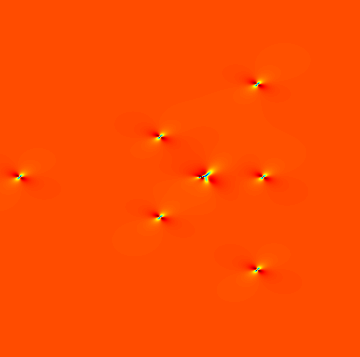}&\includegraphics[height=40mm,width=40mm]{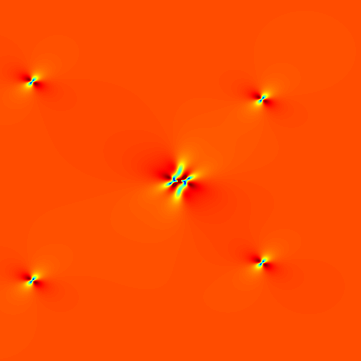}&\includegraphics[height=40mm,width=40mm]{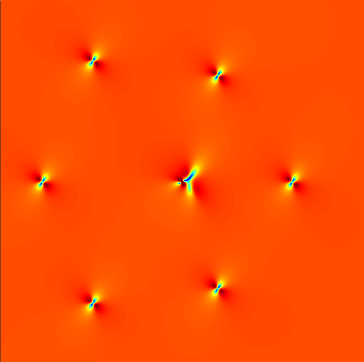}&\rotatebox{0}{\textbf{$t$}}\\
\textbf{$\left|u_4\right|$} & \includegraphics[height=40mm,width=40mm]{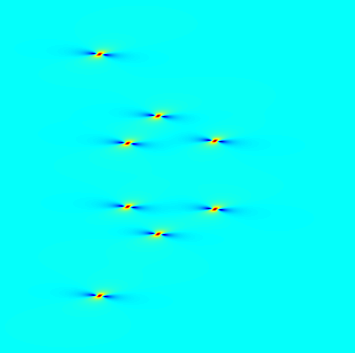}&\includegraphics[height=40mm,width=40mm]{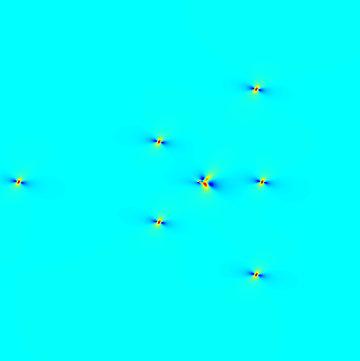}&\includegraphics[height=40mm,width=40mm]{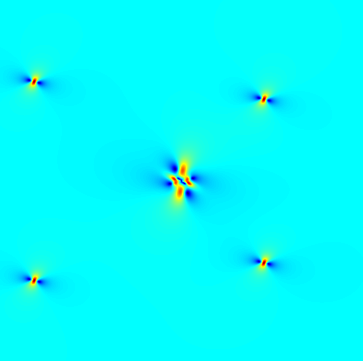}&\includegraphics[height=40mm,width=40mm]{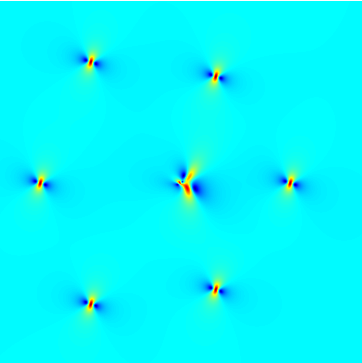}&\rotatebox{0}{\textbf{$t$}}\\
& $x$ & $x$ & $x$ & $x$
\end{tabular}
}
\caption{Predicted $3$rd type $(0,0,2,0)$-th order rogue waves of the four-component NLS equation from Theorem \ref{Rogue wave patterns-4com}. Each column corresponds to a rogue wave with a single large parameter $a_m$, whose value is indicated on top, and all other internal parameters are set to zero. Top row: predicted $\left(\bar{x}_0, \bar{t}_0\right)$ locations by formulae \eqref{prediction of locations of RW-4com}. Second row: predicted $\left|u_1(x, t)\right|$.  Third row: predicted $\left|u_2(x, t)\right|$.  Fourth row: predicted $\left|u_3(x, t)\right|$. Bottom row: predicted $\left|u_4(x, t)\right|$. First column: the $(x, t)$ intervals are $-25 \leq x \leq 25$, $-40 \leq t \leq 40$. Second column: the $(x, t)$ intervals are $-50 \leq x \leq 40$, $-35 \leq t \leq 35$. Third column: the $(x, t)$ intervals are $-30 \leq x \leq 30$, $-20 \leq t \leq 20$. Fourth column: the $(x, t)$ intervals are $-30 \leq x \leq 30$, $-15 \leq t \leq 15$.}
\label{3rd-type_RW_N2_P-4com}
\end{figure}

For $[u_{1, \mathcal{N}_1}(x, t), u_{2, \mathcal{N}_1}(x, t), u_{3, \mathcal{N}_1 }(x, t), u_{4, \mathcal{N}_1 }(x, t)]$, their corresponding predicted rogue wave patterns are shown in the last four rows of Fig. \ref{3rd-type_RW_N2_P-4com}, with the first row being the locations of the rogue waves. Each column is separated when one of the parameters $(a_2, a_3,a_4, a_6)$ is large.

As depicted in Fig. \ref{3rd-type_RW_N2_P-4com}, the large-$a_2$ solution exhibits a skewed double-triangle, corresponding to the double-triangle root structure of $W_2^{[2,5,2]}(z)$. The large-$a_3$ solution exhibits a skewed triple-triangle, corresponding to the triple-triangle root structure of $W_2^{[3,5,2]}(z)$. The large-$a_4$ solution exhibits a deformed rectangle, corresponding to the rectangle-shaped root structure of $W_2^{[4,5,2]}(z)$. The large-$a_6$ solution exhibits a deformed hexagon, corresponding to the hexagon-shaped root structure of $W_2^{[6,5,2]}(z)$. %As far as we know, this {\color{red}triple-triangle} is a new type of pattern.

\begin{figure}[h]
\centering
\renewcommand\arraystretch{0.5}
\setlength\tabcolsep{0pt}
\resizebox{\linewidth}{!}{
\begin{tabular}{m{0.6cm}<{\centering}m{4.1cm}<{\centering}m{4.1cm}<{\centering}m{4.1cm}<{\centering}m{4.1cm}<{\centering}m{0.6cm}<{\centering}r}
&\textbf{$a_2=50$}  & \textbf{$a_3= 200$} & \textbf{$a_4=500 $}& \textbf{$a_6=5000 $}&\\
\textbf{$\left|u_1\right|$} & \includegraphics[height=40mm,width=40mm]{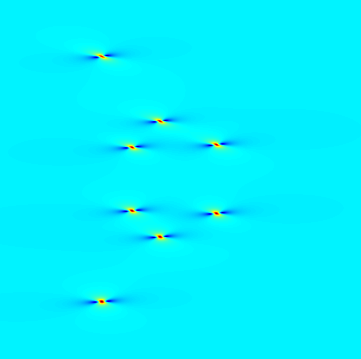}&\includegraphics[height=40mm,width=40mm]{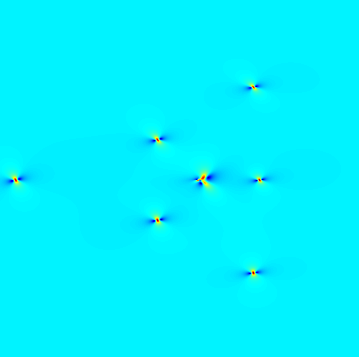}&\includegraphics[height=40mm,width=40mm]{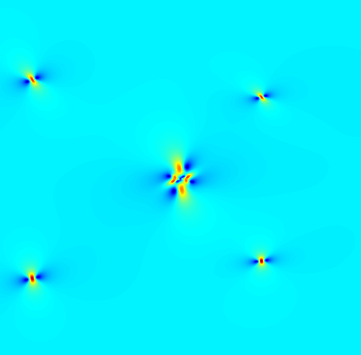}&\includegraphics[height=40mm,width=40mm]{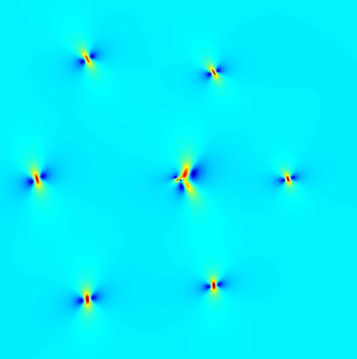}&\rotatebox{0}{\textbf{$t$}}\\
\textbf{$\left|u_2\right|$} & \includegraphics[height=40mm,width=40mm]{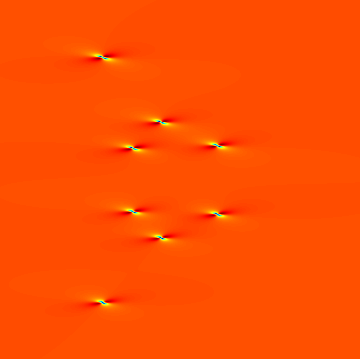}&\includegraphics[height=40mm,width=40mm]{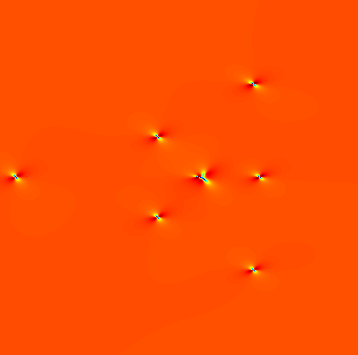}&\includegraphics[height=40mm,width=40mm]{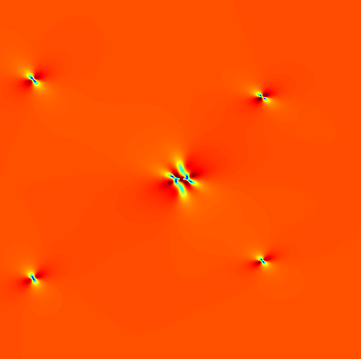}&\includegraphics[height=40mm,width=40mm]{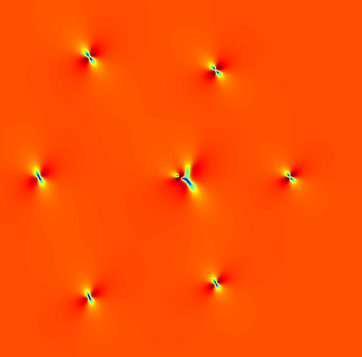}&\rotatebox{0}{\textbf{$t$}}\\
\textbf{$\left|u_3\right|$} & \includegraphics[height=40mm,width=40mm]{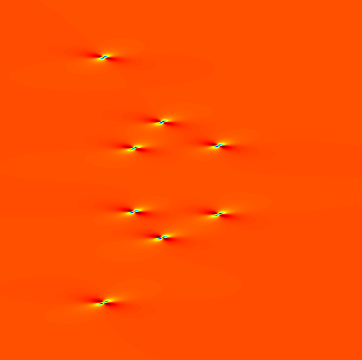}&\includegraphics[height=40mm,width=40mm]{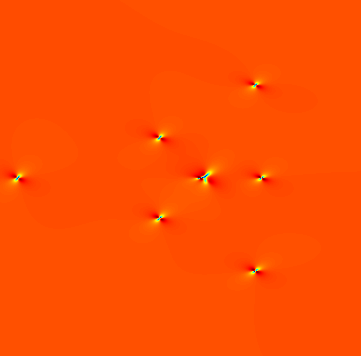}&\includegraphics[height=40mm,width=40mm]{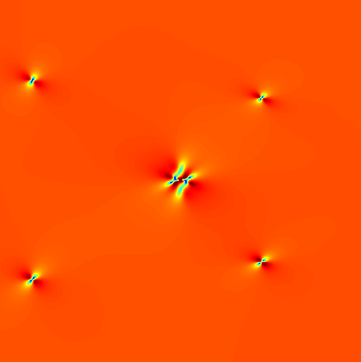}&\includegraphics[height=40mm,width=40mm]{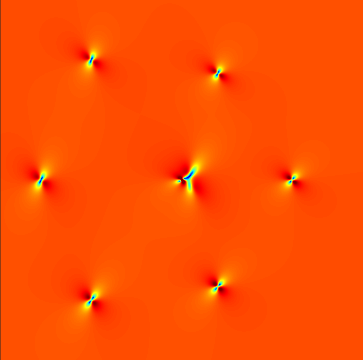}&\rotatebox{0}{\textbf{$t$}}\\
\textbf{$\left|u_4\right|$} & \includegraphics[height=40mm,width=40mm]{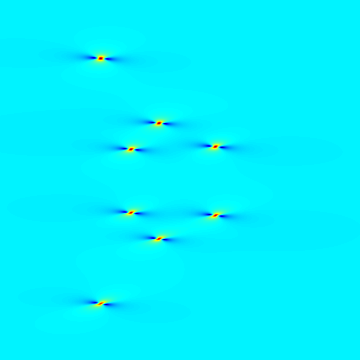}&\includegraphics[height=40mm,width=40mm]{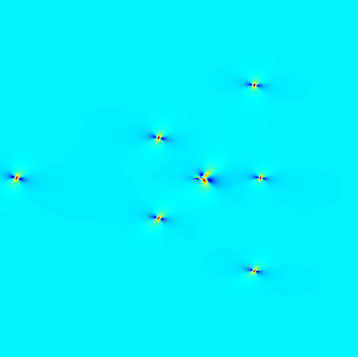}&\includegraphics[height=40mm,width=40mm]{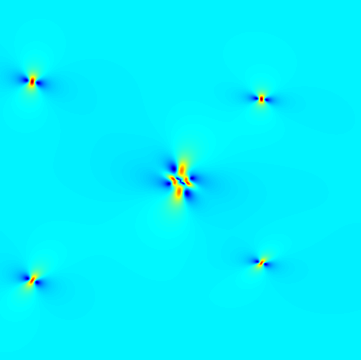}&\includegraphics[height=40mm,width=40mm]{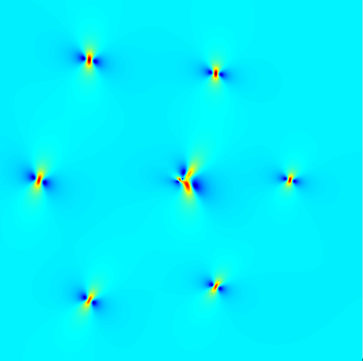}&\rotatebox{0}{\textbf{$t$}}\\
& $x$ & $x$ & $x$ & $x$
\end{tabular}
}
\caption{True $3$rd type $(0,0,2,0)$-th order rogue waves of the four-component NLS equation with the same parameters as Fig. \ref{3rd-type_RW_N2_P-4com}.
% Each column corresponds to a rogue wave with a single large parameter $a_m$, whose value is indicated on top, and all other internal parameters are set to zero. Top row: true $\left|u_1(x, t)\right|$. Middle row:true $\left|u_2(x, t)\right|$. Bottom row: true $\left|u_3(x, t)\right|$.
The $(x, t)$ interval for each column is the same as the corresponding column in Fig. \ref{3rd-type_RW_N2_P-4com}.
% First column: the $(x, t)$ intervals are $-25 \leq x \leq 25$, $-40 \leq t \leq 40$. Second column: the $(x, t)$ intervals are $-50 \leq x \leq 40$, $-35 \leq t \leq 35$. Third column: the $(x, t)$ intervals are $-30 \leq x \leq 30$, $-20 \leq t \leq 20$. Fourth column: the $(x, t)$ intervals are $-30 \leq x \leq 30$, $-15 \leq t \leq 15$.
}
\label{3rd-type_RW_N2_T-4com}
\end{figure}

Comparing the actual rogue waves with the predicted rogue waves (see Figs. \ref{3rd-type_RW_N2_P-4com} and \ref{3rd-type_RW_N2_T-4com}), we can observe that each of the rogue waves matches perfectly in terms of position and rogue wave shape. %Although there are a few rogue waves that will be slightly misplaced, this is due to our approximation error.
Notice that the predicted pattern looks very different from the root structure of $W_2^{[m,5,2]}(z)$. This is caused by $\bar{\Delta}_3$, which leads to a nonlinear transformation from the root structure. When we take $|a_m|$ very large, the term $\bar{\Delta}_3$ can be neglected, and our pattern becomes more similar to a certain linear transformation of the root structure of $W_2^{[m,5,2]}(z)$.
The numerical results also match very well, and as they are very similar to the previous error analysis, we omit the details.

\subsubsection{Fourth-type four-component NLS equation rogue wave solution}
In this circumstance, we consider $(0,0,0,2)$-th order rogue waves. For brevity, we only let one of the internal parameters $(a_2, a_3, a_4, a_6)$ be large and the others are set to $0$. The very large parameter is one of
\begin{equation}
    a_2=50,\quad a_3=200,\quad a_4=500,\quad a_5=5000.
\end{equation}
According to Theorem \ref{Rogue wave patterns-4com}, the position $(\bar{x}_0, \bar{t}_0)$ of each fundamental rogue wave
$$
u_{1, \mathcal{N}_4}(x, t), \quad u_{2, \mathcal{N}_4}(x, t), \quad u_{3, \mathcal{N}_4 }(x, t), \quad u_{4, \mathcal{N}_4 }(x, t)
$$
can be predicted by \eqref{prediction of locations of RW-4com}. The possible $(N_1, N_2, N_3, N_4)$-th order rogue wave would appear in the inner region, and the $(N_1, N_2, N_3, N_4)$ values for these five rogue wave solutions are obtained from Theorems \ref{Rogue wave patterns-4com} and \ref{root sturcture of jump 5} as
$$
(N_1,N_2,N_3,N_4) = (0,0,0,0), \quad(0,0,0,0),\quad(0,0,1,1),\quad(0,0,0,0),
$$
respectively. Note that $(0,0,0,0)$ means that there are no lower-order rogue waves in the center region. For the same reason as previous cases, the internal parameters in these predicted lower $(N_1, N_2, N_3, N_4)$-th order rogue waves of the center region are all taken to be zero.

\begin{figure}[htp]
\centering
\renewcommand\arraystretch{0.5}
\setlength\tabcolsep{0pt}
\resizebox{\linewidth}{!}{
\begin{tabular}{m{0.6cm}<{\centering}m{4.1cm}<{\centering}m{4.1cm}<{\centering}m{4.1cm}<{\centering}m{4.1cm}<{\centering}m{0.6cm}<{\centering}r}
&\textbf{$a_2=50$}  & \textbf{$a_3= 200$} & \textbf{$a_4=500 $}& \textbf{$a_6=5000 $}&\\
\rotatebox{90}{\text{ predicted locations }} & \includegraphics[height=40mm,width=40mm]{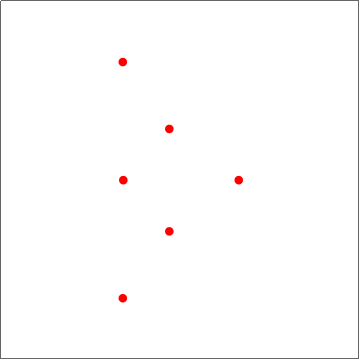}&\includegraphics[height=40mm,width=40mm]{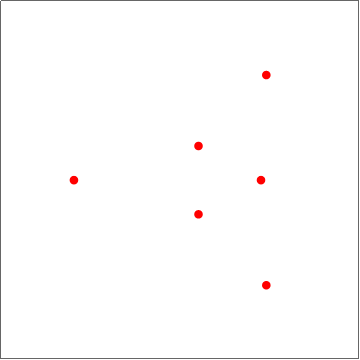}&\includegraphics[height=40mm,width=40mm]{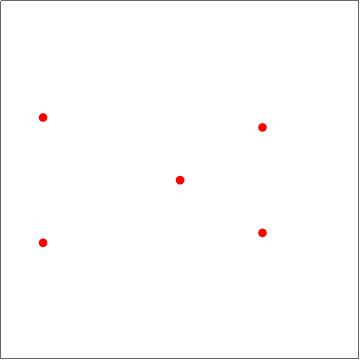}&\includegraphics[height=40mm,width=40mm]{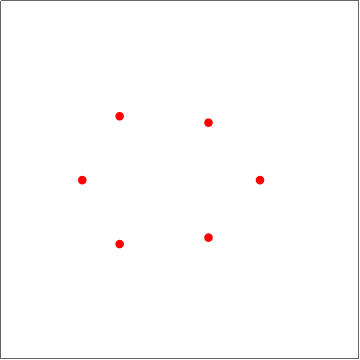}&\rotatebox{0}{\textbf{$t$}}\\
\textbf{$\left|u_1\right|$} & \includegraphics[height=40mm,width=40mm]{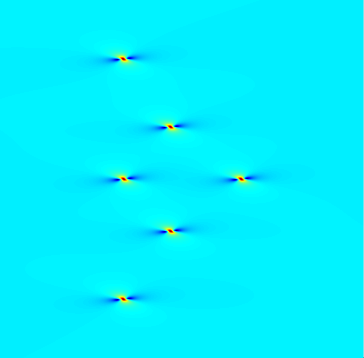}&\includegraphics[height=40mm,width=40mm]{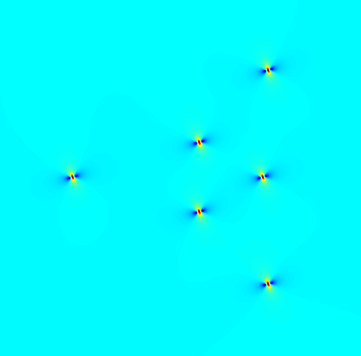}&\includegraphics[height=40mm,width=40mm]{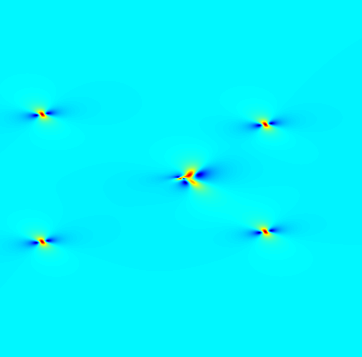}&\includegraphics[height=40mm,width=40mm]{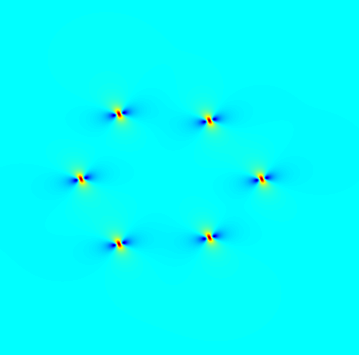}&\rotatebox{0}{\textbf{$t$}}\\
\textbf{$\left|u_2\right|$} & \includegraphics[height=40mm,width=40mm]{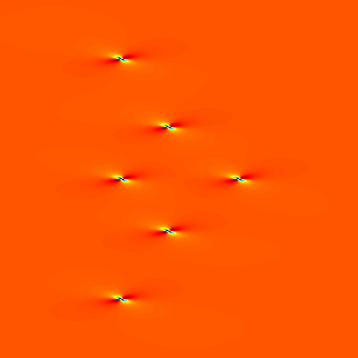}&\includegraphics[height=40mm,width=40mm]{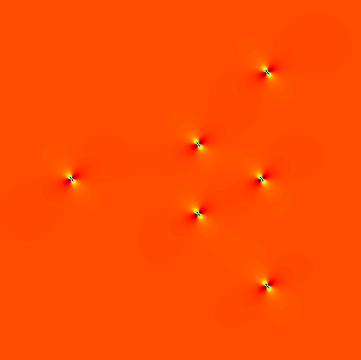}&\includegraphics[height=40mm,width=40mm]{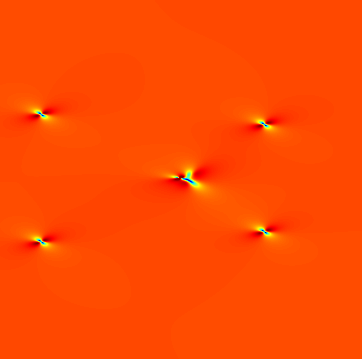}&\includegraphics[height=40mm,width=40mm]{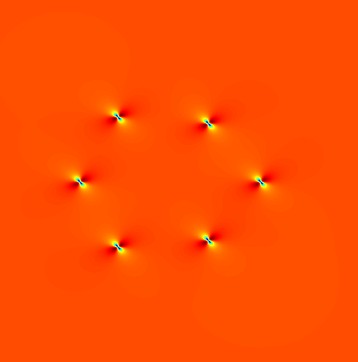}&\rotatebox{0}{\textbf{$t$}}\\
\textbf{$\left|u_3\right|$} & \includegraphics[height=40mm,width=40mm]{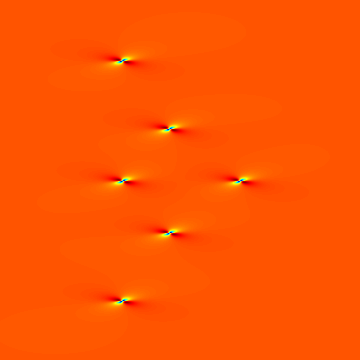}&\includegraphics[height=40mm,width=40mm]{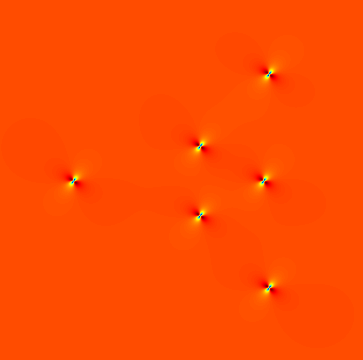}&\includegraphics[height=40mm,width=40mm]{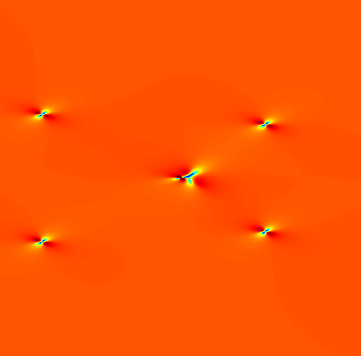}&\includegraphics[height=40mm,width=40mm]{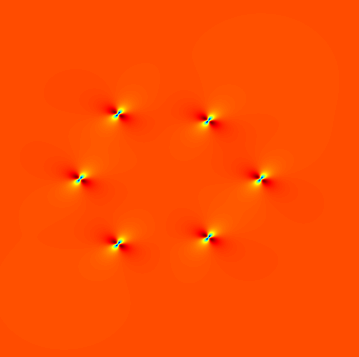}&\rotatebox{0}{\textbf{$t$}}\\
\textbf{$\left|u_4\right|$} & \includegraphics[height=40mm,width=40mm]{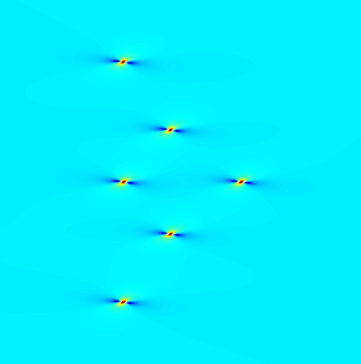}&\includegraphics[height=40mm,width=40mm]{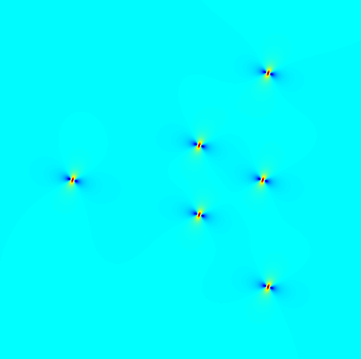}&\includegraphics[height=40mm,width=40mm]{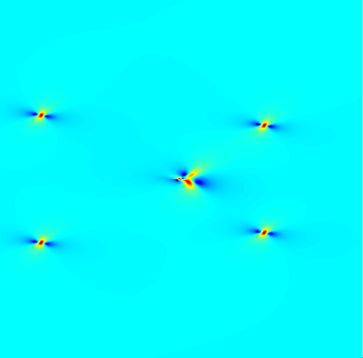}&\includegraphics[height=40mm,width=40mm]{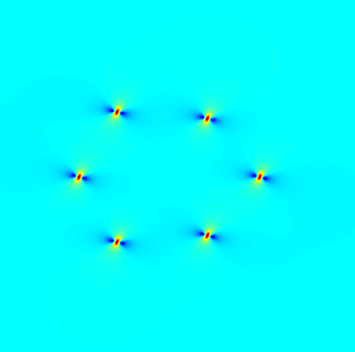}&\rotatebox{0}{\textbf{$t$}}\\
& $x$ & $x$ & $x$ & $x$
\end{tabular}
}
\caption{Predicted $4$th type $(0,0,0,2)$-th order rogue waves of the four-component NLS equation from Theorem \ref{Rogue wave patterns-4com}. Each column corresponds to rogue waves with a single large parameter $a_m$, whose value is indicated on top, and all other internal parameters are set to zero. Top row: predicted $\left(\bar{x}_0, \bar{t}_0\right)$ locations by formulae \eqref{prediction of locations of RW-4com}. Second row: predicted $\left|u_1(x, t)\right|$.  Third row: predicted $\left|u_2(x, t)\right|$.  Fourth row: predicted $\left|u_3(x, t)\right|$. Bottom row: predicted $\left|u_4(x, t)\right|$. First column: the $(x, t)$ intervals are $-25 \leq x \leq 25$, $-35 \leq t \leq 35$. Second column: the $(x, t)$ intervals are $-55 \leq x \leq 35$, $-25 \leq t \leq 25$. Third column: the $(x, t)$ intervals are $-25 \leq x \leq 25$, $-25 \leq t \leq 25$. Fourth column: the $(x, t)$ intervals are $-30 \leq x \leq 30$, $-20 \leq t \leq 20$.}
\label{4th-type_RW_N2_P-4com}
\end{figure}

For $[u_{1, \mathcal{N}_2}(x, t), u_{2, \mathcal{N}_2}(x, t), u_{3, \mathcal{N}_2 }(x, t)]$, their corresponding predicted rogue wave patterns are shown in the last three rows of Fig. \ref{4th-type_RW_N2_P-4com}, with the first row being the predicted locations of the rogue waves. It can be seen from Fig. \ref{4th-type_RW_N2_P-4com} that the first to fourth columns are skewed double-triangles, skewed triple-triangles, rectangles and  hexagons, respectively.

\begin{figure}[h]
\centering
\renewcommand\arraystretch{0.5}
\setlength\tabcolsep{0pt}
\resizebox{\linewidth}{!}{
\begin{tabular}{m{0.6cm}<{\centering}m{4.1cm}<{\centering}m{4.1cm}<{\centering}m{4.1cm}<{\centering}m{4.1cm}<{\centering}m{0.6cm}<{\centering}r}
&\textbf{$a_2=50$}  & \textbf{$a_3= 200$} & \textbf{$a_4=500 $}& \textbf{$a_6=5000 $}&\\
\textbf{$\left|u_1\right|$} & \includegraphics[height=40mm,width=40mm]{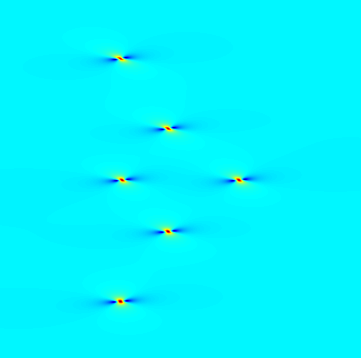}&\includegraphics[height=40mm,width=40mm]{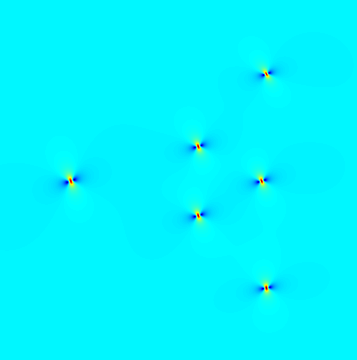}&\includegraphics[height=40mm,width=40mm]{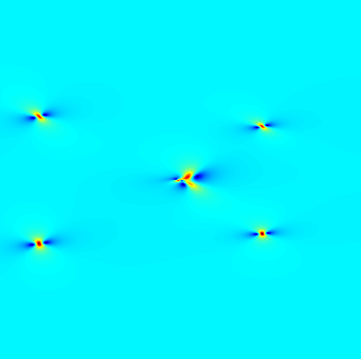}&\includegraphics[height=40mm,width=40mm]{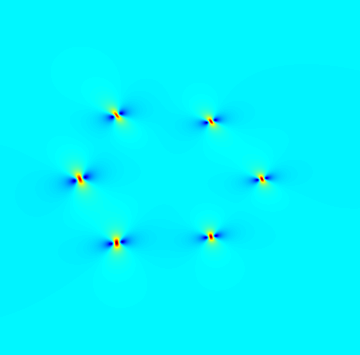}&\rotatebox{0}{\textbf{$t$}}\\
\textbf{$\left|u_2\right|$} & \includegraphics[height=40mm,width=40mm]{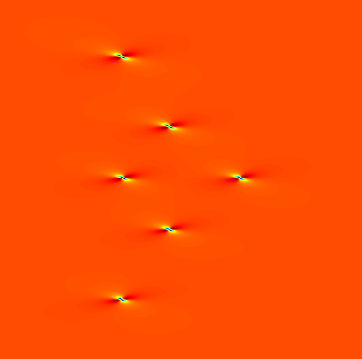}&\includegraphics[height=40mm,width=40mm]{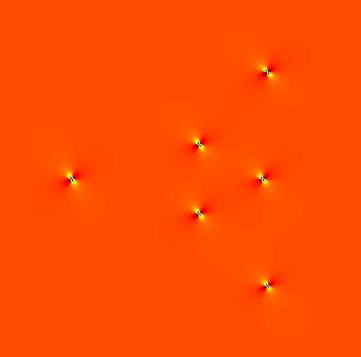}&\includegraphics[height=40mm,width=40mm]{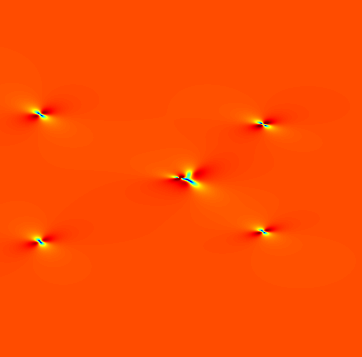}&\includegraphics[height=40mm,width=40mm]{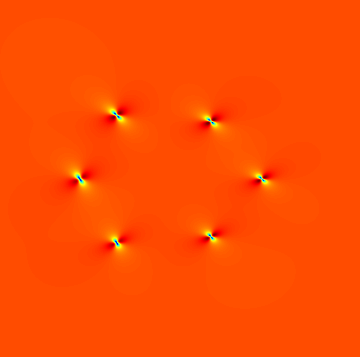}&\rotatebox{0}{\textbf{$t$}}\\
\textbf{$\left|u_3\right|$} & \includegraphics[height=40mm,width=40mm]{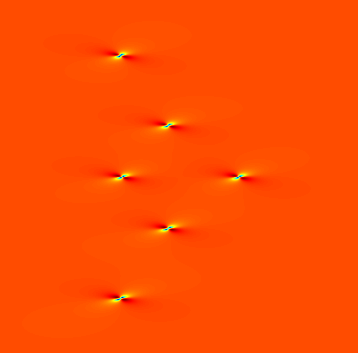}&\includegraphics[height=40mm,width=40mm]{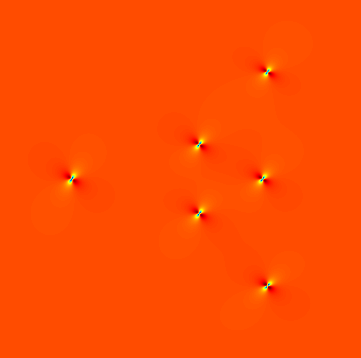}&\includegraphics[height=40mm,width=40mm]{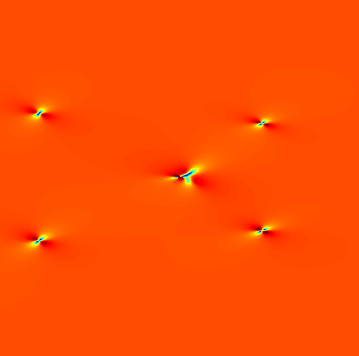}&\includegraphics[height=40mm,width=40mm]{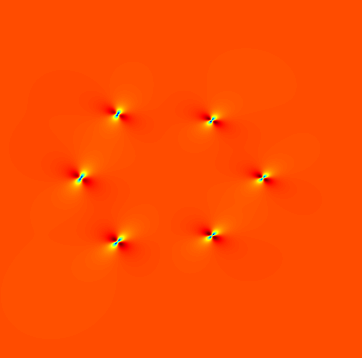}&\rotatebox{0}{\textbf{$t$}}\\
\textbf{$\left|u_4\right|$} & \includegraphics[height=40mm,width=40mm]{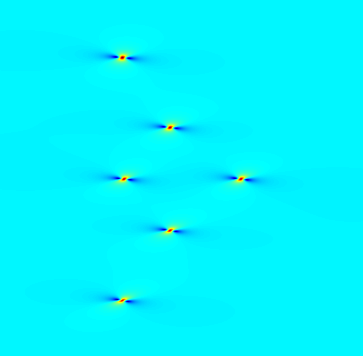}&\includegraphics[height=40mm,width=40mm]{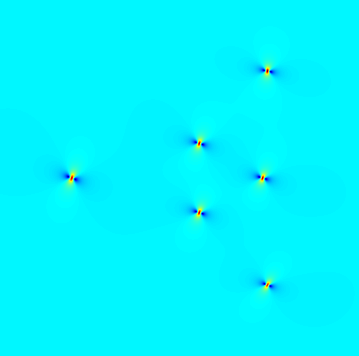}&\includegraphics[height=40mm,width=40mm]{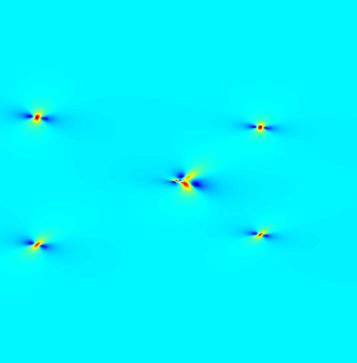}&\includegraphics[height=40mm,width=40mm]{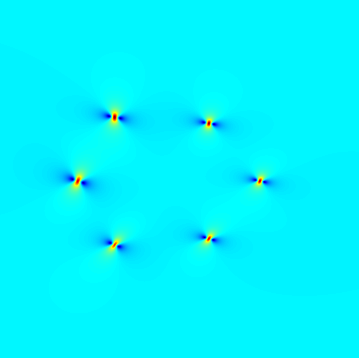}&\rotatebox{0}{\textbf{$t$}}\\
& $x$ & $x$ & $x$ & $x$
\end{tabular}
}
\caption{True $4$th type $(0,0,0,2)$-th order rogue waves of the four-component NLS equation with the same parameters as Fig. \ref{4th-type_RW_N2_P-4com}.
% Each column corresponds to rogue waves with a single large parameter $a_m$, whose value is indicated on top, and all other internal parameters are set to zero. Top row: true $\left|u_1(x, t)\right|$. Second row:true $\left|u_2(x, t)\right|$. Third row:true $\left|u_3(x, t)\right|$. Bottom row: true $\left|u_4(x, t)\right|$.
The $(x, t)$ interval for each column is the same as the corresponding column in Fig. \ref{4th-type_RW_N2_P-4com}.
% First column: the $(x, t)$ intervals are $-25 \leq x \leq 25$, $-35 \leq t \leq 35$. Second column: the $(x, t)$ intervals are $-55 \leq x \leq 35$, $-25 \leq t \leq 25$. Third column: the $(x, t)$ intervals are $-25 \leq x \leq 25$, $-25 \leq t \leq 25$. Fourth column: the $(x, t)$ intervals are $-30 \leq x \leq 30$, $-20 \leq t \leq 20$.
}
\label{4th-type_RW_N2_T-4com-T}
\end{figure}

Comparing the true rogue waves with predicted ones (see Figs. \ref{4th-type_RW_N2_P-4com} and \ref{4th-type_RW_N2_T-4com-T}), we can observe that each of the rogue waves matches perfectly in terms of position and rogue wave shape. The results of the numerical analysis also match very well, but we omit the details because they are very similar to the previous error analysis.

\section{Proof of the main results} \label{section-Proof of the main results}
%\begin{proof}

\noindent {\bf Proof of Theorem \ref{Rogue wave patterns-3com}}.  We will only provide the proof for $i=1$ as the proofs are similar in other cases.
Assume $\left|a_m\right|$ is large and the rest parameters are $O(1)$ in the 1st type rogue wave solutions of the three-component NLS equation.
We first consider the case when $\left(x,t\right)$ is far away from the origin and $(x^2+t^2)^{1/2}=O(\left|a_m\right|^{1/m})$. In this circumstance, we have
\begin{equation} \label{Simplified Schur polynomial-3com}
\begin{aligned}
S_j\left(\boldsymbol{x}^{+}(\mathbf{n})+\nu \boldsymbol{s}\right)&=S_j\left(x_1^{+}, x_2^{+}, x_3^{+},\nu s_4, x_5^{+}, x_6^{+}, x_7^{+}, \nu s_8, \cdots, x_m^{+}+\nu s_m, \cdots\right) \sim S_j(\mathbf{v}),
\end{aligned}\end{equation}
where
$$
\mathbf{v}=\left(p_0 x+2p_0p_1 \mathrm{i} t, 0, \cdots, 0, a_m, 0, \cdots\right).
$$
According to Remark \ref{Values of s_r}, we have $s_1=s_2=s_3=s_5=s_6=s_7=0$.
By the definition of Schur polynomials, we have the relation
\begin{equation} \label{Schur polynomial VS p1-3com}
S_j(\mathbf{v})=a_m^{j / m} p_j^{[m]}(z),
\end{equation}
where
\begin{equation}
z=a_m^{-1 / m}\left(\alpha_1 x+\beta_1 \mathrm{i} t\right)=a_m^{-1 / m}\left(p_0 x + 2 p_0 p_1 \mathrm{i} t\right).
\end{equation}
Then it follows that
\begin{equation} \label{Schur polynomial VS WH x+ -3com}
 \det_{1 \leq i, j \leq N}\left[S_{4 i-j}\left(\boldsymbol{x}^{+}(\mathbf{n})+\nu_j \boldsymbol{s}\right)\right] \sim (c_N^{[m,4,3]})^{-1} a_m^{3N(N+1) / 2m} W_{N}^{[m,4,3]}(z)
\end{equation}
and
\begin{equation}
 \det_{1 \leq i, j \leq N}\left[S_{4 i-j}\left(\boldsymbol{x}^{-}(\mathbf{n})+\nu_j \boldsymbol{s}^*\right)\right] \sim (c_N^{[m,4,3]})^{-1} \left(a_m^*\right)^{3N(N+1) / 2m} W_{N}^{[m,4,3]}(z^*). \end{equation}
Here, $S_{j}\equiv0$ when $j<0$.

Next, we rewrite the function $\tau_{\mathbf{n}}$ into the following form  by Laplace expansion
\begin{equation} \label{Laplace expansion of tau function-3com}
  \begin{gathered}
\tau_{\mathbf{n}}=\sum_{0 \leq \nu_1<\nu_2<\cdots<\nu_N \leq 4 N-1}
 \det_{1 \leq i, j \leq N}\left[\left(h_0\right)^{\nu_j} S_{4 i-1-\nu_j}\left(\boldsymbol{x}^{+}(\mathbf{n})+\nu_j s\right)\right] \\
\times \det_{1 \leq i, j \leq N}\left[\left(h_0^*\right)^{\nu_j} S_{4 i-1-\nu_j}\left(\boldsymbol{x}^{-}(\mathbf{n})+\nu_j s^*\right)\right],
\end{gathered}
\end{equation}
where $h_0=p_1 /\left(p_0+p_0^*\right)$.

It is clear that the highest order term in $a_m$ in this $\tau_{\mathbf{n}}$ comes from the index choices of $\nu_j=j-1$. Therefore, we have
\begin{equation}
\tau_{\mathbf{n}}\sim|\alpha|^2\left|a_m\right|^{3 N(N+1) / m}\left|W_{N}^{[m,4,3]}(z)\right|^2,
\end{equation}
where $\alpha=h_0^{N(N-1)/2} (c_N^{[m,4,3]})^{-1}$. From the asymptotic analysis above, we conclude that the leading-order term of $\tau_{\mathbf{n}}$ is independent of $\mathbf{n}$. Consequently, when  $(x,t)$ is not close to  $(\check{x}_0,\check{t}_0)$, which is related to the roots of $W_{N}^{[m,4,3]}(z)$ by
$$z_0=a_{m}^{-1/m} \left(p_0 \check{x}_0 +2p_0p_1\mathrm{i} \check{t}_0\right),$$ we have
\begin{equation}
   \frac{\tau_{\mathbf{n}_1}}{\tau_{\mathbf{n}_0}} \sim 1, \quad \frac{\tau_{\mathbf{n}_2}}{\tau_{\mathbf{n}_0}} \sim 1, \quad
   \frac{\tau_{\mathbf{n}_3}}{\tau_{\mathbf{n}_0}} \sim 1, \quad |a_m| \gg 1.
\end{equation}

However, when  $(x,t)$ is close to  $(\check{x}_0,\check{t}_0)$, the coefficient of the term with highest order in $a_m$ vanishes. To deal with this case, we have to consider lower order terms in $a_m$, which require more precise asymptotics. In this circumstance, i.e., $(x,t)$ is near  $(\check{x}_0,\check{t}_0)$, we find that
\begin{eqnarray*}
&&S_j\left(\boldsymbol{x}^{+}(\mathbf{n})+\nu \boldsymbol{s}\right)
\\
&=&S_j\left(x_1^{+}, x_2^{+}, x_3^{+},\nu s_4, x_5^{+}, x_6^{+}, x_7^{+},\nu s_8, \cdots, x_m^{+}+\nu s_m, \cdots\right)
\\
&=&\left[S_j(\hat{\mathbf{v}})+\hat{x}_2^{+}\left(\check{x}_0, \check{t}_0\right) S_{j-2}(\hat{\mathbf{v}})\right]\left[1+O\left(a_m^{-2 / m}\right)\right], \quad\left|a_m\right| \gg 1,
\end{eqnarray*}
where
\begin{eqnarray*}
\hat{\mathbf{v}}&=&\left(p_0 x+2p_0p_1 \mathrm{i} t+ n_1 \theta_{11} + n_2 \theta_{12}+ n_3 \theta_{13}, 0, \cdots, 0, a_m, 0, \cdots\right),\\
\hat{x}_2^{+}(x, t)&=&p_2 x+\left(2 p_0 p_2+p_1^2\right)(\mathrm{i} t),
\end{eqnarray*}
 $p_2 = p_1^2$
 and $a_1$ in $x_1^+$ is set to 0. Similar to \eqref{Schur polynomial VS p1-3com}, we can get
\begin{equation}
S_j(\hat{\mathbf{v}})=a_m^{j / m} p_j^{[m]}(\hat{z}),
\end{equation}
where
\begin{equation}
\hat{z}=a_m^{-1 / m}\left(p_0 x + 2 p_0 p_1 \mathrm{i} t+ n_1 \theta_{11} + n_2 \theta_{12}+ n_3 \theta_{13}\right).
\end{equation}
In this case, there are two index choices of $\nu_j$ that will produce leading-order terms in $a_m$ for $\tau_{\mathbf{n}}$. One of them is $\nu=(0,1,\cdots,N-1)$ while the other is $\nu=(0,1,\cdots,N-2,N)$.

(1) For the first choice of index, i.e., $\nu_j=j-1$, there are two parts that will provide leading-order terms. The first part stems from $S_j(\hat{\mathbf{v}})$, and we find that the dominant term involving $\boldsymbol{x}^{+}(\mathbf{n})$ is expressed as
%from the asymptotic expression in the determinants,
%we will obtain similar form like what we get above but with a different $\hat{z}$, that is
\begin{equation}%\label{}
  \alpha~ a_m^{\frac{3N(N+1)}{ 2m}} W_{N}^{[m,4,3]}(\hat{z})\left[1+O\left(a_m^{-2 / m}\right)\right].
\end{equation}
Then, we expand $W_{N}^{[m,4,3]}(\hat{z})$ around $z_0$, and noting $W_{N}^{[m,4,3]}(z_0)=0$, we obtain
\begin{equation}%\label{}
  W_{N}^{[m,4,3]}(\hat{z})=a_m^{-1 / m}\left[p_0\left(x-\check{x}_0\right)+ 2p_0p_1 \mathrm{i} \left(t-\check{t}_0\right)+ n_1 \theta_{11} + n_2 \theta_{12}+ n_3 \theta_{13}\right]\left[W_{N}^{[m,4,3]}\right]^{\prime}\left(z_0\right)\left[1+O\left(a_m^{-1 / m}\right)\right].
\end{equation}
As a result, the corresponding leading-order term in $a_m$ is
\begin{equation}\label{1 index-part 1}
  \alpha a_m^{\frac{3N(N+1)-2}{2m}}
\left[p_0\left(x-\check{x}_0\right)+ 2p_0p_1 \mathrm{i} \left(t-\check{t}_0\right)+ n_1 \theta_{11} + n_2 \theta_{12}+ n_3 \theta_{13}\right]  \left[W_{N}^{[m,4,3]}\right]^{\prime}\left(z_0\right)
 \left[1+O\left(a_m^{-1 / m}\right)\right].
\end{equation}
The other leading-order term results from the determinants containing $\hat{x}_2^{+} (\check{x}_0,\check{t}_0) S_{j-2}$, that is,
\begin{eqnarray}
  &&\hat{x}_2^{+} (\check{x}_0,\check{t}_0) h_0^{N(N-1)/2} \sum_{j=1}^N \det_{1 \leq i \leq N}\left[S_{4 i-1}(\hat{\mathbf{v}}), \cdots,  S_{4 i-(j-1)}(\hat{\mathbf{v}}), S_{4 i-j-2}(\hat{\mathbf{v}}), S_{4 i-(j+1)}(\hat{\mathbf{v}}), \cdots, S_{4 i-N}(\hat{\mathbf{v}})\right] \nonumber
  \\
  &&   \times \left[1+O\left(a_m^{-1 / m}\right)\right] .  \label{1 index-part 2}
\end{eqnarray}
Combing \eqref{1 index-part 1} and \eqref{1 index-part 2} yields the leading-order term in $a_m$ \cite{yang2022rogue} of the first determinant in \eqref{Laplace expansion of tau function-3com} containing $\boldsymbol{x}^{+}(\mathbf{n})$ corresponding to the index choice $\nu_j=j-1$, that is,
%
%first determinant involving $x^{+}(n,k,l)$ from the first index choices  will produce such a leading term in one of a determinant :
\begin{equation*}
  \alpha a_m^{[3N(N+1)-2] / 2m}\left[p_0\left(x-\check{x}_0\right)+2p_0p_1 \mathrm{i}\left(t-\check{t}_0\right)+ n_1 \theta_{11} + n_2 \theta_{12}+ n_3 \theta_{13}+\Delta_1\right]\left[W_{N}^{[m,4,3]}\right]^{\prime}\left(z_0\right)\left[1+O\left(a_m^{-1 / m}\right)\right]
\end{equation*}
where
\begin{equation}%\label{}
 \Delta_1=\frac{\hat{x}_2^{+}(\check{x}_0,\check{t}_0)}{a_m^{1 / m}} \frac{\sum_{j=1}^N \det_{1 \leq i \leq N}\left[p_{4 i-1}^{[m]}\left(z_0\right), \cdots, p_{4 i-j+1}^{[m]}\left(z_0\right), p_{4 i-j-2}^{[m]}\left(z_0\right), p_{4 i-j-1}^{[m]}\left(z_0\right), \cdots, p_{4 i-N}^{[m]}\left(z_0\right)\right]}{\left[W_{N}^{[m,4,3]}\right]^{\prime}\left(z_0\right)}
\end{equation}
and $\Delta_1=O(1)$ as
$$ \hat{x}_2^{+}(\check{x}_0,\check{t}_0) = p_2 \check{x}_0+\left(2 p_0 p_2+p_1^2\right)(\mathrm{i} \check{t}_0) =O(|a_m^{1 / m}|).$$
Further, we can absorb the $\Delta_1$ into $(\check{x}_0,\check{t}_0)$ \cite{yang2022rogue}  and obtain
\begin{equation}%\label{}
  \alpha a_m^{\frac{3N(N+1)-2}{2m}}\left[p_0\left(x-\hat{x}_0\right)+2p_0p_1 \mathrm{i}\left(t-\hat{t}_0\right)+ n_1 \theta_{11} + n_2 \theta_{12}+ n_3 \theta_{13}\right]\left[W_{N}^{[m,4,3]}\right]^{\prime}\left(z_0\right)\left[1+O\left(a_m^{-1 / m}\right)\right].
\end{equation}
where $\hat{x}_0 $ and $\hat{t}_0$ are given in \eqref{prediction of locations of RW-3com}.

Similarly, the second determinant in \eqref{Laplace expansion of tau function-3com} containing $\boldsymbol{x}^{-}(\mathbf{n})$ corresponding to the index choice $\nu_j=j-1$
contributes the term
\begin{equation}%\label{}
  \alpha^{*} (a_m^{*})^{\frac{3N(N+1)-2}{2m}}\left[p_0^{*}\left(x-\hat{x}_0\right)-2p_0^{*}p_1^{*} \mathrm{i}\left(t-\hat{t}_0\right)- n_1 \theta_{11}^{*} - n_2 \theta_{12}^{*}- n_3 \theta_{13}^{*}\right]\left[W_{N}^{[m,4,3]}\right]^{\prime}\left(z_0^{*}\right)\left[1+O\left(a_m^{-1 / m}\right)\right].
\end{equation}

(2) For the second choice of index, i.e., $\nu = (0, 1, \cdots, N-2,N)$, the dominant term in $a_m$ can be calculated in a similar way as \eqref{Schur polynomial VS WH x+ -3com}, that is,
\begin{equation}
h_0^{\frac{N(N-1)+2}{2}} a_m^{\frac{3N(N+1)-2}{2m}} \det_{1 \leq i \leq N}\left[p_{4 i-1}^{[m]}\left(z_0\right), p_{4 i-2}^{[m]}\left(z_0\right), \cdots, p_{4 i-(N-1)}^{[m]}\left(z_0\right), p_{4 i-N-1}^{[m]}\left(z_0\right)\right]\left[1+O\left(a_m^{-1 / m}\right)\right].
\end{equation}
 Since $p_{j-1}^{[m]}(z)=\left[p_j^{[m]}\right]^{\prime}(z)$, the above term can be expressed as
$$h_0 \alpha a_m^{\frac{3N(N+1)-2}{2m}}\left[W_{N}^{[m,4,3]}\right]^{\prime}\left(z_0\right)\left[1+O\left(a_m^{-1 / m}\right)\right] .$$
Similarly, its conjugate counterpart reads
$$
h_0^* \alpha^*\left(a_m^*\right)^{\frac{3N(N+1)-2}{2m}}\left[W_{N}^{[m,4,3]}\right]^{\prime}\left(z_0^*\right)\left[1+O\left(a_m^{-1 / m}\right)\right].
$$
Summarizing the above two contributions, we conclude that
\begin{eqnarray}
\tau_{\mathbf{n}}(x,t)&=&
|\alpha|^2\left|\left[W_{N}^{[m,4,3]}\right]^{\prime}\left(z_0\right)\right|^2\left|a_m\right|^{[3N(N+1)-2] / m} \times
\Big(\left[p_1\left(x-\hat{x}_0\right)+2 \mathrm{i} p_0 p_1\left(t-\hat{t}_0\right)+ n_1 \theta_{11} + n_2 \theta_{12}+ n_3 \theta_{13}\right]  \nonumber \\
&& \left[p_1^*\left(x-\hat{x}_0\right)-2 \mathrm{i} p_0^* p_1^*\left(t-\hat{t}_0\right)- n_1 \theta_{11}^{*} - n_2 \theta_{12}^{*}- n_3 \theta_{13}^{*}\right]+\left|h_0\right|^2\Big) \nonumber
\\
&&
\times
\left[1+O\left(a_m^{-1 / m}\right)\right]. \label{asymptotics of tau fucntion-3com}
\end{eqnarray}
Finally, under the assumption that all nonzero roots of the generalized Wronskian-Hermite polynomials $W_N^{[m,k,l]}$ are simple, the above leading-order term in $a_m$ for $\tau_{\mathbf{n}}(x, t)$ is non-zero. Hence, using \eqref{asymptotics of tau fucntion-3com}, we conclude that, near $(\hat{x}_0, \hat{t}_0)$, the $N$-th order rogue wave is approximated by a fundamental rogue wave of the three-component NLS equation given in Theorem \ref{Rogue wave patterns-3com} with error $O(\left|a_m\right|^{-1/m})$.

In order to study the patterns of the 1st type rogue waves  of the three-component NLS rogue waves under the condition $|a_m| \gg 1$ in the inner region with $x^2+t^2=O(1)$,
we first use similar method as that in \cite{ohta2012general} to rewrite the determinant $\tau_{\mathbf{n}}$ as a $5 N \times 5 N$ determinant
	\begin{equation}\label{three expansion form}
     \tau_{\mathbf{n}}=\left|\begin{array}{cc}
	 \mathbf{O}_{N \times N} & \Phi_{N \times 4 N} \\
	 -\Psi_{4 N \times N} & \mathbf{I}_{4 N \times 4 N}
	 \end{array}\right|,
	\end{equation}
where
		$$
	\Phi_{i, j}=\left(\frac{p_1}{p_0+p_0^*}\right)^{j-1} S_{4 i-j}\left[\boldsymbol{x}^{+}(\mathbf{n})+(j-1) \boldsymbol{s}\right], \quad \Psi_{i, j}=\left(\frac{p_1^*}{p_0+p_0^*}\right)^{i-1} S_{4 j-i}\left[\boldsymbol{x}^{-}(\mathbf{n})+(i-1) \boldsymbol{s}^*\right] .
	$$
It is clear that each element in \eqref{three expansion form} is a polynomial in $a_m$. To express these polynomials explicitly, we define $\boldsymbol{y}^{\pm}$ to be the vector $\boldsymbol{x}^{\pm}$ without the $a_m$ term, i.e.,
	\begin{equation}\label{transformation between x and y}
	\boldsymbol{x}^{+}=\boldsymbol{y}^{+}+\left(0, \cdots, 0, a_m, 0, \cdots\right), \quad \boldsymbol{x}^{-}=\boldsymbol{y}^{-}+\left(0, \cdots, 0, a_m^*, 0, \cdots\right).
	\end{equation}
Then we can expand the Schur polynomials $S_j\left(\boldsymbol{x}^{\pm}+\nu \boldsymbol{s}\right)$ by
	\begin{equation}\label{S expand}
		S_j\left(\boldsymbol{x}^{+}+\nu \boldsymbol{s}\right)=\sum_{l=0}^{[j / m]} \frac{a_m^l}{l !} S_{j-l m}\left(\boldsymbol{y}^{+}+\nu \boldsymbol{s}\right), \quad S_j\left(\boldsymbol{x}^{-}+\nu \boldsymbol{s}^*\right)=\sum_{l=0}^{[j / m]} \frac{\left(a_m^*\right)^l}{l !} S_{j-l m}\left(\boldsymbol{y}^{-}+\nu \boldsymbol{s}^*\right),
	\end{equation}
where $\left[ a\right] $ refers to the largest inter less than or equal to $a$.
To determine the highest order term in $a_m$ of $\tau_{\mathbf{n}}$, a straightforward way is to keep only the highest order of $a_m$ in each element. However, it turns out the resulting determinant will vanish. To tackle this issue, we can use similar argument as that in \cite{yang2021rogue} to perform row and column operations. Notice that we have totally three cases to consider, i.e., $m \equiv j \mod 4, j=1,2,3$.
Since the proof for  $
j=2$ is different from those in the NLS equation \cite{yang2021general} and the Manakov system \cite{yang2021rogue}, we first focus on the proof of this case. As the proofs for the cases $j=1$ and $3$ are similar to \cite{yang2021rogue}, we only provide a brief proof for $j=1$.

For $m \equiv 2 \mod 4$, i.e., $m=4r+2 \, (r\geq0)$, according to the block structure of the determinant $\tau_{\mathbf{n}}$, we can perform row operations on the matrix $\Phi_{N\times 4N}$. For convenience, we define $\hat{S}_j=S_{j}\left(\boldsymbol{y}^{+}+\nu \boldsymbol{s}\right)$ and omit $\left(p_1/(p_0+p_0^*)\right)^{j-1}$ in the following representation because they are the same in each column and have no affect on the row operations. Then, we can substitute \eqref{S expand} into $\Phi_{N \times 4 N}$ and rewrite it into the form

\begin{equation*}
\Phi_{N \times 4 N}  \sim \left[\begin{array}{ccc}
\hat{S}_{3} & \hat{S}_{2} & \cdots \\

\hat{S}_{7} & \hat{S}_{6} & \cdots\\

\vdots & \vdots & \ddots \\

\hat{S}_{m-3} & \hat{S}_{m-4} & \cdots \\

a_m \hat{S}_{1}+\hat{S}_{m+1} & a_m \hat{S}_{0}+\hat{S}_{m} & \cdots\\

a_m \hat{S}_{5}+\hat{S}_{m+5} & a_m \hat{S}_{4}+\hat{S}_{m+4} & \cdots \\

\vdots & \vdots & \ddots \\

a_m \hat{S}_{m-1}+\hat{S}_{2m-1} & a_m \hat{S}_{m-2}+\hat{S}_{2m-2} & \cdots\\

\dfrac{a_m^2}{2!} \hat{S}_{3}+a_m \hat{S}_{m+3}+\hat{S}_{2m+3}& \dfrac{a_m^2}{2!} \hat{S}_{2}+a_m \hat{S}_{m+2}+\hat{S}_{2m+2} & \cdots \\

%\dfrac{a_m^2}{2!} \hat{S}_{7}+a_m \hat{S}_{m+7}+\hat{S}_{2m+7}& a^3 z^0 +a^2 z^m& \cdots & \cdots & \cdots& \cdots & \cdots  & \cdots& \cdots\\

\vdots & \vdots & \ddots\\

\dfrac{a_m^2}{2!} \hat{S}_{m-3}+a_m \hat{S}_{2m-3}+\hat{S}_{3m-3} & \dfrac{a_m^2}{2!} \hat{S}_{m-4}+a_m \hat{S}_{2m-4}+\hat{S}_{3m-4} & \cdots\\

\dfrac{a_m^3}{3!} \hat{S}_{1}+\dfrac{a_m^2}{2!} \hat{S}_{m+1}+ O(a_m)
%\text{lower order terms}
%a_m\hat{S}_{2m+1}+\hat{S}_{3m+1}
& \dfrac{a_m^3}{3!} \hat{S}_{0}+\dfrac{a_m^2}{2!} \hat{S}_{m}+ O(a_m)& \cdots \\

\vdots & \vdots & \ddots\\

\dfrac{a_m^3}{3!} \hat{S}_{m-1}+\dfrac{a_m^2}{2!} \hat{S}_{2m-1}+ O(a_m)%a_m\hat{S}_{2m+4r-1}+\hat{S}_{3m+4r-1}
& \dfrac{a_m^3}{3!} \hat{S}_{m-2}+\dfrac{a_m^2}{2!} \hat{S}_{2m-2}+ O(a_m) & \cdots \\

\dfrac{a_m^4}{4!} \hat{S}_{3}+\dfrac{a_m^3}{3!} \hat{S}_{m+3}+\dfrac{a_m^2}{2!} \hat{S}_{2m+3}+ O(a_m)
%a_m\hat{S}_{3m+3}+\hat{S}_{4m+3}
& \dfrac{a_m^4}{4!} \hat{S}_{2}+\dfrac{a_m^3}{3!} \hat{S}_{m+2}+\dfrac{a_m^2}{2!} \hat{S}_{2m+2}+ O(a_m)& \cdots \\

\vdots & \vdots & \ddots\\

\dfrac{a_m^4}{4!} \hat{S}_{m-3}+\dfrac{a_m^3}{3!} \hat{S}_{2m-3}+\dfrac{a_m^2}{2!} \hat{S}_{3m-3}+ O(a_m)
%a_m\hat{S}_{3m+4r+1}+\hat{S}_{4m+4r+1}
& \dfrac{a_m^4}{4!} \hat{S}_{m-4}+\dfrac{a_m^3}{3!} \hat{S}_{2m-4}+\dfrac{a_m^2}{2!} \hat{S}_{3m-4}+ O(a_m)& \cdots \\

\vdots & \vdots & \ddots\\
\end{array}\right] .
\end{equation*}

In this case, we can notice that the coefficients of the highest $a_m$ power terms in the first column are proportional to
\begin{equation}\label{S sequence (1st,4r+2)}
   \hat{S}_{3},\hat{S}_{7},\cdots, \hat{S}_{m-3} ; \hat{S}_{1},\hat{S}_{5},\cdots, \hat{S}_{m-1}
\end{equation}
and repeating. To be more precise, the first $r$ rows are a sequence starting with $\hat{S}_{3}$, and the subscripts of $\hat{S}$ increase by $4$. The next $r+1$ rows, i.e., rows $r+1$ to $2r+1$, are a sequence starting with $\hat{S}_{1}$, and the subscripts increase by $4$ as well. After that, the subsequent $r$ rows are the sequence starting with a multiple of $\hat{S}_{3}$, followed by $r+1$ rows starting with a multiple of $\hat{S}_{1}$,  and so on and so forth. Each element in the second and higher columns maintains the same form as the elements in the first column, except that the subscripts decreasing by 1, where $\hat{S}_{j}\equiv0$ for $j<0$.

Notice that each $2r+1~(=m/2)$ row circulates a multiple of the sequence \eqref{S sequence (1st,4r+2)} and $N_0 \equiv N \bmod m$, i.e., $N=km+N_0$. Hence, we define the first $m/2$ rows of $\Phi_{N \times 4 N}$ as the first block matrix, the next $m/2$ rows as the second block matrix, and so on. The first $km$ rows consist of $2k$ blocks, and each of these blocks contains two parts. The last remaining $N_0$ rows are called the remaining block matrix, i.e., the $(2k+1)$-th block matrix.

The first round of the row operation is to use the first block to eliminate the highest power term of $a_m$ in each subsequent block, and leaving the lower power terms of $a_m$. This can be achieved by multiplying each row of the first part of the first block matrix by $-a_m^2/(2n-2)!$ and multiplying each row of the second part of the first block matrix by $-a_m^2/(2n-1)!$ and adding them to the corresponding row of the $n$-th block matrix. The resulting $\Phi_{N \times 4 N}$ is

\begin{equation} \label{row operation-matrix 2}
\Phi_{N \times 4 N} \sim \left[\begin{array}{ccc}
\hat{S}_{3} & \hat{S}_{2} & \cdots \\

\hat{S}_{7} & \hat{S}_{6} & \cdots\\

\vdots & \vdots & \ddots \\

\hat{S}_{m-3} & \hat{S}_{m-4} & \cdots \\

a_m \hat{S}_{1}+\hat{S}_{m+1} & a_m \hat{S}_{0}+\hat{S}_{m} & \cdots\\

a_m \hat{S}_{5}+\hat{S}_{m+5} & a_m \hat{S}_{4}+\hat{S}_{m+4} & \cdots \\

\vdots & \vdots & \ddots \\

a_m \hat{S}_{m-1}+\hat{S}_{2m-1} & a_m \hat{S}_{m-2}+\hat{S}_{2m-2} & \cdots\\

a_m \hat{S}_{m+3}+\hat{S}_{2m+3}& a_m \hat{S}_{m+2}+\hat{S}_{2m+2} & \cdots \\

\vdots & \vdots & \ddots\\

a_m \hat{S}_{2m-3}+\hat{S}_{3m-3} & a_m \hat{S}_{2m-4}+\hat{S}_{3m-4} & \cdots\\

\left(\dfrac{1}{2!}-\dfrac{1}{3!}\right)a_m^2 \hat{S}_{m+1}+ O(a_m)
%\text{lower order terms}
%a_m\hat{S}_{2m+1}+\hat{S}_{3m+1}
& \left(\dfrac{1}{2!}-\dfrac{1}{3!}\right)a_m^2 \hat{S}_{m}+ O(a_m)& \cdots \\

\vdots & \vdots & \ddots\\

\left(\dfrac{1}{2!}-\dfrac{1}{3!}\right)a_m^2 \hat{S}_{2m-1}+ O(a_m)%a_m\hat{S}_{2m+4r-1}+\hat{S}_{3m+4r-1}
& \left(\dfrac{1}{2!}-\dfrac{1}{3!}\right)a_m^2 \hat{S}_{2m-2}+ O(a_m) & \cdots \\

\dfrac{a_m^3}{3!} \hat{S}_{m+3}+\dfrac{a_m^2}{2!} \hat{S}_{2m+3}+ O(a_m)
%a_m\hat{S}_{3m+3}+\hat{S}_{4m+3}
& \dfrac{a_m^3}{3!} \hat{S}_{m+2}+\dfrac{a_m^2}{2!} \hat{S}_{2m+2}+ O(a_m)& \cdots \\

\vdots & \vdots & \ddots\\

\dfrac{a_m^3}{3!} \hat{S}_{2m-3}+\dfrac{a_m^2}{2!} \hat{S}_{3m-3}+ O(a_m)
%a_m\hat{S}_{3m+4r+1}+\hat{S}_{4m+4r+1}
& \dfrac{a_m^3}{3!} \hat{S}_{2m-4}+\dfrac{a_m^2}{2!} \hat{S}_{3m-4}+ O(a_m)& \cdots \\

\vdots & \vdots & \ddots\\
\end{array}\right] .
\end{equation}		

The second round of the row operation is to use the second block to eliminate the highest power term of $a_m$ in each subsequent block, and leaving the lower power terms of $a_m$. This results in

\begin{equation} \label{row operation-matrix 3}
\Phi_{N \times 4 N} \sim \left[\begin{array}{ccc}
\hat{S}_{3} & \hat{S}_{2} & \cdots \\

\hat{S}_{7} & \hat{S}_{6} & \cdots\\

\vdots & \vdots & \ddots \\

\hat{S}_{m-3} & \hat{S}_{m-4} & \cdots \\

a_m \hat{S}_{1}+\hat{S}_{m+1} & a_m \hat{S}_{0}+\hat{S}_{m} & \cdots\\

a_m \hat{S}_{5}+\hat{S}_{m+5} & a_m \hat{S}_{4}+\hat{S}_{m+4} & \cdots \\

\vdots & \vdots & \ddots \\

a_m \hat{S}_{m-1}+\hat{S}_{2m-1} & a_m \hat{S}_{m-2}+\hat{S}_{2m-2} & \cdots\\

a_m \hat{S}_{m+3}+\hat{S}_{2m+3}& a_m \hat{S}_{m+2}+\hat{S}_{2m+2} & \cdots \\

\vdots & \vdots & \ddots\\

a_m \hat{S}_{2m-3}+\hat{S}_{3m-3} & a_m \hat{S}_{2m-4}+\hat{S}_{3m-4} & \cdots\\

\left(\dfrac{1}{2!}-\dfrac{1}{3!}\right)a_m^2 \hat{S}_{m+1}+ O(a_m)
%\text{lower order terms}
%a_m\hat{S}_{2m+1}+\hat{S}_{3m+1}
& \left(\dfrac{1}{2!}-\dfrac{1}{3!}\right)a_m^2 \hat{S}_{m}+ O(a_m)& \cdots \\

\vdots & \vdots & \ddots\\

\left(\dfrac{1}{2!}-\dfrac{1}{3!}\right)a_m^2 \hat{S}_{2m-1}+ O(a_m)%a_m\hat{S}_{2m+4r-1}+\hat{S}_{3m+4r-1}
& \left(\dfrac{1}{2!}-\dfrac{1}{3!}\right)a_m^2 \hat{S}_{2m-2}+ O(a_m) & \cdots \\

\left(\dfrac{1}{2!}-\dfrac{1}{3!}\right)a_m^2 \hat{S}_{2m+3}+ O(a_m)
%a_m\hat{S}_{3m+3}+\hat{S}_{4m+3}
& \left(\dfrac{1}{2!}-\dfrac{1}{3!}\right)a_m^2 \hat{S}_{2m+2}+ O(a_m)& \cdots \\

\vdots & \vdots & \ddots\\

\left(\dfrac{1}{2!}-\dfrac{1}{3!}\right)a_m^2 \hat{S}_{3m-3}+ O(a_m)
%a_m\hat{S}_{3m+4r+1}+\hat{S}_{4m+4r+1}
& \left(\dfrac{1}{2!}-\dfrac{1}{3!}\right)a_m^2 \hat{S}_{3m-4}+ O(a_m)& \cdots \\

\vdots & \vdots & \ddots\\
\end{array}\right] .
\end{equation}	

We can continue to perform these row operation to $\Phi_{ N \times 4N}$, which have $2k$ rounds in total. Similar column operations can be applied to the matrix $\Psi_{4 N \times N}$.

At the end of these operations, we arrive at the situation where the determinant \eqref{three expansion form} does not vanish when we keep only the highest order term in $a_m$ for each element. The difference with the previous work in \cite{yang2021rogue} is that we cannot generate the lower triangular block matrix or upper triangular block matrix after keeping the highest order terms and row switchings. This indicates that the size of determinant $\tau_{\mathbf{n}}$ in unchanged. Moreover, it can be shown that $\tau_{\mathbf{n}}$ reduces to the form
	\begin{equation}\label{rewrite form-3com(1st,4r+2)}
		\tau_{\mathbf{n}}=\beta_1 \left|a_m\right|^{\gamma_1}\left|\begin{array}{cc}
	 \mathbf{O}_{N \times N} & \hat{\Phi}_{N \times 4 N} \\
	 -\hat{\Psi}_{4 N \times N} & \mathbf{I}_{4 N \times 4 N}
	 \end{array}\right|\left[1+O\left(a_m^{-1}\right)\right],
	\end{equation}
where $\beta_1 \not = 0, \gamma_1>0$  are constants, and

% By applying similar argument as in \cite{yang2021rogue}, it can be shown that $\tau_{\mathbf{n},r}$ reduces to
% the form
% 	\begin{equation}\label{rewrite form}
% 		\tau_{\mathbf{n},r}=\beta \left|a_m\right|^{\gamma}\left|\begin{array}{cc}
% 			\mathbf{O}_{{\overline{N}_1}\times\overline{N}_1} & \widehat{\Phi}_{\overline{N}_1 \times \widehat{N}} \\
% 			-\widehat{\Psi}_{\widehat{N} \times \overline{N}_1} & \mathbf{I}_{\widehat{N} \times \widehat{N}}
% 		\end{array}\right|\left[1+O\left(a_m^{-1}\right)\right],
% 	\end{equation}
% where $\beta \not = 0, \gamma>0$  are constants,
% $\overline{N}_1=\displaystyle{\sum_{n=1}^4}N_{n,4}$,  $\displaystyle{\widehat{N}=\max_{1\leq i \leq 4}\left(5N_i-i\right)}$,
	\begin{equation}
	\begin{aligned}
		&\hat{\Phi}=\left(\begin{array}{ll}
			\hat{\Phi}_{N_{1} \times 4N}^{(1)} \\
			\hat{\Phi}_{N_{3} \times 4N}^{(3)}
		\end{array}\right), \quad \hat{\Psi}=\left(\begin{array}{ll}
			\hat{\Psi}_{4N \times N_{1}}^{(1)} &
			\hat{\Psi}_{4N \times N_{3}}^{(3)}
		\end{array}\right),
\\
		&\hat{\Phi}_{i, j}^{(I)}=\left(h_0\right)^{-(j-1)} S_{4 i+1-I-j}\left[\boldsymbol{y}^{+}(\mathbf{n})+\left(j-1\right) \boldsymbol{s}\right],
 \\
		&\hat{\Psi}_{i, j}^{(J)}=\left(h_0^*\right)^{-(i-1)} S_{4 j+1-J-i}\left[\boldsymbol{y}^{-}(\mathbf{n})+\left(i-1\right) \boldsymbol{s}^*\right].
\end{aligned}
\end{equation}

Since the rogue wave solutions are independent of the constants $\beta_1$ and $\gamma_1$, we can rewrite \eqref{rewrite form-3com(1st,4r+2)} into a $2\times 2$ block determinant \cite{yang2022pattern}
\begin{equation}\label{reduced solution-3com(1st,4r+2)}
	\tau_{\mathbf{n}}=\operatorname{det}\left(\begin{array}{cc}
		\tau_{\mathbf{n}}^{[1,1]}  &\tau_{\mathbf{n}}^{[1,3]} \\
		\tau_{\mathbf{n}}^{[3,1]} & \tau_{\mathbf{n}}^{[3,3]}
	\end{array}\right)\left[1+O\left(a_m^{-1}\right)\right]
\end{equation}
where
\begin{equation}
\tau_{\mathbf{n}}^{[I, J]}=\left(m_{4 i-I, 4 j-J}^{(\mathbf{n}, I, J)}\right)_{1 \leq i \leq N_I, 1 \leq j \leq N_J}, \quad 1 \leq I, J \leq 3,
\end{equation}
and
\begin{equation}\label{element-3com(1st,4r+2)}
	m_{i, j}^{(\mathbf{n}, I, J)}=\sum_{v=0}^{\min (i, j)}\left[\frac{\left|p_{1}\right|^{2}}{\left(p_{0}+p_{0}^{*}\right)^{2}}\right]^{v} S_{i-v}\left(\boldsymbol{y}^{+}(\mathbf{n})+v \boldsymbol{s}\right) S_{j-v}\left(\boldsymbol{y}^{-}(\mathbf{n})+v \boldsymbol{s}^{*}\right) .
\end{equation}
Note that the determinant $\tau_{\mathbf{n}}$ is still of order $N$, but the degree of $\tau_{\mathbf{n}}$ with respect to $x$ or $t$ is reduced, so this is still a lower-order rogue wave. Moreover, in this case, $\tau_{\mathbf{n}}$ in the inner region is always approximately a $2 \times 2$ block matrix regardless of the values of $N$ and $m$, i.e., $N_2 = 0$ in \eqref{definition of N-theorem}. As a result of this, when  $x^2+t^2=O(1)$ and $|a_m| \gg 1$, the determinant in \eqref{three expansion form} is approximately a $\left(N_{1}, 0,N_{3}\right)$-th order rogue wave of the three-components NLS equation
$$[u_{1, \widehat{\mathcal{N}}_1}(x,
	t), \quad u_{2, \widehat{\mathcal{N}}_1}(x, t), \quad
	u_{3, \widehat{\mathcal{N}}_1}(x, t)]$$
%$$\left[u_{1}(x, t), \quad u_{2}(x,
%t), \quad u_{3}(x, t)\right]$$
where $ \widehat{\mathcal{N}}_1=\left(N_{1}, 0,N_{3}\right)$, $u_{j, \widehat{\mathcal{N}}_1} \, (j=1,2,3)$ is given in Theorem \ref{RW solutions of vector NLS} with $N_j = N_{j,3}$, and the internal parameters
\begin{eqnarray*}
  \left(\hat{a}_{1,n}, \hat{a}_{2,n}, \hat{a}_{3,n}, \hat{a}_{5,n}, \hat{a}_{6,n}\ldots, \hat{a}_{4 N_{n,3}-n,n}\right), \quad n=1,3,
  \end{eqnarray*}
are related to those in the original rogue wave as
	$$
	\hat{a}_{j, 1}=\hat{a}_{j, 3}=a_j, \quad j=1,2,3,5,6,7 \cdots,m-1,m+1,\cdots
	$$
and
$$
	\hat{a}_{m, 1}=\hat{a}_{m, 3}=0.
	$$
From \eqref{reduced solution-3com(1st,4r+2)}, we deduce that the approximation error of this lower-order rogue wave is $O\left(\left|a_m\right|^{-1}\right)$ .
% It is important to note that the approximate rogue wave expression in the inner region

Next, we consider the case $m\equiv1 \mod 4$, i.e., $m=4r+1 \, (r>1)$. Notice that the coefficients of the highest $a_m$ power terms in the first column of $\Psi_{N \times 4N}$ are proportional to
$$
\hat{S}_{3},\hat{S}_{7},\cdots, \hat{S}_{4r-1},\hat{S}_{2},\hat{S}_{6},\cdots, \hat{S}_{4r-2},\hat{S}_{1},\hat{S}_{5},\cdots, \hat{S}_{4r-3},\hat{S}_{0},\hat{S}_{4},\cdots, \hat{S}_{4r}
$$
and repeating. Similar to the previous case, we can think of the first $m$ rows as the first block matrix of $\Phi_{N \times 4N}$, the next $m$ rows as the second block matrix, and so on. On account of $N=k m + N_0$, the remaining $N_0$ rows are called the remaining block matrix, i.e., the $(k+1)$-th block matrix. Each block matrix can be divided into four parts, for example, the first column of these four parts are sequences starting with
$\hat{S}_{3}$, $\hat{S}_{2}$, $\hat{S}_{1}$ and $\hat{S}_{0}$ respectively. %, {\color{red}and the subscripts is a arithmetic progression with a common difference of 4.}

Then, using similar argument as in \cite{yang2021general}, we arrive at the situation where the determinant \eqref{three expansion form} does not vanish when we keep only the highest order term in $a_m$ for each element.
In this case, after row and column swapping, upper and lower triangular block matrices will be generated. After expanding these block matrices, $\tau_{\mathbf{n}}$ reduces to
the form
	\begin{equation}\label{rewrite form-3com(1st,4r+1)}
		\tau_{\mathbf{n}}=\beta_2 \left|a_m\right|^{\gamma_2}\left|\begin{array}{cc}			\mathbf{O}_{{\bar{N}_3}\times\bar{N}_3} & \widehat{\Phi}_{\bar{N}_3 \times \hat{N}} \\
			-\widehat{\Psi}_{\hat{N} \times \bar{N}_3} & \mathbf{I}_{\hat{N} \times \hat{N}}
		\end{array}\right|\left[1+O\left(a_m^{-1}\right)\right],
	\end{equation}
where $\beta_2 \not = 0, \gamma_2>0$  are constants, $\bar{N}_3=\displaystyle{\sum_{n=1}^3}N_{n,3}$,  $\displaystyle{\hat{N}=\max_{1\leq i \leq 3}\left(4N_{i,3}-i+1\right)}$,
	\begin{equation}
	\begin{aligned}
		&\widehat{\Phi}=\left(\begin{array}{lll}
			\widehat{\Phi}_{N_{1,3} \times \widehat{N}}^{(1)} \\
			\widehat{\Phi}_{N_{2,3} \times \widehat{N}}^{(2)}\\
			\widehat{\Phi}_{N_{3,3} \times \widehat{N}}^{(3)}
		\end{array}\right), \quad \widehat{\Psi}=\left(\begin{array}{lll}
			\widehat{\Psi}_{\widehat{N} \times N_{1,3}}^{(1)} & \widehat{\Psi}_{\widehat{N} \times N_{2,3}}^{(2)}&
			\widehat{\Psi}_{\widehat{N} \times N_{3,3}}^{(3)}
		\end{array}\right)
\\
		&\widehat{\Phi}_{i, j}^{(I)}=\left(h_0\right)^{-(j-1)} S_{4 i+1-I-j}\left[\boldsymbol{y}^{+}(\mathbf{n})+\left(j-1+\nu_0\right) \boldsymbol{s}\right]
 \\
		&\widehat{\Psi}_{i, j}^{(J)}=\left(h_0^*\right)^{-(i-1)} S_{4 j+1-J-i}\left[\boldsymbol{y}^{-}(\mathbf{n})+\left(i-1+\nu_0\right) \boldsymbol{s}^*\right]
\end{aligned}
\end{equation}
and $\nu_0=N-\bar{N}_1$. Finally, using similar argument as in \cite{yang2022pattern}, we find that $\tau_{\mathbf{n}}$ can be asymptotically reduced
to a $(N_{1,3},N_{2,3},N_{3,3})$-th order rogue wave of the three-components NLS equation in the inner region.
Notice that the internal parameters
\begin{eqnarray*}
  \left(\hat{a}_{1,n}, \hat{a}_{2,n}, \hat{a}_{3,n}, \hat{a}_{5,n}, \hat{a}_{6,n}\ldots, \hat{a}_{4 N_{n,3}-n,n}\right), \quad n=1,2,3,
  \end{eqnarray*}
are related to those in the original rogue wave as
	$$
	\hat{a}_{j, 1}=\hat{a}_{j, 2}=\hat{a}_{j, 3}=a_j+\displaystyle{\left(N-\bar{N}_3\right)} s_j, \quad j=1,2,3,5,6,7 \cdots.
	$$

As pointed before, the proofs of our 1st type and 3rd type rogue waves of the three-component NLS are very similar, so we omit the proof of 3rd type. However, there are some differences in the proof of 2nd type rogue waves in the inner region. This is explained as follows.

% In order to study the patterns of the 2-nd type rogue waves  of the three-component NLS rogue waves under the condition $a_m \gg 1$ in the inner region with $x^2+t^2=O(1)$,
We rewrite the determinant $\tau_{\mathbf{n}}$ as a $5 N \times 5 N$ determinant as in \eqref{three expansion form}. Note that $m \equiv j \mod 4, j=1,2,3$, and the different case is still $j=2$, i.e., $m=4r+2 \,(r\geq0)$. In this case, we can substitute \eqref{S expand} into \eqref{three expansion form} to expand each element into a polynomial in $a_m$. Similar to the proof of 1st type, we can rewrite $\Phi_{N\times4N}$ as follows
\begin{equation}
\Phi_{N \times 4 N} \sim \left[\begin{array}{ccc}
\hat{S}_{2} & \hat{S}_{1} & \cdots \\

\hat{S}_{6} & \hat{S}_{5} & \cdots\\

\vdots & \vdots & \ddots \\

\hat{S}_{m-4} & \hat{S}_{m-5} & \cdots \\

a_m \hat{S}_{0}+\hat{S}_{m} & \hat{S}_{m-1} & \cdots\\

a_m \hat{S}_{4}+\hat{S}_{m+4} & a_m \hat{S}_{3}+\hat{S}_{m+3} & \cdots \\

\vdots & \vdots & \ddots \\

a_m \hat{S}_{m-2}+\hat{S}_{2m-2} & a_m \hat{S}_{m-3}+\hat{S}_{2m-3} & \cdots\\

\dfrac{a_m^2}{2!} \hat{S}_{2}+a_m \hat{S}_{m+2}+\hat{S}_{2m+2}& \dfrac{a_m^2}{2!} \hat{S}_{1}+a_m \hat{S}_{m+1}+\hat{S}_{2m+1} & \cdots \\

%\dfrac{a_m^2}{2!} \hat{S}_{7}+a_m \hat{S}_{m+7}+\hat{S}_{2m+7}& a^3 z^0 +a^2 z^m& \cdots & \cdots & \cdots& \cdots & \cdots  & \cdots& \cdots\\

\vdots & \vdots & \ddots\\

\dfrac{a_m^2}{2!} \hat{S}_{m-4}+a_m \hat{S}_{2m-4}+\hat{S}_{3m-4} & \dfrac{a_m^2}{2!} \hat{S}_{m-5}+a_m \hat{S}_{2m-5}+\hat{S}_{3m-5} & \cdots\\

\dfrac{a_m^3}{3!} \hat{S}_{0}+\dfrac{a_m^2}{2!} \hat{S}_{m}+ O(a_m)
& \dfrac{a_m^2}{2!} \hat{S}_{m-1}+ O(a_m)& \cdots \\

\vdots & \vdots & \ddots\\

\dfrac{a_m^3}{3!} \hat{S}_{m-2}+\dfrac{a_m^2}{2!} \hat{S}_{2m-2}+ O(a_m)
& \dfrac{a_m^3}{3!} \hat{S}_{m-3}+\dfrac{a_m^2}{2!} \hat{S}_{2m-3}+ O(a_m) & \cdots \\

% \dfrac{a_m^4}{4!} \hat{S}_{2}+\dfrac{a_m^3}{3!} \hat{S}_{m+2}+\dfrac{a_m^2}{2!} \hat{S}_{2m+2}+ O(a_m)
% & \dfrac{a_m^4}{4!} \hat{S}_{1}+\dfrac{a_m^3}{3!} \hat{S}_{m+1}+\dfrac{a_m^2}{2!} \hat{S}_{2m+1}+ O(a_m) & \cdots \\

% \vdots & \vdots & \ddots\\

% \dfrac{a_m^5}{5!} \hat{S}_{0}+\dfrac{a_m^4}{4!} \hat{S}_{m}+\dfrac{a_m^3}{3!} \hat{S}_{2m}+ O(a_m^2)
% & \dfrac{a_m^4}{4!} \hat{S}_{m-1}+\dfrac{a_m^3}{3!} \hat{S}_{2m-1}+\dfrac{a_m^2}{2!} \hat{S}_{3m-1}+ O(a_m) & \cdots \\
% \dfrac{a_m^4}{4!} \hat{S}_{3}+\dfrac{a_m^3}{3!} \hat{S}_{m+3}+\dfrac{a_m^2}{2!} \hat{S}_{2m+3}+ O(a_m)
% %a_m\hat{S}_{3m+3}+\hat{S}_{4m+3}
% & \dfrac{a_m^4}{4!} \hat{S}_{2}+\dfrac{a_m^3}{3!} \hat{S}_{m+2}+\dfrac{a_m^2}{2!} \hat{S}_{2m+2}+ O(a_m)& \cdots \\

% \vdots & \vdots & \ddots\\

% \dfrac{a_m^4}{4!} \hat{S}_{m-3}+\dfrac{a_m^3}{3!} \hat{S}_{2m-3}+\dfrac{a_m^2}{2!} \hat{S}_{3m-3}+ O(a_m)
% %a_m\hat{S}_{3m+4r+1}+\hat{S}_{4m+4r+1}
% & \dfrac{a_m^4}{4!} \hat{S}_{m-4}+\dfrac{a_m^3}{3!} \hat{S}_{2m-4}+\dfrac{a_m^2}{2!} \hat{S}_{3m-4}+ O(a_m)& \cdots \\

\vdots & \vdots & \ddots\\
\end{array}\right] .
\end{equation}

It can be seen that the matrix $\Phi_{N\times4N}$ can be divided into a number of blocks. We use the same method as before, that is, we use the preceding blocks to eliminate the highest-order terms in $a_m$ of the subsequent blocks in turn. After the above operations, we find that only the coefficient of the highest-power term in $(r+1)$-th row is $\hat{S}_{0}$. This inspires us to eliminate one row and one column through some operations.

We first keep only the highest remaining power of $a_m$ in the $(r+1)$-th row of $\Phi_{N\times4N}$. Then, from the original determinant $\tau_{\mathbf{n}}$, we can expand it according to the $(r+1)$-th row, and obtain %ing the remaining $\Phi_{N\times4N}$ as
\begin{equation}
\Phi_{N \times 4 N} \sim \left[\begin{array}{ccc}
\hat{S}_{1} & \hat{S}_{0} & \cdots \\

\hat{S}_{5} & \hat{S}_{4} & \cdots\\

\vdots & \vdots & \ddots \\

\hat{S}_{m-5} & \hat{S}_{m-6} & \cdots \\

% a_m \hat{S}_{0}+\hat{S}_{m} & \hat{S}_{m-1} & \cdots\\

a_m \hat{S}_{3}+\hat{S}_{m+3} & a_m \hat{S}_{2}+\hat{S}_{m+2} & \cdots \\

\vdots & \vdots & \ddots \\

a_m \hat{S}_{m-3}+\hat{S}_{2m-2} & a_m \hat{S}_{m-4}+\hat{S}_{2m-3} & \cdots\\

a_m \hat{S}_{m+1}+\hat{S}_{2m+1}& a_m \hat{S}_{m}+\hat{S}_{2m} & \cdots \\

\vdots & \vdots & \ddots\\

a_m \hat{S}_{2m-5}+\hat{S}_{3m-5} & a_m \hat{S}_{2m-6}+\hat{S}_{3m-6} & \cdots\\

\left(\dfrac{1}{2!}-\dfrac{1}{3!}\right)a_m^2 \hat{S}_{m-1}+ O(a_m)
& \left(\dfrac{1}{2!}-\dfrac{1}{3!}\right)a_m^2 \hat{S}_{m-2}+ O(a_m)& \cdots \\

\vdots & \vdots & \ddots\\

\left(\dfrac{1}{2!}-\dfrac{1}{3!}\right)a_m^2 \hat{S}_{2m-3}+ O(a_m)
& \left(\dfrac{1}{2!}-\dfrac{1}{3!}\right)a_m^2\hat{S}_{2m-4}+ O(a_m) & \cdots \\

% \left(\dfrac{1}{2!}-\dfrac{1}{3!}+\dfrac{1}{4!}\right)a_m^2 \hat{S}_{2m+1}+ O(a_m)
% & \left(\dfrac{1}{2!}-\dfrac{1}{3!}+\dfrac{1}{4!}\right)a_m^2 \hat{S}_{2m}+ O(a_m) & \cdots \\
% \vdots & \vdots & \ddots\\
% ? & ? & \ddots\\

\vdots & \vdots & \ddots\\
\end{array}\right] .
\end{equation}

Similar treatment can be applied to the matrix $\Psi_{4N\times N}$. It can be observed that we have a similar situation to the inner region of 1st type rogue wave with $m=4r+2$. Finally, we can rewrite \eqref{rewrite form-3com(1st,4r+2)} into a $2\times 2$ block determinant
\begin{equation}\label{reduced solution-3com(2nd,4r+2)}
	\tau_{\mathbf{n}}=\operatorname{det}\left(\begin{array}{cc}
		\tau_{\mathbf{n}}^{[1,1]}  &\tau_{\mathbf{n}}^{[1,3]} \\
		\tau_{\mathbf{n}}^{[3,1]} & \tau_{\mathbf{n}}^{[3,3]}
	\end{array}\right)\left[1+O\left(a_m^{-1}\right)\right]
\end{equation}
where
\begin{equation}
\tau_{\mathbf{n}}^{[I, J]}=\left(m_{4 i-I, 4 j-J}^{(\mathbf{n}, I, J)}\right)_{1 \leq i \leq N_I, 1 \leq j \leq N_J}, \quad 1 \leq I, J \leq 3,
\end{equation}
and
\begin{equation}\label{element-3com(1st,4r+2)}
	m_{i, j}^{(\mathbf{n}, I, J)}=\sum_{v=0}^{\min (i, j)}\left[\frac{\left|p_{1}\right|^{2}}{\left(p_{0}+p_{0}^{*}\right)^{2}}\right]^{v} S_{i-v}\left(\boldsymbol{x}_I^{+}(\mathbf{n})+v \boldsymbol{s}\right) S_{j-v}\left(\boldsymbol{x}_J^{-}(\mathbf{n})+v \boldsymbol{s}^{*}\right) .
\end{equation}

Note that the determinant $\tau_{\mathbf{n}}$ is always $(N-1)\times(N-1)$ and $\tau_{\mathbf{n}}$ in the inner region is always approximately a $2 \times 2$ block matrix regardless of the values of $N$ and $m$, i.e., $N_2 = 0$ in \eqref{definition of N-theorem}. Moreover, we remark that the internal parameters
\begin{eqnarray*}
  \left(\hat{a}_{1,n}, \hat{a}_{2,n}, \hat{a}_{3,n}, \hat{a}_{5,n}, \hat{a}_{6,n}\ldots, \hat{a}_{4 N_{n,2}-n,n}\right), \quad n=1,3,
  \end{eqnarray*}
are related to those in the original rogue wave as
	$$
	\hat{a}_{j, 1}=\hat{a}_{j, 3}=a_j+s_j, \quad j=1,2,3,5,6,7 \cdots,m-1,m+1,\cdots
	$$
and
$$
	\hat{a}_{m, 1}=\hat{a}_{m, 3}=s_m.
$$

This completes the proof of Theorem \ref{Rogue wave patterns-3com} for the inner region.

%\newpage

\vspace{2em}

\noindent {\bf Proof of Theorem \ref{Rogue wave patterns-4com}}.
Since the proofs are similar for different $i \in \{1,2,3,4\}$,  it suffices to present the proof for $i=1$.

Assume $|a_m|$ is large and other parameters are $O(1)$. We first consider the situation when $(x,t)$ is located in the outer region, i.e., $\sqrt{x^2+t^2}=O\left(\left|a_m\right|^{1 / m}\right)$.
Since the proof is very simlar to Theorem \ref{Rogue wave patterns-3com}, we only show the differences.

To begin with, we have
\begin{equation}
S_j\left(\boldsymbol{x}^{+}(\mathbf{n})+\nu \boldsymbol{s}\right)=S_j\left(x_1^{+}, x_2^{+}, x_3^{+}, x_4^{+},\nu s_5,  x_6^{+}, x_7^{+}, x_8^{+}, x_9^{+},  \nu s_{10}, \cdots, x_m^{+}+\nu s_m, \cdots\right) \sim S_j(\mathbf{v}),
\end{equation}
where
\begin{equation}
\mathbf{v}=\left(p_0 x+2 p_0 p_1 \mathrm{i} t, 0, \cdots, 0, a_m, 0, \cdots\right).
\end{equation}
This relation is the same as \eqref{Simplified Schur polynomial-3com}, but the values of $p_0$ and $p_1$ are different from the three-component NLS equation.
%, but only keep the same relation.
Then, after some calculations similar to the proof of Theorem \ref{Rogue wave patterns-3com}, we find that the highest order term in $a_m$ for $\tau_{\mathbf{n}}$ is
\begin{equation}
\tau_{\mathbf{n}} \sim|\alpha|^2\left|a_m\right|^{4 N(N+1) / m}\left|W_N^{[m, 5,4]}(z)\right|^2,
\end{equation}
where
$$\alpha={h_0}^{N(N-1) / 2} (c_N^{[m, 5,4]})^{-1}, \quad h_0=p_1 /\left(p_0+p_0^*\right), \quad z=a_m^{-1 / m}\left(p_0 x+2 p_0 p_1 \mathrm{i} t\right).$$
Note that the order of $a_m$ is changed from $3 N(N+1) / m$ in the three-component case to $4 N(N+1) / m$.
Thus, the solutions
$$u_{1, \mathcal{N}_1}(x, t), \quad u_{2, \mathcal{N}_1}(x, t), \quad u_{3, \mathcal{N}_1}(x, t), \quad u_{4, \mathcal{N}_1}(x, t)$$ are the plane-wave backgrounds,
except at or near $\left(\tilde{x}_0, \tilde{t}_0\right)$,  where
 \begin{equation} \label{root relation-4com}
 z_0=a_m^{-1 / m}\left(p_0 \tilde{x}_0+2 p_0 p_1 \mathrm{i} \tilde{t}_0\right)
\end{equation}
 is a root of $W_N^{[m, 5,4]}(z)$.

In what follows, we show that, when $(x, t)$ is contained in a small neighborhood of $\left(\tilde{x}_0, \tilde{t}_0\right)$ given by \eqref{root relation-4com}, the underlying rogue wave is approximately a fundamental rogue wave. % that is located within $O(1)$ distance from $\left(\tilde{x}_0, \tilde{t}_0\right)$.
Denote by
$$
\hat{x}_2^{+}(x, t)=p_2 x+\left(2 p_0 p_2+p_1^2\right)(\mathrm{i} t),
$$
which contains the dominant terms of $x_2^{+}(x, t)$ in \eqref{values of x+} with the index `$I $'  removed. %, when $(x, t)$ is in the outer region.
Then, for $(x, t)$ in the neighborhood of $\left(\tilde{x}_0, \tilde{t}_0\right)$, we have a more refined asymptotics for $S_j\left(\boldsymbol{x}^{+}(\mathbf{n})+\nu \boldsymbol{s}\right)$
\begin{equation}
S_j\left(\boldsymbol{x}^{+}(\mathbf{n})+\nu \boldsymbol{s}\right)=\left[S_j(\hat{\mathbf{v}})+\hat{x}_2^{+}\left(\tilde{x}_0, \tilde{t}_0\right) S_{j-2}(\hat{\mathbf{v}})\right]\left[1+O\left(a_m^{-2 / m}\right)\right], \quad\left|a_m\right| \gg 1,
\end{equation}
where
\begin{equation}
    \hat{\mathbf{v}}=\left(p_0 x+2 p_0 p_1 \mathrm{i} t+n_1 \theta_{11}+n_2 \theta_{12}+n_3 \theta_{13}+n_4 \theta_{14}, 0, \cdots, 0, a_m, 0, \cdots\right).
\end{equation}
Here, the normalization of $a_1=0$ has been utilized. Next, we rewrite $\tau_{\mathbf{n}}$ in a similar form as  \eqref{Laplace expansion of tau function-3com} by means of Laplace expansion. Further, the contribution from the first index choice of $\nu_j=j-1$ can be expressed as
\begin{equation}
\alpha a_m^{\frac{ {2 N(N+1)-1}}{m}} \left[p_0\left(x-\tilde{x}_0\right)+2 p_0 p_1 \mathrm{i}\left(t-\tilde{t}_0\right)+ \sum_{k=1}^{4}n_k \theta_{1k}+\bar{\Delta}_1\right]\left[W_N^{[m, 5,4]}\right]^{\prime}\left(z_0\right)\left[1+O\left(a_m^{-1 / m}\right)\right]
\end{equation}
where
\begin{equation}
\bar{\Delta}_1=\frac{\hat{x}_2^{+}\left(\tilde{x}_0, \tilde{t}_0\right)}{a_m^{1 / m}} \frac{\sum_{j=1}^N \det_{1 \leq i \leq N}\left[p_{5 i-1}^{[m]}\left(z_0\right), \cdots, p_{5 i-j-2}^{[m]}\left(z_0\right), \cdots, p_{5 i-N}^{[m]}\left(z_0\right)\right]}{\left[W_N^{[m, 5,4]}\right]^{\prime}\left(z_0\right)}
\end{equation}
and $\bar{\Delta}_1=O(1)$ as $\hat{x}_2^{+}\left(\tilde{x}_0, \tilde{t}_0\right)=O\left(\left|a_m^{1 / m}\right|\right)$. By absorbing $\bar{\Delta}_1$ into $\left(\tilde{x}_0, \tilde{t}_0\right)[28]$, we obtain
\begin{equation}
\alpha a_m^{\frac{2 N(N+1)-1}{ m}}\left[p_0\left(x-\bar{x}_0\right)+2 p_0 p_1 \mathrm{i}\left(t-\bar{t}_0\right)+\sum_{k=1}^{4}n_k \theta_{1k}\right]\left[W_N^{[m, 5,4]}\right]^{\prime}\left(z_0\right)\left[1+O\left(a_m^{-1 / m}\right)\right] .
\end{equation}
where $\bar{x}_0$ and $\bar{t}_0$ are given in Theorem \ref{Rogue wave patterns-4com}.

For the second index choice, i.e., $\nu=(0,1, \cdots, N-2, N)$, the dominant terms in $a_m$ can be calculated in a similar way as (76), that is,
\begin{equation}
h_0^{\frac{N(N-1)+2}{2}} a_m^{\frac{2 N(N+1)-1}{ m}} \det_{1 \leq i \leq N}\left[p_{5 i-1}^{[m]}\left(z_0\right), p_{5 i-2}^{[m]}\left(z_0\right), \cdots, p_{5 i-(N-1)}^{[m]}\left(z_0\right), p_{5 i-N-1}^{[m]}\left(z_0\right)\right]\left[1+O\left(a_m^{-1 / m}\right)\right] .
\end{equation}
Since $p_{j-1}^{[m]}(z)=\left[p_j^{[m]}\right]^{\prime}(z)$, the above term can be expressed as
\begin{equation}
h_0 \alpha a_m^{\frac{2 N(N+1)-1}{ m}}\left[W_N^{[m, 5,4]}\right]^{\prime}\left(z_0\right)\left[1+O\left(a_m^{-1 / m}\right)\right] .
\end{equation}
\\
Summarizing the above two contributions, we conclude that
\begin{eqnarray}
&&\tau_{\mathbf{n}}(x, t)
\\
&=&|\alpha|^2\left|\left[W_N^{[m, 5,4]}\right]^{\prime}\left(z_0\right)\right|^2\left|a_m\right|^{\frac{2 N(N+1)-1}{ m}} \times \Big(\left[p_1\left(x-\bar{x}_0\right)+2 \mathrm{i} p_0 p_1\left(t-\bar{t}_0\right)+n_1 \theta_{11}+n_2 \theta_{12}+n_3 \theta_{13}+n_4 \theta_{14}\right]   \nonumber\\
&& {\left[p_1^*\left(x-\bar{x}_0\right)-2 \mathrm{i} p_0^* p_1^*\left(t-\bar{t}_0\right)-n_1 \theta_{11}^*-n_2 \theta_{12}^*-n_3 \theta_{13}^*-n_4 \theta_{14}^*\right]+\left|h_0\right|^2\Big) }
\times\left[1+O\left(a_m^{-1 / m}\right)\right].
\end{eqnarray}
Thus, the proof for outer region is completed.

 In order to study the patterns of the 1st type rogue waves  of the four-component NLS rogue waves under the condition $|a_m| \gg 1$ in the inner region with $x^2+t^2=O(1)$,
we first rewrite the determinant $\tau_{\mathbf{n}}$ as a $6 N \times 6 N$ determinant
	\begin{equation}\label{four expansion form}
     \tau_{\mathbf{n}}=\left|\begin{array}{cc}
	 \mathbf{O}_{N \times N} & \Phi_{N \times 5 N} \\
	 -\Psi_{5 N \times N} & \mathbf{I}_{5 N \times 5 N}
	 \end{array}\right|,
	\end{equation}
where
		$$
	\Phi_{i, j}=\left(\frac{p_1}{p_0+p_0^*}\right)^{j-1} S_{5 i-j}\left[\boldsymbol{x}^{+}(\mathbf{n})+(j-1) \boldsymbol{s}\right], \quad \Psi_{i, j}=\left(\frac{p_1^*}{p_0+p_0^*}\right)^{i-1} S_{5 j-i}\left[\boldsymbol{x}^{-}(\mathbf{n})+(i-1) \boldsymbol{s}^*\right] .
	$$
	
Then, we can apply \eqref{transformation between x and y} and \eqref{S expand} to express each element in \eqref{four expansion form} into a  polynomial in $a_m$ explicitly.
Notice that have totally four cases to consider, i.e., $m \equiv j \mod 5, j=1,2,3,4$. Since the proofs for all cases are similar, it suffices to provide the proof for $j=1$. To determine the highest order term in $a_m$ of $\tau_{\mathbf{n}}$,
% a straightforward way is to keep only the highest order of $a_m$ in each element. However, it turns out the resulting determinant will vanish. To tackle this issue,
we can use similar argument as that in \cite{yang2021rogue,yang2022rogue} to perform row and column operations.
%As the procedure for the row and column operations is similar to Manakov system \cite{yang2022rogue} and the three-component NLS of $j=1,3$
%case, we omit the details.
% After applying similar argument as in \cite{yang2021rogue},
After these operations, $\tau_{\mathbf{n}}$ can be reduced to
the form
	\begin{equation}\label{rewrite form}
		\tau_{\mathbf{n}}=\beta \left|a_m\right|^{\gamma}\left|\begin{array}{cc}
			\mathbf{O}_{{\overline{N}_4}\times\overline{N}_4} & \widehat{\Phi}_{\overline{N}_4 \times \widehat{N}} \\
			-\widehat{\Psi}_{\widehat{N} \times \overline{N}_4} & \mathbf{I}_{\widehat{N} \times \widehat{N}}
		\end{array}\right|\left[1+O\left(a_m^{-1}\right)\right],
	\end{equation}
where $\beta \not = 0, \gamma>0$  are constants, $\overline{N}_4=\displaystyle{\sum_{n=1}^4}N_{n,4}$,  $\displaystyle{\widehat{N}=\max_{1\leq i \leq 4}\left(5N_{i,4}-i + 1\right)}$,
	\begin{equation}
	\begin{aligned}
		&\widehat{\Phi}=\left(\begin{array}{llll}
			\widehat{\Phi}_{N_{1,4} \times \widehat{N}}^{(1)} \\
			\widehat{\Phi}_{N_{2,4} \times \widehat{N}}^{(2)}\\
			\widehat{\Phi}_{N_{3,4} \times \widehat{N}}^{(3)}\\
			\widehat{\Phi}_{N_{4,4} \times \widehat{N}}^{(4)}
		\end{array}\right), \quad \widehat{\Psi}=\left(\begin{array}{llll}
			\widehat{\Psi}_{\widehat{N} \times N_{1,4}}^{(1)} & \widehat{\Psi}_{\widehat{N} \times N_{2,4}}^{(2)}&
			\widehat{\Psi}_{\widehat{N} \times N_{3,4}}^{(3)}&
			\widehat{\Psi}_{\widehat{N} \times N_{4,4}}^{(4)}
		\end{array}\right)
\\
		&\widehat{\Phi}_{i, j}^{(I)}=\left(h_0\right)^{-(j-1)} S_{5 i-I}\left[\boldsymbol{y}^{+}(\mathbf{n})+\left(j-1+\nu_0\right) \boldsymbol{s}\right]
 \\
		&\widehat{\Psi}_{i, j}^{(J)}=\left(h_0^*\right)^{-(i-1)} S_{5 j-J}\left[\boldsymbol{y}^{-}(\mathbf{n})+\left(i-1+\nu_0\right) \boldsymbol{s}^*\right]
\end{aligned}
\end{equation}
and $\nu_0=N-\overline{N}_4$.
Since the rogue wave solutions are independent of the constants $\beta$ and $\gamma$, we can rewrite \eqref{rewrite form} into a $4\times 4$ block determinant
%, which has the same structure with our rouge wave solution but with a lower order of $\Sigma_1^4$:
\begin{equation}\label{reduced solution-4com}
	\tau_{\mathbf{n}}=\operatorname{det}\left(\begin{array}{cccc}
		\tau_{\mathbf{n}}^{[1,1]} & \tau_{\mathbf{n}}^{[1,2]} &\tau_{\mathbf{n}}^{[1,3]}
		&\tau_{\mathbf{n}}^{[1,4]} \\
		\tau_{\mathbf{n}}^{[2,1]} & \tau_{\mathbf{n}}^{[2,2]} & \tau_{\mathbf{n}}^{[2,3]} & \tau_{\mathbf{n}}^{[2,4]}\\
		\tau_{\mathbf{n}}^{[3,1]} & \tau_{\mathbf{n}}^{[3,2]} &\tau_{\mathbf{n}}^{[3,3]}
		&\tau_{\mathbf{n}}^{[3,4]} \\
		\tau_{\mathbf{n}}^{[4,1]} & \tau_{\mathbf{n}}^{[4,2]} &\tau_{\mathbf{n}}^{[4,3]}
		&\tau_{\mathbf{n}}^{[4,4]}
	\end{array}\right)\left[1+O\left(a_m^{-1}\right)\right]
\end{equation}
where
\begin{equation}
	\tau_{\mathbf{n}}^{[I, J]}=\left(m_{5 i-I, 5 j-J}^{(\mathbf{n},  I, J)}\right)_{1 \leq i \leq N_{I,4}, 1 \leq j \leq N_{J,4}}
\end{equation}
and
\begin{equation}\label{element}
	m_{i, j}^{(\mathbf{n}, I, J)}=\sum_{\nu=0}^{\min (i, j)}\left[\frac{\left|p_1\right|^2}{\left(p_0+p_0^*\right)^2}\right]^\nu S_{i-\nu}\left(\boldsymbol{y}^{+}(\mathbf{n})+\nu_0 \boldsymbol{s}+\nu \boldsymbol{s}\right) S_{j-\nu}\left(\boldsymbol{y}^{-}(\mathbf{n})+\nu_0 \boldsymbol{s}^*+\nu \boldsymbol{s}^*\right) .
\end{equation}

Finally, the determinant in \eqref{reduced solution-4com} becomes a $\left(N_{1,4}, N_{2,4},N_{3,4},N_{4,4}\right)$-th order rogue wave of the four-components NLS equation, and the internal parameters
\begin{eqnarray*}
  \left(\bar{a}_{1,n}, \bar{a}_{2,n}, \bar{a}_{3,n}, \bar{a}_{4,n}, \bar{a}_{6,n}\ldots, \bar{a}_{5 N_{n,4}-n,n}\right), \quad n=1,2,3,4, %\\
  \end{eqnarray*}
are related to those in the original rogue wave as
	$$
	\bar{a}_{j, 1}=\bar{a}_{j, 2}=\bar{a}_{j, 3}=\bar{a}_{j, 4}=a_j+\displaystyle{\left(N-\overline{N}_4\right)} s_j, \quad j=1,2,3,4,6,7 \cdots.
	$$
%which is the same as the relation {\color{red}$\cdots$} in Theorem {\color{red}$\cdots$}.
From \eqref{reduced solution-4com}, we deduce that the approximation error of this lower-order rogue wave is $O\left(\left|a_m\right|^{-1}\right)$ . This completes the proof of Theorem \ref{Rogue wave patterns-4com} for the inner region.

\section{Conclusion} \label{section-Conclusion}

In summary, we have constructed rogue waves of the vector (or $M$-component) NLS equation \eqref{vector NLS} and analyzed their patterns for $M=3,4$. These solutions are expressed in terms of Gram-type determinants of $K\times K$ block matrices ($K=1,2,\cdots, M$)  with index jumps of $M+1$ via Kadomtsev-Petviashvili hierarchy reduction technique. One crucial step in this process is solving a system of algebraic equations (see Lemma \ref{multiple roots} and its proof). The rogue wave patterns corresponding to $M=3,4$ and $K=1$ have been investigated comprehensively. We find that when specific internal parameters are large enough, these patterns are described by new polynomial hierarchies, i.e., the generalized Wronskian-Hermite polynomials, in contrast with the scalar NLS equation and the Manakov system. Since the Yablonskii-Vorob'ev polynomial hierarchy and Okamoto polynomial hierarchies are special cases of the generalized Wronskian-Hermite polynomials, our results have unified rogue wave patterns of the scalar NLS equation and the vector NLS equation  for $M=2,3,4$. It is worth noting that the case $M=3$ presents a unique feature as, in certain cases, the sizes of the Gram-type determinants cannot be reduced in the approximation of inner regions.

%Our graphical analysis reveals that the rogue wave patterns for $M=3,4$ exhibit diverse new shapes, and similar to the Manakov system \cite{yang2022rogue},
The rogue wave patterns for $M=3,4$ exhibit very rich structures similar to the Manakov system \cite{yang2022rogue}, these patterns are, in general, distorted from root structures of the generalized Wronskian-Hermite polynomials. The predicted rogue wave patterns have been compared with true solutions, and excellent agreement is achieved. As pointed out in \cite{yang2021universal,yang2022rogue}, universal rogue wave patterns, which depend on the index jumps, exist in integrable systems. We expect that the patterns uncovered in the present paper will appear in many other systems and thus are universal, as long as the corresponding Schur polynomials have index jumps of $4$ or $5$.

\section*{Acknowledgements}

 B.F. Feng was partially supported by National Science Foundation (NSF) under Grant No. DMS-1715991 and U.S. Department of Defense (DoD), Air Force for Scientific Research (AFOSR) under grant No. W911NF2010276.
C.F. Wu was supported by the National Natural Science Foundation of China (Grant Nos. 11701382 and 11971288) and Guangdong Basic and Applied Basic Research Foundation, China (Grant No. 2021A1515010054). We would like to thank Mr. Yuke Wang for drawing some of the figures.

\section*{Appendix A}

%\begin{proof}
In this appendix, we provide the proof of Lemma \ref{multiple roots}.
Assume $\xi$ is a root of $\mathcal{R}_M(z)=0$ of  multiplicity $M$ with $\Im (\xi)\not= 0 $, %where $x,y$ are real and $y \not=0$,
 then we have

\begin{equation} \label{system of equations}
\mathcal{R}^{(n)}_M(\xi) =0, \quad n=0,1,2,\dots, M-1,
%  \begin{aligned}
%     \mathcal{R}_M(\xi) &= \sum_{j=1}^M\frac{r_j }{(\xi+  k_j)^2} + 2=  0, \\
%  \mathcal{R}^{(n)}_M(\xi) &= (-1)^n (n+1)!  \sum_{j=1}^M\frac{r_j }{(\xi+  k_j)^{n+2}} = 0, \quad n=1,2,\dots, M-1.
%  \end{aligned}
\end{equation}
where
\begin{equation*}%\label{}
  \mathcal{R}^{(m)}_M(\xi) = (-1)^m (m+1)!  \sum_{j=1}^M\frac{r_j }{(\xi+  k_j)^{m+2}}, \quad m \geq 1.
\end{equation*}
The system of equations \eqref{system of equations} is linear in $r_j$, $j=1,2,\dots, M$, so we can solve for them and obtain
\begin{equation} \label{solution in r_j}
( \xi + k_j )^{M+1} =  -\dfrac{1}{2} \prod_{\substack{i=1 \\ i \not=j}}^M (k_j-k_i) r_j.
\end{equation}
Denote by
\begin{equation}\label{change of variable in r_j}
  \xi= x+ \mathrm{i}  y, \quad -\dfrac{1}{2} \prod_{\substack{i=1 \\ i \not=j}}^M (k_j-k_i) r_j = \lambda_j^{M+1} \exp(\mathrm{i} \theta_j \pi), \quad j=1,2,\dots, M,
\end{equation}
where $\lambda_j >0,  x, y$ are real, $y \not= 0$ and
\begin{equation}%\label{}
  \theta_j =
  \begin{cases}
  0, \quad \text{ if }  -\dfrac{1}{2} \prod_{\substack{i=1 \\ i \not=j}}^M (k_j-k_i) r_j >0,
  \\
  1, \quad \text{ if }  -\dfrac{1}{2} \prod_{\substack{i=1 \\ i \not=j}}^M (k_j-k_i) r_j <0,
  \end{cases}
\end{equation}
then we deduce from \eqref{solution in r_j} that, for each $k_j$, there exits $l_j \in \{0,1,\dots, M\}$ such that
\begin{equation} \label{k_j-1}
  x+k_j+ \mathrm{i} y =
  \begin{cases}
  \lambda_j \exp[2l_j \pi \mathrm{i}/(M+1)], \quad \text{ if } \theta_j=0,
  \\
  \lambda_j\exp[(2 l_j +1) \pi \mathrm{i}/(M+1)], \quad \text{ if } \theta_j=1,
  \end{cases}
\end{equation}
Comparing both sides of \eqref{k_j-1} gives
\begin{equation} \label{k_j-2}
   y =
  \begin{cases}
  \lambda_j \sin[2l_j \pi /(M+1)], \quad \text{ if } \theta_j=0,
  \\
   \lambda_j \sin[(2 l_j +1) \pi /(M+1)], \quad \text{ if } \theta_j=1.
  \end{cases}
\end{equation}
%As $M>1$ and the $k_j$'s are distinct, we have  $y\not = 0$.
This implies that all the corresponding $ \sin[2l_j \pi /(M+1)]$ or $\sin[(2 l_j +1) \pi /(M+1)], j=1,2,\dots, M,$ should have the same sign. Without loss of generality, we may assume $y>0$. %On the other hand, there are only $M$
Note that the set
\begin{equation}%\label{}
 \{1,  \exp[ \pi \mathrm{i}/(M+1)], \exp[ 2\pi \mathrm{i}/(M+1)], \dots, \exp[2M \pi \mathrm{i}/(M+1)], \exp[(2 M +1) \pi \mathrm{i}/(M+1)]\}
\end{equation}
contains exactly $M$ elements with positive imaginary parts, which are
\begin{equation}%\label{}
   \exp[ \pi \mathrm{i}/(M+1)], \quad \exp[ 2\pi \mathrm{i}/(M+1)], \dots, \quad \exp[M \pi \mathrm{i}/(M+1)].%, \exp[(2 M +1) \pi \mathrm{i}/(M+1)\
\end{equation}
Since the $k_j$'s are distinct, it then follows that
\begin{equation} %\label{}
  x+k_j+ \mathrm{i} y = \lambda_j \exp[\sigma_j \pi \mathrm{i}/(M+1)]
\end{equation}
where $(\sigma_1, \sigma_2, \dots,\sigma_M)$ can be any permutation of the set $\{1,2,\dots,M\}$. Without loss of generality, we may take
\begin{equation}%\label{}
 \sigma_j = j,
\end{equation}
where $j=1,2,\dots, M$. In this circumstance, we have $\theta_j=[1+(-1)^{j+1}]/2$ and
\begin{eqnarray} %\label{}
  x &=&  \lambda_j \cos[j \pi /(M+1)] -  k_j,
   \\
  y &=& \lambda_j \sin[j \pi /(M+1)],
\end{eqnarray}
 and hence
\begin{eqnarray}
  \lambda_j  &=& \lambda_1 \frac{\sin[  \pi /(M+1)] }{\sin[j \pi /(M+1)]}, \label{lambda-j}
  \\
  k_j &=& k_1 +  \lambda_j \cos[j \pi /(M+1)] - \lambda_1 \cos[ \pi /(M+1)],
  \\
      &=& k_1 +  \lambda_1 \left( \sin[  \pi /(M+1)]  \cot[j \pi /(M+1)] -  \cos[ \pi /(M+1)]\right)
\end{eqnarray}
where $j=1,2,\dots, M$. Further,  we find from \eqref{change of variable in r_j} and \eqref{lambda-j} that
\begin{eqnarray}%\label{}
  r_j %&=& (-1)^j \left( \frac{\sin[  \pi /(M+1)] }{\sin[j \pi /(M+1)]} \right)^M \prod_{i=2}^M (k_1-k_i) \prod_{\substack{i=1 \\ i \not=j}}^M (k_j-k_i)^{-1} r_1
%  \\
  = 2 (-1)^{j+1}  \prod_{\substack{i=1 \\ i \not=j}}^M (k_j-k_i)^{-1}  \left(\lambda_1 \frac{\sin[  \pi /(M+1)] }{\sin[j \pi /(M+1)]} \right)^{M+1} .
\end{eqnarray}
As $\mathcal{R}_M(z)$ is a rational function with real coefficients, it is clear that $\xi^*$ is a root of $\mathcal{R}_M(z)=0$ of  multiplicity $M$ as well.
This completes the proof.

\section*{Appendix B}
In this appendix, we apply Hirota's bilinear method to derive rogue wave solutions of the vector NLS
equation \eqref{vector NLS} presented in Theorem \ref{RW solutions of vector NLS} based on the KP reduction technique. For convenience, we only consider the case when $\tau_{\mathbf{n}}$ given in \eqref{tau-block matrix-theorem} consists of $M \times M$ block matrices, i.e., $K=M$, as other cases can be treated in a similar manner. In such case, we have $I_j = j \,(j=1,2,\dots, M)$ in \eqref{tau-block matrix-theorem}.
%Since the proofs are very similar for $M=3$ and $4$, it suffices to provide a brief proof for $M=4$.

We first transform the vector NLS equation \eqref{vector NLS} into a set of bilinear equations
\begin{equation}\label{bilinear equations_vector NLS}
\begin{aligned}
 &\left(D_x^2+\sum_{j=1}^M \sigma_j \rho_j^2\right) f \cdot f=\sum_{j=1}^M \sigma_j \rho_j^2 g_j g_j^*, \\
&\left(\mathrm{i} D_t+D_x^2+2 \mathrm{i} k_j D_x\right) g_j \cdot f=0, \quad j =1,2,\cdots,M,
\end{aligned}
\end{equation}
 under the non-zero boundary condition at $\pm \infty$ by the variable transformation
 \begin{equation}\label{transformation}
   u_j=\rho_j \frac{g_j}{f} \mathrm{e}^{\mathrm{i}\left(k_j x + w_j t\right)},  \quad j =1,2,\cdots,M,
 \end{equation}
where $w_j=\sum_{j=1}^{M}\sigma_j \rho_j^2 - k_j^2,$  $f$ is a real-valued function, $g_j$ is a complex-valued function,  and $D$ is the Hirota's bilinear operator \cite{hirota2004direct} defined by
$$
D_x^m D_t^n f \cdot g=\left.\left(\frac{\partial}{\partial x}-\frac{\partial}{\partial x^{\prime}}\right)^m\left(\frac{\partial}{\partial t}-\frac{\partial}{\partial t^{\prime}}\right)^n\left[f(x, t) g\left(x^{\prime}, t^{\prime}\right)\right]\right|_{x^{\prime}=x, t^{\prime}=t} .
$$
Next we define
\begin{equation*}
  \begin{aligned}
m^{\mathbf{n}} &=\frac{1}{p+q}\sum_{j=1}^{M}\left(-\frac{p-\mathrm{i} k_j}{q+\mathrm{i} k_j}\right)^{n_j} e^{\xi+\eta}, \\
\xi &=p x+p^2 y+\sum_{j=1}^M\frac{1}{p-\mathrm{i} k_j} v_j +\xi_0(p), \\
\eta &=q x-q^2 y+\sum_{j=1}^M\frac{1}{q+\mathrm{i} k_j} v_j +\eta_0(q),
\end{aligned}
\end{equation*}
where $\mathbf{n}=\left(n_1, n_2, \ldots, n_M\right)$ with $n_j$ being integers, $p, q, v_j$   are arbitrary complex constants,  $j=1,2,\dots, M$,  and $ \xi_0(p), \eta_0(q)$ are arbitrary functions of $p$ and $q$ respectively.
Let $\mathcal{A}_i$ and $\mathcal{B}_j$ be differential operators of order $i$ and $j$, respectively, defined by
$$
\mathcal{A}_i(p)=\frac{1}{i !}\left[f_1(p) \partial_p\right]^i, \quad \mathcal{B}_j(q)=\frac{1}{i !}\left[f_2(q) \partial_q\right]^j,
$$
where $ f_1(p), f_2(q) $ are arbitrary functions of $p$ and $q$ respectively.
Then it can be calculated that \cite{ohta2012general} the determinant
\begin{eqnarray*}
%\begin{aligned}
\tau_{\mathbf{n}} = \det_{1 \leq  \nu, \mu  \leq N}\left( m_{i_\nu, j_\mu}^{\mathbf{n}}  \right)%| _{2N\times 2N}
\end{eqnarray*}
where $\left(i_1, i_2, \cdots, i_N\right)$ and $\left(j_1, j_2, \cdots, j_N\right)$ are arbitrary sequences of indices, and the matrix element $m_{i j}^{\mathbf{n}}$ is defined as
\begin{equation} \label{phi equation}
m_{i j}^{\mathbf{n}}=\mathcal{A}_i \mathcal{B}_j m^{\mathbf{n}},
\end{equation}
would satisfy the bilinear equations %in the KP hierarchy
%{\color{red}
\begin{equation}\label{KP equations}
\begin{aligned}
&\left(\frac{1}{2} D_x D_{v_j}-1\right) \tau_{\mathbf{n}} \cdot \tau_{\mathbf{n}}=-\tau_{\mathbf{n}_{j,1}} \tau_{\mathbf{n}_{j,-1}}, \quad j= 1,2,\cdots, M,\\
&\left(D_x^2-D_y+2 \mathrm{i} k_j D_x\right) \tau_{\mathbf{n}_{j,1}} \cdot \tau_{\mathbf{n}}=0, \quad j= 1,2,\cdots, M,
\end{aligned}
\end{equation}
where $$ \mathbf{n}_{j,i}=\mathbf{n}+i \times \mathbf{n}_{j}, \quad \mathbf{n}_j=\sum_{l=1}^M \delta_{j l} \boldsymbol{e}_l, $$%}
$\boldsymbol{e}_l$ is the standard unit vector in $\mathbb{R}^M$ and $\delta_{j l}$ is the Kronecker delta.

In what follows, we will establish the reductions from the bilinear equations \eqref{KP equations} in the
KP hierarchy to the bilinear equations \eqref{bilinear equations_vector NLS},   thereby obtaining rogue wave solutions of the vector NLS equation \eqref{vector NLS}. This procedure consists of several steps.

\begin{itemize}
  \item [i)] {\it Dimension reduction}

  Note that
  \begin{equation}%\label{}
\left(2 \partial_x+\sum_{k=1}^M\sigma_k \rho_k^2 \partial_{v_k}\right) m_{i j}^{\mathbf{n}}=\mathcal{A}_i \mathcal{B}_j\left[\mathcal{G}_M(p)+\mathcal{H}_M(q)\right] m^{\mathbf{n}},
  \end{equation}
  where
  \begin{equation}
\mathcal{G}_M(p)= \sum_{j=1}^M\frac{\sigma_j \rho_j^2}{p-\mathrm{i} k_j} +2 p, \quad \mathcal{H}_M(q)= \sum_{j=1}^M\frac{\sigma_j \rho_j^2}{q+\mathrm{i} k_j} +2 q.
\end{equation}
  This implies that
  \begin{equation}\label{dimension reduction1}
    \left(2 \partial_x+\sum_{k=1}^M\sigma_k \rho_k^2 \partial_{v_k}\right) m_{i j}^{\mathbf{n}}=\sum_{\mu=0}^i \frac{1}{\mu !}\left[\left(f_1 \partial_p\right)^\mu \mathcal{G}_M(p)\right] m_{i-\mu, j}^{\mathbf{n}}+\sum_{l=0}^j \frac{1}{l !}\left[\left(f_2 \partial_q\right)^l \mathcal{H}_M(q)\right] m_{i, j-l}^{\mathbf{n}}.
  \end{equation}
  Then we can use the method introduced in \cite{yang2021general} to find $f_1(p)$ and  $f_2(q)$ such that
  \begin{equation}\label{select f}
    \left(f_1 \partial_p\right)^{M+1} \mathcal{G}_M(p)=\mathcal{G}_M(p), \quad  \left(f_2 \partial_q\right)^{M+1} \mathcal{H}_M(q)=\mathcal{H}_M(q).
  \end{equation}
Choosing $q_0 = p_0^*$ and using \eqref{select f} and the assumption that $p_0$ is a root of $\mathcal{G}'_M(p)=0$ of multiplicity $M$, the equation \eqref{dimension reduction1} reduces to
\begin{eqnarray}\label{dimension reduction2}
    &&\left(2 \partial_x+\sum_{k=1}^M\sigma_k \rho_k^2 \partial_{v_k}\right) m_{i j}^{\mathbf{n}} \Big|_{p=p_0,q=q_0}  \nonumber
    \\
    =&& \mathcal{G}_M(p_0) \sum_{\substack{\mu=0 \\ \mu \equiv 0 (\text{mod } (M+1))}}^i \frac{1}{\mu !} m_{i-\mu, j}^{\mathbf{n}}+\mathcal{H}_M(q_0) \sum_{\substack{l=0 \\ l \equiv 0 (\text{mod } (M+1))}}^i \frac{1}{l !} m_{i, j-l}^{\mathbf{n}} \Big|_{p=p_0,q=q_0}  .
  \end{eqnarray}
Let $N=N_1+N_2+\cdots+N_M$, where $N_j, j=1,2,\cdots,M,$ are positive integers, and define the determinant $\tau_{\mathbf{n}} $ by
\begin{equation}
\tau_{\mathbf{n}}=\det\left(\begin{array}{llll}
\tau_{\mathbf{n}}^{[1,1]} & \tau_{\mathbf{n}}^{[1,2]}&\cdots&\tau_{\mathbf{n}}^{[1,M]} \\
\tau_{\mathbf{n}}^{[2,1]} & \tau_{\mathbf{n}}^{[2,2]}&\cdots&\tau_{\mathbf{n}}^{[2,M]} \\ \vdots & \vdots&\ddots &\vdots \\\tau_{\mathbf{n}}^{[M,1]} & \tau_{\mathbf{n}}^{[M,2]}&\cdots&\tau_{\mathbf{n}}^{[M,M]}
\end{array}\right),
\end{equation}
where
\begin{equation}
\tau_{\mathbf{n}}^{[I, J]}=\operatorname{mat}_{1 \leq i \leq N_I, 1 \leq j \leq N_J}\left(\left.m_{(M+1) i-I, (M+1) j-J}^{\mathbf{n}}\right|_{p=p_0, q=q_0, \xi_0=\xi_{0, I}, \eta_0=\eta_{0, J}}\right), \quad 1 \leq I, J \leq M,
\end{equation}
and $m_{i,j}^{\mathbf{n}}$ is given by \eqref{phi equation}.\\
  With \eqref{dimension reduction2}, we can use similar argument as in \cite{ohta2012general} to show that the determinant $\tau_{\mathbf{n}}$ satisfies the dimensional reduction condition
 \begin{equation}\label{dimensional reduction condition}
\left(2 \partial_x+\sum_{k=1}^M\sigma_k \rho_k^2 \partial_{v_k}\right)\tau_{\mathbf{n}} =N \left[\mathcal{G}_M(p_0)+\mathcal{H}_M(q_0)\right] \tau_{\mathbf{n}}.
  \end{equation}
  Therefore, we can use \eqref{dimensional reduction condition} to eliminate the variables $v_j, j =1,2,\cdots,M$, from the higher dimensional bilinear system \eqref{KP equations}. As a result of this, we have
  \begin{equation} \label{bilinear equations_after dimension reduction}
\begin{aligned}
& \left(D_x^2+\sum_{j=1}^M \sigma_j \rho_j^2\right) \tau_{\mathbf{n}} \cdot \tau_{\mathbf{n}} = \sum_{j=1}^M
 \sigma_j \rho_j^2 \tau_{\mathbf{n}_{j,1}} \tau_{\mathbf{n}_{j,-1}}
 \\
&\left(\mathrm{i} D_t+D_x^2+2 \mathrm{i} k_j D_x\right) \tau_{\mathbf{n}_{j,1}} \cdot \tau_{\mathbf{n}}=0,\quad j=1,2,\cdots,M,
\end{aligned}
\end{equation}
where $t = - \mathrm{i} y$.

  \item [ii)] {\it Complex conjugate reduction}

  Impose the parameter constraint
   \begin{equation*}%\label{}
     \xi_{0,I} = \eta_{0,I}^*,
   \end{equation*}
  and in view of $p_0 = q_0^*$, we have $\left[f_1\left(p_0\right)\right]^*=$ $f_2\left(q_0\right)$. It then follows that
  \begin{equation}\label{complex conjugacy condition}
    \tau_{\mathbf{n}} = \tau_{-\mathbf{n}}^*.
  \end{equation}
  Define
  \begin{equation} \label{solution_differential operator}
    f=\tau_{\mathbf{n}_0}, \quad g_j=\tau_{\mathbf{n}_j}, \quad j=1,2,\cdots,M,
  \end{equation}

%and
%$$
%h_1=\tau_{-1,0,0,0}, \quad h_2=\tau_{0,-1,0,0}, \quad h_3=\tau_{0,0,-1,0}, \quad h_4=\tau_{0,0,0,-1},
%$$
then the complex conjugacy condition \eqref{complex conjugacy condition} implies that $f$  is real. %$h_j^*=g_j,j=1,2,3,4$.
Therefore, from \eqref{bilinear equations_after dimension reduction} and \eqref{complex conjugacy condition}, we conclude that the functions $f$ and $g_j$ satisfy the bilinear system \eqref{bilinear equations_vector NLS}, thereby yielding rational solutions to the vector NLS equation \eqref{vector NLS} via the transformation \eqref{transformation}.

% \text {. }

  \item [iii)]{\it  Introduction of free parameters}

  We apply the method proposed in \cite{yang2021general} to introduce free parameters in the following form
\begin{equation}%\label{}
  \xi_{0,I}=\sum_{n=1}^{\infty}  a_{n,I}  \ln ^{n} \mathcal{U}(p),
\end{equation}
where $\mathcal{U}(p)$ is defined by the relation
$$ f_1(p) =   \dfrac{\mathcal{U}(p)}{ \mathcal{U}'(p)},$$
and the $a_{n,I}$'s are free complex constants.

  \item [iv)] {\it Simplification of solutions}

  With the aid of the generator $\mathcal{D}$ of the differential operators $\left(p \partial_p\right)^k\left(q \partial_q\right)^l$ given as
  \begin{equation} \label{generator of differential operators}
    \mathcal{D}=\sum_{k=0}^{\infty} \sum_{l=0}^{\infty} \frac{\kappa^k}{k !} \frac{\lambda^l}{l !}\left(p \partial_p\right)^k\left(q \partial_q\right)^l=\exp \left(\kappa p \partial_p+\lambda q \partial_q\right)=\exp \left(\kappa \partial_{\ln p}+\lambda \partial_{\ln q}\right),
  \end{equation}
  we are able to simplify the solutions expressed by \eqref{solution_differential operator} using differential operators into the form of Schur polynomials as presented in Theorem \ref{RW solutions of vector NLS}. Since the computations are very similar to those in the three-wave system by Yang and Yang \cite{yang2021general}, we omit the details.

\end{itemize}

Thus the proof of Theorem \ref{RW solutions of vector NLS} is completed.

\section*{Appendix C}

In the first part of this appendix, we provide the values of $N_1, N_2,N_3,N_4$ that appear in Theorem \ref{root sturcture of jump 5} in the following lemma.

\begin{lemma} \label{values of N_i-root structure}
The values of $N_1, N_2,N_3,N_4$ involved in Theorem \ref{root sturcture of jump 5} are characterized as follows.

\begin{itemize}
  \item When $m \equiv 1 \mod 5$, we have

  \begin{equation*}
l=4: \left(N_1, N_2,N_3,N_4\right)= \begin{cases}\left(N_0, 0,0,0\right), & 0 \leq N_0 \leq\left[\frac{m}{5}\right]
\\ \left(\left[\frac{m}{5}\right], N_0-\left[\frac{m}{5}\right],0,0\right), & \left[\frac{m}{5}\right]+1 \leq N_0 \leq 2\left[\frac{m}{5}\right]
\\ \left(\left[\frac{m}{5}\right],\left[\frac{m}{5}\right], N_0-2\left[\frac{m}{5}\right]\right), &  2\left[\frac{m}{5}\right]+1 \leq N_0 \leq 3\left[\frac{m}{5}\right]
\\ \left(\left[\frac{m}{5}\right],\left[\frac{m}{5}\right],\left[\frac{m}{5}\right], N_0-3\left[\frac{m}{5}\right]\right), &  3\left[\frac{m}{5}\right]+1 \leq N_0 \leq 4\left[\frac{m}{5}\right]
\\
\left(m-1-N_0, m-1-N_0,m-1-N_0,m-1-N_0\right), & \text 4\left[\frac{m}{5}\right]+1 \leq N_0 \leq m-1\end{cases}
\end{equation*}
\begin{equation*}
l=3: \left(N_1, N_2,N_3,N_4\right)= \begin{cases}\left(0,N_0, 0,0\right), & 0 \leq N_0 \leq\left[\frac{m}{5}\right]
\\ \left(0,\left[\frac{m}{5}\right], N_0-\left[\frac{m}{5}\right],0\right), & \left[\frac{m}{5}\right]+1 \leq N_0 \leq 2\left[\frac{m}{5}\right]
\\ \left(0,\left[\frac{m}{5}\right],\left[\frac{m}{5}\right], N_0-2\left[\frac{m}{5}\right]\right), &  2\left[\frac{m}{5}\right]+1 \leq N_0 \leq 3\left[\frac{m}{5}\right]
\\
\left(\left[\frac{m}{5}\right]-1,\left[\frac{m}{5}\right]-1,\left[\frac{m}{5}\right]-1,N_0 -3\left[\frac{m}{5}\right]-1\right), & \text 3\left[\frac{m}{5}\right]+1 \leq N_0 \leq 4\left[\frac{m}{5}\right]+1
\\
\left(m-1-N_0,m-1-N_0,m-1-N_0,m-N_0\right), & \text 4\left[\frac{m}{5}\right]+2 \leq N_0 \leq m-1
\end{cases}
\end{equation*}
\begin{equation*}
l=2: \left(N_1, N_2,N_3,N_4\right)= \begin{cases}\left(0, 0,N_0,0\right), & 0 \leq N_0 \leq\left[\frac{m}{5}\right]
\\
\left(0,0,\left[\frac{m}{5}\right],N_0-\left[\frac{m}{5}\right]\right), & \left[\frac{m}{5}\right]+1 \leq N_0 \leq 2\left[\frac{m}{5}\right]
\\
\left(\left[\frac{m}{5}\right]-1,\left[\frac{m}{5}\right]-1, N_0-2\left[\frac{m}{5}\right]-1,0\right), &  2\left[\frac{m}{5}\right]+1 \leq N_0 \leq 3\left[\frac{m}{5}\right]+1
\\
\left(\left[\frac{m}{5}\right]-1,\left[\frac{m}{5}\right]-1,\left[\frac{m}{5}\right],N_0 -3\left[\frac{m}{5}\right]-1\right), & \text 3\left[\frac{m}{5}\right]+2 \leq N_0 \leq 4\left[\frac{m}{5}\right]+1
\\
\left(m-1-N_0,m-1-N_0,m-N_0,m-N_0\right), & \text 4\left[\frac{m}{5}\right]+2 \leq N_0 \leq m-1
\end{cases}
\end{equation*}
\begin{equation*}
l=1: \left(N_1, N_2,N_3,N_4\right)= \begin{cases}\left(0, 0,0,N_0\right), & 0 \leq N_0 \leq\left[\frac{m}{5}\right]
\\
\left(\left[\frac{m}{5}\right]-1, N_0-\left[\frac{m}{5}\right]-1,0,0\right), & \left[\frac{m}{5}\right]+1 \leq N_0 \leq 2\left[\frac{m}{5}\right]+1
\\
\left(\left[\frac{m}{5}\right]-1,\left[\frac{m}{5}\right], N_0-2\left[\frac{m}{5}\right]-1,0\right), &  2\left[\frac{m}{5}\right]+2 \leq N_0 \leq 3\left[\frac{m}{5}\right]+1
\\
\left(\left[\frac{m}{5}\right]-1,\left[\frac{m}{5}\right],\left[\frac{m}{5}\right],N_0 -3\left[\frac{m}{5}\right]-1\right), & \text 3\left[\frac{m}{5}\right]+2 \leq N_0 \leq 4\left[\frac{m}{5}\right]+1
\\
\left(m-1-N_0,m-N_0,m-N_0,m-N_0\right), & \text 4\left[\frac{m}{5}\right]+2 \leq N_0 \leq m-1
\end{cases}
\end{equation*}

  \item  When $m \equiv 2 \mod 5$, we have
  \begin{equation*}
l=4: \left(N_1, N_2,N_3,N_4\right)= \begin{cases}\left(N_0, 0,0,0\right), & 0 \leq N_0 \leq\left[\frac{m}{5}\right]
\\ \left(\left[\frac{m}{5}\right],0, N_0-\left[\frac{m}{5}\right],0\right), & \left[\frac{m}{5}\right]+1 \leq N_0 \leq 2\left[\frac{m}{5}\right]
\\ \left(N_0-2\left[\frac{m}{5}\right]-1,\left[\frac{m}{5}\right],0,\left[\frac{m}{5}\right]\right)&  2\left[\frac{m}{5}\right]+1 \leq N_0 \leq 3\left[\frac{m}{5}\right]
\\ \left(\left[\frac{m}{5}\right],\left[\frac{m}{5}\right],N_0-3\left[\frac{m}{5}\right]-1,\left[\frac{m}{5}\right]\right), &  3\left[\frac{m}{5}\right]+1 \leq N_0 \leq 4\left[\frac{m}{5}\right]
\\
\left(m-1-N_0, m-1-N_0,m-1-N_0,m-1-N_0\right), & \text 4\left[\frac{m}{5}\right]+1 \leq N_0 \leq m-1
\end{cases}
\end{equation*}
\begin{equation*}
l=3: \left(N_1, N_2,N_3,N_4\right)= \begin{cases}\left(0,N_0, 0,0\right), & 0 \leq N_0 \leq\left[\frac{m}{5}\right]
\\
\left(0,\left[\frac{m}{5}\right], 0,N_0-\left[\frac{m}{5}\right]\right), & \left[\frac{m}{5}\right]+1 \leq N_0 \leq 2\left[\frac{m}{5}\right]+1
\\
\left( N_0-2\left[\frac{m}{5}\right]-1,\left[\frac{m}{5}\right],0,\left[\frac{m}{5}\right]+1\right), &  2\left[\frac{m}{5}\right]+2 \leq N_0 \leq 3\left[\frac{m}{5}\right]+1
\\
\left(\left[\frac{m}{5}\right],\left[\frac{m}{5}\right],N_0 -3\left[\frac{m}{5}\right]-1,\left[\frac{m}{5}\right]+1\right), & \text 3\left[\frac{m}{5}\right]+2 \leq N_0 \leq 4\left[\frac{m}{5}\right]+1
\\
\left(m-1-N_0,m-1-N_0,m-1-N_0,m-N_0\right), & \text 4\left[\frac{m}{5}\right]+2 \leq N_0 \leq m-1
\end{cases}
\end{equation*}
\begin{equation*}
l=2: \left(N_1, N_2,N_3,N_4\right)= \begin{cases}\left(0, 0,N_0,0\right), & 0 \leq N_0 \leq\left[\frac{m}{5}\right]
\\
\left(N_0-\left[\frac{m}{5}\right]-1,0,0,\left[\frac{m}{5}\right]\right), & \left[\frac{m}{5}\right]+1 \leq N_0 \leq 2\left[\frac{m}{5}\right]+1
\\
\left(\left[\frac{m}{5}\right],0,N_0-2\left[\frac{m}{5}\right]-1, \left[\frac{m}{5}\right]\right), &  2\left[\frac{m}{5}\right]+2 \leq N_0 \leq 3\left[\frac{m}{5}\right]+1
\\
\left(\left[\frac{m}{5}\right]-1,\left[\frac{m}{5}\right]-1,N_0 -3\left[\frac{m}{5}\right]-2,\left[\frac{m}{5}\right]\right), & \text 3\left[\frac{m}{5}\right]+2 \leq N_0 \leq 4\left[\frac{m}{5}\right]+2
\\
\left(m-1-N_0,m-1-N_0,m-N_0,m-N_0\right), & \text 4\left[\frac{m}{5}\right]+3 \leq N_0 \leq m-1
\end{cases}
\end{equation*}
\begin{equation*}
l=1: \left(N_1, N_2,N_3,N_4\right)= \begin{cases}\left(0, 0,0,N_0\right), & 0 \leq N_0 \leq\left[\frac{m}{5}\right]+1
\\
\left( N_0-\left[\frac{m}{5}\right]-1,0,0,\left[\frac{m}{5}\right]+1\right), & \left[\frac{m}{5}\right]+2 \leq N_0 \leq 2\left[\frac{m}{5}\right]+1
\\
\left(\left[\frac{m}{5}\right],0, N_0-2\left[\frac{m}{5}\right]-1,\left[\frac{m}{5}\right]+1\right), &  2\left[\frac{m}{5}\right]+2 \leq N_0 \leq 3\left[\frac{m}{5}\right]+1
\\
\left(\left[\frac{m}{5}\right]-1,\left[\frac{m}{5}\right],N_0 -3\left[\frac{m}{5}\right]-2,\left[\frac{m}{5}\right]\right), & \text 3\left[\frac{m}{5}\right]+2 \leq N_0 \leq 4\left[\frac{m}{5}\right]+2
\\
\left(m-1-N_0,m-N_0,m-N_0,m-N_0\right), & \text 4\left[\frac{m}{5}\right]+3 \leq N_0 \leq m-1
\end{cases}
\end{equation*}

\item  When $m \equiv 3 \mod 5$, we have
  \begin{equation*}
l=4: \left(N_1, N_2,N_3,N_4\right)= \begin{cases}\left(N_0, 0,0,0\right), & 0 \leq N_0 \leq\left[\frac{m}{5}\right]
\\ \left(\left[\frac{m}{5}\right],0,0, N_0-\left[\frac{m}{5}\right]\right), & \left[\frac{m}{5}\right]+1 \leq N_0 \leq 2\left[\frac{m}{5}\right]+1
\\ \left(\left[\frac{m}{5}\right],N_0-2\left[\frac{m}{5}\right]-1,0,\left[\frac{m}{5}\right]\right)&  2\left[\frac{m}{5}\right]+2 \leq N_0 \leq 3\left[\frac{m}{5}\right]+1
\\ \left(\left[\frac{m}{5}\right],N_0-3\left[\frac{m}{5}\right]-2,\left[\frac{m}{5}\right],\left[\frac{m}{5}\right]\right), &  3\left[\frac{m}{5}\right]+2 \leq N_0 \leq 4\left[\frac{m}{5}\right]+2
\\
\left(m-1-N_0, m-1-N_0,m-1-N_0,m-1-N_0\right), & \text 4\left[\frac{m}{5}\right]+3 \leq N_0 \leq m-1
\end{cases}
\end{equation*}
\begin{equation*}
l=3: \left(N_1, N_2,N_3,N_4\right)= \begin{cases}\left(0,N_0, 0,0\right), & 0 \leq N_0 \leq\left[\frac{m}{5}\right]
\\
\left(N_0-\left[\frac{m}{5}\right]-1,0,\left[\frac{m}{5}\right], 0\right), & \left[\frac{m}{5}\right]+1 \leq N_0 \leq 2\left[\frac{m}{5}\right]+1
\\
\left( \left[\frac{m}{5}\right],0,\left[\frac{m}{5}\right],N_0-2\left[\frac{m}{5}\right]-1\right), &  2\left[\frac{m}{5}\right]+2 \leq N_0 \leq 3\left[\frac{m}{5}\right]+2
\\
\left(\left[\frac{m}{5}\right],N_0 -3\left[\frac{m}{5}\right]-2,\left[\frac{m}{5}\right],\left[\frac{m}{5}\right]+1\right), & \text 3\left[\frac{m}{5}\right]+3 \leq N_0 \leq 4\left[\frac{m}{5}\right]+2
\\
\left(m-1-N_0,m-1-N_0,m-1-N_0,m-N_0\right), & \text 4\left[\frac{m}{5}\right]+3 \leq N_0 \leq m-1
\end{cases}
\end{equation*}
\begin{equation*}
l=2: \left(N_1, N_2,N_3,N_4\right)= \begin{cases}\left(0, 0,N_0,0\right), & 0 \leq N_0 \leq\left[\frac{m}{5}\right]+1
\\
\left(N_0-\left[\frac{m}{5}\right]-1,0,\left[\frac{m}{5}\right]+1,0\right), & \left[\frac{m}{5}\right]+2 \leq N_0 \leq 2\left[\frac{m}{5}\right]+1
\\
\left(\left[\frac{m}{5}\right],0, \left[\frac{m}{5}\right]+1,N_0-2\left[\frac{m}{5}\right]-1\right), &  2\left[\frac{m}{5}\right]+2 \leq N_0 \leq 3\left[\frac{m}{5}\right]+1
\\
\left(\left[\frac{m}{5}\right],N_0 -3\left[\frac{m}{5}\right]-2,\left[\frac{m}{5}\right]+1,\left[\frac{m}{5}\right]+1\right), & \text 3\left[\frac{m}{5}\right]+2 \leq N_0 \leq 4\left[\frac{m}{5}\right]+2
\\
\left(m-1-N_0,m-1-N_0,m-N_0,m-N_0\right), & \text 4\left[\frac{m}{5}\right]+3 \leq N_0 \leq m-1
\end{cases}
\end{equation*}
\begin{equation*}
l=1: \left(N_1, N_2,N_3,N_4\right)= \begin{cases}\left(0, 0,0,N_0\right), & 0 \leq N_0 \leq\left[\frac{m}{5}\right]+1
\\
\left( 0,N_0-\left[\frac{m}{5}\right]-1,0,\left[\frac{m}{5}\right]+1\right), & \left[\frac{m}{5}\right]+2 \leq N_0 \leq 2\left[\frac{m}{5}\right]+1
\\
\left(\left[\frac{m}{5}\right],N_0-2\left[\frac{m}{5}\right]-2,0, \left[\frac{m}{5}\right]\right), &  2\left[\frac{m}{5}\right]+2 \leq N_0 \leq 3\left[\frac{m}{5}\right]+2
\\
\left(\left[\frac{m}{5}\right]-1,N_0 -3\left[\frac{m}{5}\right]-3,\left[\frac{m}{5}\right],\left[\frac{m}{5}\right]\right), & \text 3\left[\frac{m}{5}\right]+3 \leq N_0 \leq 4\left[\frac{m}{5}\right]+3
\\
\left(m-1-N_0,m-N_0,m-N_0,m-N_0\right), & \text 4\left[\frac{m}{5}\right]+4 \leq N_0 \leq m-1
\end{cases}
\end{equation*}

\item  When $m \equiv 4 \mod 5$, we have
  \begin{equation*}
l=4: \left(N_1, N_2,N_3,N_4\right)= \begin{cases}\left(N_0, 0,0,0\right), & 0 \leq N_0 \leq\left[\frac{m}{5}\right]
\\ \left(N_0-\left[\frac{m}{5}\right]-1,\left[\frac{m}{5}\right],0,0\right), & \left[\frac{m}{5}\right]+1 \leq N_0 \leq 2\left[\frac{m}{5}\right]+1
\\ \left(N_0-2\left[\frac{m}{5}\right]-2,\left[\frac{m}{5}\right],\left[\frac{m}{5}\right],0\right)&  2\left[\frac{m}{5}\right]+2 \leq N_0 \leq 3\left[\frac{m}{5}\right]+2
\\ \left(N_0-3\left[\frac{m}{5}\right]-3,\left[\frac{m}{5}\right],\left[\frac{m}{5}\right],\left[\frac{m}{5}\right]\right), &  3\left[\frac{m}{5}\right]+3 \leq N_0 \leq 4\left[\frac{m}{5}\right]+3
\\
\left(m-1-N_0, m-1-N_0,m-1-N_0,m-1-N_0\right), & \text 4\left[\frac{m}{5}\right]+3 \leq N_0 \leq m-1
\end{cases}
\end{equation*}
\begin{equation*}
l=3: \left(N_1, N_2,N_3,N_4\right)= \begin{cases}\left(0,N_0, 0,0\right), & 0 \leq N_0 \leq\left[\frac{m}{5}\right]+1
\\
\left(N_0-\left[\frac{m}{5}\right]-1,\left[\frac{m}{5}\right]+1,0, 0,\right), & \left[\frac{m}{5}\right]+2 \leq N_0 \leq 2\left[\frac{m}{5}\right]+1
\\
\left(N_0-2\left[\frac{m}{5}\right]-2, \left[\frac{m}{5}\right],\left[\frac{m}{5}\right]+1,0\right), &  2\left[\frac{m}{5}\right]+2 \leq N_0 \leq 3\left[\frac{m}{5}\right]+2
\\
\left(N_0 -3\left[\frac{m}{5}\right]-3,\left[\frac{m}{5}\right],\left[\frac{m}{5}\right],\left[\frac{m}{5}\right]+1\right), & \text 3\left[\frac{m}{5}\right]+3 \leq N_0 \leq 4\left[\frac{m}{5}\right]+2
\\
\left(m-1-N_0,m-1-N_0,m-1-N_0,m-N_0\right), & \text 4\left[\frac{m}{5}\right]+3 \leq N_0 \leq m-1
\end{cases}
\end{equation*}
\begin{equation*}
l=2: \left(N_1, N_2,N_3,N_4\right)= \begin{cases}\left(0, 0,N_0,0\right), & 0 \leq N_0 \leq\left[\frac{m}{5}\right]+1
\\
\left(0,N_0-\left[\frac{m}{5}\right]-1,\left[\frac{m}{5}\right]+1,0\right), & \left[\frac{m}{5}\right]+2 \leq N_0 \leq 2\left[\frac{m}{5}\right]+2
\\
\left(N_0-2\left[\frac{m}{5}\right]-2,\left[\frac{m}{5}\right]+1, \left[\frac{m}{5}\right]+1,0\right), &  2\left[\frac{m}{5}\right]+3 \leq N_0 \leq 3\left[\frac{m}{5}\right]+2
\\
\left(N_0 -3\left[\frac{m}{5}\right]-3,\left[\frac{m}{5}\right],\left[\frac{m}{5}\right]+1,\left[\frac{m}{5}\right]+1\right), & \text 3\left[\frac{m}{5}\right]+3 \leq N_0 \leq 4\left[\frac{m}{5}\right]+3
\\
\left(m-1-N_0,m-1-N_0,m-N_0,m-N_0\right), & \text 4\left[\frac{m}{5}\right]+4 \leq N_0 \leq m-1
\end{cases}
\end{equation*}
\begin{equation*}
l=1: \left(N_1, N_2,N_3,N_4\right)= \begin{cases}\left(0, 0,0,N_0\right), & 0 \leq N_0 \leq\left[\frac{m}{5}\right]+1
\\
\left( 0,0,N_0-\left[\frac{m}{5}\right]-1,\left[\frac{m}{5}\right]+1\right), & \left[\frac{m}{5}\right]+2 \leq N_0 \leq 2\left[\frac{m}{5}\right]+2
\\
\left(0,N_0-2\left[\frac{m}{5}\right]-2,\left[\frac{m}{5}\right]+1, \left[\frac{m}{5}\right]+1\right), &  2\left[\frac{m}{5}\right]+3 \leq N_0 \leq 3\left[\frac{m}{5}\right]+3
\\
\left(N_0 -3\left[\frac{m}{5}\right]-3,\left[\frac{m}{5}\right]+1,\left[\frac{m}{5}\right]+1,\left[\frac{m}{5}\right]+1\right), & \text 3\left[\frac{m}{5}\right]+4 \leq N_0 \leq 4\left[\frac{m}{5}\right]+3
\\
\left(m-1-N_0,m-N_0,m-N_0,m-N_0\right), & \text 4\left[\frac{m}{5}\right]+4 \leq N_0 \leq m-1.
\end{cases}
\end{equation*}
\end{itemize}

\end{lemma}

In the second part of this appendex, we will prove Theorems \ref{root sturcture of jump 4} and \ref{root sturcture of jump 5} for root structures of the generalized Wronskian-Hermite polynomials $W_N^{[m,k,l]}$.  We only provide the proof for $k=5$, $l=4$ and $m \equiv 1,2,3,4\mod5$, as other cases can be proved in a similar manner.

	Firstly, we define a new class of special Schur polynomials $S_j^{[m]}(z ; a)$ and the polynomials $\widehat{W}_N^{[m,5,4]}(z ; a)$ as
	%and then construct the new polynomials which is similar to our $W_N^{[m,5,4]}$, but with the added parameters $a$ as
	
	\begin{eqnarray}
		\sum_{j=0}^{\infty} S_j^{[m]}(z ; a) \epsilon^j&=&\exp \left[z \epsilon+a \epsilon^m\right],
	%\end{equation}
%	\begin{equation}
\\
		\widehat{W}_N^{[m,5,4]}(z ; a)&=&c_N^{[m,5,4]}
        \left|\begin{array}{cccc}
			S_4^{[m]}(z ; a) & S_3^{[m]}(z ; a) & \cdots & S_{5-N}^{[m]}(z ; a) \\
			S_9^{[m]}(z ; a) & S_8^{[m]}(z ; a) & \cdots & S_{10-N}^{[m]}(z ; a) \\
			\vdots & \vdots & \vdots & \vdots \\
			S_{5 N-1}^{[m]}(z ; a) & S_{5 N-2}^{[m]}(z ; a) & \cdots & S_{4 N}^{[m]}(z ; a)
		\end{array}\right|, \label{exprsion of W hat N}
	\end{eqnarray}
	where $a$  is a parameter, $c_N^{[m,5,4]}$ is a constant defined in \eqref{def of cN}, and $S_j^{[m]}(z ; a) \equiv 0 $  when $ j<0.$ Compared with the generalized Wronskian-Hermite polynomials $W_N^{[m,k,l]}$, the polynomials $\widehat{W}_N^{[m,5,4]}(z ; a)$ have a new parameter $a$. Note that the polynomials $S_j^{[m]}(z ; a)$ are related to   $p_j^{[m]}(z)$ by
	%By a simple scaling transformation,  we can observe that the relation between $S_j^{[m]}(z ; a)$ and $p_j^{[m]}(z)$ are
	\begin{equation}\label{relation between S and P-proof for root structures}
		S_j^{[m]}(z ; a)=a^{j / m} p_j^{[m]}(\hat{z}), \quad \hat{z}=a^{-1 / m} z.
	\end{equation}
	In addition, we have
	%Furthermore, we can extend the relation to
	%$\widehat{W}_N^{[m,5,4]}(z ; a)$ and  %$W_N^{[m,5,4]}(z)$ as
	\begin{equation}\label{relation between W and W hat-proof for root structures}
		\widehat{W}_N^{[m,5,4]}(z ; a)=a^{2N(N+1)/m} W_N^{[m,5,4]}(\hat{z}).
	\end{equation}
	From \eqref{relation between S and P-proof for root structures} and \eqref{relation between W and W hat-proof for root structures}, we find that each term in the polynomial  $\widehat{W}_N^{[m]}(z ; a)$ is a constant multiple  of $a^s z^k$ and $k+ms=2N(N+1)$. This indicates that when the power of $a$ is
	larger, the power of $z$ is lower. Thus, to find the lowest order of $z$, it suffices to find the highest order of $a$. To this end, we rewrite the polynomials $S_j^{[m]}(z ; a)$ as
	\begin{equation}\label{the expression of S_j(z,a)}
		S_j^{[m]}(z ; a)=\sum_{n=0}^{[j / m]} \frac{a^n}{n !(j-n m) !} z^{j-n m}
	\end{equation}
	and substitute \eqref{the expression of S_j(z,a)} into the determinant \eqref{exprsion of W hat N}. Note that coefficients of each $a^sz^k$ in each row are proportional to each other, thus we can ignore them. In particular, for $m=5r+1$, where $r$ is positive integer, we get 	
	%For $m=5j+1$
	\begin{equation}\label{matrix of W hat N in root structure}
		\widehat{W}_N^{[m,5,4]}(z ; a) \sim c_N\left|\begin{array}{cccccccc}
			z^4 & z^3 & \cdots \\
			
			\vdots & \vdots & \ddots\\
			
			z^{5r-1} & z^{5r-2} & \cdots \\
			
			a z^3 + \cdots & a z^2 + \cdots & \cdots  \\
			
			\vdots & \vdots & \ddots\\
			
			a z^{5r-2}+z^{5r-2+m} & a z^{5r-3}+z^{5r-3+m} & \cdots \\
			
			a^2 z^2 +a z^{2+m} + \cdots & a^2 z^1 +a z^{1+m} + \cdots & \cdots  \\
			
			\vdots & \vdots & \ddots\\
			
			a^2 z^{5r-3} +a z^{5r-3+m} + \cdots & a^2 z^{5r-4} +a z^{5r-4+m} + \cdots & \cdots \\
			
			a^3 z^1 +a^2 z^{1+m} + \cdots & a^3 z^0 +a^2 z^m + \cdots & \cdots  \\
			
			\vdots & \vdots & \ddots\\
			a^3 z^{5r-4} +a^2 z^{5r-4+m} + \cdots & a^3 z^{5r-5} +a^2 z^{5r-5+m} + \cdots & \cdots \\
			a^4 z^0 + a^3 z^m + \cdots & a^3 z^{5r} + \cdots  & \cdots \\
			\vdots & \vdots & \ddots\\
			a^4 z^{5r}+a^3 z^{5r+m} + \cdots  & a^4 z^{5r-1}+a^3 z^{5r-1+m}  + \cdots & \cdots \\
			\vdots & \vdots & \ddots
		\end{array}\right| .
	\end{equation}
	
	Next, we perform row operations to \eqref{matrix of W hat N in root structure}, which consist of several steps.
	\begin{enumerate}
		\item Note that the coefficients of  the highest order terms in $a$ in the first column of \eqref{matrix of W hat N in root structure} are periodic and one period is given by
		\begin{equation}\label{blocks}
			z^4 \cdots z^{5r-1};\quad z^3 \cdots z^{5r-2};\quad z^2 \cdots z^{5r-3};\quad z^1 \cdots z^{5r-4};\quad z^0 \cdots z^{5r}.
		\end{equation}According to this periodicity,
	we divide the determinants \eqref{matrix of W hat N in root structure} into  $\left[N/m\right] $  block matrices of size $ m  \times N $ and one $N_0 \times N$ block matrix, where $N_0 \equiv N \mod m$. In addition, we divide  the first column in each block  into five parts, which have distinct initial powers in $z$, and the difference of the powers in $z$ of consecutive terms in each part is $5$. %from we can see five different series with different initial powers but the same addition $5$, such as $z^4, z^9 \cdots z^{5j-1}$, and $4$ is its initial power.
		   We denote the number of parts starting with power $j$ by $N_{5-j}, j=1, \dots, 4$. %Correspondingly, we label $z^3 \cdots z^{5j-2}$ as the part two with number $N_2$, and so on.
		\item We are only concerned with the first column of \eqref{matrix of W hat N in root structure}, since other columns have similar structures. According to the above discussions,  %the coefficients of highest $a$ have
	 %periodicity,
	 we may use each part of the first block %which multiply with a certain power of $a$,
	 to cancel the highest order terms in $a$ of the corresponding parts for the subsequent blocks. %For the remain $N_0$ rows we can only use few rows of the first block to cancel its highest $a$.
		After the first round row operations, the coefficients of the highest order terms in $a$ in the first column of the second block become
		\begin{equation}
			z^{4+m} \cdots z^{5r-1+m}; z^{3+m} \cdots z^{5r-2+m}; z^{2+m} \cdots z^{5r-3+m}; z^{1+m} \cdots z^{5r-4+m}; z^m \cdots z^{5r+m},
		\end{equation}
	and from the third to the last blocks, the corresponding coefficients change to $z^{i+m}$ from $z^i$. In the second round, we can use the second block to cancel the highest order terms in $a$ of the blocks below.
	Then we continue this process until %after which we use the third block to do the same thing and until
	the last block.
		At the end of these operations, we can exchange the rows in each block such that the highest order terms in $a$ of the first column are
		$$
		z^0, z^1, z^2, \cdots, z^{5r};
		z^m, z^{m+1}, z^{m+2},\cdots, z^{m+5r};
		\cdots;
		z^{km}, z^{km+1},z^{km+2},\cdots,z^{km+5r};\cdots.
		$$
			and the determinant \eqref{matrix of W hat N in root structure} becomes
		$$
		\widehat{W}_{N \times  N}^{[m,5,4]} \sim \left(\begin{array}{ccc}
			\mathbf{L}_{\left(N-N_0\right) \times\left(N-N_0\right)} & \mathbf{0}_{\left(N-N_0\right) \times  N_0}  \\
			\mathbf{M}_{N_0 \times\left(N-N_0\right)} & \widehat{W}_{N_0 \times  N_0}
		\end{array}\right)
		$$ where $\mathbf{L}_{\left(N-N_0\right) \times\left(N-N_0\right)}$ is a lower
		triangular matrix whose diagonal entries are all $1$.
		\item Therefore, to calculate the lowest power of $z$ of $\widehat{W}_N^{[m,5,4]}(z ; a)$, it suffices to compute the power of the reduced $N_0 \times N_0$ determinant $\widehat{W}_{N_0 \times  N_0}$ and the final result is $\Gamma$ as given in \eqref{gamma-4com}.
	\end{enumerate}
	
	Next, we derive the factorization of $W_N^{[m,5,4]}(z)$ provided in Theorem \ref{root sturcture of jump 5}. Since the multiplicity of the zero root of $W_N^{[m,5,4]}(z)$ is $\Gamma$, we can write
	\begin{equation}\label{multiplicity of the zero root fot root structure}
	W_N^{[m,5,4]}(z)=z^{\Gamma} q_N^{[m]}(z).
   \end{equation}
	%where $q_N^{[m]}(z)$ is a polynomial of $z$ with a nonzero constant term.
	Note that
	\begin{equation}
\label{pj of root structures}
	p_j^{[m]}(\omega z)=\omega^j p_j^{[m]}(z),
\end{equation}
	where $\omega$ is any of the $m$-th root of 1, i.e., $\omega^m=1$. %This symmetry of $p_j^{[m]}(z)$ leads to the symmetry of $W_N^{[m,5,4]}$ as
	From \eqref{multiplicity of the zero root fot root structure} and \eqref{pj of root structures}, we immediately have
	$$
	W_N^{[m,5,4]}(\omega z)=\omega^{2N(N+1)} W_N^{[m,5,4]}(z)
	$$
	and
    %The symmetry  of the polynomial $W_N^{[m,5,4]}(z)$ induces a symmetry for $q_N^{[m]}(z)$ as

	$$
	q_N^{[m]}(\omega z)=\omega^{2N(N+1)-\Gamma} q_N^{[m]}(z) .
	$$
	Since $2N(N+1)-\Gamma$ is a multiple of $m$, we have $\omega^{2N(N+1)-\Gamma}=1$, and hence
	$$
	q_N^{[m]}(\omega z)=q_N^{[m]}(z) .
	$$
		This completes the proof.

\begin{remark}	
We note that the row operations performed above are similar to those in the proof of rogue patterns in the inner region. For some special cases, such as $m=4r+2$ in the three-components NLS equation, the proof needs some modifications similar to the proof of Theorem \ref{root sturcture of jump 4}.

\end{remark}

\bibliographystyle{abbrv}
\bibliography{References}

\end{document}